\documentclass[11pt]{article}
\usepackage{amsfonts,amssymb,graphicx}
\usepackage{fullpage}
\usepackage[tight]{subfigure}
\usepackage{rotating}
\usepackage{units}
\usepackage{sidecap}
\pdfoutput=1
\usepackage{ifpdf}
\ifpdf
  \usepackage{color}
  \definecolor{darkblue}{rgb}{0.3,0.3,0.6}
    \definecolor{darkgreen}{rgb}{0,0.6,0}
  \usepackage[pdftitle={Deformations, Moduli Stabilisation and Gauge Kinetic Functions},pdfauthor={Gabriele Honecker, Isabel Koltermann, Wieland Staessens},pdfsubject={String theory},pdfkeywords={string theory},colorlinks=true,linkcolor=black,urlcolor=darkblue,filecolor=darkblue,citecolor=darkblue,menucolor=darkblue,breaklinks=true,bookmarks=true,anchorcolor=black,unicode=true]{hyperref}
\else
  
\fi
\usepackage[numbers,compress]{natbib}
\usepackage{a4wide}
\usepackage{longtable}
\usepackage{graphicx,cancel}
\usepackage{subfigure}
\usepackage[centertags]{amsmath}
\usepackage{multirow}
\usepackage{slashed}
\usepackage{pdflscape}
\usepackage{rotating}
\usepackage{tikz}
\usetikzlibrary{arrows,positioning}
\usepackage{stmaryrd}
\usepackage{color}
\usepackage[normalem]{ulem}
 
\newcommand{\bCentering}{\centering}
\newcommand{\bCaption}{\caption}
\newcommand{\sgn}{{\rm sgn}}
\newcommand{\unity}{{\footnotesize\mbox{1\!\!I}}}

\def\muc{\multicolumn}

\def\Z{\mathbb{Z}}

\def\R{\mathbb{R}}
\def\C{\mathbb{C}}
\def\P{\mathbb{P}}
\def\unity{1\!\!{\rm I}}
\def\ov{\overline}
\def\N{\mathbf{N}}
\def\Sym{\mathbf{Sym}}
\def\Anti{\mathbf{Anti}}
\def\Adj{\mathbf{Adj}}

\def\ov{\overline}
\def\1{{\bf 1}}
\def\2{{\bf 2}}
\def\3{{\bf 3}}
\def\4{{\bf 4}}
\def\6{{\bf 6}}
\def\8{{\bf 8}}
\def\OR{\Omega\mathcal{R}}

\def\pp{\uparrow\uparrow}

\def\targ#1#2{\genfrac{[}{]}{0pt}{}{#1}{#2}}
\def\tarh#1#2{\genfrac{(}{)}{0pt}{}{#1}{#2}}
\def\tari#1#2{\genfrac{\{}{\}}{0pt}{}{#1}{#2}}
\def\targ2#1#2{\genfrac{}{}{0pt}{}{#1}{#2}}

\definecolor{blus}{rgb}{0.1,0.1,0.8}
\definecolor{GreenYellow}{cmyk}{0.15,0,0.69,0}
\definecolor{Yellow}{cmyk}{0,0,1,0}
\definecolor{Goldenrod}{cmyk}{0,0.10,0.84,0}
\definecolor{Dandelion}{cmyk}{0,0.29,0.84,0}
\definecolor{Apricot}{cmyk}{0,0.32,0.52,0}
\definecolor{Peach}{cmyk}{0,0.50,0.70,0}
\definecolor{Melon}{cmyk}{0,0.46,0.50,0}
\definecolor{YellowOrange}{cmyk}{0,0.42,1,0}
\definecolor{Orange}{cmyk}{0,0.61,0.87,0}
\definecolor{BurntOrange}{cmyk}{0,0.51,1,0}
\definecolor{Bittersweet}{cmyk}{0,0.75,1,0.24}
\definecolor{RedOrange}{cmyk}{0,0.77,0.87,0}
\definecolor{Mahogany}{cmyk}{0,0.85,0.87,0.35}
\definecolor{Maroon}{cmyk}{0,0.87,0.68,0.32}
\definecolor{BrickRed}{cmyk}{0,0.89,0.94,0.28}
\definecolor{Red}{cmyk}{0,1,1,0}
\definecolor{OrangeRed}{cmyk}{0,1,0.50,0}
\definecolor{RubineRed}{cmyk}{0,1,0.13,0}
\definecolor{WildStrawberry}{cmyk}{0,0.96,0.39,0}
\definecolor{Salmon}{cmyk}{0,0.53,0.38,0}
\definecolor{CarnationPink}{cmyk}{0,0.63,0,0}
\definecolor{Magenta}{cmyk}{0,1,0,0}
\definecolor{VioletRed}{cmyk}{0,0.81,0,0}
\definecolor{Rhodamine}{cmyk}{0,0.82,0,0}
\definecolor{Mulberry}{cmyk}{0.34,0.90,0,0.02}
\definecolor{RedViolet}{cmyk}{0.07,0.90,0,0.34}
\definecolor{Fuchsia}{cmyk}{0.47,0.91,0,0.08}
\definecolor{Lavender}{cmyk}{0,0.48,0,0}
\definecolor{Thistle}{cmyk}{0.12,0.59,0,0}
\definecolor{Orchid}{cmyk}{0.32,0.64,0,0}
\definecolor{DarkOrchid}{cmyk}{0.40,0.80,0.20,0}
\definecolor{Purple}{cmyk}{0.45,0.86,0,0}
\definecolor{Plum}{cmyk}{0.50,1,0,0}
\definecolor{Violet}{cmyk}{0.79,0.88,0,0}
\definecolor{RoyalPurple}{cmyk}{0.75,0.90,0,0}
\definecolor{BlueViolet}{cmyk}{0.86,0.91,0,0.04}
\definecolor{Periwinkle}{cmyk}{0.57,0.55,0,0}
\definecolor{CadetBlue}{cmyk}{0.62,0.57,0.23,0}
\definecolor{CornflowerBlue}{cmyk}{0.65,0.13,0,0}
\definecolor{MidnightBlue}{cmyk}{0.98,0.13,0,0.43}
\definecolor{NavyBlue}{cmyk}{0.94,0.54,0,0}
\definecolor{RoyalBlue}{cmyk}{1,0.50,0,0}
\definecolor{Blue}{cmyk}{1,1,0,0}
\definecolor{Cerulean}{cmyk}{0.94,0.11,0,0}
\definecolor{Cyan}{cmyk}{1,0,0,0}
\definecolor{ProcessBlue}{cmyk}{0.96,0,0,0}
\definecolor{SkyBlue}{cmyk}{0.62,0,0.12,0}
\definecolor{Turquoise}{cmyk}{0.85,0,0.20,0}
\definecolor{TealBlue}{cmyk}{0.86,0,0.34,0.02}
\definecolor{Aquamarine}{cmyk}{0.82,0,0.30,0}
\definecolor{BlueGreen}{cmyk}{0.85,0,0.33,0}
\definecolor{Emerald}{cmyk}{1,0,0.50,0}
\definecolor{JungleGreen}{cmyk}{0.99,0,0.52,0}
\definecolor{SeaGreen}{cmyk}{0.69,0,0.50,0}
\definecolor{Green}{cmyk}{1,0,1,0}
\definecolor{ForestGreen}{cmyk}{0.91,0,0.88,0.12}
\definecolor{PineGreen}{cmyk}{0.92,0,0.59,0.25}
\definecolor{LimeGreen}{cmyk}{0.50,0,1,0}
\definecolor{YellowGreen}{cmyk}{0.44,0,0.74,0}
\definecolor{SpringGreen}{cmyk}{0.26,0,0.76,0}
\definecolor{OliveGreen}{cmyk}{0.64,0,0.95,0.40}
\definecolor{RawSienna}{cmyk}{0,0.72,1,0.45}
\definecolor{Sepia}{cmyk}{0,0.83,1,0.70}
\definecolor{Brown}{cmyk}{0,0.81,1,0.60}
\definecolor{Tan}{cmyk}{0.14,0.42,0.56,0}
\definecolor{Gray}{cmyk}{0,0,0,0.50}
\definecolor{Black}{cmyk}{0,0,0,1}
\definecolor{White}{cmyk}{0,0,0,0}

\definecolor{mygr}{rgb}{0,0.6,0}
\definecolor{mygrey}{rgb}{0,0.1,0.2}
\definecolor{myblue}{rgb}{0,0.5,0.9}
\definecolor{myblue2}{rgb}{0,0.5,0.5}
\definecolor{myblue3}{rgb}{0,0.7,0.9}
\definecolor{myorange}{rgb}{1,0.5,0}
\definecolor{mypurple}{rgb}{0.6,0,1}
\definecolor{mygolden}{rgb}{1,0.8,0.2}
\definecolor{mycyan}{rgb}{0,1,1}
\definecolor{mymagenta}{rgb}{1,0,1}
\definecolor{mykiwi}{rgb}{0.8,1,0.5}
\definecolor{mybrown}{cmyk}{0.14, 0.42, 0.56, 0.2}
\definecolor{myturq}{cmyk}{0.99, 0, 0.2, 0.4}
\definecolor{myaubergine}{cmyk}{0.2, 0.7, 0, 0.4}
\definecolor{CycleGreen}{cmyk}{0.56,0,0.8,0}
\definecolor{CycleBrown}{cmyk}{0, 0.4, 0.9, 0.2}

\newcommand{\bCaptionfonts}{\small}
\makeatletter
\long\def\@makecaption#1#2{%
  \vskip\abovecaptionskip
  \sbox\@tempboxa{{\bCaptionfonts #1: #2}}%
  \ifdim \wd\@tempboxa >\hsize
    {\bCaptionfonts #1: #2\par}
  \else
    \hbox to\hsize{\hfil\box\@tempboxa\hfil}%
  \fi
  \vskip\belowcaptionskip}
\makeatother

\makeatletter
\let\ORIGINALlatex@openbib@code=\@openbib@code
\renewcommand{\@openbib@code}{\ORIGINALlatex@openbib@code\setlength{\itemsep}{1ex plus.5ex minus.5ex}\setlength{\parsep}{0pt}}
\makeatother

\def\mathtab#1#2#3{\begin{table}[th]\bCentering$#1$\bCaption{#3}\label{tab:#2}\end{table}}
\def\mathtabfix#1#2#3{\begin{table}[th]\bCentering\resizebox{\linewidth}{!}{$#1$}\bCaption{#3}\label{tab:#2}\end{table}}

\allowdisplaybreaks[4]

\begin{document}
\begin{center}
\begin{flushright}
{\small MITP/16-103\\ 
IFT-UAM/CSIC-17-010\\
\today}

\end{flushright}

\vspace{20mm}
{\Large\bf Deformations, Moduli Stabilisation and Gauge Couplings at One-Loop}

\vspace{12mm}
{\large Gabriele Honecker${}^{a,\heartsuit}$, Isabel Koltermann${}^{a,\clubsuit}$,  and Wieland Staessens${}^{b,c,\spadesuit}$
}

\vspace{8mm}
{
\it $^a$PRISMA Cluster of Excellence, MITP 
 \& Institut f\"ur Physik  (WA THEP), \\Johannes Gutenberg-Universit\"at, 55099 Mainz, Germany\\
$^b$ Instituto de F\'isica Te\'orica UAM-CSIC, Cantoblanco, 28049 Madrid, Spain\\
$^c$ Departamento de F\'{\i}sica Te\'orica, 
Universidad Aut\'onoma de Madrid, 
28049 Madrid, Spain
\;$^{\heartsuit}${\tt Gabriele.Honecker@uni-mainz.de},~ $^{\clubsuit}${\tt kolterma@uni-mainz.de}\\ $^{\spadesuit}${\tt wieland.staessens@csic.es}}

\vspace{12mm}{\bf Abstract}\\[2ex]\parbox{140mm}{
We investigate deformations of $\Z_2$ orbifold singularities on the toroidal orbifold $T^6/(\Z_2\times \Z_6)$ with discrete torsion in the framework of Type IIA orientifold model building with intersecting D6-branes wrapping special Lagrangian cycles. To this aim, we employ the hypersurface formalism developed previously for the orbifold $T^6/(\Z_2\times \Z_2)$ with discrete torsion and adapt it to the $(\Z_2\times \Z_6\times \OR)$ point group by modding out the remaining $\Z_3$ subsymmetry and the  orientifold projection $\OR$. We first study the local behaviour of the $\Z_3 \times \OR$ invariant deformation orbits under non-zero deformation and then develop methods to assess the deformation effects on the fractional three-cycle volumes globally. We confirm that D6-branes supporting $USp(2N)$ or $SO(2N)$ gauge groups do not constrain any deformation, while deformation parameters associated to cycles wrapped by D6-branes with $U(N)$ gauge groups are constrained by D-term supersymmetry breaking.
These features are exposed in global prototype MSSM, Left-Right symmetric and Pati-Salam models first constructed in~\cite{Ecker:2015vea,Ecker:2014hma}, for which we here count the number of stabilised moduli and study flat directions changing the values of some gauge couplings. 
\\
Finally, we confront the behaviour of tree-level gauge couplings under non-vanishing deformations along flat directions with the one-loop gauge threshold corrections at the orbifold point
and discuss phenomenological implications, in particular on possible LARGE volume scenarios and the corresponding value of the string scale $M_{\text{string}}$, for the same global D6-brane models. 
}
\end{center}

\thispagestyle{empty}
\clearpage 

\tableofcontents

\setlength{\parskip}{1em plus1ex minus.5ex}
\section{Introduction}\label{S:intro}

Since the dawn of string phenomenology, toroidal orbifolds have played a prominent r\^ole in string model building~\cite{Dixon:1985jw,Dixon:1986jc,Bailin:1999nk,Blumenhagen:2005mu,Nilles:2008gq,Maharana:2012tu,Blumenhagen:2013fgp}: they provide for exactly solvable conformal field theories, allow for supersymmetric compactifications and are capable of accommodating the necessary ingredients to construct chiral gauge theories. In the context of Type IIA orientifold model building with intersecting D6-branes, factorisable toroidal orbifolds come with factorisable {\it special Lagrangian} ({\it sLag}) three-cycles as underlying building blocks for such chiral gauge theories, see e.g.~\cite{Cvetic:2001tj,Cvetic:2001nr,Blumenhagen:2002wn,Honecker:2003vq,MarchesanoBuznego:2003axu,Honecker:2004kb,Honecker:2004np,Blumenhagen:2004xx,Blumenhagen:2005mu,Blumenhagen:2005tn,Gmeiner:2005vz,Blumenhagen:2006ci,Gmeiner:2007zz,Gmeiner:2008xq,Forste:2010gw,Ibanez:2012zz,Honecker:2012qr,Ecker:2014hma,Ecker:2015vea}.\footnote{Type IIA orientifold compactifications on non-factorisable toroidal orbifolds~\cite{Blumenhagen:2004di} have only recently been considered for model building purposes~\cite{Forste:2014bfa,Seifert:2015fwr,Berasaluce-Gonzalez:2016kqb,Seifert:2017}.} More precisely, these fractional three-cycles are wrapped by (stacks of coincident) D6-branes, which support (non)-Abelian gauge theories on their worldvolumes. Consequently, the parameters characterising the gauge theory are related to geometric data associated to the three-cycles. The square of the tree-level gauge coupling for instance scales inversely proportional to the volume of the three-cycle wrapped by the corresponding D6-brane~\cite{Aldazabal:2000cn,Ibanez:2012zz}. 

At the singular orbifold point, all exceptional three-cycles located at orbifold singularities have vanishing volumes, and the volume of a fractional three-cycle is simply (a fraction of)  the volume of the bulk three-cycle inherited from the ambient six-torus. However, a thorough study of the four-dimensional effective field theory emerging from a Type IIA orientifold compactification requires to consider a region in moduli space where the orbifold singularities have been resolved or deformed.\footnote{In addition to phenomenological considerations, the known prescriptions for identifying dual string theoretic descriptions via mirror symmetry to Type IIB orientifolds or via M-theory to $E_8 \times E_8$ heterotic compactifications, see e.g.~\cite{Greene:1990ud,Kachru:1995wm,Hunt:1996zh,Brunner:1997bf},  are to our best knowledge only valid for smooth Calabi-Yau backgrounds. 
} The resolution or deformation of such singular points will have undeniable geometric and physical consequences for the D6-branes wrapping them. In first instance, one has to verify whether the {\it sLag} condition of the corresponding fractional three-cycle is preserved under the deformation or not. 
Whenever the deformation violates the {\it sLag} condition, supersymmetry is broken via the appearance of a Fayet-Iliopoulos D-term in the four-dimensional effective field theory; the deformation modulus is then bound to be stabilised at the singular orbifold point. If a fractional three-cycle remains {\it sLag} under a particular deformation, its volume - and thereby also the associated inverse of the tree-level gauge coupling squared at $M_{\text{string}}$ - is expected to alter with an increasing deformation along this flat direction. 
One can of course also deform a singularity in the toroidal orbifold at which none of the D6-branes are located, in which case the associated deformation modulus can take any vacuum expectation value ({\it vev}) without affecting the physics of the chiral gauge theory at leading order.    

When resolving orbifold singularities on a singular Calabi-Yau variety, one usually turns to the toolbox of algebraic geometry and toric geometry, see e.g.~\cite{Aspinwall:1994ev}, which would offer us the necessary techniques to resolve exceptional two- and four-cycles through blow-ups. Toric singularities and blow-up resolutions of divisor four-cycles happen to be part of the modus operandi for constructing chiral gauge theories on the Type IIB side~\cite{Aldazabal:2000sa,Uranga:2000ck,Verlinde:2005jr,Buican:2006sn,Blumenhagen:2007sm,Blumenhagen:2008zz,Balasubramanian:2009tv,Balasubramanian:2012wd,Cicoli:2012vw,Cicoli:2013mpa,Cicoli:2013zha} using fractional D3-branes located at the singularities or D7-branes wrapping the resolved four-cycles. However, in the case of Type IIA model building with fractional D6-branes on orbifolds with discrete torsion, the orbifold singularities have to be deformed rather than blown up, which forces us to consider different tools from algebraic geometry: by viewing two-tori as elliptic curves in the weighted projective space $\P_{112}^2$, a factorisable toroidal orbifold with discrete torsion can be described as a hypersurface in a weighted projective space, with its topology being a double cover of $\P^1 \times \P^1 \times \P^1$. Building on this hypersurface formalism first sketched in~\cite{Vafa:1994rv} for the $T^6/(\Z_2 \times \Z_2)$ orbifold with discrete torsion and extended to its $T^6/(\Z_2 \times \Z_2 \times \OR)$ and $T^6/(\Z_2 \times \Z_6' \times \OR)$ orientifold versions with underlying isotropic square~\cite{Blaszczyk:2014xla,Koltermann:2016oyv}  or hexagonal~\cite{Blaszczyk:2015oia,Koltermann:2015oyv,Koltermann:2016oyv}  two-tori, respectively,
we focus here on the so far most fertile patch in the Type IIA orientifold landscape with rigid D6-branes~\cite{Ecker:2014hma,Ecker:2015vea,Ecker:2016iqw,Honecker:2016gyz}, the $T^6/(\Z_2 \times \Z_6 \times \OR)$ orientifold with discrete torsion and  one rectangular and two hexagonal underlying two-tori. In this case, the $\Z_2^{(1)}$-twisted sector conceptually differs from the $\Z_2^{(2)}$- and $\Z_2^{(3)}$-twisted sectors, necessitating separate discussions for the respective deformations and making the deformations of this toroidal orbifold more intricate than the other previously discussed orbifolds with discrete torsion. 

Upon embedding a toroidal orbifold with discrete torsion as a hypersurface in a weighted projective space with carefully chosen weights, (a subset of) {\it sLag} three-cycles can be constructed as the fixed loci under the anti-holomorphic involution contained in the orientifold projection $\OR$, in a similar spirit as in~\cite{Brunner:1999jq,Blumenhagen:2002wn,Blumenhagen:2002vp,Palti:2009bt}. The deformations in the hypersurface formalism allow for the description of exceptional and fractional three-cycles, besides the bulk three-cycles, by which the set of {\it sLags} three-cycles on the deformed toroidal orbifold can be immensely extended, all corresponding to calibrated submanifolds~\cite{Harvey:1982xk,McLean:1996nm,Joyce:2001xt,Joyce:2001nm} with respect to the same holomorphic volume three-form~$\Omega_3$. It is exactly the presence of these fractional {\it sLag} three-cycles that makes toroidal orbifolds with discrete torsion so appealing for D6-brane model building. Contrarily to a bulk {\it sLag} three-cycle, a fractional {\it sLag} three-cycle is not necessarily accompanied by an open string moduli space~\cite{McLean:1996nm}, as it is (at least in the absence of an additional $\Z_3$ symmetry completely) projected out by the $\Z_2\times \Z_2$ point group. The absence of open string deformation moduli ensures that the non-Abelian gauge group supported by a stack of D6-branes cannot be spontaneously broken by the displacement of a D-brane in that stack. On a more formal level, knowledge about the moduli space of {\it sLag} three-cycles is vital in the search for the mirror manifold~\cite{Strominger:1996it,Gross:1997hn,Gross:1999rm,Joyce:2001xt,Joyce:2001nm,Morrison:2015qca} of the deformed toroidal orbifold. The absence of an open string moduli space for fractional three-cycles is expected to complicate this search, which makes studying the geometric characteristics of fractional {\it sLag} three-cycles and uncovering their relations to the closed string moduli space all the more essential.  

In this article, a first step in revealing those relations for the $T^6/(\Z_2 \times \Z_6 \times \OR)$ orientifold with discrete torsion is taken by studying the functional dependence of the fractional three-cycle volumes on the complex structure (deformation) moduli, whose {\it vev}s measure the volumes of the exceptional three-cycles. Through this connection, the viability of a non-zero deformation is assessed by virtue of the preserved {\it sLag} conditions of the fractional three-cycles away from the orbifold point, as mentioned before. The physical implications of these deformations for D6-brane model building are discussed in terms of potential Fayet-Iliopoulos terms and/or altering tree-level gauge coupling strength in the effective four-dimensional gauge theories resulting from the orientifold compactifications with D6-branes.  

Also K\"ahler moduli are expected to have a substantial influence on the effective four-dimensional gauge theories, as exhibited through their presence in the one-loop threshold corrections to the gauge couplings at the singular orbifold point, see e.g.~\cite{Lust:2003ky,Akerblom:2007np,Blumenhagen:2007ip,Gmeiner:2009fb,Honecker:2011sm,Honecker:2011hm} in the context of D6-branes. These gauge threshold corrections can be sizeable for specific anisotropic choices of two-torus volumes~\cite{Honecker:2012qr}, given by the vacuum expectation values of the (CP-even part of the) K\"ahler moduli. In the class of models under consideration, these sizeable gauge threshold corrections are able to lift the degeneracy of the tree-level gauge coupling strengths for distinct fractional D6-brane stacks wrapping the same bulk three-cycle. With a lifted degeneracy of the gauge couplings already at the singular orbifold point at one-loop, it becomes more conceivable to  construct global intersecting D6-brane models with e.g.~a very strongly coupled hidden gauge group, whose gaugino condensate forms a natural source for spontaneous supersymmetry breaking. Clearly, establishing the full moduli-dependence of the one-loop correction to the gauge coupling represents a {\it conditio sine qua non} for string model builders, both at and away from the singular orbifold point. 
 
This article is organised as follows: in section~\ref{S:DefOrbS}, we briefly review the hypersurface formulation for describing {\it local} deformations of $T^6/(\Z_2 \times \Z_2 \times \OR)$ 
singularities as discussed in~\cite{Vafa:1994rv,Blaszczyk:2014xla,Blaszczyk:2015oia,Koltermann:2015oyv} and then go on to discuss additional constraints imposed by the extra $\Z_3$ symmetry of the $T^6/(\Z_2 \times \Z_6 \times \OR)$ orientifold.  Special attention will be devoted to the {\it sLag} cycles used for particle physics model building.
In section~\ref{S:DefModuliGlobalModels}, additional subtleties in {\it global} deformations of $T^6/(\Z_2 \times \Z_6 \times \OR)$ singularities are discussed, and several 
prototype examples of global D6-brane models with particle physics spectra are examined.
Section~\ref{S:GKFDefModuli} is devoted to the computation of the one-loop corrections to the gauge couplings at the orbifold point and the phenomenological implications of their specific geometric moduli dependences.
Finally, section~\ref{S:conclu} contains our conclusions and outlook.
Additional technical details useful for the computation of deformations and one-loop corrections are relegated to appendices~\ref{A:MT},~\ref{A:CorrectionTerms} and~\ref{A:Tables-MatterSector}.

\section{Deforming Orbifold Singularities in the Hypersurface Formalism}\label{S:DefOrbS}

To start, we first briefly review the construction of {\it fractional} three-cycles as sums of toroidal and exceptional three-cycles stuck at orbifold singularities in section~\ref{Ss:RemZ2Z6},
in particular on the orientifold of phenomenological interest $T^6/(\Z_2 \times \Z_6 \times \OR)$. 
Then, we move on to reviewing  {\it Lagrangian} ({\it Lag}) lines on two-tori of rectangular and hexagonal shape in the hypersurface formalism in section~\ref{Ss:LCHF}. 
In section~\ref{Ss:IntersectSummary}, we first discuss deformations of $\Z_2 \times \Z_2$ singularities 
on \mbox{$T^6= (T^2)^3$} and afterwards impose relations among deformations due to the specific additional $\Z_3$ symmetry of the $\Z_2 \times \Z_6$ action. As a final element,
in section~\ref{Ss:DefsLagS} we discuss the general procedures allowing for the quantitative study of  {\it special Lagrangian} ({\it sLag}) three-cycles on deformations of
 $T^6/(\Z_2 \times \Z_6 \times \OR)$ in the hypersurface formalism.

\subsection{Reminiscing about three-cycles on the $T^6/(\Z_2 \times \Z_6 \times \OR)$ orientifold}\label{Ss:RemZ2Z6}

The action of the orbifold group $\Z_2 \times \Z_6$ on the factorisable six-torus $T^6 = T_{(1)}^2\times T_{(2)}^2 \times T_{(3)}^2$ consists of a rotation of the complex coordinates $z_k$ parametrising the respective two-torus $T_{(k)}^2$ with $k\in \{1,2,3\}$:
\begin{equation} 
\theta^m \omega^n: z_k \rightarrow e^{2 \pi i (m a_k + n b_k)} z_k, \qquad \qquad \text{ with } \vec{a} = \frac{1}{2} (1,-1,0), \quad \vec{b} = \frac{1}{6}(0,1,-1).
\end{equation}
Note that the point group $\Z_2 \times \Z_6$ is generated by the elements $\theta$ and $\omega$, with $\theta$ generating the $\Z_2$-factor acting on the four-torus $T_{(1)}^2\times T_{(2)}^2$ and $\omega$ generating the $\Z_6$-part acting on the four-torus $T_{(2)}^2\times T_{(3)}^2$. As a direct product of two Abelian factors containing each $\Z_2$ as a (sub)group, the orbifold group allows for a global discrete torsion factor $\eta=\pm 1$~\cite{Vafa:1994rv,Blumenhagen:2005tn,Forste:2010gw}, whose presence alters the amount of two- and three-cycles supported in the orbifold twisted sectors, as indicated in table~\ref{tab:Z2Z6HodgeNumbers} listing the Hodge numbers per sector. In the absence of discrete torsion ($\eta=1$), the $\Z_2^{(i)}$ singularities can be resolved through a blow-up in the respective twisted sector. In the presence of discrete torsion $(\eta = -1)$, one has to resort to deformations of the $\Z_2^{(i)}$ singularities, yielding exceptional three-cycles located at the former $\Z_2^{(i)}$ fixed loci. The three-cycles in the $\Z_2^{(i)}$-twisted sectors turn out to be useful tools with regard to particle physics phenomenology and D6-brane model building~\cite{Ecker:2014hma,Ecker:2015vea}, encouraging us to focus for the remainder of the article on the orbifold with discrete torsion.  
\mathtabfix{
\begin{array}{|c||c||c|c|c|c|c|c|c||c|}
\hline 
\multicolumn{10}{|c|}{\text{\bf Hodge Numbers $(h^{11},h^{21})$ per sector for the factorisable orbifold $T^6/(\Z_2 \times \Z_6)$}}\\
\hline \hline
\multicolumn{2}{|c||}{} & \Z_2^{(1)} &  \Z_2^{(2)} &  \Z_2^{(3)} & \Z_3 & \multicolumn{3}{|c||}{\Z_6^{(\prime)}}& \text{\bf Hodge Numbers} \\ 
\hline \hline
\eta & \text{\bf Untwisted} &  \omega^3 &   \theta \omega^3 & \theta   & \omega^2 & \omega & \theta \omega & \theta  \omega^2 & \left(\begin{array}{c} h^{11} \\ h^{21} \end{array}\right)\\
\hline \hline
\eta = +1 & \left(\begin{array}{c} 3 \\ 1\end{array} \right)  &  \left(\begin{array}{c} 6\\ 0\end{array} \right) &  \left(\begin{array}{c} 8\\0 \end{array} \right) &  \left(\begin{array}{c} 8 \\ 0\end{array} \right) & \left(\begin{array}{c} 8 \\2 \end{array} \right) & \left(\begin{array}{c} 2 \\  0\end{array} \right) & \left(\begin{array}{c} 8 \\ 0  \end{array} \right) & \left(\begin{array}{c} 8 \\ 0 \end{array} \right) & \left(\begin{array}{c} 51 \\ 1+ 2 \end{array} \right)  \\
\hline \hline
\eta = -1 &  \left(\begin{array}{c} 3 \\ 1 \end{array} \right)  &  \left(\begin{array}{c}0 \\ 6 \end{array} \right) &  \left(\begin{array}{c} 0\\ 4\end{array} \right) &  \left(\begin{array}{c}0 \\ 4\end{array} \right) & \left(\begin{array}{c}8 \\2 \end{array} \right) & \left(\begin{array}{c} 0\\2 \end{array} \right) & \left(\begin{array}{c} 4 \\ 0 \end{array} \right) & \left(\begin{array}{c} 4\\0 \end{array} \right) & \left(\begin{array}{c} 19\\ 15 + 2 \times 2 \end{array} \right)  \\
\hline
\end{array}
}{Z2Z6HodgeNumbers}{Hodge numbers $(h^{11},h^{21})$ per sector for the factorisable toroidal orbifold $T^6/(\Z_2 \times \Z_6)$ with lattice configuration $SU(2)^2\times SU(3)\times SU(3)$. In the absence of discrete torsion, the Hodge numbers match those of the orbifold $T^6/(\Z_2\times \Z_2)$, namely $(h^{11},h^{21})=(51,3)$,
but with a different distribution over twisted sectors. Considering the orbifolds with discrete torsion leads to a reduction of the initial 51 three-cycles on $T^6/(\Z_2\times \Z_2)$ to only 19 three-cycles on $T^6/(\Z_2\times \Z_6)$ due to the additional $\Z_3$-action.}

A first observation regarding the $\Z_6$-action deals with the shape of the two-tori $T_{(2)}^2 \times T_{(3)}^2$ whose underlying lattice is constrained to be (up to overall rescaling per two-torus)
the root lattice of $SU(3)\times SU(3)$, i.e.~both lattices are hexagonal, and the complex structures of these two-tori are fixed. Only the first two-torus, whose lattice configuration corresponds to
(up to overall scaling) the root lattice of $SU(2)^2$, has an unfrozen complex structure modulus, matching the Hodge number $h^{21}_{\rm bulk} = 1$ for the $\Z_2 \times \Z_6$ orbifold. The comparison with the orbifold $T^6/(\Z_2\times \Z_2)$ with discrete torsion shows that the additional $\Z_3$-action reduces the number of three-cycles in the $\Z_2^{(i)}$ twisted sectors by triple identifications, and thereby also enforces the simultaneous deformation of the associated $\Z_2\times \Z_2$ singularities, as we will discuss in detail in section~\ref{Sss:HFZ3}. For now, we restrict ourselves to counting the (orbits of) singularities appearing in the various twisted sectors of the orbifold and to indicating how they relate to the Hodge numbers in table~\ref{tab:Z2Z6HodgeNumbers}:
\begin{itemize}
\item Three $\Z_2^{(i=1,2,3)}$-twisted sectors generated by $(\omega^3, \omega^3 \theta, \theta)$ respectively, where each $\Z_2^{(i)}$ sector comes with 16 fixed two-tori or fixed lines labelled by the points $(\alpha \beta)$ with $\alpha, \beta \in \{1,2,3,4\}$ along $T^4_{(i)} \equiv T^2_{(j)} \times T^2_{(k)}$ as depicted in figure~\ref{Fig:T2LatticesSquareHexa}. 
In the $\Z_2^{(1)}$-twisted sector, the fixed point $(11)$ on $T_{(2)}^2\times T_{(3)}^2$ remains invariant under the orbifold action by $\omega$, while the other fixed points recombine into orbifold-invariant orbits consisting of three fixed points each. More explicitly, the $\Z_6$-action rotates the fixed points as $2\rightarrow 3 \rightarrow 4 \rightarrow 2$ on $T^2_{(2)}$ and as $2\rightarrow 4 \rightarrow 3 \rightarrow 2$ on $T^2_{(3)}$, which implies the following five orbits of $\Z_2^{(1)}$ fixed points: $[(21),(31),(41)]$, $[(12),(14),(13)]$, $[(33),(42),(24)]$, $[(22),(34),(43)]$ and $[(44),(23),(32)]$. On the orbifold $T^6/(\Z_2\times \Z_6)$ without discrete torsion, these $\Z_2^{(i)}$ fixed point orbits contribute to the $h^{11}$ K\"ahler moduli, while they contribute to the $h^{21}$ complex structure moduli  for the orbifold set-up with discrete torsion when tensored with a one-cycle on $T_{(1)}^2$. In the $\Z_2^{(j=2,3)}$-twisted sectors, the four fixed points $(\alpha 1)$ on $T_{(1)}^2 \times T_{(k=3,2)}^2$ are invariant under the $\Z_6$-action, while the other twelve fixed points recombine into four $\Z_6$-invariant orbits of three fixed points each: $[(\alpha 2),(\alpha 3),(\alpha 4)]$.  
On the orbifold without discrete torsion, all fixed point orbits contribute to the counting of $h^{11}$, while on the orbifold with discrete torsion only the non-trivial $\Z_6$-invariant orbits tensored with a one-cycle on $T_{(j)}^2$ (which is also rotated under $\Z_6$) contribute to $h^{21}$.
\item One $\Z_3$-twisted sector generated by $\omega^2$ with nine fixed two-tori labelled by the fixed points $(a b)$ with $a,b \in \{1,5,6\}$ on $T_{(2)}^2\times T_{(3)}^2$. The fixed points are subject to the $\Z_2\times \Z_2$ orbifold action mapping $5\leftrightarrow 6$ and $1\circlearrowleft$, such the $\Z_3$ fixed points along $T^4_{(1)}$ recombine into four invariant orbits: $(11)$, $[(15),(16)]$, $[(51),(61)]$ and $[(55),(56),(65),(66)]$. 
As detailed in~\cite{Forste:2010gw}, the $\Z_3$-twisted sector does not feel the discrete torsion phase $\eta=\pm 1$, such that in any case, each fixed point orbit supports two two-cycles per $T^4_{(1)}/\Z_3$ singularity, and three-cycles arise from tensoring the $\Z_2 \times \Z_2$ quadruplet $[(55),(56),(65),(66)]$ on $T^4_{(1)}$ with one-cycles on $T^2_{(1)}$.
\item One $\Z_6$-twisted and two $\Z_6'$-twisted sectors generated by $(\omega, \theta \omega, \theta \omega^2)$, respectively. The $\Z_6$-twisted sector associated to $\omega$ comes with one fixed two-torus or fixed line located at the singularity (11) on $T_{(2)}^2\times T_{(3)}^2$. 
As detailed in~\cite{Forste:2010gw}, the discrete torsion phase acts non-trivially in this sector, which accounts for $h^{11}=2$ in the case of $\eta=+1$ and $h^{21}=2$ in the case of $\eta=-1$ in analogy to the other $\Z_2^{(1)}$ twisted sector.\\
The other two $\Z_6^{\prime}$ actions have a different structure: the one generated by $\theta \omega$ yields twelve fixed points labelled by $(\alpha a 1)$, and the last one generated by $\theta \omega^2$ comes with twelve fixed points labelled by $(\alpha 1 a)$, where $\alpha \in \{1,2,3,4\}$ and $a\in \{1,5,6\}$ for both cases. Under the $\Z_2\times \Z_2$ orbifold action, the twelve $\Z_6^{\prime}$ fixed points in the $\theta \omega$ sector recombine into eight orbits $[(\alpha 1 1)]_{\theta \omega}$, $[(\alpha 5 1), (\alpha 6 1)]_{\theta \omega}$. In the absence of discrete torsion $(\eta=+1)$, each orbit supports a two-cycle and its dual four-cycle, while in the presence of
non-trivial discrete torsion $(\eta=-1)$, only the non-trivial orbits $[(\alpha 5 1), (\alpha 6 1)]_{\theta \omega}$ support each one two-cycle and its dual four-cycle, see~\cite{Forste:2010gw} for details. The second $\Z_6^{\prime}$-twisted sector is obtained by permutation of two-torus indices, $T^2_{(2)} \leftrightarrow T^2_{(3)}$.
\end{itemize}

Now that we have a clear understanding of untwisted and twisted sectors and how they contribute to the Hodge numbers, we can infer the different types of orbifold-invariant three-cycles supported on the orbifold $T^6/(\Z_2 \times \Z_6)$ with discrete torsion:
\begin{itemize}
\item[\bf (1)] {\bf Bulk three-cycles} are orbifold-invariant products of three one-cycles, where each one-cycle extends along a different two-torus $T_{(i)}^2$. The homology class of each one-cycle is specified by two integer-valued co-prime torus wrapping numbers $(n^i,m^i)$ w.r.t. the basis one-cycles $\pi_{2i-1}, \pi_{2i}$
of each two-torus $T_{(i)}^2$, see figure~\ref{Fig:T2LatticesSquareHexa} for the conventional choice of basis used in this article. 
The orbifold-invariant products of the basis three-cycles combine into four basis bulk three-cycles $(\rho_1, \rho_2, \rho_3, \rho_4)$, matching the Betti-number $b^{\rm bulk}_3 = 2 (h^{21}_{\rm bulk} + 1) = 4$ counting the number of basis three-cycles inherited from the factorisable six-torus $(T^2)^3$ after  $\Z_2\times \Z_6$ identifications. Generic bulk three-cycles can then be expressed in terms of these four basis bulk three-cycles: 
\begin{equation}\label{Eq:Bulk-Wrapping}
\begin{array}{rcl}
\Pi^{\rm bulk} &=& n^1 (n^2 n^3- m^2 m^3) \rho_1 + n^1 (n^2 m^3 + m^2 n^3+m^2m^3) \rho_2  \\
  && +\, m^1 (n^2 n^3- m^2 m^3) \rho_3 + m^1 (n^2 m^3 + m^2 n^3+m^2m^3) \rho_4
 .
\end{array}
\end{equation}
\item[\bf(2)] {\bf Exceptional three-cycles} are orbifold-invariant products of a one-cycle on the $\Z_2^{(i)}$-invariant two-torus $T_{(i)}^2$ with an exceptional divisor $e_{\alpha \beta}^{(i)}$ located at the $\Z_2^{(i)}$ fixed points $(\alpha \beta)$ along the four-torus $T^4_{(i)}$. The $\Z_2^{(i)}$ fixed points can be resolved by gluing in a two-sphere per singularity. The $\Z_6$-invariant  products of the basis one-cycles with the exceptional divisors yield twelve basis exceptional cycles $({\boldsymbol \epsilon}_{\lambda}^{(1)}, \tilde {\boldsymbol \epsilon}_{\lambda }^{(1)})$ in the $\Z_2^{(1)}$-twisted sector with $\lambda\in\{0,1,2,3,4,5\}$ and $2 \times 8$ basis exceptional cycles $({\boldsymbol \epsilon}_{\alpha}^{(k)}, \tilde {\boldsymbol \epsilon}_{\alpha }^{(k)})$ in the $\Z_2^{(k=2,3)}$-twisted sectors with $\alpha \in \{1,2,3,4\}$. The dimensionality of the full set of basis exceptional three-cycles expected from all $\Z_2^{(i)}$-sectors combined matches the Betti-number $b^{\Z_2}_3 = 2 h_{\Z_2}^{21} = 2 \cdot \left( 6 + 2 \times 4  \right)  = 28$.\\ Furthermore, the $\Z_6$- and $\Z_3$-twisted sectors also yield $b^{\Z_6+\Z_3}_3 = 2 \cdot (2+2)=8$ exceptional three-cycles located at the $\omega$ and $\omega^2$ fixed points along $T_{(2)}^2\times T_{(3)}^2$ as detailed above. These latter basis three-cycles need to be taken into account when searching for a unimodular basis of the full three-cycle lattice, but they do not contribute to the standard CFT constructions of Type IIA/$\OR$ orientifold models~\cite{Blumenhagen:1999md,Blumenhagen:1999ev,Forste:2000hx}, in particular they expected to contribute to the open string one-loop annulus amplitude~\cite{Blumenhagen:2002wn,Blumenhagen:2006ci}, such that they require no further attention from our part, and we shall only focus on the exceptional three-cycles that can be expressed in terms of the $\Z_2^{(i)}$ exceptional basis three-cycles.  
\item[\bf(3)] {\bf Fractional three-cycles} are linear combinations of some bulk three-cycle and several exceptional three-cycles. When a bulk three-cycle passes through the $\Z_2^{(i)}$-fixed points and represents its own $\Z_2$-orbifold image, one has to add the appropriate set of exceptional three-cycles (weighted with appropriate sign factors)
in order to form a closed fractional three-cycle. As such, a fractional three-cycle can be expressed as
\begin{equation}\label{Eq:Pi-frac}
\Pi^{\rm frac} = \frac{1}{4} \Pi^{\rm bulk} +  \frac{1}{4} \sum_{i=1}^3 \Pi^{\Z_2^{(i)}} =\frac{1}{4} \Pi^{\rm bulk} + \frac{1}{4} \sum_{i=1}^3 \sum_{\lambda} \left(  x^{(i)}_\lambda {\boldsymbol \epsilon}_{\lambda}^{(i)} + y^{(i)}_\lambda \tilde {\boldsymbol \epsilon}_{\lambda}^{(i)}  \right),
\end{equation} 
where the integer-valued exceptional wrapping numbers $\left(x^{(i)},  y^{(i)}\right)$ are constructed from the torus wrapping numbers $(n^i,m^i)$ along the two-torus $T_{(i)}^2$ weighted by sign factors associated to the discrete $\Z_2^{(i)}$-eigenvalues $\pm 1$ and to (-1) exponentiated by  the discrete Wilson-lines $(\tau^{j},\tau^{k}) \in \{0,1\}$. The explicit form of $\left(x^{(i)},  y^{(i)}\right)$ is constrained by the position of the two-cycle on $T_{(i)}^4$ set by the discrete shift parameter $(\sigma^{j}, \sigma^{k}) \in \{0,1\}$, as detailed in table 36 of~\cite{Ecker:2014hma}. 
The sum over $\lambda$ runs over at most four different values in the $\Z_2^{(1)}$-twisted sector with $\lambda \in \{ 0,1,2,3,4,5\}$ and two values in the $\Z_2^{(2,3)}$-twisted sectors with $\lambda \in \{1,2,3,4 \}$
for the orbifold $T^6/(\Z_2 \times \Z_6)$.
\end{itemize}
A detailed discussion of three-cycles on the orbifold $T^6/(\Z_2\times\Z_6)$ can be found in~\cite{Forste:2010gw}, while their prospects for intersecting D6-brane model building have been thoroughly investigated in~\cite{Ecker:2014hma,Ecker:2015vea}. For instance, phenomenologically viable models with three chiral generations were identified in abundance on this orbifold - a feature that can be traced back to the $\Z_3$-factor in the orbifold group, in analogy to other orbifolds with a $\Z_3$-factor within the point group~\cite{Honecker:2004kb,Gmeiner:2007zz,Gmeiner:2008xq,Honecker:2012qr,Honecker:2013kda}.

Type IIA string compactifcations on $T^6/(\Z_2\times \Z_6)$ preserve ${\cal N}=2$ supersymmetry in the closed string sector, which can be broken to a phenomenologically more appealing ${\cal N}=1$ supersymmetry by including an orientifold projection consisting of the worldsheet parity ${\Omega}$, a left-moving fermion number projection $(-)^{F_L}$ and an anti-holomorphic involution ${\cal R}$. The fixed planes under the involution ${\cal R}$ combine into four inequivalent orbits under the $\Z_6$-action, corresponding to the O6-planes $\OR$ and $\OR\Z_2^{(i=1,2,3)}$, respectively, as listed in table~\ref{tab:O6PlanesZ2Z6} for the {\bf aAA}-lattice configuration.\footnote{The anti-holomorphic involution ${\cal R}$ also constrains the shape of the two-torus lattices and limits the orientation of each lattice w.r.t.~the orientifold-invariant direction to two invariant orientations: {\bf A} or {\bf B} for a hexagonal lattice and {\bf a} or {\bf b} for a rectangular lattice. Through a non-supersymmetric rotation of the lattices, the {\it a priori} six independent lattice configurations can be reduced~\cite{Ecker:2014hma} to two physically distinct ones: the {\bf aAA} and {\bf bAA} lattices. From the model building perspective with intersecting D6-branes, the {\bf aAA}-lattice configuration turned out~\cite{Ecker:2014hma,Ecker:2015vea} to be the most fruitful background allowing for global three-generation
MSSM, Left-Right (L-R) symmetric and Pati-Salam (PS) models.} Each of the O6-planes carries RR-charge whose sign is denoted by $\eta_{\OR (\Z_2^{(i)})} \in \{\pm 1\}$, and worldsheet consistency of the Klein-bottle relates them to the discrete torsion parameter~\cite{Blumenhagen:2005tn,Forste:2010gw}:
\begin{equation}
\eta = \eta_{\OR} \prod_{i=1}^3 \eta_{\OR \Z_2^{(i)}}.
\end{equation}
 This implies that at least one of the O6-planes is exotic, with the sign of the RR-charges opposite w.r.t. the other O6-planes. Anticipating the phenomenologically appealing global models discussed in sections~\ref{S:DefModuliGlobalModels} and~\ref{S:GKFDefModuli}, we select the $\OR\Z_2^{(3)}$-plane as the exotic O6-plane, i.e. \mbox{$\eta_{\OR \Z_2^{(3)}}=-1$}.\footnote{The choice $\eta_{\OR\Z_2^{(2)}}=-1$ is equivalent upon permutation of $T^2_{(2)} \leftrightarrow T^2_{(3)}$, while there exists a second inequivalent choice of $\eta_{\OR}=-1$ allowing for supersymmetric solutions to the RR tadpole cancellation conditions.}
 The absence of twisted sector contributions in the tree-channel for the Klein bottle and M\"obius strip amplitudes indicates that the sum over all O6-planes corresponds topologically to a (fraction of a) pure bulk three-cycle. As a consequence, D6-branes wrapping fractional three-cycles should be chosen such that the sum of their bulk three-cycles cancel the RR-charges of the O6-planes, while the sum of the $\Z_2^{(i)}$-twisted exceptional three-cycle part should vanish among itself for each $i \in \{1,2,3\}$, in order to ensure vanishing RR tadpoles. Note that the basis bulk and exceptional three-cycles decompose into $\OR$-even and $\OR$-odd three-cycles under the orientifold projection, depending on the choice of the exotic O6-plane, as can be deduced from table~\ref{tab:Z2Z6OrientifoldExceptionalCycles}.
 
 \mathtab{
\begin{array}{|c|c|c|c|}
\hline
\multicolumn{4}{|c|}{\text{\bf ${\cal R}$-invariant planes on $T^6/(\Z_2\times \Z_6\times \OR)$}}\\
\hline \hline
\text{O6-plane} & \text{Torus wrapping numbers} & \text{RR-charge} & \text{Global Models}\\
\hline
\OR & (1,0;1,0;1,0) & \eta_{\OR} & + 1\\
\OR\Z_2^{(1)} & (1,0;-1,2;1,-2) & \eta_{\OR\Z_2^{(1)}} & +1 \\
\OR\Z_2^{(2)} & (0,1;1,0;1,-2) & \eta_{\OR\Z_2^{(2)}} & +1 \\
\OR\Z_2^{(3)} & (0,1;1,-2;1,0) & \eta_{\OR\Z_2^{(3)}}& -1 \\
\hline
\end{array}
}{O6PlanesZ2Z6}{O6-planes on the {\bf aAA} lattice of $T^6/(\Z_2\times \Z_6\times \OR)$ with discrete torsion ($\eta=-1$). The last column indicates the regular/exotic sign of the RR-charges for the global models discussed in sections~\ref{S:DefModuliGlobalModels} and~\ref{S:GKFDefModuli}.}

\mathtab{
\begin{array}{|c|c||c|c|c||c|c|c|}
\hline
\multicolumn{8}{|c|}{\text{\bf Orientifold images of basis three-cycles on}\, T^6/(\Z_2 \times \Z_6\times \OR) \,  \text{\bf with } \eta = -1  }\\
\hline
\hline
 \multicolumn{2}{|c||}{ \text{\bf Bulk cycles}}& \multicolumn{3}{|c||}{ \Z_2^{(1)} \text{\bf twisted sector}} &  \multicolumn{3}{|c|}{ \Z_2^{(l)} \text{\bf twisted sector with}\; l = 2,3} \\
\hline
\hline
\OR (\rho_\alpha)& \alpha & \OR({\boldsymbol \epsilon}^{(1)}_{\lambda}) & \OR(\tilde{\boldsymbol \epsilon}^{(1)}_{\lambda}) & \lambda &  \OR({\boldsymbol \epsilon}^{(l)}_{\alpha}) & \OR(\tilde{\boldsymbol \epsilon}^{(l)}_{\alpha}) & \alpha \\
\hline
\hline
 \rho_1 &1 & - \eta_{(1)} {\boldsymbol \epsilon}^{(1)}_{\lambda} 
& \eta_{(1)} \, \tilde{\boldsymbol \epsilon}^{(1)}_{\lambda}
& 0,1,2,3 
& - \eta_{(l)} \, {\boldsymbol \epsilon}^{(l)}_{\alpha}  
&\eta_{(l)} \left( \tilde{\boldsymbol \epsilon}^{(l)}_{\alpha} -{\boldsymbol \epsilon}^{(l)}_{\alpha} \right)
& 1,2,3,4\\
\rho_1 - \rho_2& 2&  - \eta_{(1)}  {\boldsymbol \epsilon}^{(1)}_{5} 
& \eta_{(1)} \, \tilde{\boldsymbol \epsilon}^{(1)}_{5}
&4 
&  
&
& \\
-\rho_3&3 &- \eta_{(1)} {\boldsymbol \epsilon}^{(1)}_{4} 
& \eta_{(1)} \, \tilde{\boldsymbol \epsilon}^{(1)}_{4}
&5
&  
&
& \\
\rho_4-\rho_3&4  & & & & & & \\
\hline
\end{array}
}{Z2Z6OrientifoldExceptionalCycles}{Orientifold images of the basis bulk and $\Z_2^{(k)}$ exceptional three-cycles for the {\bf aAA} lattice configuration, depending on the choice of the exotic O6-plane orbit with sign factor $\eta_{(k)} \equiv \eta_{\OR} \eta_{\OR\Z_2^{(k)}}$.}

In order for the D6-branes to preserve the same ${\cal N}=1$ supersymmetry, they are required to wrap {\it special Lagranigian} ({\it sLag}) three-cycles $\Pi$:
\begin{eqnarray}
&&{\cal J}_{(1,1)}\big|_{\Pi} =0, \label{Eq:LagCon}\\
&&{\rm Re}(\Omega_3)\big|_{\Pi} > 0, \qquad {\rm Im}(\Omega_3)\big|_{\Pi} =0. \label{Eq:SpecCon}
\end{eqnarray}  
Three-cycles satisfying condition (\ref{Eq:LagCon}) where the pullback of the K\"ahler (1,1)-form ${\cal J}_{(1,1)}$ w.r.t.  the~three-cycle worldvolume vanishes, are called {\it Lagrangian}  ({\it Lag}) cycles. It is straightforward to check that the (factorisable) bulk three-cycles satisfy this condition. Three-cycles satisfying condition (\ref{Eq:SpecCon}) are calibrated w.r.t.~the (real part of the) holomorphic volume form $\Omega_3$, deserving the epithet {\it special}. At the orbifold point, the condition (\ref{Eq:SpecCon}) reduces to constraints on the torus wrapping numbers and the bulk complex structure moduli. Deforming the background away from the orbifold point can yield an exceptional three-cycle with non-vanishing volume, which no longer satisfies the special condition, implying that supersymmetry can only be maintained when the volume of such an exceptional three-cycle vanishes; 
in other words the twisted complex structure modulus is stabilised at vanishing vacuum expectation value ({\it vev}). Explicit examples of this phenomenon will be discussed in section~\ref{S:DefModuliGlobalModels}.

For the sake of completeness regarding the discussion of geometric moduli on the \mbox{$T^6/(\Z_2 \times \Z_6 \times \OR)$} orientifold with discrete torsion, 
we also list the counting of K\"ahler moduli and closed string vectors on the {\bf aAA} lattice in table~\ref{tab:Z2Z6HodgeSplitting}.
The counting on the inequivalent {\bf bAA} lattice can be found in table~46 of~\cite{Forste:2010gw}.
\mathtabfix{
\begin{array}{|c||c||c|c|c|c|c|c|c||c|}
\hline 
\multicolumn{10}{|c|}{\text{\bf Hodge Numbers $(h^{11}_+,h^{11}_-)$ per sector for the factorisable orientifold $T^6/(\Z_2 \times \Z_6 \times \OR)$ with $\eta=-1$}}\\
\hline \hline
\multicolumn{2}{|c||}{} & \Z_2^{(1)} &  \Z_2^{(2)} &  \Z_2^{(3)} & \Z_3 & \multicolumn{3}{|c||}{\Z_6^{(\prime)}}& \text{\bf Hodge Numbers} \\ 
\hline \hline
\eta & \text{\bf Untwisted} &  \omega^3 &   \theta \omega^3 & \theta   & \omega^2 & \omega & \theta \omega & \theta  \omega^2 & \left(\begin{array}{c} h^{11}_+ \\ h^{11}_- \end{array}\right)
\\\hline \hline
\eta = +1 & \left(\begin{array}{c} 0 \\ 3 \end{array}\right) &  \left(\begin{array}{c} 1 \\ 5 \end{array}\right) & 
\left(\begin{array}{c} 0 \\ 8 \end{array}\right) & \left(\begin{array}{c} 0 \\ 8 \end{array}\right) & 
 \left(\begin{array}{c} 0 \\ 8 \end{array}\right) &  \left(\begin{array}{c} 0 \\ 2 \end{array}\right) & 
 \left(\begin{array}{c} 0 \\ 8 \end{array}\right) & \left(\begin{array}{c} 0 \\ 8 \end{array}\right) & 
 \left(\begin{array}{c} 1 \\ 50 \end{array}\right)
\\\hline \hline
\eta = -1 & \left(\begin{array}{c} 0 \\ 3 \end{array}\right) &   \left(\begin{array}{c} 0 \\ 0 \end{array}\right) & 
\left(\begin{array}{c} 0 \\ 0 \end{array}\right) & \left(\begin{array}{c} 0 \\ 0 \end{array}\right) & 
 \left(\begin{array}{c} 0 \\ 8 \end{array}\right) &  \left(\begin{array}{c} 0 \\ 0 \end{array}\right) & 
 \left(\begin{array}{c} 2(1+\eta_{(2)})  \\ 2(1-\eta_{(2)})  \end{array}\right)
&  \left(\begin{array}{c} 2(1+\eta_{(3)})  \\ 2(1-\eta_{(3)})  \end{array}\right)
&   \left(\begin{array}{c} 4 + 2(\eta_{(2)}+\eta_{(3)}) \\ 15 - 2(\eta_{(2)}+\eta_{(3)})  \end{array}\right)
\\\hline
\end{array}
}{Z2Z6HodgeSplitting}{Splitting of the Hodge number $h^{11}$ into $\OR$-even and $\OR$-odd part on the {\bf aAA} lattice.
$h^{11}_-$ counts the number of K\"ahler moduli on the orientifold, while $h^{11}_+$ counts the number of closed string vectors.
The models in sections~\protect\ref{S:DefModuliGlobalModels} and~\protect\ref{S:GKFDefModuli} obey $\eta_{(2)}=1=-\eta_{(3)}$.
}
It is noteworthy that for the phenomenologically interesting choice of exotic $\OR\Z_2^{(3)}$-plane, i.e. $\eta_{(2)}=-\eta_{(3)}=1$, the $\theta\omega$-twisted sector does not contain any K\"ahler moduli, i.e.~$h^{11}_-=0$.
The orientifold projection thus removes the geometric moduli  in this sector required for resolving the $\Z_6^{\prime}$ singularities, and a full resolution and deformation of the 
toroidal orbifold background to a smooth Calabi-Yau threefold is not possible in the presence of the $\OR$ orientifold action on Type IIA string theory.

\subsection{Lagrangian lines on the elliptic curve in the hypersurface formalism}\label{Ss:LCHF}
At the orbifold point, the geometric engineering method for D6-brane models on $T^6/\Z_{2N}$ or $T^6/(\Z_2 \times \Z_{2M})$ backgrounds reviewed above formally uses 
exceptional divisors at $\Z_2$-singularities
and their topological intersection numbers, even though their volumes are set to zero, or in other words the associated twisted complex structure moduli 
have vanishing {\it vev}s. 
When moving away from the orbifold point into the Calabi-Yau moduli space by deforming the $\Z_2$-singularities, we have to use an extended toolbox of algebraic geometry and embed the orbifold as a hypersurface in an ambient toric space. The first step in this process consists in reformulating the two-tori as elliptic curves in the weighted complex projective space $\P_{112}^2$ and describing {\it Lag} lines on the elliptic curves.   
 
\begin{figure}[h]
\begin{center}
\begin{tabular}{c@{\hspace{0.4in}}c}
\includegraphics[scale=1.0]{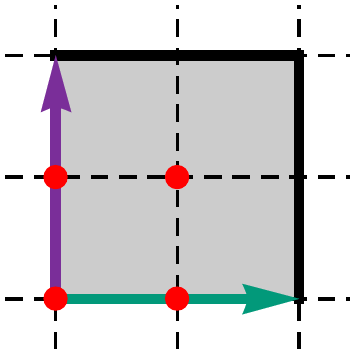} \begin{picture}(0,0) \put(-35,5){$\color{myturq} \pi_1$} \put(-110,75){$\color{myaubergine}  \pi_2$} \put(-100,5){$\color{red} 1$} \put(-100,40){$\color{red} 4$}  \put(-50,40){$\color{red} 3$}  \put(-50,5){$\color{red} 2$}  \end{picture}& \qquad \quad \includegraphics[scale=1.0]{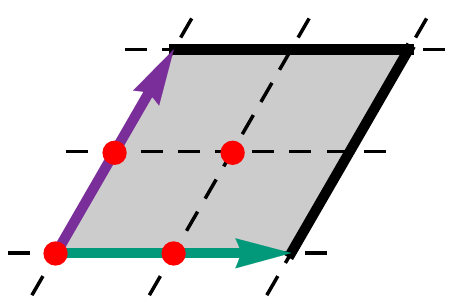} \begin{picture}(0,0) \put(-55,4){$\color{myturq} \pi_1$} \put(-115,65){$\color{myaubergine}  \pi_2$} \put(-122,4){$\color{red} 1$} \put(-120,35){$\color{red} 4$}  \put(-68,35){$\color{red} 2$}  \put(-86,4){$\color{red} 3$}  \end{picture} \\
(a) & (b)
\end{tabular}
\caption{Lattice configuration for a square (a) and hexagonal (b) two-torus $T^2_{(i)}$. The grey areas indicate a fundamental cell in the two-torus lattice, while the coloured vectors represent the basis one-cycles $\pi_{2i-1},\pi_{2i}$ spanning the fundamental cell (with the two-torus index $i$ suppressed in the following). 
The red points correspond to $\Z_2$ fixed points on $T^2$, whose labels match the indices of the roots of $F(x,v)$ in equation (\ref{Eq:GenFEC}). An anti-holomorphic involution keeping the one-cycle $\pi_1$ fixed also leaves all fixed points on the square two-torus invariant, while such an involution permutes the fixed points 2 and 4 on the hexagonal lattice and leaves the other two points 1 and 3 invariant.\label{Fig:T2LatticesSquareHexa}}
\end{center}
\end{figure}
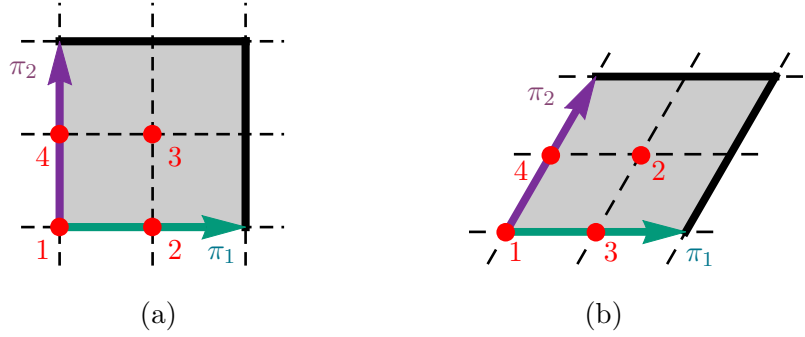

Thus, we introduce the coordinates $(x,v,y)$  with weights $(1,1,2)$ as the homogeneous coordinates of the projective space $\P_{112}^2$ and describe a two-torus as a hypersurface within $\P_{112}^2$. More explicitly, a two-torus corresponds to an elliptic curve in $\P_{112}^2$, which forms the zero locus of a polynomial $f$ of degree 4:
\begin{equation}
f \equiv -y^2 +  F(x,v) = 0, \qquad F(x,v) = 4 v x^3 - g_2 v^3 x - g_3 v^4, \label{Eq:TEHypersurface}
\end{equation}
 where we choose the Weierstrass form for the elliptic curve. There exists a $\Z_2$ reflection symmetry acting only on $y\rightarrow -y$, yet its fixed points correspond to the roots of the polynomial $F(x,v)$. By expanding $F(x,v)$ in terms of its (finite) roots $\epsilon_2$, $\epsilon_3$ and $\epsilon_4$,
 \begin{equation}\label{Eq:GenFEC}
 F(x,v) = 4v ( x - \epsilon_2 v) ( x - \epsilon_3 v) ( x - \epsilon_4 v),
\end{equation}  
the coefficients $g_2$ and $g_3$ are easily related to the roots: $g_2 = - 4 \left( \epsilon_2 \epsilon_3 + \epsilon_2 \epsilon_4 + \epsilon_3 \epsilon_4  \right)$ and $g_3 = 4 \epsilon_2 \epsilon_3 \epsilon_4$, with the roots satisfying the condition $\epsilon_2 + \epsilon_3 + \epsilon_4 = 0$. The fourth root $\epsilon_1$ located at $x = \infty$ (in the $v=1$ patch) represents the $\Z_2$ fixed point at the origin.
The coefficients $g_2$ and $g_3$ are on the other hand uniquely determined by the torus lattice and its complex structure parameter $\tau$, such that we can limit ourselves to those torus lattices relevant for the orbifold $T^6/(\Z_2\times \Z_6)$:
\begin{itemize}
\item[(1)] {\bf a}-type lattices or untilted (rectangular) tori with ${\rm Re}(\tau) = 0$: generically the roots $\epsilon_{\alpha}$ are all real and can be ordered as $\epsilon_4<\epsilon_3<\epsilon_2$. A square torus with $\tau = i$ represents a special case for which $g_3=0$, $\epsilon_2 = - \epsilon_4 = 1$ and $\epsilon_3=0$. 
\item[(2)] {\bf b}-type lattices or tilted tori with ${\rm Re}(\tau) \neq 0$: generically the roots $\epsilon_2$ and $\epsilon_4$ are related by complex conjugation, $\epsilon_2 = \ov{\epsilon_4} \equiv \xi $, while $\epsilon_3 = - 2 {\rm Re}(\xi) $ is a real parameter. A hexagonal torus with $\tau = e^{i \frac{\pi}{3}}$
(cf. figure~\ref{Fig:T2LatticesSquareHexa}\,(b))
 forms a special case with $g_2 = 0$ and $\xi = e^{2\pi i/3} $, for which the elliptic curve exhibits an additional $\Z_3$ symmetry $\frac{x}{v} \rightarrow e^{2\pi i/3} \frac{x}{v}$. This $\Z_3$ symmetry is in correspondence with a $\Z_3$ subgroup acting on the hexagonal two-torus lattice, suggesting that the two-tori $T_{(2)}^2$ and $T_{(3)}^2$ are perfectly described by this type of elliptic curve. 
\end{itemize} 
A pictorial representation of a square untilted and a hexagonal torus lattice with their respective roots is given in figure~\ref{Fig:T2LatticesSquareHexa}. 
The two-torus $T_{(1)}^2$ is not affected by the $\Z_3$ subgroup of the $\Z_2\times \Z_6$ point group, hence its torus lattice can in principle be either untilted ({\bf a}-type lattice) or tilted ({\bf b}-type lattice). As a tilted two-torus $T_{(1)}^2$ does not provide for any (known) phenomenologically appealing global intersecting D6-brane models~\cite{Ecker:2014hma,Ecker:2015vea}, we confine ourselves to an untilted $T_{(1)}^2$ and simplify the set-up even more by choosing a square two-torus when studying deformations.
This simplification is justified by the fact that for the choice of an exotic $\OR\Z_2^{(3)}$-plane, the (bulk) RR tadpole cancellation conditions can only be solved in a supersymmetric way if all D6-branes extend along $\pi_1$ on the {\bf a}-type lattice, see section~\ref{S:DefModuliGlobalModels} for several examples. Such configurations are supersymmetric for any value of the complex structure parameter Im$(\tau)>0$ on $T^2_{(1)}$.  
 
The full map between a two-torus and an elliptic curve is given by Weierstrass' elliptic function $\wp(z)$, mapping bijectively the holomorphic coordinate $z$ on the two-torus with modular parameter $\tau$ to the elliptic curve with coefficients $g_2$ and $g_3$. It is easy to see that the Weierstrass' elliptic function $\wp(z)$ satisfies the hypersurface equation (\ref{Eq:TEHypersurface}) through the identification $\wp(z) = x/v$, $\wp'(z) = y/v^2$, which reduces to a differential equation on $\wp(z)$.

One-cycles on a two-torus $T_{(i)}^2$ were introduced in the previous section parameterised by the torus wrapping numbers $(n^i,m^i)$ w.r.t.~the basis one-cycles. In order to discuss {\it Lag} lines on an elliptic curve, we introduce an anti-holomorphic involution $\sigma$ acting on the homogeneous coordinates as follows:
\begin{equation}\label{Eq:sigma-for-Lags}
\sigma: \quad \left(\begin{array}{c} x \\ v \end{array} \right) \longmapsto  A \left(\begin{array}{c} \ov x \\ \ov v \end{array} \right), \qquad \qquad y \longmapsto e^{i \beta} \ov{y},
\end{equation}
 with $A \in GL_2(\C)$. For this action to be an involution, the matrix $A$ has to satisfy the condition $A \ov A = \unity$. The involution also has to be a symmetry of the elliptic curve, which boils down to the following condition $\sigma \left( F(x,v) \right) = e^{2i \beta} \ov F(\ov x, \ov v)$. Solving both conditions allows to extract the unequivocal forms of the various anti-holomorphic involutions~\cite{Blaszczyk:2014xla}. Afterwards, one can determine the fixed loci for each individual anti-holomorphic involution, which will constitute only a subset of all {\it Lag} lines on the elliptic curve~\cite{Blaszczyk:2014xla,Blaszczyk:2015oia,Koltermann:2015oyv}. Fortunately for us, the {\it Lag} lines defined as fixed loci under $\sigma$ are in one-to-one correspondence with the torus one-cycles used as building blocks for global intersecting D6-brane models, as can be checked explicitly by virtue of the Weierstrass' elliptic function $\wp(z)$. Distinguishing between square and hexagonal lattices leads to the following classification of 
 {\it Lag} lines:
 \begin{itemize}
 \item[(1)] untilted square torus: we distinguish between four one-cycles {\bf aX} (with {\bf X} $=$ {\bf I}, {\bf II}, {\bf III}, {\bf IV}) passing through two roots $\epsilon_{\alpha}$
 and one-cycles {\bf cX} (with {\bf X} $=$ {\bf I}, {\bf II}) not passing through any of the roots. The first type of one-cycles will serve as fractional cycles, while the latter type of one-cycles remain bulk cycles once the orbifold $T^6/(\Z_2\times \Z_6)$ is modelled as a hypersurface in an ambient toric space. A full overview of these {\it Lag} lines on the square torus and the relations to the roots $\epsilon_{\alpha}$ is offered in table \ref{tab:LagSquareT2}; their positions in the $x$-plane are depicted in figure \ref{Fig:TorusCyclesxplane} (a) for the $v=1$ coordinate patch.   
 \begin{table}[h]
\begin{center}
\begin{tabular}{|c|c|c|c|c|}
\hline 
\multicolumn{5}{|c|}{\bf Lagrangian lines on square untilted torus $T^2$}\\
\hline \hline 
$(n^i,m^i)$ & displacement & condition in $x$ & label & picture\\
\hline
\raisebox{-20pt}{\multirow{3}{*}{$\pm (1,0)$}} & \raisebox{-8pt}{0} &  \raisebox{-8pt}{$\epsilon_2 \leq x $} &  \raisebox{-8pt}{{\color{myorange} \bf aI}}  &\raisebox{-20pt}{\begin{picture}(0,0) \put(-14,2){\includegraphics[scale=0.6]{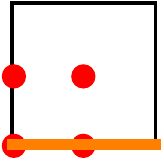}} \end{picture}} \\
\cline{2-5}
&  \raisebox{-8pt}{1} &  \raisebox{-8pt}{$ \epsilon_4 \leq x \leq \epsilon_3$} &  \raisebox{-8pt}{$\color{SkyBlue} \bf aIII$ } & \raisebox{-20pt}{\begin{picture}(0,0) \put(-18,2){ \includegraphics[scale=0.6]{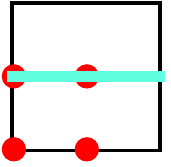}}\end{picture}}   \\
\cline{2-5} &  \raisebox{-8pt}{ continuous} &  \raisebox{-8pt}{$|x-\epsilon_4|^2 = 2 \epsilon_4^2 + \epsilon_2 \epsilon_3$} & \raisebox{-8pt}{ {\color{mygr} \bf cI}} &   \raisebox{-20pt}{\begin{picture}(0,0) \put(-14,2){\includegraphics[scale=0.6]{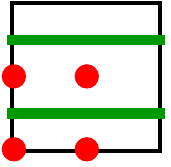}}\end{picture}}   \\
\hline \hline
\raisebox{-20pt}{\multirow{3}{*}{$\pm (0,1)$}} &  \raisebox{-8pt}{0} & \raisebox{-8pt}{ $ x \leq \epsilon_4  $}  & \raisebox{-8pt}{ {\color{mybrown} \bf aII} }&  \raisebox{-20pt}{\begin{picture}(0,0) \put(-14,2){\includegraphics[scale=0.6]{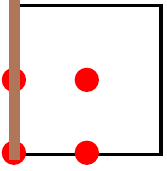}}\end{picture}}    \\
\cline{2-5}
& \raisebox{-8pt}{1} & \raisebox{-8pt}{ $ \epsilon_3 \leq x \leq \epsilon_2  $}  & \raisebox{-8pt}{ {\color{mypurple} \bf aIV} }&  \raisebox{-20pt}{\begin{picture}(0,0) \put(-14,2){\includegraphics[scale=0.6]{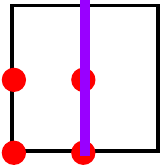}}\end{picture}}    \\
\cline{2-5} &  \raisebox{-8pt}{continuous} & \raisebox{-8pt}{$|x-\epsilon_2|^2 = 2 \epsilon_2^2 + \epsilon_4 \epsilon_3$}  &  \raisebox{-8pt}{{\color{myblue} \bf cII} } &  \raisebox{-20pt}{\begin{picture}(0,0) \put(-18,2){ \includegraphics[scale=0.6]{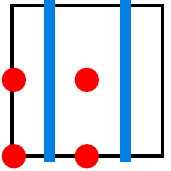}}\end{picture}}  \\
\hline
\end{tabular}
\caption{Overview of {\it Lag} lines on an elliptic curve corresponding to a square untilted two-torus. The first column lists the torus wrapping numbers $(n^i,m^i)$ describing how the one-cycles wrap on the two-torus lattice, while the second column indicates a potential displacement along one of the basis one-cycles (we distinguish between no displacement, a displacement over one-half of a basis one-cycle or a continuous displacement in between the latter two options). The third column presents the equation for the {\it Lag} line in the homogenous coordinate $x$ (considering the $v=1$ patch) in line with figure \ref{Fig:TorusCyclesxplane} (a). The last column gives a graphical representation of the {\it Lag} one-cycle on the two-torus lattice. \label{tab:LagSquareT2}}
\end{center}
\end{table}  
 \item[(2)] hexagonal torus: here we can identify the one-cycles  {\bf bX} (with {\bf X} $=$ {\bf I}, {\bf II}, {\bf III}, {\bf IV}) passing through two roots $\epsilon_{\alpha}$ and corresponding to $\sigma$-involution invariant directions. Due to the $\Z_3$ symmetry, we also find the by $\pm \frac{2\pi}{3}$ rotated images of these one-cycles, tripling the number of individual one-cycles. A full list of {\it Lag} lines is given in table \ref{tab:LagHexaT2}, while figure \ref{Fig:TorusCyclesxplane} (b) shows their position in the $x$-plane for the $v=1$ coordinate patch and clearly exhibits the $\Z_3$ symmetry.  
 \end{itemize} 
\begin{table}[h]
\begin{center}
\begin{tabular}{|c|c|c|c|@{\hspace{0.3in}}c@{\hspace{0.3in}}|}
\hline 
\multicolumn{5}{|c|}{\bf Lagrangian lines on hexagonal torus $T^2$}\\
\hline \hline 
$(n^i,m^i)$ & displacement & condition in $x$ & label & picture  \\
\hline
 \raisebox{-18pt}{\multirow{2}{*}{$\pm (1,0)$}} &  \raisebox{-8pt}{0} & \raisebox{-8pt}{$1\leq x$}  & \raisebox{-8pt}{{$\color{myblue3} \bf bI^0$}}  & \raisebox{-22pt}{\begin{picture}(0,0) \put(-24,2){\includegraphics[scale=0.6]{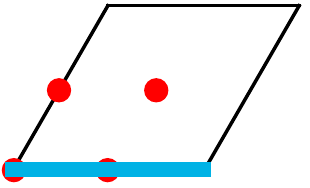}} \end{picture}} \\
\cline{2-5}
& \raisebox{-8pt}{1} & \raisebox{-8pt}{$|x -1|^2 = 3, {\rm Re}(x)\leq -1/2 $} & \raisebox{-8pt}{{$\color{myblue3} \bf bIII^0$}}   & \raisebox{-22pt}{\begin{picture}(0,0) \put(-24,2){\includegraphics[scale=0.6]{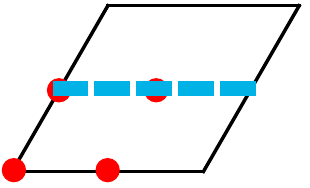}} \end{picture}}  \\
\hline 
 \raisebox{-18pt}{\multirow{2}{*}{$\pm (-1,2)$} }& \raisebox{-8pt}{0 }& \raisebox{-8pt}{$x\leq 1$}  &\raisebox{-8pt}{ {$\color{myblue3} \bf bII^0$} } & \raisebox{-22pt}{\begin{picture}(0,0) \put(-24,2){\includegraphics[scale=0.6]{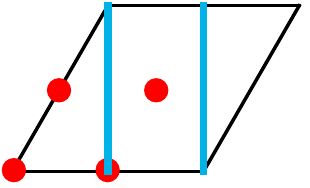}} \end{picture}} \\
\cline{2-5}
& \raisebox{-8pt}{1} & \raisebox{-8pt}{ $|x -1|^2 = 3, -1/2  \leq {\rm Re}(x)  $}  & \raisebox{-8pt}{{$\color{myblue3} \bf bIV^0$} }  & \raisebox{-22pt}{\begin{picture}(0,0) \put(-24,2){\includegraphics[scale=0.6]{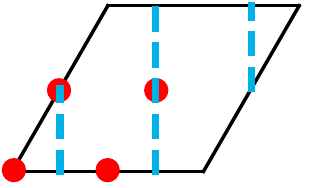}} \end{picture}} \\
\hline \hline
 \raisebox{-18pt}{\multirow{2}{*}{$\pm (0,1)$}} &\raisebox{-8pt}{ 0} &\raisebox{-8pt}{ $1\leq  \xi^2 x$  } &\raisebox{-8pt}{ {$\color{myorange} \bf bI^-$}}  & \raisebox{-22pt}{\begin{picture}(0,0) \put(-24,2){\includegraphics[scale=0.6]{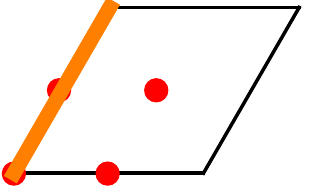}} \end{picture}} \\
\cline{2-5}
& \raisebox{-8pt}{1} & \raisebox{-8pt}{$|\xi^2 x -1|^2 = 3, {\rm Re}(\xi^2 x)\leq -1/2 $}  &\raisebox{-8pt}{ {$\color{myorange} \bf bIII^-$} }  &  \raisebox{-22pt}{\begin{picture}(0,0) \put(-24,2){\includegraphics[scale=0.6]{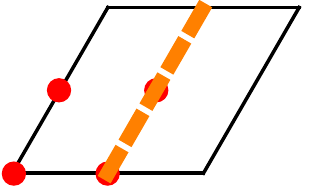}} \end{picture}}  \\
\hline 
 \raisebox{-18pt}{\multirow{2}{*}{$\pm (2,-1)$}} & \raisebox{-8pt}{0} &\raisebox{-8pt}{ $\xi^2 x\leq 1$} &\raisebox{-8pt}{ {$\color{myorange} \bf bII^-$}}  &  \raisebox{-22pt}{\begin{picture}(0,0) \put(-24,2){\includegraphics[scale=0.6]{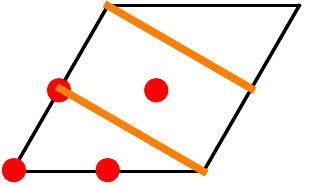}} \end{picture}} \\
\cline{2-5}
& \raisebox{-8pt}{1} & \raisebox{-8pt}{$ | \xi^2 x -1|^2 = 3, -1/2  \leq {\rm Re}(\xi^2 x)  $}   & \raisebox{-8pt}{{$\color{myorange} \bf bIV^-$} }  & \raisebox{-22pt}{\begin{picture}(0,0) \put(-24,2){\includegraphics[scale=0.6]{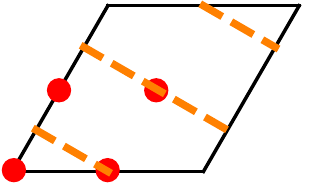}} \end{picture}}  \\
\hline \hline
 \raisebox{-18pt}{\multirow{2}{*}{$\pm (1,-1)$} }&\raisebox{-8pt}{ 0} & \raisebox{-8pt}{$1\leq  \xi x$ }  &\raisebox{-8pt}{ {$\color{mypurple} \bf bI^+$}}  &  \raisebox{-22pt}{\begin{picture}(0,0) \put(-24,2){\includegraphics[scale=0.6]{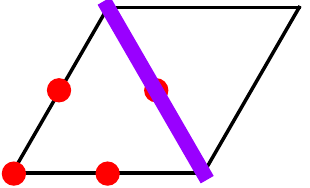}} \end{picture}} \\
\cline{2-5}
& \raisebox{-8pt}{1} &\raisebox{-8pt}{ $|\xi x -1|^2 = 3, {\rm Re}(\xi x)\leq -1/2 $} & \raisebox{-8pt}{{$\color{mypurple} \bf bIII^+$}}   &  \raisebox{-22pt}{\begin{picture}(0,0) \put(-24,2){\includegraphics[scale=0.6]{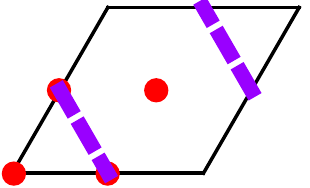}} \end{picture}}   \\
\hline 
 \raisebox{-18pt}{\multirow{2}{*}{$\pm (1,1)$}} &\raisebox{-8pt}{ 0} &\raisebox{-8pt}{ $\xi x\leq 1$} &\raisebox{-8pt}{ {$\color{mypurple} \bf bII^+$} } &  \raisebox{-22pt}{\begin{picture}(0,0) \put(-24,2){\includegraphics[scale=0.6]{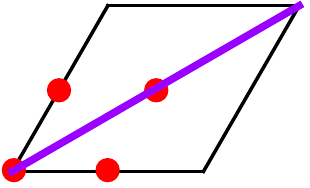}} \end{picture}} \\
\cline{2-5}
& \raisebox{-8pt}{1} & \raisebox{-8pt}{$|\xi x -1|^2 = 3, -1/2  \leq {\rm Re}(\xi x)  $ }  &\raisebox{-8pt}{ {$\color{mypurple} \bf bIV^+$}}   &  \raisebox{-22pt}{\begin{picture}(0,0) \put(-24,2){\includegraphics[scale=0.6]{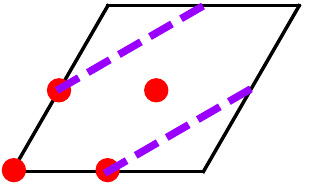}} \end{picture}}  \\
\hline 
\end{tabular}
\caption{Overview of {\it Lag} one-cycles on an elliptic curve corresponding to a hexagonal two-torus. The first column lists the torus wrapping numbers $(n^i,m^i)$ describing how the one-cycles wrap on the two-torus lattice, while the second column indicates a potential displacement along one of the basis one-cycles. The third column presents the equation for the {\it Lag} line in the homogenous coordinate $x$ (considering the $v=1$ patch) in line with figure \ref{Fig:TorusCyclesxplane} (b). The last column gives a graphical representation of the {\it Lag} one-cycle on the two-torus lattice.\label{tab:LagHexaT2}}
\end{center}
\end{table} 
 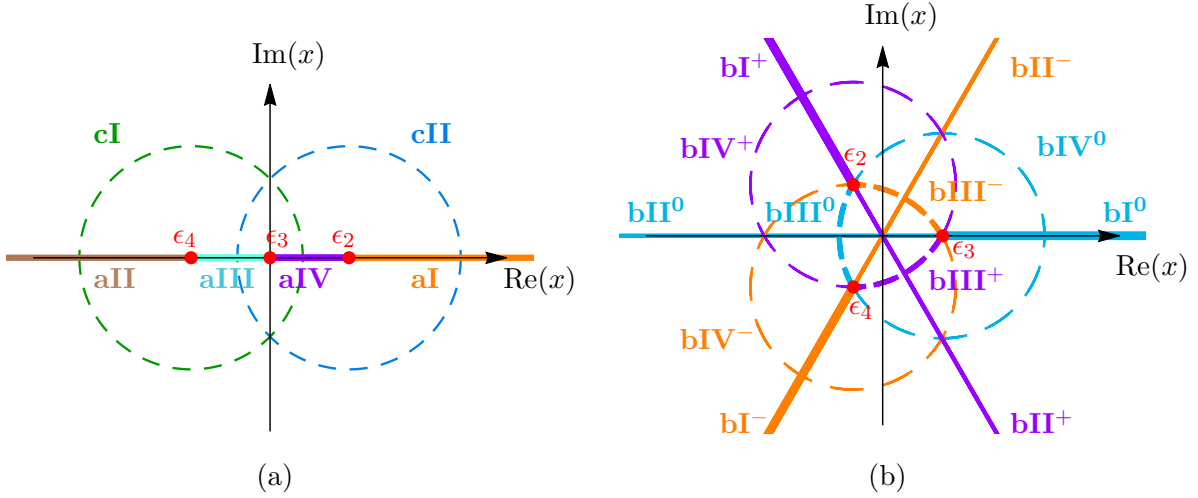
\begin{figure}[h]
 \begin{center}
 \begin{tabular}{c@{\hspace{0.4in}}c}
 \includegraphics[width=7cm]{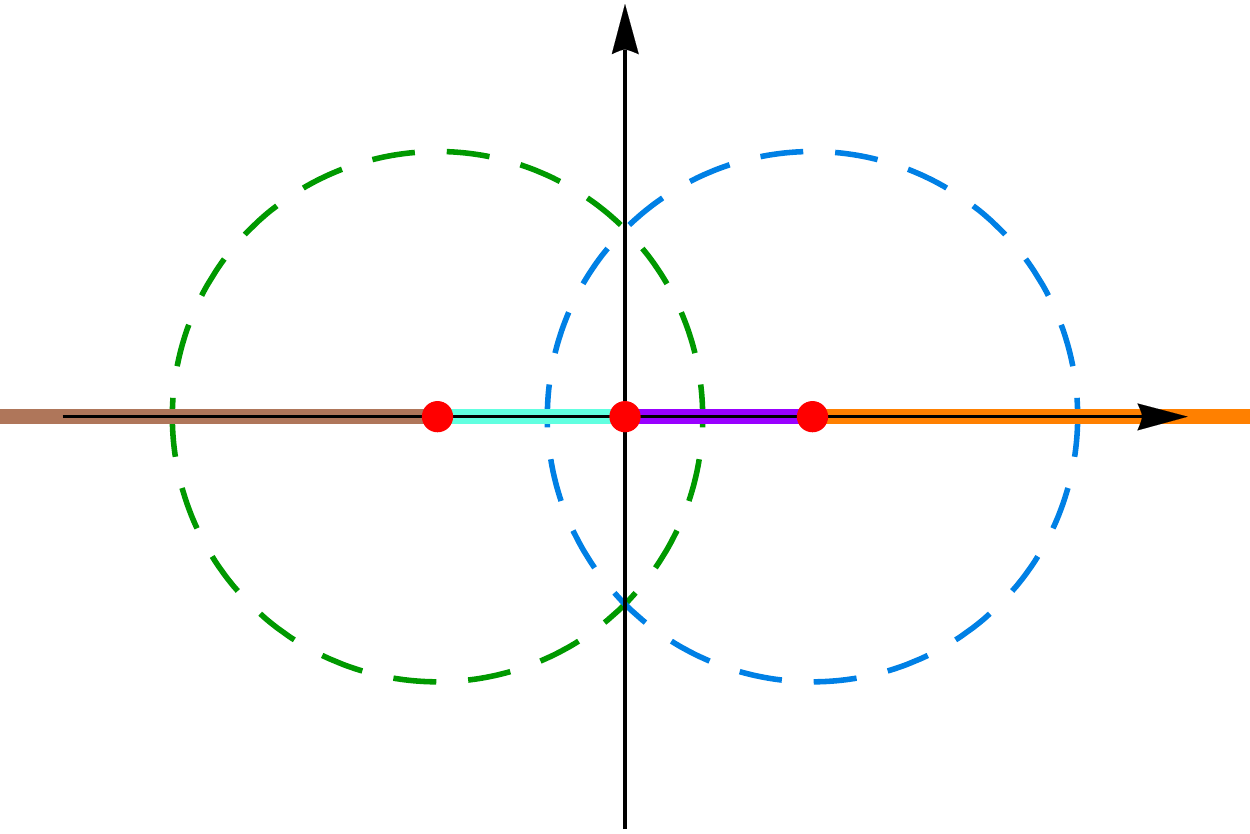} \begin{picture}(0,0) \put(-170,110){\color{mygr} \bf cI} \put(-50,110){\color{myblue} \bf cII}  \put(-170,55){\color{mybrown} \bf aII}   \put(-50,55){\color{myorange} \bf aI}  \put(-100,55){\color{mypurple} \bf aIV} \put(-130,55){\color{SkyBlue} \bf aIII} \put(-80,72){\color{red} $\epsilon_2$}  \put(-105,72){\color{red} $\epsilon_3$} \put(-140,72){\color{red} $\epsilon_4$} 
 \put(-110,140){${\rm Im}(x)$}   \put(-15,55){${\rm Re}(x)$}  \end{picture} &  \includegraphics[width=7cm]{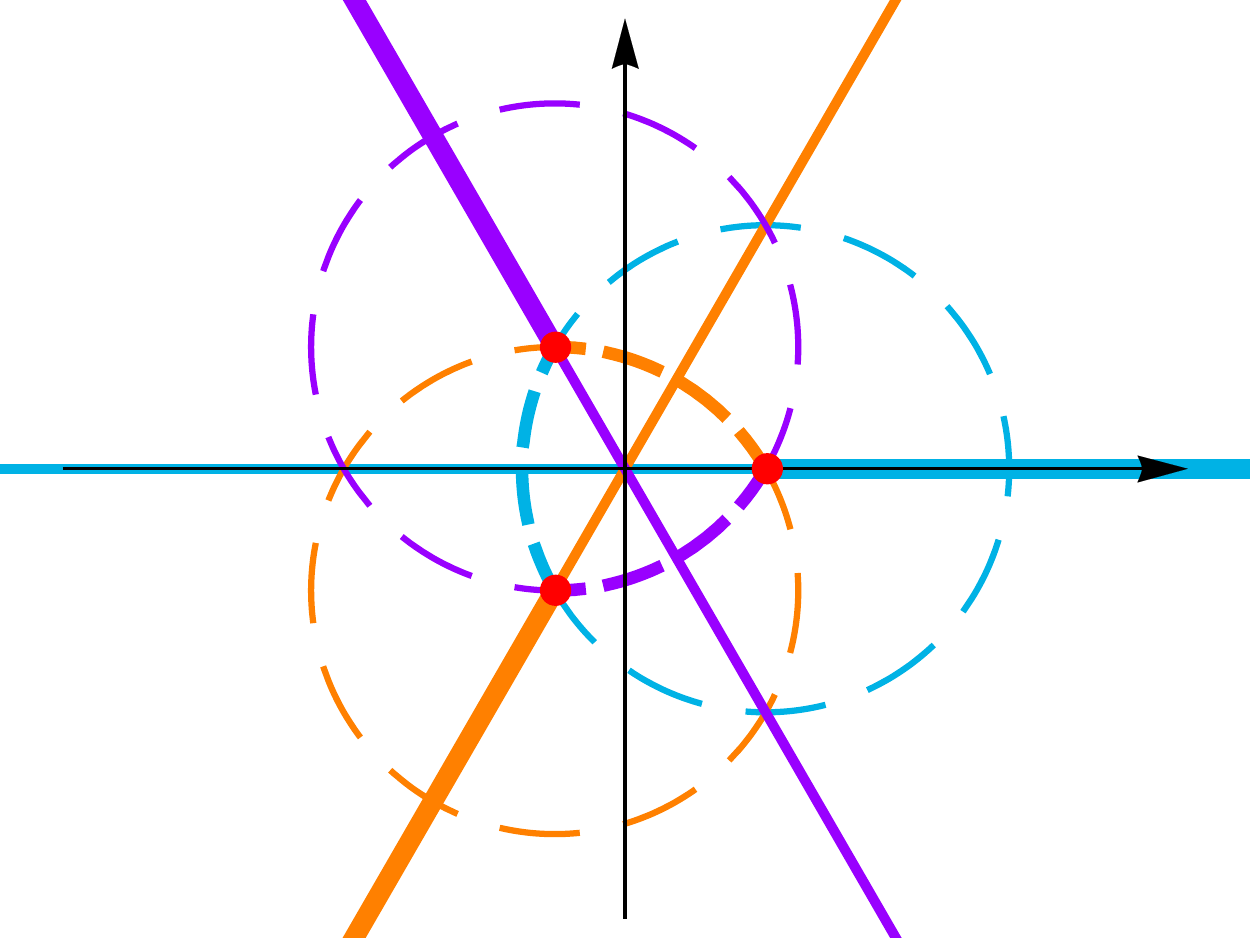} \begin{picture}(0,0) \put(-77,68){\color{red} $\epsilon_3$}  \put(-118,103){\color{red} $\epsilon_2$} \put(-116,45){\color{red} $\epsilon_4$} 
 \put(-110,155){${\rm Im}(x)$}   \put(-15,60){${\rm Re}(x)$} \put(-20,80){\color{myblue3} ${\bf bI^0}$} \put(-200,80){\color{myblue3} ${\bf bII^0}$} \put(-148,80){\color{myblue3} ${\bf bIII^0}$}   \put(-45,105){\color{myblue3} ${\bf bIV^0}$}  \put(-165,0){\color{myorange} ${\bf bI^-}$} \put(-55,135){\color{myorange} ${\bf bII^-}$}  \put(-86,88){\color{myorange} ${\bf bIII^-}$} \put(-180,32){\color{myorange} ${\bf bIV^-}$}    \put(-165,135){\color{mypurple} ${\bf bI^+}$}   \put(-55,0){\color{mypurple} ${\bf bII^+}$}  \put(-86,54){\color{mypurple} ${\bf bIII^+}$}   \put(-180,104){\color{mypurple} ${\bf bIV^+}$}  \end{picture} \\
 (a) & (b)
 \end{tabular}
 \caption{Graphic representation of {\it Lag} lines on an elliptic curve embedded in $\P^2_{112}$ corresponding to a square untilted two-torus (a) or a hexagonal two-torus (b). The {\it Lag} lines are drawn in the $x$-plane with $v=1$, following the notation in the overview tables \ref{tab:LagSquareT2} and \ref{tab:LagHexaT2}.
   \label{Fig:TorusCyclesxplane}}
 \end{center}
 \end{figure}
 
 The holomorphic one-form $\Omega_1$ defined on the elliptic curve descends from the following holomorphic two-form defined on the ambient space $\P^2_{112}$:
 \begin{equation}
 \Omega_1 = \frac{1}{\pi i} \int_\gamma \frac{x dv \wedge dy - v dx \wedge dy  + y dx\wedge dv}{f},
 \end{equation}  
where we divided by the polynomial $f$ defined on the l.h.s. of equation~\eqref{Eq:TEHypersurface}
to obtain a well-defined scale-invariant two-form on $\P^2_{112}$. The integral is taken over a curve $\gamma$ around the singular region $f=0$, such that we can apply Cauchy's residue theorem in a suitable patch to obtain the expression:
\begin{equation}\label{Eq:CYOneForm}
\Omega_1 = \frac{vdx - xdv}{y\big|_{f=0}}.
\end{equation}
The expression for $y$ in terms of $x$ and $v$ follows by imposing the hypersurface equation $f=0$ and choosing one branch of the square root. With this prescription, the holomorphic one-form (\ref{Eq:CYOneForm}) allows us to uncover the calibration form for each of the {\it Lag} lines identified above. Roughly speaking, {\it Lag} cycles with {\bf X}=even are calibrated w.r.t.~${\rm Re}(\Omega_1)$, while {\it Lag} lines with {\bf X}=odd have ${\rm Im}(\Omega_1)$ as  calibration form.

\clearpage
\subsection{$T^6/(\Z_2 \times \Z_6 \times \OR)$ in the hypersurface formalism}\label{Ss:IntersectSummary}
Describing deformations of the orbifold $T^6/(\Z_2 \times \Z_6)$ requires us to adopt a hypersurface formalism in which $T^6/(\Z_2 \times \Z_6)$ with discrete torsion is embedded into an appropriate toric space. In a first step, we will review in section~\ref{Sss:HFTZ2Z2N} how the deformations of $T^6/(\Z_2 \times \Z_2)$ with discrete torsion thrive in the hypersurface formalism, after which we mod out the remaining $\Z_3$ symmetry within the $\Z_2 \times \Z_6$ point group in section~\ref{Sss:HFZ3} to end up with the hypersurface formalism describing the deformations of $T^6/(\Z_2 \times \Z_6)$. Finally, we impose additional constraints due to the orientifold involution $\OR$.
\subsubsection{Hypersurface formalism for $T^6/(\Z_2 \times \Z_{2})$ with discrete torsion}\label{Sss:HFTZ2Z2N}
The factorisable orbifold $T^6/(\Z_2 \times \Z_2)$ with discrete torsion can be mapped \cite{Vafa:1994rv} to the direct product of three distinct elliptic curves modded out by the orbifold group $\Z_2 \times \Z_2$, whose action only keeps the $\Z_2 \times \Z_2$-invariant subring of the ring of polynomials. The $\Z_2 \times \Z_2$-invariant ``monomials" are subject to a single equation describing a hypersurface in the toric space parametrised by the coordinates $(x_i,v_i,y)_{i=1,2,3}$ with weights $q_i$ according to the weight diagram in table \ref{tab:ScalingToricZ2Z2}. 
\begin{table}[h]
\begin{center}
\begin{tabular}{|c||c|c|c|c|c|}
\hline \multicolumn{6}{|c|}{\bf Scaling charges $q_i$ for $T^6/(\Z_2 \times \Z_2)$ toric space }\\
\hline \hline
weights $q_i$& $(x_1,v_1)$ & $(x_2,v_2)$ & $(x_3,v_3)$ & $y$& $f(x_i,v_i,y)$\\
\hline
$q_1$ & 1 & 0 & 0 & 2 & 4\\
$q_2$ & 0 & 1 & 0 & 2 & 4\\
$q_3$ & 0 & 0 & 1 & 2 & 4\\
\hline
\end{tabular}
\caption{Overview of the weights $q_i$ for the coordinates $(x_i,v_i,y)_{i=1,2,3}$  parametrising the toric space, as well as for the polynomial $f$ allowing for the embedding of $T^6/(\Z_2 \times \Z_2)$ as a hypersurface in this toric space.  \label{tab:ScalingToricZ2Z2}}
\end{center}
\end{table}

Indeed, the orbifold $T^6/(\Z_2 \times \Z_2)$ with discrete torsion or its deformations are described by the zero locus of the polynomial $f(x_i,v_i,y)$ whose most general form reads:
\begin{align}
\begin{split}
f \equiv& -y^2 + F_{(1)}(x_1,v_1) F_{(2)}(x_2,v_2) F_{(3)}(x_3,v_3) \\
&- \sum_{i\neq j \neq k \neq i} \sum_{\alpha,\beta=1}^4 \varepsilon^{(i)}_{\alpha\beta} F_{(i)}(x_i,v_i) \delta F_{(j)}^{\alpha} (x_j,v_j) \delta F_{(k)}^{\beta} (x_k,v_k) \\
&+ \sum_{\alpha,\beta,\gamma =1}^4 \varepsilon_{\alpha\beta\gamma} \delta F_{(1)}^{\alpha} (x_1,v_1) \delta F_{(2)}^{\beta} (x_2,v_2) \delta F_{(3)}^{\gamma} (x_3,v_3) 
\, .
\label{Eq:T6Z2Z2HypersurfaceEquation}
 \end{split}
\end{align}
A few comments regarding this polynomial are in order:
\begin{itemize}
\item The polynomials $F_{(i)}(x_i,v_i)$ correspond to the homogeneous polynomials of degree four defining the two-torus $T_{(i)}^2$ as in equation (\ref{Eq:TEHypersurface}) 
or its rewritten form~\eqref{Eq:GenFEC}
and encode information about the complex structure of $T^2_{(i)}$. As each two-torus has its own complex structure, we have {\it a priori} three bulk complex structure moduli in total
(with the number later on being reduced by imposing an extra $\Z_3$ symmetry, cf. section~\ref{Sss:HFZ3}).  
Setting all the deformation parameters to zero, $\varepsilon_{\alpha \beta}^{(i)} = 0 = \varepsilon_{\alpha \beta \gamma}$, corresponds to the orbifold point in the complex structure moduli space, with the roots of $F_{(i)}(x_i,v_i)$ determining the positions of the $\Z_2^{(k\neq i)}$ fixed points. 
\item The deformation polynomials $\delta F_{(i)}^{\alpha}(x_i,v_i)$ are also homogeneous polynomials of degree four and have the same roots as $F_{(i)}(x_i,v_i)$ up to the $\alpha^{\rm th}$ zero. Thus, $\delta F_{(i)}^{\alpha}(x_i,v_i)$ allows to deform the $\Z_2$-fixed point associated to the $\alpha^{\rm th}$ root.
\item Deformations of the form $\delta F_{(i)}^\alpha F_{(j)} F_{(k)}$ with $(ijk)$ a cyclic permutation of $(123)$ are not explicitly considered in equation~\eqref{Eq:T6Z2Z2HypersurfaceEquation} as they correspond to deformations of the complex structure for two-torus $T_{(i)}^2$, up to $PSL(2,\C)$ transformations acting on $(x_i,v_i)$.\footnote{The $PSL(2,\C)$ transformations are able to eliminate three complex parameters, leaving exactly one independent parameter per two-torus. \label{Ref:PSLT2}
} Hence, their coefficients correspond to untwisted moduli and their CFT counter-parts are given by the three truly marginal operators from the untwisted sector in the associated ${\cal N}=(2,2)$ super-conformal field theory, following the construction prescriptions in~\cite{Dixon:1986qv}.
\item The parameter $\varepsilon_{\alpha \beta}^{(i)}$ allows for the deformation of the $\Z_2^{(i)}$ singularity with index $(\alpha \beta)$ on the four-torus $T^4_{(i)}\equiv T^2_{(j)} \times T^2_{(k)}$ with $(ijk)$ some permutation of $(123)$. Counting the number of distinct deformation parameters, one finds $3 \times 4 \times 4 = 48$ parameters in total, one for each $\Z_2^{(i)}$ singularity. The number 48 matches exactly the number of truly marginal operators in the $\Z_2^{(i)}$ twisted sectors of the corresponding ${\cal N}=(2,2)$ SCFT.
\item Within the set of possible deformations of $T^6/(\Z_2 \times \Z_2)$, one also observes the deformations associated to the parameters $\varepsilon_{\alpha \beta \gamma}$. The total number of these parameters amounts to $4\times 4\times 4 = 64$ and is in one-to-one correspondence with the number of $\Z_2\times \Z_2$ fixed points on the orbifold. However, these parameters do not represent independent deformations, but rather depend on the complex structure deformations $\varepsilon_{\alpha \beta}^{(i)}$ from the $\Z_2^{(i)}$ twisted sectors, such that (at most) 64 conifold singularities remain and cannot be deformed away. This reflection is supported~\cite{Vafa:1994rv} by the absence of truly marginal operators in the ${\cal N}=(2,2)$ SCFT corresponding to $\varepsilon_{\alpha \beta \gamma}$-deformations.\footnote{Contrary to the blow-up procedure, where blowing up the co-dimension two singularities in the $\Z_2^{(i)}$ twisted sectors also eliminates the co-dimension three singularities on $T^6/(\Z_2\times\Z_2)$ without discrete torsion, the deformation procedure does not automatically lead to the resolution of the 64 $\Z_2\times \Z_2$ fixed points on $T^6/(\Z_2\times\Z_2)$ with discrete torsion.} 
Thus, determining whether or not conifold singularities are present can only be assessed through the geometric description of the deformed orbifold in the hypersurface formalism in terms of the independent deformation parameters $\varepsilon_{\alpha \beta}^{(i)}$.  
\end{itemize}

In order to determine the holomorphic volume form $\Omega_3$ on $T^6/(\Z_2 \times \Z_2)$ with discrete torsion,
 we extend the philosophy from above that allowed us to identify the appropriate hypersurface equation (\ref{Eq:T6Z2Z2HypersurfaceEquation}). More explicitly, we consider the wedge product of three one-forms $\Omega_1$, one for each two-torus $T_{(i)}^2$ as defined in equation~(\ref{Eq:CYOneForm}), and mod out the $\Z_2 \times \Z_2$ symmetry to obtain the following (simplified) expression (in the $v_i=1$ patch):
\begin{equation}\label{Eq:Holo3FormZ2Z2}
\Omega_3 = \frac{dx_1\wedge dx_2 \wedge dx_3}{y(x_i)\big|_{f=0}},
\end{equation}
up to an overall normalisation constant and possible phase. The expression for $y$ in terms of $x_i$ follows by imposing the hypersurface equation $f\stackrel{!}{=}0$ 
on the defining equation~\eqref{Eq:T6Z2Z2HypersurfaceEquation} and fixing a branch cut for the square root. 

A thorough analysis of {\it Lag} lines on the (deformed) orbifold $T^6/(\Z_2 \times \Z_2)$ on square tori was the subject of~\cite{Blaszczyk:2014xla} and further comments 
and extensions to products of three hexagonal tori can be found in \cite{Blaszczyk:2015oia,Koltermann:2015oyv,Koltermann:2016oyv}. As is well known, the orbifold $T^6/(\Z_2 \times \Z_2)$ with discrete torsion does not support exceptional two-cycles, implying the absence of twisted sector blow-up modes. If one wants to blow up the $\Z_2^{(i)}$ singularities rather than deform them, one ought to look at the version of the orbifold without discrete torsion, see e.g.~\cite{Aspinwall:1988rp,Aspinwall:1994ev}. Physical implications of resolving the orbifold singularities through blow-ups have been investigated in~\cite{Denef:2005mm,Lust:2006zh,Cvetic:2007ju,Blaszczyk:2011hs}.

\subsubsection{Hypersurface formulation for orbifolds with additional $\Z_3 \times \OR$ action}\label{Sss:HFZ3}
Setting up the hypersurface formalism for $T^6/(\Z_2\times \Z_6)$ with discrete torsion now consists in acting with the $\Z_3$-subgroup 
generated by $2\vec{b}=\frac{1}{3}(0,1,-1)$ on the hypersurface formulation of $T^6/(\Z_2\times \Z_2)$ with discrete torsion, such that the resulting hypersurface polynomial is invariant under the $\Z_3$-symmetry.\footnote{Notice that in~\cite{Blaszczyk:2015oia,Koltermann:2015oyv} a different $T^6/(\Z_2 \times \Z_6')$ orbifold with the $\Z_6'$ factor generated by 
$\vec{b}^{\, \prime}=\frac{1}{6}(-2,1,1)$ was considered. The analysis in that case with $\Z_2 \times \Z_6'$ point group was simpler due to all three two-tori being of hexagonal shape and all three $\Z_2^{(i)}$-twisted sectors being equivalent (up to a relative sign factor in the orientifold projection if one of the $\OR\Z_2^{(j)}$-planes is chosen as the exotic one).
} 
The $\Z_3$-action generated by $\omega^2$ will in the first place restrict the form of the homogeneous polynomials $F_2$ and $F_3$, while the form of $F_1$ remains generic as in equation (\ref{Eq:GenFEC}). Anticipating the torus lattice configurations for the global models in section \ref{S:DefModuliGlobalModels}, we consider the first two-torus to be a square untilted two-torus, such that the homogeneous polynomials are given by: 
\begin{equation}
\begin{array}{rcl}
F_{(1)} (x_1, v_1) &=& 4 v_1 x_1 (x_1^2 -  v_1^2),\\
F_{(2)} (x_2, v_2) &=& 4 v_2 (x_2^3 - v_2^3),\\ 
F_{(3)} (x_3, v_3) &=& 4 v_3 (x_3^3 - v_3^3).\\ 
\end{array}
\end{equation} 
Evidently, the $\Z_3$-action also constrains the form of the deformation polynomials $\delta F_{(j=2,3)}^\alpha$, while the deformation polynomials $\delta F_{(1)}^\alpha$ are shaped by the untilted square lattice choice for the two-torus $T_{(1)}^2$:
\begin{equation}\label{Eq:DefPolyZ2Z6}
\begin{array}{rcl@{\hspace{0.4in}}rcl}
\delta F_{(1)}^1 &=&  x_1^2 (x_1^2 - v_1^2), & \delta F_{(j)}^1 &=& x_j (x_j^3 - v_j^3), \qquad\qquad \text{for} \quad j=2,3 \\
\delta F_{(1)}^2 &=&  v_1 x_1(x_1 + v_1)^2, & \delta F_{(j)}^2 &=& v_j^2 (v_j - x_j) (v_j - \xi x_j) , \\
\delta F_{(1)}^3 &=& - v_1^2 (x_1^2 - v_1^2), & \delta F_{(j)}^3 &=& v_j^2 (v_j - \xi x_j) (v_j - \xi^2 x_j) , \\
\delta F_{(1)}^4 &=& -  v_1 x_1 (x_1 -  v_1)^2, & \delta F_{(j)}^4 &=& v_j^2 (v_j - x_j) (v_j - \xi^2 x_j) . \\
\end{array}
\end{equation}
By virtue of the Weierstrass' elliptic function, one can easily deduce that the $\Z_3$-subgroup also acts on the homogeneous coordinates $x_i$ as follows:\footnote{Note that the holomorphic three-form defined in equation (\ref{Eq:Holo3FormZ2Z2}) remains invariant under the $\Z_3$-symmetry.}
\begin{equation}\label{Eq:Z3onx}
\omega^2: (x_1,x_2,x_3) \mapsto (x_1, \xi x_2, \xi^2 x_3), \qquad \text{ with } \xi = e^{2\pi i/3},
\end{equation}
which leaves the homogeneous polynomials $F_{(i)}(x_i,v_i)$ invariant, but forces the deformation polynomials to transform as follows:
\begin{equation}
\begin{array}{ll}
\omega^2: & \delta F_{(1)}^\alpha \mapsto  \delta F_{(1)}^\alpha \qquad \forall\, \alpha \in\{1,2,3,4\},\\
& \delta F_{(2)}^1 \mapsto \xi \delta F_{(2)}^1, \hspace{5mm}  \delta F_{(2)}^2 \mapsto \delta F_{(2)}^3 \mapsto \delta F_{(2)}^4 \mapsto \delta F_{(2)}^2, \\
& \delta F_{(3)}^1 \mapsto \xi^2 \delta F_{(3)}^1, \quad \delta F_{(3)}^2 \mapsto \delta F_{(3)}^4 \mapsto \delta F_{(3)}^3 \mapsto \delta F_{(3)}^2. \\
\end{array}
\end{equation}
Keeping in mind these transformation properties, we can deduce that the linear combination $\delta F_{(j)}^2+\delta F_{(j)}^3+\delta F_{(j)}^4 = 3 v_j^4$ represents a  $\Z_3$-invariant polynomial for $j=2,3$. However, a simple coordinate transformation on $x_j$ eliminates any deformation of the type $3 v_j^4 F_{(1)} F_{(k\neq j)}$, leaving the complex structure of the two-torus $T_{(j=2,3)}^2$ unaltered. The non-invariance of $\delta F_{(j=2,3)}^1$ under $\Z_3$ also excludes any type of deformation of the form $\delta F_{(j=2,3)}^1 F_{(1)} F_{(k=3,2)}$. This observation agrees with the considerations in section~\ref{Ss:RemZ2Z6} that the complex structures of the two-tori $T_{(j=2,3)}^2$ are frozen to hexagonal shape by the $\Z_3$-action. 
\newline
The deformation polynomials $\delta F_{(1)}^\alpha$ are left invariant by the $\Z_3$-action, such that deformations of the type $\delta F_{(1)}^\alpha F_{(2)} F_{(3)}$ do exist, up to $PSL(2,\C)$ transformations acting on $(x_1,v_1)$, and they represent one untwisted complex structure modulus, in line with $h^{21}_{\rm bulk} = 1$. Recall from footnote~\ref{Ref:PSLT2} on page~\pageref{Ref:PSLT2} that the three complex parameters of the $PSL(2,\C)$ symmetry allow to reduce the four deformations to a single independent deformation.

Deformations of the $\Z_2^{(i)}$ singularities are performed through polynomials of the form $F_{(i)} \delta F_{(j)}^\alpha \delta F_{(k)}^\beta$, which should also be $\Z_3$-invariant for consistency. In this sense, the $\Z_3$-action will put restrictions on the deformation parameters $\varepsilon_{\alpha \beta}^{(i)}$ and reduce the number of independent deformation moduli. For the $\Z_2^{(1)}$-twisted sector, we find six independent deformation parameters which concur with the Hodge number $h^{21}_{\Z_2^{(1)}} = 6$ from table \ref{tab:Z2Z6HodgeNumbers}:
\begin{itemize}
\item \vspace{-0.2in} $\varepsilon_{11}^{(1)} \equiv \varepsilon_{0}^{(1)} $ is left untouched and deforms the singularity $(11)$ on the four-torus $T^2_{(2)} \times T^2_{(3)}$;
\item $\xi^2\,\varepsilon_{21}^{(1)} =\varepsilon_{31}^{(1)} = \xi\, \varepsilon_{41}^{(1)} \equiv \varepsilon_{1}^{(1)} $ deforms the singular orbit $[(21),(31),(41)]$ on $T^2_{(2)} \times T^2_{(3)}$;
\item $\xi^2\,\varepsilon_{12}^{(1)} =\varepsilon_{13}^{(1)} = \xi\, \varepsilon_{14}^{(1)} \equiv \varepsilon_{2}^{(1)} $ deforms the singular orbit $[(12),(13),(14)]$ on $T^2_{(2)} \times T^2_{(3)}$;
\item  $\varepsilon_{33}^{(1)} =\varepsilon_{42}^{(1)} = \varepsilon_{24}^{(1)} \equiv \varepsilon_{3}^{(1)} $ deforms the singular orbit $[(33),(42),(24)]$ on $T^2_{(2)} \times T^2_{(3)}$;
\item  $\varepsilon_{22}^{(1)} =\varepsilon_{34}^{(1)} = \varepsilon_{43}^{(1)} \equiv \varepsilon_{4}^{(1)} $ deforms the singular orbit $[(22),(34),(43)]$ on $T^2_{(2)} \times T^2_{(3)}$;
\item  $\varepsilon_{44}^{(1)} =\varepsilon_{23}^{(1)} = \varepsilon_{32}^{(1)} \equiv \varepsilon_{5}^{(1)} $ deforms the singular orbit $[(44),(23),(32)]$ on $T^2_{(2)} \times T^2_{(3)}$.
\end{itemize}

The $\Z_2^{(j=2,3)}$ sectors are equivalent as a result of the $T_{(2)}^2\leftrightarrow T_{(3)}^2$ exchange symmetry, such that we can treat them jointly. In the $\Z_2^{(j)}$-sector we find four independent deformation parameters, which match the Hodge numbers $h^{21}_{\Z_2^{(2)}} = h^{21}_{\Z_2^{(3)}}=4$ in table \ref{tab:Z2Z6HodgeNumbers}:
\begin{itemize}
\item \vspace{-0.2in} $\forall\, \alpha \in \{1,2,3,4\}$ we find $\varepsilon_{\alpha 1}^{(j)} = 0$, which excludes any type of deformation of the fixed points $(\alpha 1)$ on $T_{(1)}^2 \times T_{(k\neq j)}^2$; 
\item $\varepsilon_{1 2}^{(j)} = \varepsilon_{1 3}^{(j)} = \varepsilon_{1 4}^{(j)} \equiv \varepsilon_{1}^{(j)}$ deforms the singular orbit $[(12),(13),(14)]$ on $T^2_{(1)} \times T^2_{(k)}$;
\item $\varepsilon_{2 2}^{(j)} = \varepsilon_{2 3}^{(j)} = \varepsilon_{2 4}^{(j)} \equiv \varepsilon_{2}^{(j)}$ deforms the singular orbit $[(22),(23),(24)]$ on $T^2_{(1)} \times T^2_{(k)}$;
\item $\varepsilon_{3 2}^{(j)} = \varepsilon_{3 3}^{(j)} = \varepsilon_{3 4}^{(j)} \equiv \varepsilon_{3}^{(j)}$ deforms the singular orbit $[(32),(33),(34)]$ on $T^2_{(1)} \times T^2_{(k)}$;
\item $\varepsilon_{4 2}^{(j)} = \varepsilon_{4 3}^{(j)} = \varepsilon_{4 4}^{(j)} \equiv \varepsilon_{4}^{(j)}$ deforms the singular orbit $[(42),(43),(44)]$ on $T^2_{(1)} \times T^2_{(k)}$.
\end{itemize}
\vspace{-0.2in}The skeptical reader might object that there exists a certain freedom in choosing the forms of the deformation polynomials, yet any consistent $\Z_3$-invariant choice of the polynomials $\delta F_{(j)}^\alpha$ should yield the same numbers, as the pairing of the $\Z_2^{(i)}$ fixed points into orbifold-invariant orbits is an eternal consequence of the $\Z_3$-action. After all, the amount of independent $\Z_2^{(i)}$ deformations has to match the Hodge numbers $h^{21}_{\Z_2}$ in the $\Z_2^{(i)}$-twisted sectors discussed in section~\ref{Ss:RemZ2Z6}.

The last type of deformations to consider have the form $ \delta F_{(1)}^\alpha \delta F_{(2)}^\beta \delta F_{(3)}^\gamma$ and are also subject to the $\Z_3$-action. The invariance of $ \delta F_{(1)}^\alpha$ under the $\Z_3$-action suggests that we should only worry about the remaining two deformation polynomials and make sure they recombine into $\Z_3$-invariant combinations with the following relations:
\begin{itemize}
\item \vspace{-0.2in} $\varepsilon_{\alpha11}$ remains unconstrained $(\forall\, \alpha \in \{1,2,3,4\})$;
\item $\xi^2\, \varepsilon_{\alpha 21} = \varepsilon_{\alpha 31} = \xi\, \varepsilon_{\alpha 41} $ $(\forall\, \alpha \in \{1,2,3,4\})$;
\item $\xi^2\, \varepsilon_{\alpha 12} = \varepsilon_{\alpha 13} = \xi\, \varepsilon_{\alpha 14} $ $(\forall\, \alpha \in \{1,2,3,4\})$;
\item $\varepsilon_{\alpha 33} = \varepsilon_{\alpha 24} = \varepsilon_{\alpha 42} $ $(\forall\, \alpha \in \{1,2,3,4\})$;
\item $\varepsilon_{\alpha 22}= \varepsilon_{\alpha 34} = \varepsilon_{\alpha 43}$ $(\forall\, \alpha \in \{1,2,3,4\})$;
\item $\varepsilon_{\alpha 44} = \varepsilon_{\alpha 23} = \varepsilon_{\alpha 32}$ $(\forall\, \alpha \in \{1,2,3,4\})$.
\end{itemize}
\vspace{-0.2in} Hence, we obtain in total $6\times 4 = 24$ $\Z_3$-invariant deformations of the type $ \delta F_{(1)}^\alpha \delta F_{(2)}^\beta \delta F_{(3)}^\gamma$ on the $T^6/(\Z_2 \times \Z_6)$ orbifold, which should be contrasted with the 64 parameters on the \mbox{$T^6/(\Z_2\times \Z_2)$} orbifold. Observe that the initial 64 $\Z_2\times \Z_2$ fixed points split up into four fixed points, which are left invariant under the $\Z_3$ action, and 60 fixed points, which are regrouped into $\Z_3$-invariant triplets. This simple counting explains the 24 allowed deformations $ \delta F_{(1)}^\alpha \delta F_{(2)}^\beta \delta F_{(3)}^\gamma$. Similar to the $\Z_2 \times \Z_2$ orbifold, the $\varepsilon_{\alpha \beta \gamma}$-deformations depend on the twisted complex structure deformation parameters $\varepsilon_{\alpha \beta}^{(i)}$, such that at most 24 conifold singularities remain upon deformation by $\varepsilon_{\alpha \beta}^{(i)}
$. In CFT language this would imply that the truly marginal operators associated to the $\varepsilon_{\alpha \beta \gamma}$-deformations do not exist in the associated ${\cal N}=(2,2)$ SCFT.  
Hence, also for this orbifold the potential presence of conifold singularities can only be assessed by investigating the hypersurface equation algebraically in the hypersurface formalism.

The last element missing to describe {\it sLags} in this hypersurface set-up is the orientifold involution $\sigma_{{\cal R}}$ which acts on the homogeneous coordinates as follows:
\begin{equation}\label{Eq:OrInZ2Z6}
\sigma_{{\cal R}}: (x_i,v_i,y) \longmapsto (\ov x_i, \ov v_i, \ov y). 
\end{equation} 
This anti-holomorphic involution,  constructed from the involution $\sigma$ defined in equation (\ref{Eq:sigma-for-Lags}) by choosing $A=\1_2$ on each two-torus $T^2_{(i)}$, 
has to be a symmetry of the hypersurface, which boils down to the condition $\sigma_{\cal R}(f) = \ov f$. At the orbifold point, this latter condition constrains the shape of the three two-tori to be either of {\bf a}-type  or {\bf b}-type  (the latter corresponding to both {\bf A}- and {\bf B}-orientation for hexagonal lattices)
as discussed in section~\ref{Ss:LCHF} and ensures that the orientifold involution is an automorphism of the torus-lattices. The orientifold involution also acts on the deformation polynomials (\ref{Eq:DefPolyZ2Z6}) as follows:
\begin{equation}
\begin{array}{ll}
\delta F_{(1)}^\alpha \mapsto \ov{\delta F_{(1)}^\alpha} & \forall\, \alpha \in \{1,2,3,4\}\\
\delta F_{(j)}^1 \mapsto \ov{\delta F_{(j)}^1}, & \quad  \delta F_{(j)}^2 \mapsto \ov{\delta F_{(j)}^4}, \qquad \delta F_{(j)}^3 \mapsto \ov{\delta F_{(j)}^3}, \qquad \delta F_{(j)}^4 \mapsto \ov{\delta F_{(j)}^2}
\quad \text{ for } j=2,3
,
\end{array}
\end{equation} 
which concurs with the $\OR$-action on the $\Z_2$ fixed points of the {\bf aAA} lattice, whose individual two-torus positions are depicted in figure \ref{Fig:T2LatticesSquareHexa}. Taking into account the action of the involution, we observe that the deformation parameters are even further reduced: the {\it a priori} complex deformation parameters are either constrained to be real, or two complex deformation parameters are identified, leaving only one independent complex deformation parameter. The latter occurs for the deformation parameters $\varepsilon_4^{(1)}$ and $\varepsilon_5^{(1)}$, for which we can introduce $\varepsilon_{4|5}^{(1)} = \frac{1}{2} \left( \varepsilon_{4+5}^{(1)}  \pm i \varepsilon_{4-5}^{(1)}   \right)$ with $\varepsilon_{4\pm5}^{(1)}  \in \R$. All the other deformation parameters  $\varepsilon_{\lambda=0,1,2,3}^{(1)}$ and $\varepsilon_{\alpha=1,2,3,4}^{(2,3)}$ are constrained to be real. A summary of the independent deformation parameters for the orientifold $T^6/(\Z_2\times \Z_6 \times \OR)$
with discrete torsion and {\bf aAA} lattice is given in table~\ref{tab:DefParZ2Z6Complete}.
\mathtabfix{
\begin{array}{|c||c|c|c|c|}
\hline \multicolumn{5}{|c|}{\text{\bf Independent deformation parameters for $T^6/(\Z_2\times \Z_6 \times \OR)$ with discrete torsion}}\\
\hline \hline
\Z_2^{(i)} & & \text{Parameter identification} & \text{Parameter range} & \text{Exceptional wrapping numbers}\\
 \hline \hline
& \varepsilon_0^{(1)} &\varepsilon_{11}^{(1)} & \R& \left(x_0^{(1)},y_0^{(1)}\right) \\
& \varepsilon_1^{(1)} & \xi^2\,\varepsilon_{21}^{(1)} =\varepsilon_{31}^{(1)} = \xi\, \varepsilon_{41}^{(1)} & \R & \left(x_1^{(1)},y_1^{(1)}\right) \\
 \Z_2^{(1)}& \varepsilon_2^{(1)} & \xi^2\,\varepsilon_{12}^{(1)} =\varepsilon_{13}^{(1)} = \xi\, \varepsilon_{14}^{(1)} & \R & \left(x_2^{(1)},y_2^{(1)}\right) \\
& \varepsilon_3^{(1)} & \varepsilon_{33}^{(1)} =\varepsilon_{42}^{(1)} = \varepsilon_{24}^{(1)}   & \R & \left(x_3^{(1)},y_3^{(1)}\right) \\
& \varepsilon_4^{(1)}, \varepsilon_5^{(1)}  & \varepsilon_{22}^{(1)} =\varepsilon_{34}^{(1)} = \varepsilon_{43}^{(1)} =\ov \varepsilon_{44}^{(1)} = \ov \varepsilon_{23}^{(1)} =\ov \varepsilon_{32}^{(1)}  & \C & \left(x_4^{(1)},y_4^{(1)}, x_5^{(1)},y_5^{(1)}\right) \\
 \hline  \hline
\Z_2^{(2)} &  \varepsilon_{\alpha=1,2,3,4}^{(2)}  & \varepsilon_{\alpha 2}^{(2)} = \varepsilon_{\alpha 3}^{(2)} = \varepsilon_{\alpha 4}^{(2)} &  \R& \left(x_\alpha^{(2)},y_\alpha^{(2)}\right) \\
 \hline \hline 
\Z_2^{(3)}   &  \varepsilon_{\alpha=1,2,3,4}^{(3)}  & \varepsilon_{\alpha 2}^{(3)} = \varepsilon_{\alpha 3}^{(3)} = \varepsilon_{\alpha 4}^{(3)} & \R & \left(x_\alpha^{(3)},y_\alpha^{(3)}\right) \\
  \hline
\end{array}
}{DefParZ2Z6Complete}{Overview of the independent deformation parameters per $\Z_2^{(i)}$-twiste sector for \mbox{$T^6/(\Z_2\times \Z_6 \times \OR)$} with discrete torsion ($\eta = -1$) and the restrictions following from the $\Z_3$-action and the $\OR$-action. The last column relates the deformation parameters to the relevant exceptional wrapping numbers introduced in equation~\protect\eqref{Eq:Pi-frac}
of section~\ref{Ss:RemZ2Z6}.}

In conclusion, the $T^6/(\Z_2\times \Z_6 \times \OR)$ orientifold with discrete torsion corresponds to the zero locus of the following polynomial $f$:
\begin{align}\label{Eq:HyperZ2Z6Full}
\begin{split}
f=& - y^2 +  v_1 x_1 (x_1^2 -  v_1^2) v_2 (x_2^3 - v_2^3) v_3 (x_3^3 - v_3^3) \\
& \qquad- F_{(1)}\left\{ \varepsilon_0^{(1)} \cdot x_2 (x_2^3 - v_2^3) \cdot x_3  (x_3^3 - v_3^3) + \varepsilon_1^{(1)} \cdot 3 v_2^3 x_2  \cdot x_3  (x_3^3 - v_3^3) \right.\\
& \qquad \qquad \quad + \varepsilon_2^{(1)} \cdot x_2  (x_2^3 - v_2^3) \cdot 3 v_3^3 x_3 + \varepsilon_3^{(1)} \cdot 3 v_2^2 v_3^2 \left(v_2^2 v_3^2 + v_2 v_3 x_2 x_3 + x_2^2 x_3^2\right)\\
& \qquad \qquad \quad \left.+ \varepsilon_{4+5}^{(1)} \cdot 3 v_2^2 v_3^2 \left(v_2 v_3 - x_2 x_3\right) \left(2 v_2 v_3 + x_2 x_3\right) + \varepsilon_{4-5}^{(1)} \cdot 3 \sqrt{3} v_2^2 v_3^2 x_2 x_3 \left(v_2 v_3 - x_2 x_3 \right) \right\}\\
& \qquad- \sum_{j,k\in\{2,3\},\, j\neq k} F_{(j)} \left\{\varepsilon_1^{(j)} \cdot x_1^2 (x_1^2- v_1^2) \cdot 3 v_k^4 +  \varepsilon_2^{(j)} \cdot v_1 x_1 (x_1 + v_1)^2 \cdot 3 v_k^4 \right. \\
& \qquad \qquad \qquad \qquad \quad  \left.+  \varepsilon_3^{(j)} \cdot v_1^2  (x_1^2 - v_1^2) \cdot 3 v_k^4 + \varepsilon_4^{(j)} \cdot v_1 x_1  (x_1 - v_1)^2 \cdot 3 v_k^4 \right\} .
\end{split}
\end{align}
In this polynomial expression, one clearly notices the difference in the $\Z_2^{(1)}$ deformations on the one hand and the $\Z_2^{(2,3)}$ deformations on the other hand, in analogy with the difference in exceptional cycles from the respective $\Z_2^{(i)}$-twisted sectors as reviewed in section~\ref{Ss:RemZ2Z6}. This will obviously imply that the effect of $\Z_2^{(1)}$ deformations on {\it sLag} three-cycles has to be studied separately from the effect of $\Z_2^{(2,3)}$ deformations. The latter deformations on the other hand are expected to follow a similar pattern due to the two-torus exchange symmetry \mbox{$T_{(2)}^2 \leftrightarrow T_{(3)}^2$}, which is reflected in the coordinate permutation $(x_2,v_2)\leftrightarrow (x_3,v_3)$ accompanied by a permutation of the twisted parameters $\varepsilon_{1}^{(1)} \leftrightarrow \varepsilon_{2}^{(1)}$.

\subsection{Deforming special Lagrangian cycles on $T^6/(\Z_2 \times \Z_6 \times \OR)$}\label{Ss:DefsLagS}
Having formulated the consistent hypersurface formalism to discuss deformations of \mbox{$T^6/(\Z_2 \times \Z_6 \times \OR)$} with discrete torsion, we can now turn our attention to the geometric properties of {\it sLag} three-cycles away from the orbifold point, after providing a concise translation of the three-cycles introduced in section~\ref{Ss:RemZ2Z6} into the hypersurface formalism.

\subsubsection{\textit{\textbf{sLags}} at the Orbifold Point}\label{Sss:sLagsOrbi}
A minimal set of {\it sLags} is defined as the fixed loci under the orientifold involution $\sigma_{\cal R}$, introduced in equation~(\ref{Eq:OrInZ2Z6}). Due to the additional $\Z_3$-symmetry in equation~(\ref{Eq:Z3onx}) on the $x_{2,3}$ coordinates, this set of {\it sLag}s can be extended by demanding that they be invariant under the action of $\sigma_{\cal R} \times \Z_3$, a group isomorphic to the symmetric group~$S_3$. 
To describe the location of the O6-planes in the hypersurface formalism, it suffices to determine the invariant solutions for the $\sigma_{\cal R}$-sector, as the $\Z_3$ element $\omega^2$ maps the O6-planes from the $\sigma_{\cal R}\omega^2$-and $\sigma_{\cal R}\omega^4$-sectors to this one.
In a coordinate patch where $v_i\neq0$, we can use the $(\C^\star)^3$-scaling symmetry of the ambient space to set $v_i=1$, such that the O6-planes form a three-dimensional subspace spanned by $\{{\rm Re}(x_1), {\rm Re}(x_2), {\rm Re}(x_3)\}$ within the complex three-dimensional space parametrised by the coordinates $\{ x_1, x_2, x_3\}$. 
Furthermore, this three-dimensional subspace corresponds to the region $y^2(x_i)\geq 0$, implying that the O6-planes are calibrated w.r.t. ${\rm Re}(\Omega_3)$,  the real part of the holomorphic three-form $\Omega_3$ defined in equation~(\ref{Eq:Holo3FormZ2Z2}). This identification of the O6-planes as a real three-dimensional subspace of the $x_i$-planes matches the geometric description in terms of the torus wrapping numbers provided by table~\ref{tab:O6PlanesZ2Z6} and will allow us to verify which {\it sLag} three-cycles are calibrated w.r.t. the same holomorphic three-form and therefore preserve the same ${\cal N}=1$ supersymmetry.

When it comes to the {\it sLag} three-cycles, we should first offer a clear dictionary between the three types of three-cycles defined in section~\ref{Ss:RemZ2Z6} and three-dimensional subspaces on the $x_i$-planes. In order to construct (factorisable) three-cycles on the $x_i$-planes, we can consider the product $N_1\otimes M_2 \otimes M_3$ consisting of {\it Lag} lines on each of the two-tori, with $N_1$ one of the {\it Lag} lines from table \ref{tab:LagSquareT2} and $M_2, M_3$ {\it Lag} lines from table~\ref{tab:LagHexaT2} (or displacements thereof). A necessary condition for the product $N_1\otimes M_2 \otimes M_3$ to be a supersymmetric three-cycle is that their relative angles w.r.t. the~O6-planes add up to 0 modulo $2\pi$.\footnote{The relative angles w.r.t.~the~$\OR$- invariant plane can be inferred from the torus wrapping numbers: $\phi^{(1)} = \arctan\left( \frac{m^1}{n^1} \right) $ determines the angle (mod $\pi$) on an {\bf A}-type square $T_{(1)}^2$ and $\phi^{(j)} = \arctan\left( \frac{\sqrt{3} m^j}{2 n^j + m^j}\right)$ on an
{\bf A}-type hexagonal $T_{(j)}^2$ with $j=2,3$, and $\sgn(n^i)$ encodes the orientation to arrive at (mod $2\pi$).} The factorisable bulk three-cycles of section~\ref{Ss:RemZ2Z6} are then constructed from {\it Lag} lines which do not cross the $\Z_2^{(i)}$ fixed points, such as lines {\bf cI} and {\bf cII} on $T_{(1)}^2$ and continuous displacements of {\bf bX}$^{0,\pm}$ with ${\bf X}\in \{{\bf I},{\bf II}, {\bf III}, {\bf IV}\}$. This type of {\it Lag} lines forms curves (circles) on each $x_i$-plane separately, as for instance shown in figure~\ref{Fig:TorusCyclesxplane}, indicating that a bulk three-cycle has typically the topology of a three-torus $T^3$. Only bulk three-cycles that lie sufficiently close to a deformed $\Z_2^{(i)}$ singularity will experience alterations to their overall three-dimensional volume, yet they will always keep their {\it sLag} property. This vanilla-like behaviour of bulk three-cycles on deformed orbifolds suggests us to  dwell on them no longer than necessary, but rather to focus on the two other types of three-cycles.

As a matter of fact, the interesting phenomena occur for exceptional and fractional three-cycles passing through deformed singularities. At the orbifold point, there is no possibility to express the exceptional divisors $e^{(i)}_{\alpha \beta}$ in terms of the homogeneous coordinates $x_i$, as their volumes are shrunk to zero. Hence, we relegate the discussion of the exceptional three-cycles to the next two subsections, where we will discuss in detail the geometry of exceptional divisors located at deformed $\Z_2^{(i)}$ fixed points for the various distinct deformation parameters. In the meantime, we develop a dictionary for the fractional three-cycles associated to the $\Z_2^{(i)}$-twisted sectors at the orbifold point and assume that for only one twisted sector at a time the deformations will be turned on. In that case, we can describe the fractional three-cycles for the $\Z_2^{(i)}$-twisted sector as a direct product of a one-cycle on $T_{(i)}^2$ and a two-cycle on $T_{(i)}^4/\Z_2^{(i)}$ (or on ${\rm Def}\left(T_{(i)}^4/\Z_2^{(i)}\right)$ in later sections). If we limit ourselves to the cycles {\bf aI}, {\bf aII}, {\bf aIII} and {\bf aIV} on $T_{(1)}^2$, the total number of two-cycles on $T_{(2)}^2\times T_{(3)}^2$ based on table~\ref{tab:LagHexaT2} is $12^2=144$. The sum of the relative angles for these two-cycles w.r.t.~the~O6-planes adds up to $\{0,\pm\frac{\pi}{6},\pm\frac{\pi}{3},\frac{\pi}{2}\}$, suggesting six different calibration angles.
The 24  two-cycles with calibration angle 0  combine with the one-cycles {\bf aI} or {\bf aIII} to form three-cycles calibrated w.r.t.~the~same three-form $\Omega_3$ as the O6-planes, while 
the 24  two-cycles with calibration angle $\frac{\pi}{2}$ combine with one-cycles {\bf aII} or {\bf aIV} to form supersymmetric three-cycles. This leads to 96 three-cycles which are further reduced to 32 
independent ones as a result of the identification of the two-cycles on $T^4_{(1)} \equiv T_{(2)}^2\times T_{(3)}^2$ under the $\Z_3$-action. Taking into account the possibility of turning on discrete Wilson lines or discrete $\Z_2^{(i)}$-eigenvalues at the singularities offers a large enough class of fractional three-cycles to construct a variety of phenomenologically interesting
global intersecting D6-brane models~\cite{Ecker:2014hma,Ecker:2015vea}.

If we look closer at the fractional three-cycles constructed from {\bf aI}, {\bf aII}, {\bf aIII}, {\bf aIV}, {\bf bI$^0$} and {\bf bII$^0$}, we notice that these one-cycles all lie along the real axis ${\rm Re}(x_i)$ in the complex $x_i$-planes in figure~\ref{Fig:TorusCyclesxplane} and have $\Z_2$ fixed points as boundaries. For the {\bf aI}, {\bf aII}, {\bf bI$^0$} and {\bf bII$^0$} cycles one should imagine the point $\epsilon_1$ 
at infinity $x_i= \pm\infty$ as the second boundary, representing the singularity at the origin of the two-tori $T^2_{(i)}$. For a fractional three-cycle associated to e.g.~the $\Z_2^{(1)}$-twisted sector, the one-cycle on $T_{(1)}^2$ has a topology of a circle $S^1$, while the two-cycle on $T^4_{(1)}/\Z_2^{(1)}$ corresponds to a two-torus pinched down at the boundary points, i.e.~at the $\Z_2^{(1)}$ fixed points lying on the zero locus $y=0$. Hence, the topology of a fractional three-cycle is simply $S^1\times T^2/\Z_2$. A more pictorial representation will be given in
sections~\ref{Sss:sLagsZ21} and~\ref{Sss:sLagsZ23}. For these kinds of fractional three-cycles, the holomorphic three-form $\Omega_3$ factorises as $\Omega_3 = \Omega_2 \wedge \Omega_1$, such that we can compute the integrals of $\Omega_2$ over the two-cycles on $T^4_{(i)}/\Z_2^{(i)}$ separately from the integrals of $\Omega_1$ over the one-cycles on $T_{(i)}^2$.

In the next two subsections we discuss the effects of deformations in the $\Z_2^{(1)}$- and $\Z_2^{(2,3)}$-twisted sectors on the exceptional and fractional three-cycles and investigate how their volumes increase or decrease due to the deformation. Due to the exchange symmetry $T_{(2)}^2\leftrightarrow T_{(3)}^2$ it suffices to discuss only one of the $\Z_2^{(2,3)}$-twisted sectors, as the other one will yield the same results. Hence, we can choose to focus on deformations in the $\Z_2^{(3)}$-twisted sector.

\subsubsection{\textbf{\textit{sLag}s} in the deformed $\Z_2^{(1)}$-twisted sector}\label{Sss:sLagsZ21}
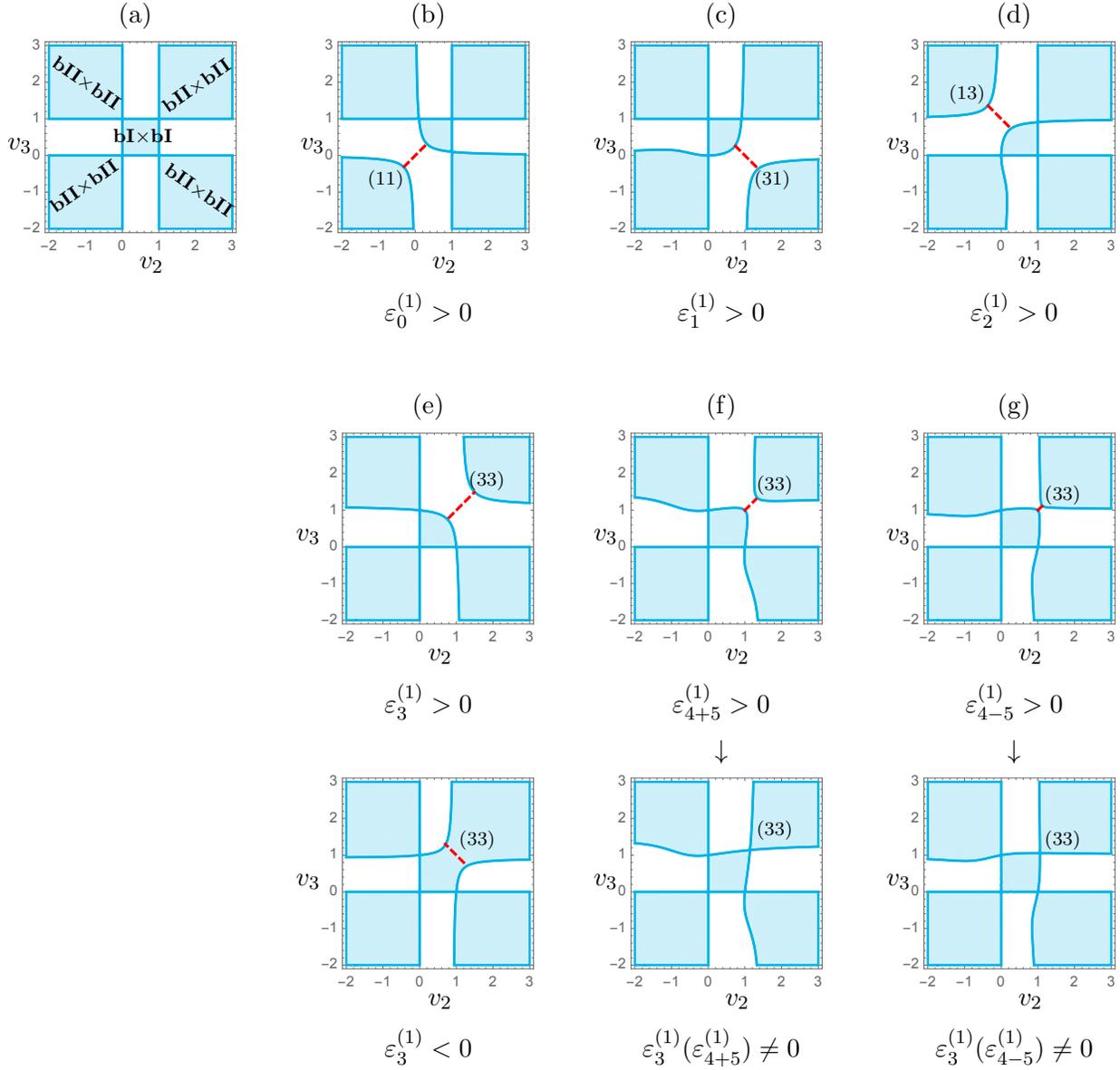
\begin{figure}[h]
 \begin{center}
\hspace*{-0.3in} \begin{tabular}{c@{\hspace{0.4in}}c@{\hspace{0.4in}}c@{\hspace{0.4in}}c}
 (a) & (b) & (c) & (d) \\
  \vspace{0.1in}\includegraphics[scale=0.6]{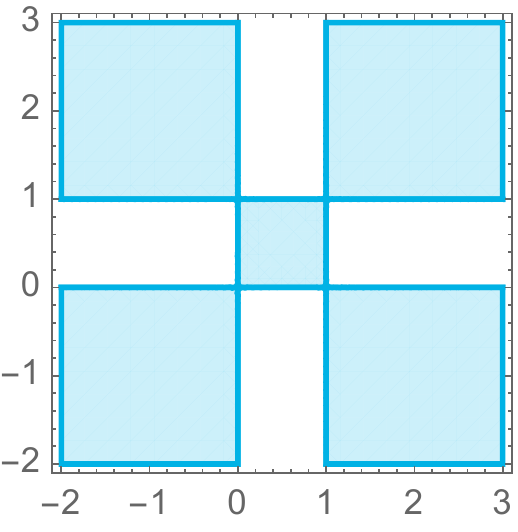} \begin{picture}(0,0)  \put(-45,-5){$v_2$}  \put(-100,45){$v_3$} \put(-56,48){\scriptsize \bf bI$\times$bI} \put(-33,60){\begin{rotate}{35}\scriptsize \bf bII$\times$bII\end{rotate}}  \put(-35,35){\begin{rotate}{-35}\scriptsize \bf bII$\times$bII\end{rotate}} \put(-80,18){\begin{rotate}{35}\scriptsize \bf bII$\times$bII \end{rotate}} \put(-82,80){\begin{rotate}{-35}\scriptsize \bf bII$\times$bII\end{rotate}} \end{picture} &  \includegraphics[scale=0.6]{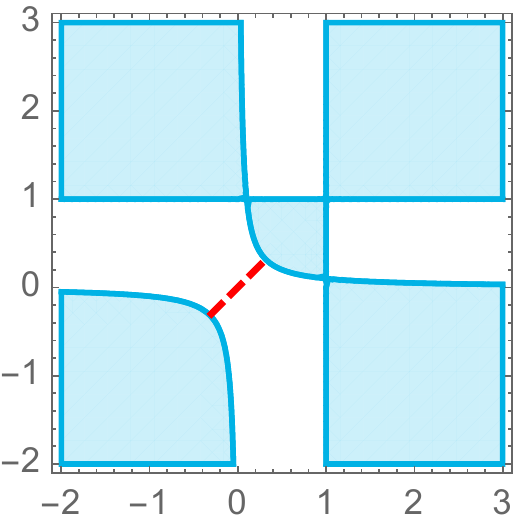} \begin{picture}(0,0)  \put(-45,-5){$v_2$}  \put(-100,45){$v_3$} \put(-72,30){\scriptsize (11)}  \end{picture}  &  \includegraphics[scale=0.6]{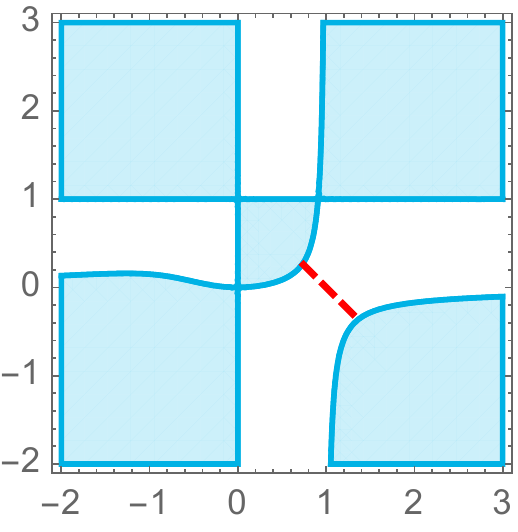} \begin{picture}(0,0)  \put(-45,-5){$v_2$}  \put(-100,45){$v_3$} \put(-33,30){\scriptsize (31)}   \end{picture}   &  \includegraphics[scale=0.6]{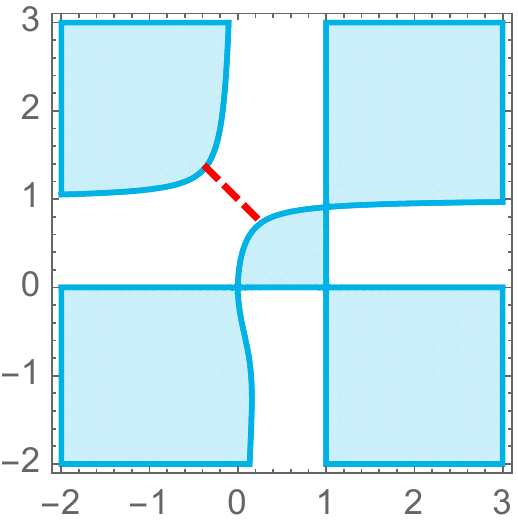}  \begin{picture}(0,0)  \put(-45,-5){$v_2$}  \put(-100,45){$v_3$}  \put(-74,66){\scriptsize (13)}    \end{picture}  \\
 &  $\varepsilon_0^{(1)}> 0$ & $\varepsilon_1^{(1)}> 0$ &  $\varepsilon_2^{(1)}> 0$  \\
 &  & &    
 \\
 & (e) & (f) & (g) \\
  \vspace{0.1in}&   \includegraphics[scale=0.6]{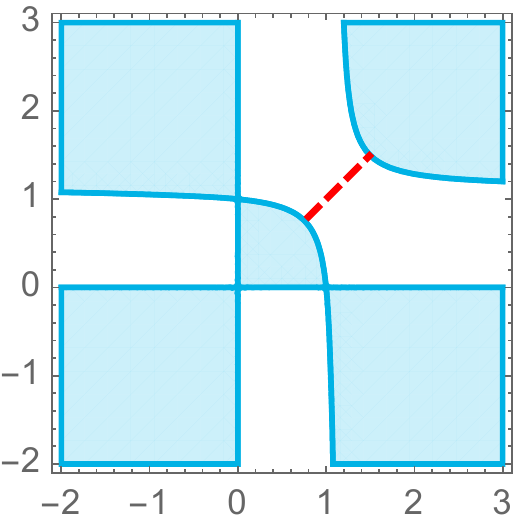}\begin{picture}(0,0)  \put(-45,-5){$v_2$}  \put(-100,45){$v_3$} \put(-28,68){\scriptsize (33)}   \end{picture} &  \includegraphics[scale=0.6]{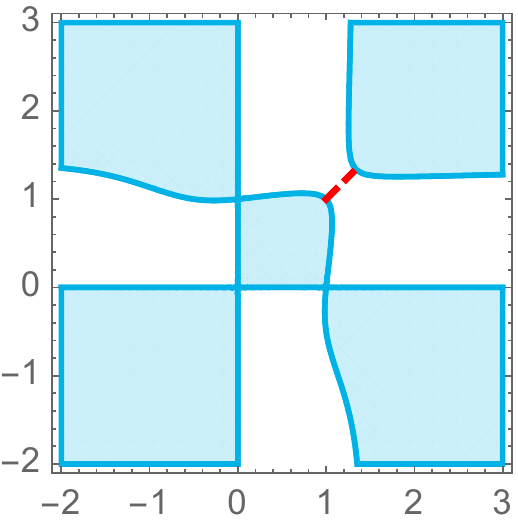} \begin{picture}(0,0)  \put(-45,-5){$v_2$}  \put(-100,45){$v_3$} \put(-32,66){\scriptsize (33)}  \end{picture}   &  \includegraphics[scale=0.6]{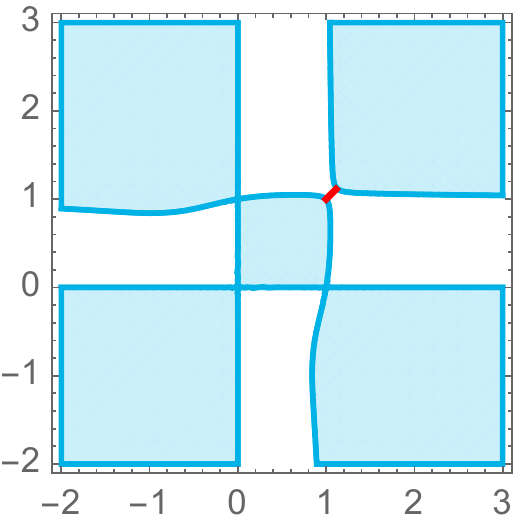} \begin{picture}(0,0)  \put(-45,-5){$v_2$}  \put(-100,45){$v_3$}  \put(-34,62){\scriptsize (33)}\end{picture}  \\
 &  $\varepsilon_3^{(1)}> 0$&  $\varepsilon_{4+5}^{(1)}> 0$&  $\varepsilon_{4-5}^{(1)}> 0$\\
 & & $\downarrow$  & $\downarrow$  \\
  \vspace{0.1in}  & \includegraphics[scale=0.6]{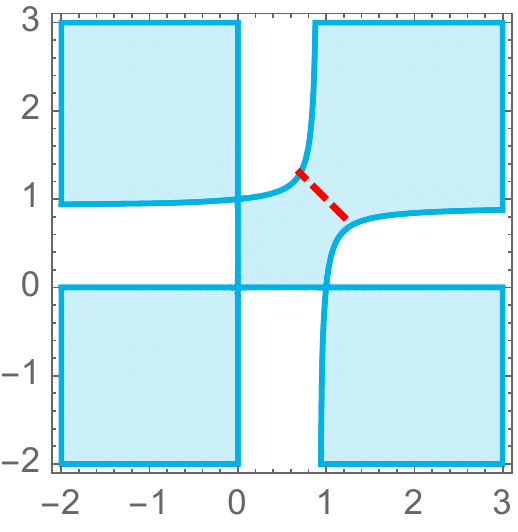}\begin{picture}(0,0)  \put(-45,-5){$v_2$}  \put(-100,45){$v_3$} \put(-32,62){\scriptsize (33)}   \end{picture} & \includegraphics[scale=0.6]{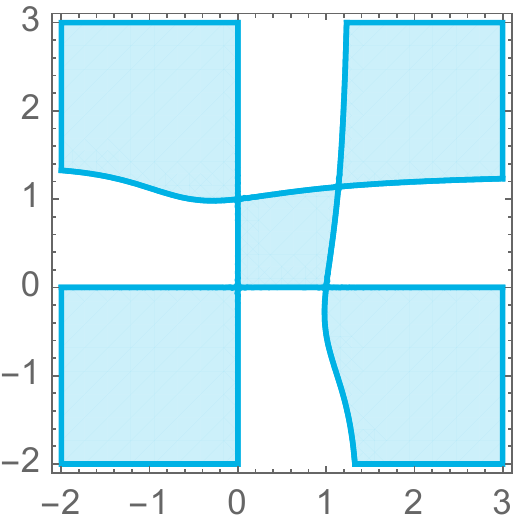} \begin{picture}(0,0)  \put(-45,-5){$v_2$}  \put(-100,45){$v_3$} \put(-32,66){\scriptsize (33)}  \end{picture}   &  \includegraphics[scale=0.6]{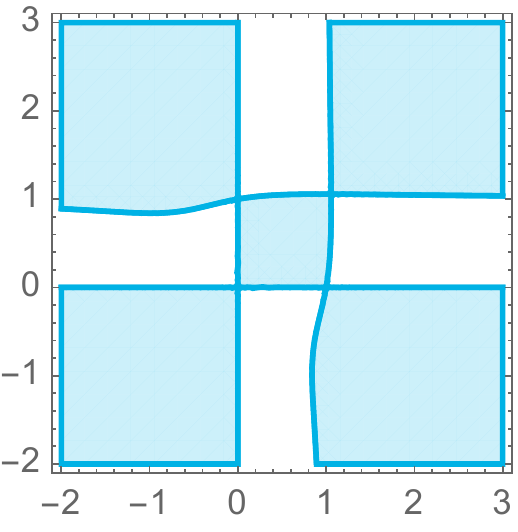} \begin{picture}(0,0)  \put(-45,-5){$v_2$}  \put(-100,45){$v_3$}  \put(-34,62){\scriptsize (33)}\end{picture} \\
 & $\varepsilon_3^{(1)}< 0$  & $\varepsilon_3^{(1)}(\varepsilon_{4+5}^{(1)}) \neq 0$ & $\varepsilon_3^{(1)}(\varepsilon_{4-5}^{(1)})\neq 0$
 \end{tabular}
 \caption{Hypersurface equation (\ref{Eq:HyperZ2Z6Full}) in the ($\R$-projected) $(v_2,v_3)$ plane for various $\Z_2^{(1)}$ deformation parameters: all deformations are set to zero in (a); $\varepsilon_0^{(1)} = 0.1$ deforms the singularity (11) in (b); $\varepsilon_1^{(1)} = 0.1$ deforms the singularity (31) in (c); $\varepsilon_2^{(1)} = 0.1$ deforms the singularity (13) in (d); $\varepsilon_3^{(1)} = 0.1$ or $\varepsilon_3^{(1)} = - 0.1$ deforms the singularity (33) in (e); $\varepsilon_{4+5}^{(1)} = 0.2$ deforms the singularity (33) in (f); $\varepsilon_{4-5}^{(1)} = 0.2$ deforms the singularity (33) in (g). 
The fact that the singularity (33) is deformed for non-vanishing deformation parameters $\varepsilon_{4\pm5}^{(1)}$ requires the introduction of a correction-term $\varepsilon_{3}^{(1)}$ as a function of $\varepsilon_{4\pm5}^{(1)}$ restoring the singular nature of the point (33) as shown in the lower parts of figures (f) and~(g).
 The contour lines correspond to zeros of $y$, while the blue-coloured regions represent {\it sLag} two-cycles calibrated w.r.t. ${\rm Re}(\Omega_2)$. The exceptional cycles with non-zero volume arising at the deformed singularities are indicated by a red dashed line.  
As this figure only depicts the $\R$-restricted subspace of the $(v_2,v_3)$ plane, we can only depict the behaviour of the singularities located on the real axis, i.e.~the points 1 ($v_i=0$) and 3 ($v_i=1$) on $T^2_{(i)}$. The $\Z_2$ fixed points 2 and 4 are complex roots and therefore do not show up in the plots. As explained in the main text, their behaviour under deformations can be depicted after performing a $\Z_3$-rotation on the homogeneous coordinates.
 \label{Fig:DefZ21hypersurfaceEq}}
 \end{center}
 \end{figure}

For a qualitative appreciation of the deformation effects in the $\Z_2^{(1)}$-twisted sector, we switch to the $x_i=1$ patch and describe the {\it sLag}s in terms of the homogeneous coordinates $v_i$. 
 The cycles {\bf bI$^0$} and {\bf bII$^0$} are still given by real hypersurfaces at the orbifold point in terms of the homogeneous coordinates $v_{i=2,3}$, though in comparison to figure~\ref{Fig:TorusCyclesxplane} their regional conditions are changed: for the {\it Lag} line {\bf bI}$^0$ we find the constraint $0 \leq v_i\leq 1$, while the {\it Lag} line {\bf bII}$^0$ consists of the union $\{-\infty \leq v_i\leq 0\} \cup \{ 1\leq v_i \leq + \infty \}$. Figure~\ref{Fig:DefZ21hypersurfaceEq} 
depicts the $T^2/\Z_2$ topology for the two-cycles constructed from {\bf bI$^0$} and {\bf bII$^0$} on ${\rm Def}\left(T_{(1)}^4/\Z_2^{(1)}\right)$, with the blue-shaded regions representing {\it sLag} two-cycles calibrated with respect to ${\rm Re}(\Omega_2)$. The blue contour-lines correspond to the zero locus $y=0$ in the $\R$-projected plane $(v_2, v_3)$, and these lines intersect at the $\Z_2^{(1)}$ fixed points (11), (13), (31) and (33). 
 The fact that we are able to depict graphically the behaviour of the aforementioned singularities follows immediately from the choice of the coordinate patch $x_i = 1$. Other singularities correspond to complex roots and therefore do not lie in the $\R$-restricted plane $(v_2, v_3)$, such that they are not depicted in figure~\ref{Fig:DefZ21hypersurfaceEq}.
The two-cycles {\bf bI}$^0 \times${\bf bI}$^0$ and {\bf bII}$^0 \times${\bf bII}$^0$ should be paired with a one-cycle {\bf aI} or {\bf aIII} on $T_{(1)}^2$ to form a {\it sLag} three-cycle calibrated with respect to ${\rm Re}(\Omega_3)$. 
The white regions in figure~\ref{Fig:DefZ21hypersurfaceEq} on the other hand represent {\it sLag} two-cycles calibrated with respect to ${\rm Im}(\Omega_2)$, namely the two-cycles {\bf bI}$^0 \times${\bf bII}$^0$ and {\bf bII}$^0 \times${\bf bI}$^0$. 
\newline
Anticipating the examples later on, the hidden stacks $h_{1,2}$ in the Left-Right (L-R) symmetric model I of section~\ref{Ss:ExLRSM} belong to the three-cycle type
{\bf aI}$\times${\bf bI}$^0 \times${\bf bI}$^0$, while the hidden stacks of the Pati-Salam (PS) II model of section~\ref{Ss:ExPS} and of the L-R symmetric II model of section~\ref{Ss:ExLRSM} are of the type {\bf aI}$\times${\bf bII}$^0 \times${\bf bII}$^0$, and the hidden stacks of the L-R symmetric IIb model are of the type {\bf aIII}$\times${\bf bII}$^0 \times${\bf bII}$^0$.

By turning on each $\Z_2^{(1)}$ deformation parameter $\varepsilon^{(1)}_\lambda$ separately in figure~\ref{Fig:DefZ21hypersurfaceEq}, we can explicitly see which singularities are deformed and which singularities are displaced by the respective deformation parameter. This information is also summarised in the upper part of table~\ref{tab:OverviewDeformperZ2Sector}, which results from determining the singular points of the hypersurface equation (\ref{Eq:HyperZ2Z6Full}). At the deformed singularity, an exceptional cycle with non-vanishing volume emerges, which is indicated by a red dashed line in figure~\ref{Fig:DefZ21hypersurfaceEq}. An interesting observation is that the point (33) gets deformed too when turning on the deformation parameters $\varepsilon_{4+5}^{(1)}$ and $\varepsilon_{4-5}^{(1)}$. This phenomenon was also observed~\cite{Blaszczyk:2015oia} for specific deformations of complex co-dimension 2 singularities on the subspace $T^4_{(i)}/\Z_2$ of the orbifold $T^6/(\Z_2\times \Z_6')$ with discrete torsion and can be resolved by turning on a correction-term $\varepsilon_3^{(1)}$ depending on the respective deformation parameter, as depicted in the lower diagrams of figures~\ref{Fig:DefZ21hypersurfaceEq} (f) and (g). A similar consideration holds for the fixed points (22) and (44) which are also deformed, but now for a non-vanishing deformation parameter $\varepsilon_3^{(1)}\neq 0$. Further details about the counter-terms can be found in appendix~\ref{A:CorrectionTerms}.      
\clearpage
\mathtabfix{
\begin{array}{|c|c||c|c|}
\hline 
\multicolumn{4}{|c|}{\text{\bf Behaviour of $\Z_2^{(i)}$ fixed points under deformations of $T^6/(\Z_2\times \Z_6 \times \OR)$ with discrete torsion}}\\
\hline \hline
\multicolumn{4}{|c|}{\text{\bf Deformations in the $\Z_2^{(1)}$-twisted sector}}\\
\hline \hline
\text{modulus} & \text{deformed singular orbit} & \text{additional deformed orbit}  & \text{displaced orbit}   \\
\hline 
 \varepsilon_0^{(1)}&  e_0^{(1)} \equiv  (11) &  & e_1^{(1)} , e_2^{(1)}  \\
 \varepsilon_1^{(1)}&e_1^{(1)} \equiv   [(31),(41),(21)] & &  e_3^{(1)}, e_4^{(1)}, e_5^{(1)}    \\ 
 \varepsilon_2^{(1)}&  e_2^{(1)} \equiv  [(13),(12),(14)] &   &e_3^{(1)}, e_4^{(1)}, e_5^{(1)}  \\ 
 \varepsilon_3^{(1)}   & e_3^{(1)} \equiv  [(33),(42),(24)] & e_4^{(1)},   e_5^{(1)}  & \\ 
 \varepsilon_{4+5}^{(1)}\, \& \,   \varepsilon_{4-5}^{(1)}  & e_4^{(1)} \equiv [(22),(34),(43)], e_5^{(1)}  \equiv  [(44),(32),(23)]      &e_3^{(1)}   & \\
  \hline \hline 
\multicolumn{4}{|c|}{\text{\bf Deformations in the $\Z_2^{(3)}$-twisted sector}}\\
\hline \hline
   \varepsilon_1^{(3)} & e_1^{(3)} \equiv  [(13),(14),(12)] &  & e_2^{(3)}, e_4^{(3)}  \\
    \varepsilon_2^{(3)} & e_2^{(3)} \equiv [(23),(24),(22)]&  & e_1^{(3)}, e_3^{(3)}   \\
     \varepsilon_3^{(3)} & e_3^{(3)} \equiv  [(33),(34),(32)] &  & e_2^{(3)}, e_4^{(3)}   \\
     \varepsilon_4^{(3)} & e_4^{(3)} \equiv [(43),(44),(42)] & & e_1^{(3)}, e_3^{(3)}  \\
     \hline
 \end{array}
}{OverviewDeformperZ2Sector}{Overview of the deformed singular $\Z_3$-orbits composed of the $\Z_2^{(i)}$ fixed points of \mbox{$T^6/(\Z_2\times \Z_6 \times \OR)$} with discrete torsion per deformation parameter $\varepsilon_{\lambda=0,\ldots,4\pm 5}^{(1)}$ and $\varepsilon_{\alpha=1,\ldots,4}^{(3)}$. The third column indicates which other singular orbits are deformed, while the last column lists which orbits remain singular but are displaced.
Figure~\ref{Fig:DefZ21hypersurfaceEq} provides a graphical representation of how the $\Z_2^{(1)}$ singularities (11), (13), (31) and (33) on $T_{(1)}^4/\Z_2^{(1)}$ behave under deformations. The other singular points belonging to the respective $\Z_3$-invariant orbits are not depicted as they do not lie in the $\R$-restricted plane. To depict these latter singularities, one has to perform an appropriate $\Z_3$-rotation on the coordinates. Similar considerations hold for the $\Z_2^{(3)}$ singularities depicted in figure~\ref{Fig:DefZ23hypersurfaceEq}.
}

For a more quantitative picture of the deformation effects, we go back to the coordinate patch in which $v_i=1$ 
where we can rewrite the hypersurface equation in the vicinity of a singular point on $T_{(1)}^4/\Z_2^{(1)}$ as a $\C^2/\Z_2$ (or $A_1$-type) singularity.
In first instance, we look at the zero locus of the hypersurface equation (\ref{Eq:HyperZ2Z6Full}), turn on each $\Z_2^{(1)}$ deformation separately and discuss their effects in a local patch around the singular point (33). This {\bf local} description can be extracted straightfordwardly for the deformation $\varepsilon_{3}^{(1)}$, which deforms the exceptional cycles ${\boldsymbol \epsilon}_{3}^{(1)}$ and $\tilde {\boldsymbol \epsilon}_{3}^{(1)}$ at the singularity (33). For the other deformations  ($\varepsilon_0^{(1)}, \varepsilon_1^{(1)}, \varepsilon_2^{(1)}$), we have to perform a M\"obius transformation as explained in appendix~\ref{A:MT}, or a complex rotation ($\varepsilon_{4+5}^{(1)}, \varepsilon_{4-5}^{(1)}$) to extract the proper local structure by placing the singularity at the point (33). 
More explicitly, there exists a M\"obius transformation $\lambda_3$~\cite{Blaszczyk:2015oia} acting on the homogeneous coordinate $x_i$ that allows to map the $\Z_2$ fixed point $\epsilon_1$ situated at the origin of a two-torus $T^2_{(i)}$ to the fixed point $\epsilon_3$ located on the real axis in the new coordinate $\tilde x_i$, such that a singular point $(\alpha \beta)$ with either $\alpha=1$ and/or $\beta=1$ can always be mapped by $\lambda_3$ to the point (33) in the new coordinates. The $\Z_2$ fixed points $\epsilon_2$ and $\epsilon_4$ on the other hand are mapped to the point $\epsilon_3$ by a $\Z_3$-transformation as can be seen in figure~\ref{Fig:TorusCyclesxplane} (b). 
With an appropriate rescaling of the homogeneous coordinates, we then find that the singularity is locally described by the following hypersurface equation:
\begin{equation}\label{Eq:HyperLocalZ21}
\tilde y^2 = \tilde x_2 \tilde x_ 3 - c_\lambda \varepsilon_{\lambda}^{(1)}, \qquad \qquad  c_a = \left\{ \begin{array}{cl}  9 & (\lambda = 0),  \\ 3 & (\lambda = 1,2),   \\ 1 & (\lambda = 3), \\  \frac{1}{2} & (\lambda = 4\pm5).   
   \end{array}\right.
\end{equation}
A first observation is of course that the co-dimension two singularity locally takes the form of a $\C^2/\Z_2$ singularity in the $(\tilde x_2, \tilde x_3)$-plane. The two-cycles {\bf bI}$^0 \times${\bf bI}$^0$ and  {\bf bII}$^0 \times${\bf bII}$^0$, with torus wrapping numbers $(n^2,m^2;n^3,m^3)=(1,0;1,0)$ and $(-1,2;1,-2)$ on $T_{(2)}^{2} \times T_{(3)}^2$, respectively, pass through the $\Z_2^{(1)}$ fixed points affected by a non-vanishing deformation parameter $\varepsilon_{\lambda=0,1,2,3}^{(1)}$.  A full fractional three-cycle calibrated with respect to ${\rm Re}(\Omega_3)$ is then constructed as described in section~\ref{Ss:RemZ2Z6} by combining these two-cycles with e.g.~the one-cycle {\bf aI} on $T_{(1)}^2$, such that the $\Z_2^{(1)}$ exceptional  contributions to a fractional three-cycle can be expressed in terms of the basis three-cycles $({\boldsymbol \epsilon}_{\lambda}^{(1)}, \tilde{{\boldsymbol \epsilon}}_{\lambda}^{(1)}) \sim \text{orbits of }(\pi_1 \otimes e_\lambda^{(1)}, \pi_2 \otimes e_\lambda^{(1)})$ as follows: 
\begin{equation}
\Pi^{\Z_{2}^{(1)}} = (-)^{\tau^{\Z_2^{(1)}}} \left( {\boldsymbol \epsilon}_{0}^{(1)} + (-)^{\tau^2} {\boldsymbol \epsilon}_{1}^{(1)} + (-)^{\tau^3} {\boldsymbol \epsilon}_{2}^{(1)} +(-)^{\tau^2 + \tau^3} {\boldsymbol \epsilon}_{3}^{(1)}  \right). 
\end{equation}
Now, by switching on one of the associated four deformation parameters, an exceptional two-cycle $e^{(1)}_{\lambda=0,1,2,3}$ with non-vanishing volume grows out of the respective fixed point along $T^4_{(1)}/\Z_6$, and it clear from this construction that the volume of an associated fractional three-cycle is also influenced by the evolution of the volume of the exceptional cycle under deformation. At the orbifold point, the two-cycles {\bf bI}$^0 \times${\bf bI}$^0$ and  {\bf bII}$^0 \times${\bf bII}$^0$ along $T^4_{(1)}/\Z_6$ can be represented by a set of 
real two-dimensional regions $\tilde x_2 \cdot \tilde x_3 \geq 0$ 
in the $(\tilde x_2, \tilde x_3)$-plane. When turning on a deformation parameter $\varepsilon_{\lambda=0,1,2,3}^{(1)}$, the volume of these two-cycles will shrink or grow depending on the sign of the deformation parameter:
\begin{itemize}
\item\vspace{-0.1in} \underline{$\varepsilon_\lambda^{(1)} > 0:$} {\bf bI}$^0 \times${\bf bI}$^0$ and {\bf bII}$^0 \times${\bf bII}$^0$ are still two separate two-cycles, both with shrinking sizes as an exceptional two-cycle $e_{\lambda}^{(1)}$ grows out of the singularity (33) in the region $\tilde y^2 <0$. This situation is represented in the upper diagram of figure~\ref{Fig:DefZ21hypersurfaceEq} (b)-(c)-(d)-(e). 
The exceptional two-cycle satisfies the algebraic condition $\tilde x_2 = \ov{\tilde x_3}$, by which the hypersurface equation~(\ref{Eq:HyperLocalZ21}) reduces to the equation for a two-sphere $S^2$ with radius $\sqrt{c_\lambda \varepsilon_\lambda^{(1)}}$. As the two-cycles {\bf bI}$^0 \times${\bf bI}$^0$ and {\bf bII}$^0 \times${\bf bII}$^0$ remain two separate {\it sLag}s, the exceptional two-cycle $e_{\lambda}^{(1)}$ has to be calibrated with respect to the same two-form ${\rm Re}(\Omega_2)$. This statement can be shown explicitly by computing the (non-vanishing) volume $\int_{e_\lambda^{(1)}} {\rm Re}(\Omega_2)$ of the exceptional two-cycle as a function of the parameter $\varepsilon_\lambda^{(1)}$ in the hypersurface formalism and taking into account that $\tilde y^2 <0$ and $\tilde x_2 = \ov{\tilde x_3}$. Extending these considerations to the three-cycles on $(T^2_{(1)} \times T^4_{(1)}/\Z_6)/\Z_2^{(3)}$, 
we conclude that the exceptional three-cycle ${\boldsymbol \epsilon}_{\lambda}^{(1)}$ is calibrated with respect to the same three-form ${\rm Re}(\Omega_3)$ as the bulk three-cycles and that there should be a relative minus sign between the bulk three-cycle $\Pi^{\rm bulk}$ and $\Pi^{\Z_2^{(1)}}$ in order for the volume of the fractional cycle to decrease upon deformation of the singularity, in line with figure~\ref{Fig:DefZ21hypersurfaceEq} (e).

The cycles {\bf bI}$^0 \times${\bf bII}$^0$ and {\bf bII}$^0 \times${\bf bI}$^0$ on the other hand have merged into one big two-cycle and are no longer {\it sLag} two-cycles separately. As these two-cycles are calibrated with respect to ${\rm Im}(\Omega_2)$, we should take a union two-cycle {\bf bI}$^0 \times${\bf bII}$^0$$\oplus${\bf bII}$^0 \times${\bf bI}$^0$ from which the exceptional cycle $e_{\lambda}^{(1)}$ is eliminated, such that the union two-cycle remains {\it sLag} with respect to ${\rm Im}(\Omega_2)$.

\item \underline{$\varepsilon_\lambda^{(1)} < 0:$} {\bf bI}$^0 \times${\bf bI}$^0$ and {\bf bII}$^0 \times${\bf bII}$^0$ are no longer separate two-cycles but melt together as shown in the lower diagram of figure~\ref{Fig:DefZ21hypersurfaceEq} (e), while an exceptional two-cycle $e_{\lambda}^{(1)}$ grows out of the singularity (33) in the region $\tilde y^2 >0$. The hypersurface equation~(\ref{Eq:HyperLocalZ21}) reproduces the topology of a $S^2$ for the algebraic condition $\tilde x_2 = - \ov{\tilde x_3}$, which implies that the exceptional two-cycle $e_{\lambda}^{(1)}$ is now calibrated with respect to ${\rm Im}(\Omega_2)$. A union two-cycle {\bf bI}$^0 \times${\bf bI}$^0$$\oplus${\bf bII}$^0 \times${\bf bII}$^0$ from which the exceptional two-cycle $e_{\lambda}^{(1)}$ is eliminated, will then correspond to  one big {\it sLag} cycle calibrated with respect to ${\rm Re}(\Omega_2)$.

Once again, the two-cycles {\bf bI}$^0 \times${\bf bII}$^0$ and {\bf bII}$^0 \times${\bf bI}$^0$ both behave differently 
with respect to the two-cycles  {\bf bI}$^0 \times${\bf bI}$^0$ and {\bf bII}$^0 \times${\bf bII}$^0$
under the deformation as their sizes shrink for increasing $|\varepsilon_\lambda^{(1)}|$. Combining them with e.g.~the one-cycle {\bf aII} 
on $T_{(1)}^2$ allows for the construction of fractional three-cycles calibrated with respect to  ${\rm Re}(\Omega_3)$, and bulk three-cycles parallel to the $\OR\Z_2^{(2)}$-and $\OR\Z_2^{(3)}$-plane, respectively. Their $\Z_2^{(1)}$ exceptional contributions $\Pi^{\Z_2^{(1)}}$ can thus be decomposed in terms of the basis three-cycles $\tilde {\boldsymbol \epsilon}_{\lambda}^{(1)}$:
\begin{equation} \label{Eq:Pi-tildes-only}
\Pi^{\Z_{2}^{(1)}} = (-)^{\tau^{\Z_2^{(1)}}} \left( \tilde{\boldsymbol \epsilon}_{0}^{(1)} + (-)^{\tau^2} \tilde{\boldsymbol \epsilon}_{1}^{(1)} + (-)^{\tau^3} \tilde{\boldsymbol \epsilon}_{2}^{(1)} +(-)^{\tau^2 + \tau^3} \tilde{\boldsymbol \epsilon}_{3}^{(1)}  \right),
\end{equation}
implying that the basis three-cycles $\tilde{\boldsymbol \epsilon}_{\lambda=0,1,2,3}^{(1)}$ are calibrated with respect to ${\rm Re}(\Omega_3)$. 
As the volumes of these fractional three-cycles shrink for a non-vanishing deformation according to figure~\ref{Fig:DefZ21hypersurfaceEq} (e), there should be a relative minus sign between $\Pi^{\rm bulk}$ and the contribution to $\Pi^{\Z_2^{(1)}}$.  
For instance, for $\varepsilon^{(1)}_3<0$ equation~\eqref{Eq:Pi-tildes-only} describes a fractional three-cycle with $\tau^{\Z_2^{(1)}} + \tau^2 + \tau^3 = 1$ mod 2.
\end{itemize}
The situation for the deformations $\varepsilon_{4+5}^{(1)}$ and $\varepsilon_{4-5}^{(1)}$ is different, as they deform singularities through which the two-cycles {\bf bI}$^0 \times${\bf bIII}$^0$, {\bf bIII}$^0 \times${\bf bI}$^0$, {\bf bIII}$^0 \times${\bf bIII}$^0$, {\bf bII}$^0 \times${\bf bIV}$^0$, {\bf bIV}$^0 \times${\bf bII}$^0$ and {\bf bIV}$^0 \times${\bf bIV}$^0$  calibrated w.r.t.~${\rm Re}(\Omega_2)$
pass. As such the singular point (33) should not be deformed for (at least) a (small) non-vanishing deformation $\varepsilon_{4+5}^{(1)}$ or $\varepsilon_{4-5}^{(1)}$, which explains the required non-vanishing correction term $\varepsilon_3^{(1)}\big(\varepsilon_{4+5}^{(1)}\big)$ or $\varepsilon_3^{(1)}\big(\varepsilon_{4-5}^{(1)}\big)$, respectively, as depicted in the lower diagrams of figures~\ref{Fig:DefZ21hypersurfaceEq} (f) and (g). A brief discussion on how to obtain these corrections terms is given in appendix~\ref{A:CorrectionTerms}. 
The (local) discussion of the singularities deformed by a non-vanishing parameter $\varepsilon_{4+5}^{(1)}$ and $\varepsilon_{4-5}^{(1)}$ follows the same pattern as the one conducted above for the other deformations $\varepsilon^{(1)}_{\lambda \in \{0,1,2,3\}}$. Nonetheless, there is an important 
difference, as the exceptional three-cycles $\big({\boldsymbol \epsilon}^{(1)}_4, \tilde{\boldsymbol \epsilon}^{(1)}_4 \big)$ are not mapped to (linear combinations of) themselves under the orientifold projection, but to 
(linear combinations of) the three-cycles $\big({\boldsymbol \epsilon}^{(1)}_5, \tilde{\boldsymbol \epsilon}^{(1)}_5 \big)$ and vice versa, as indicated in table~\ref{tab:Z2Z6OrientifoldExceptionalCycles}. This implies that,
within the context of Type IIA/$\OR$ orientifolds,  the $\Z_3$-invariant orbits $e_4^{(1)}$ and $e_5^{(1)}$ from table \ref{tab:OverviewDeformperZ2Sector} are always deformed simultaneously for a single non-zero deformation $\varepsilon^{(1)}_{4+5}$ or $\varepsilon^{(1)}_{4-5}$.

\begin{figure}[h]
\begin{center}
\begin{tabular}{c@{\hspace{0.6in}}c@{\hspace{0.6in}}c}
\includegraphics[scale=0.5]{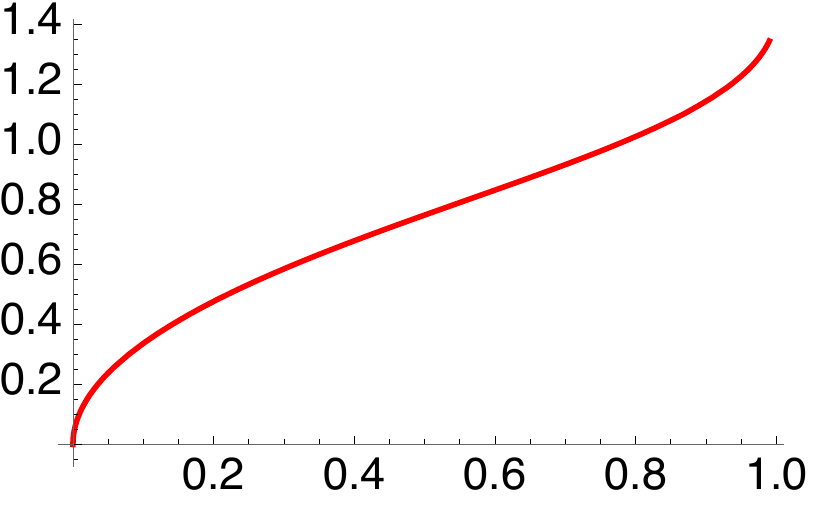} \begin{picture}(0,0) \put(0,0){$\varepsilon_0^{(1)}$} \put(-130,20){\begin{rotate}{90}Vol$_{\text{norm}}(e_0)$\end{rotate}} \end{picture} & \includegraphics[scale=0.52]{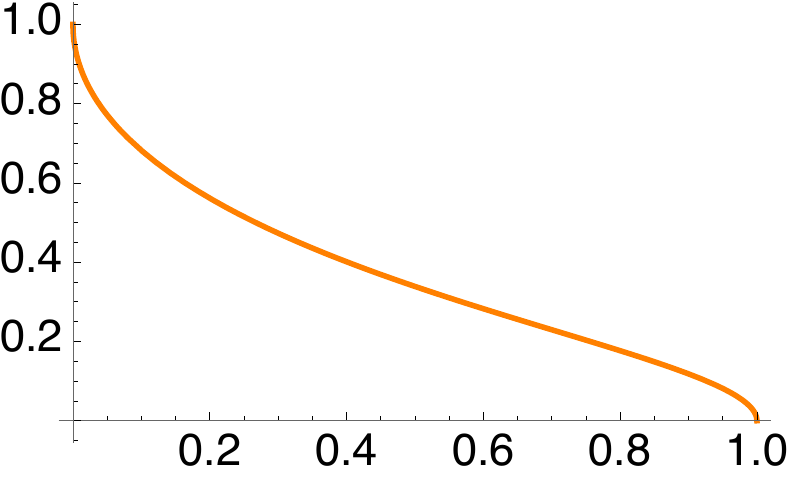} \begin{picture}(0,0) \put(0,0){$\varepsilon_0^{(1)}$} \put(-130,10){\begin{rotate}{90}Vol$_{\rm norm}(\Pi^{\text{frac}}_-)$\end{rotate}} \end{picture} & \includegraphics[scale=0.54]{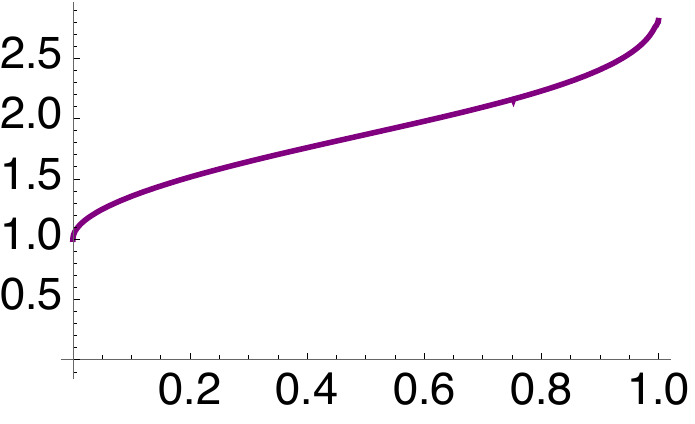} \begin{picture}(0,0) \put(0,0){$\varepsilon_0^{(1)}$} \put(-125,10){\begin{rotate}{90}Vol$_{\rm norm}(\Pi^{\text{frac}}_+)$\end{rotate}} \end{picture}
\end{tabular}
\caption{Normalised volume of the exceptional cycle $e^{(1)}_0$ on $T^4_{(1)}/\Z_6$ (left) and the fractional cycles
 $\Pi^{\text{frac}}_{\pm} =\frac{ \Pi^{\text{bulk}} \pm {\boldsymbol \epsilon}^{(1)}_0}{2}$ on $T^2_{(1)} \times T^4_{(1)}/\Z_6$ (middle and right) 
as a function of the deformation parameter~$\varepsilon^{(1)}_0$, with the representant of the bulk cycle $\Pi^{\text{bulk}}$ chosen as the cycle {\bf bIII}$^0\times${\bf bIII}$^0$ along $T^4_{(1)}/\Z_6$.\label{Fig:DeformedCycleVolumeZ21e0}}
\end{center}
\end{figure}
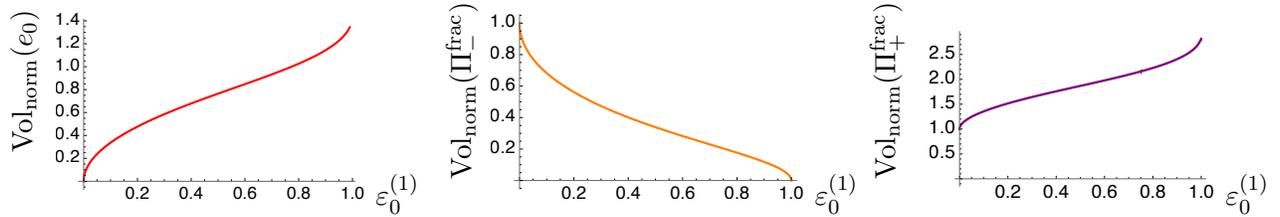

Exposing the {\bf global} behaviour of the exceptional three-cycle volumes for each deformation separately requires us to impose the algebraic condition $x_2 = \pm \ov x_3$ on the full hypersurface equation (\ref{Eq:HyperZ2Z6Full}) and to extract a real hypersurface equation allowing for a geometric description of an exceptional three-cycle in the hypersurface formalism. Let us work this out explicitly for three-cycles with a bulk orbit parallel to {\bf bI}$^0\times${\bf bI}$^0$ on $T_{(1)}^4$. Consistency with the plots in figure~\ref{Fig:DefZ21hypersurfaceEq} indicates that exceptional three-cycles calibrated w.r.t.~Re$(\Omega_3)$ are subject to the constraint $y^2 \big({\rm Re}(x_1),{\rm Re}(x_2),{\rm Im}(x_2),\varepsilon_\lambda^{(1)}\big)\leq 0$, which allows to define the integration domain for the volume of the respective exceptional three-cycles. Consider first the deformation $\varepsilon_0^{(1)}$ of the complex co-dimension 2 singularity at the origin $(11)$ on $T_{(1)}^4/\Z_6$, which can be placed along the real axes in the $(x_2,x_3)$-planes by virtue of the M\"obius transformation $\lambda_3$. Imposing subsequently the algebraic condition  $x_2 =\ov x_3$ yields a real hypersurface equation reminiscent of a $\Z_2$-singularity on $T^4/\Z_2$, implying that the $\Z_3$-action does not influence the geometrical properties of this exceptional three-cycle. Depicting the volume of the exceptional cycle $e_0^{(1)}$ as a function of the deformation parameter $\varepsilon_0^{(1)}$ fully confirms this statement, as can be seen explicitly from the left plot of figure~\ref{Fig:DeformedCycleVolumeZ21e0}. For small deformations, the exceptional cycle volume exhibits a square-root like dependence on $\varepsilon_0^{(1)}$, characteristic for deformed exceptional two-cycles on $\C^2/\Z_2$. For larger values of the deformation parameter, the exceptional cycle volume goes over into a more linear-like behaviour, before it evolves into a quadratic dependence for very large values of $\varepsilon_0^{(1)}$, enforced by the topology of the ambient $T^4$.\footnote{The volumes of the exceptional cycle and the fractional three-cycles are normalised to the volume of the fractional cycle at the orbifold point, i.e. Vol$(\Pi^{\rm frac})=1$ for vanishing deformation $\varepsilon_{\lambda}^{(1)}$ with \mbox{$\lambda \in\{0,1,2,3,4 \pm 5\}$}, throughout the paper. In this section we compute the volumes for the fractional cycles with bulk orbit parallel to {\bf aI}$\times${\bf bI}$^0\times${\bf bI}$^0$, such that the integration contours lie completely along the real lines ${\rm Re}(x_i)\geq 1$, with $i\in \{1,2,3\}$.}  The middle panel of figure~\ref{Fig:DeformedCycleVolumeZ21e0} shows the $\varepsilon_0^{(1)}$-dependence of the fractional three-cycle volume with bulk orbit parallel to {\bf aI}$\times${\bf bI}$^0\times${\bf bI}$^0$, which shrinks to zero as the deformation parameter goes to one. Hence, this plot depicts the global behaviour of the fractional three-cycle $\Pi^{\rm frac}_- = \frac{1}{2} \left( \Pi^{\rm bulk} - {\boldsymbol \epsilon}_0^{(1)}\right)$. On the right panel of figure~\ref{Fig:DeformedCycleVolumeZ21e0}, we depict the $\varepsilon_0^{(1)}$-dependence of the volume of the fractional three-cycle $\Pi^{\rm frac}_+ = \frac{1}{2} \left( \Pi^{\rm bulk} + {\boldsymbol \epsilon}_0^{(1)}\right)$, where the bulk orbit is once more parallel to {\bf aI}$\times${\bf bI}$^0\times${\bf bI}$^0$. For this latter fractional three-cycle we observe that its volume grows for increasing values of the deformation parameter, with the same functional behaviour as the exceptional cycle volume. Closer inspection of the behaviour of the bulk cycle volume under deformation reveals that the correct representant in the homology class of bulk cycles corresponds to the three-cycle {\bf aI}$\times${\bf bIII}$^0\times${\bf bIII}$^0$, which happens to lie furthest away from the deformed singularity $(11)$, and its volume is therefore the least affected by the deformation. One can confirm this explicitly by adding the exceptional cycle volume to (twice) the volume of the fractional cycle $\Pi^{\text{frac}}_-$ and comparing the volume-dependence of the resulting bulk cycle to the volume-dependence of the bulk three-cycle {\bf aI}$\times${\bf bIII}$^0\times${\bf bIII}$^0$ under deformation.

\begin{figure}[h]
\begin{center}
\begin{tabular}{c@{\hspace{0.8in}}c@{\hspace{0.8in}}c}
\includegraphics[scale=0.5]{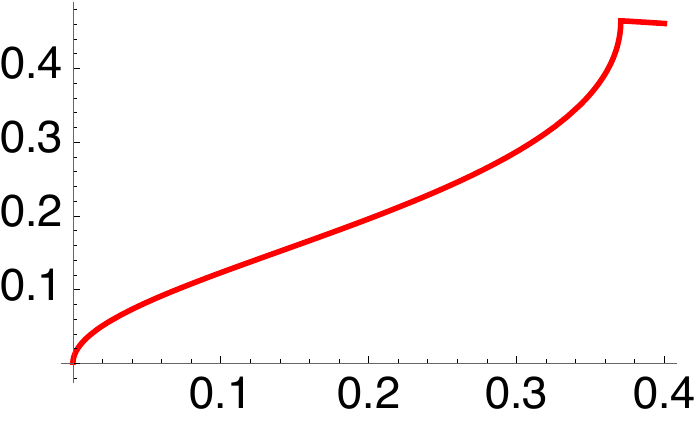} \begin{picture}(0,0) \put(0,0){$\varepsilon_3^{(1)}$} \put(-115,15){\begin{rotate}{90}Vol$_{\rm norm}(e_3)$\end{rotate}} \end{picture} & 
\includegraphics[scale=0.52]{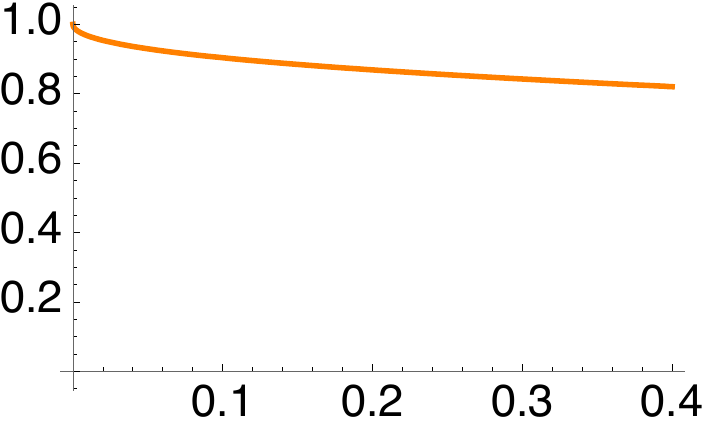} \begin{picture}(0,0) \put(0,0){$\varepsilon_3^{(1)}$} \put(-120,10){\begin{rotate}{90}Vol$_{\rm norm}(\Pi^{\rm frac}_-)$\end{rotate}} \end{picture} 
&
 \includegraphics[scale=0.54]{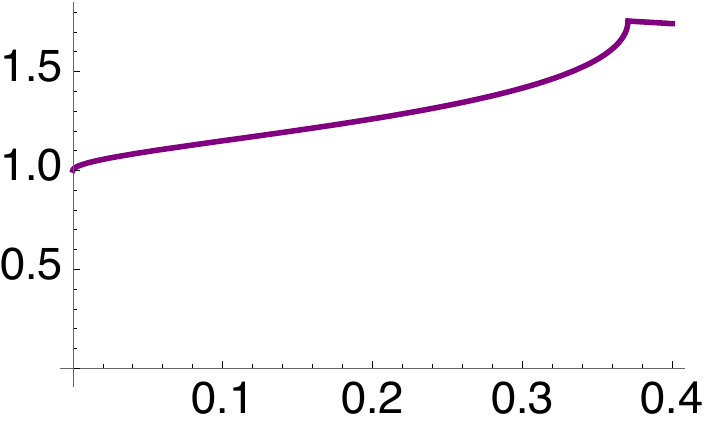} \begin{picture}(0,0) \put(0,0){$\varepsilon_3^{(1)}$} \put(-120,10){\begin{rotate}{90}Vol$_{\rm norm}(\Pi^{\rm frac}_+)$\end{rotate}} \end{picture}
\end{tabular}
\caption{Normalised volume of (one $S^2$ in) the resolved exceptional cycle $e^{(1)}_3$ on $T^4_{(1)}/\Z_6$ and the fractional cycles
 $\Pi^{\text{frac}}_\pm =\frac{ \Pi^{\text{bulk}} \pm {\boldsymbol \epsilon}^{(1)}_3}{2}$ on $T^2_{(1)} \times T^4_{(1)}/\Z_6$
as a function of the deformation parameter $\varepsilon^{(1)}_3$, with the representant of the bulk cycle $\Pi^{\text{bulk}}$ lying along the cycle ${\cal C}^0\times {\cal C}^0$ on $T^4_{(1)}/\Z_6$ defined in the main text. Note that a factor 1/3 has to be taken into account
when computing the exceptional cycle volume for the fractional three-cycle with bulk orbit {\bf aI}$\times${\bf bI}$^0\times${\bf bI}$^0$, as the fractional cycle only wraps one of the three $\Z_2^{(1)}$ singularities in the orbit $e_3^{(1)}$, namely the singularity (33), per $\Z_3$-image.
\label{Fig:DeformedCycleVolumeZ21e3}}
\end{center}
\end{figure}
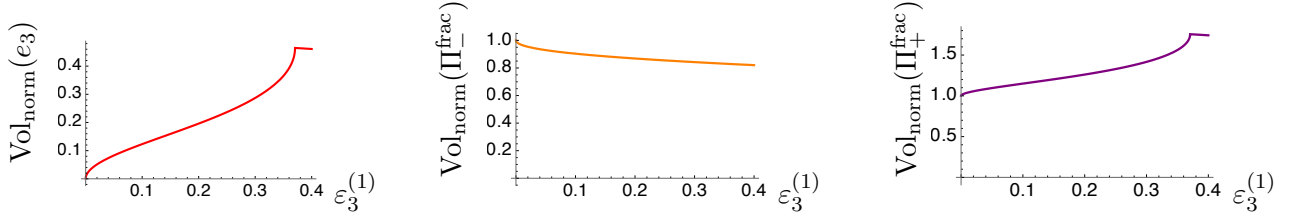

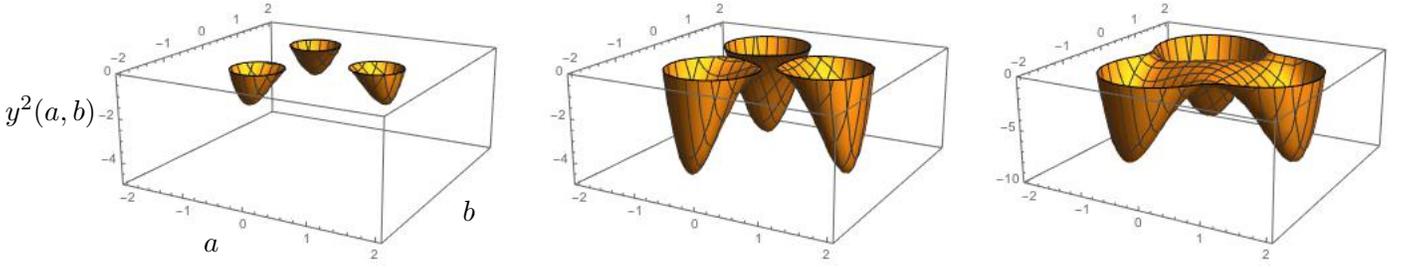
\begin{figure}[h]
\begin{center}
\begin{tabular}{c@{\hspace{-0.2in}}c@{\hspace{-0.2in}}c}
\hspace*{-0.4in}\includegraphics[scale=0.5]{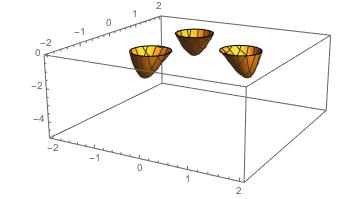} \begin{picture}(0,0) \put(-205,55){$y^2(a,b)$} \put(-130,5){$a$} \put(-32,17){$b$} \end{picture}
& \includegraphics[scale=0.5]{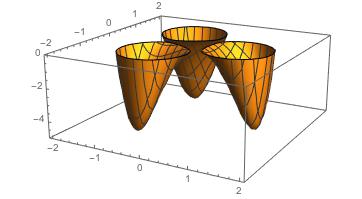} &\includegraphics[scale=0.5]{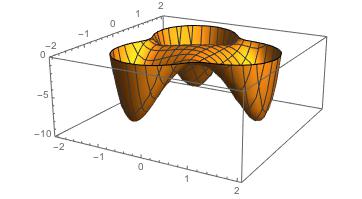}
\end{tabular}
\caption{Three-dimensional plots of the hypersurface equation $y^2(a,b)\leq 0$ describing the exceptional cycle $e_3^{(1)}$ on $T^4_{(1)}/\Z_6$ as a function of the coordinates $a={\rm Re}(x_2)$ and $b={\rm Im}(x_2)$ for deformation parameters $\varepsilon_3^{(1)} = 0.15$ (left), $\varepsilon_3^{(1)} = 0.30$ (middle) and $\varepsilon_3^{(1)} = 0.5$ (right). In the last plot, the three $S^2$'s resolving the $\Z_2^{(1)}$ singularities $(33)$, $(42)$ and $(24)$ merge together into a single exceptional cycle.\label{Fig:EvolutionExcCycleE3}}
\end{center}
\end{figure}

\begin{figure}[h]
\begin{center}
\begin{tabular}{c@{\hspace{0.8in}}c}
\includegraphics[scale=0.8]{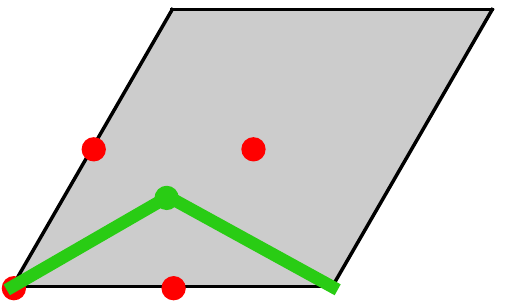}   \begin{picture}(0,0)  \put(-122,-8){$\color{red} 1$} \put(-110,35){$\color{red} 4$}  \put(-72,35){$\color{red} 2$}  \put(-84,-8){$\color{red} 3$}  \put(-62,16){$\color{CycleGreen} {\cal C}^0$}\end{picture}
&\includegraphics[scale=0.4]{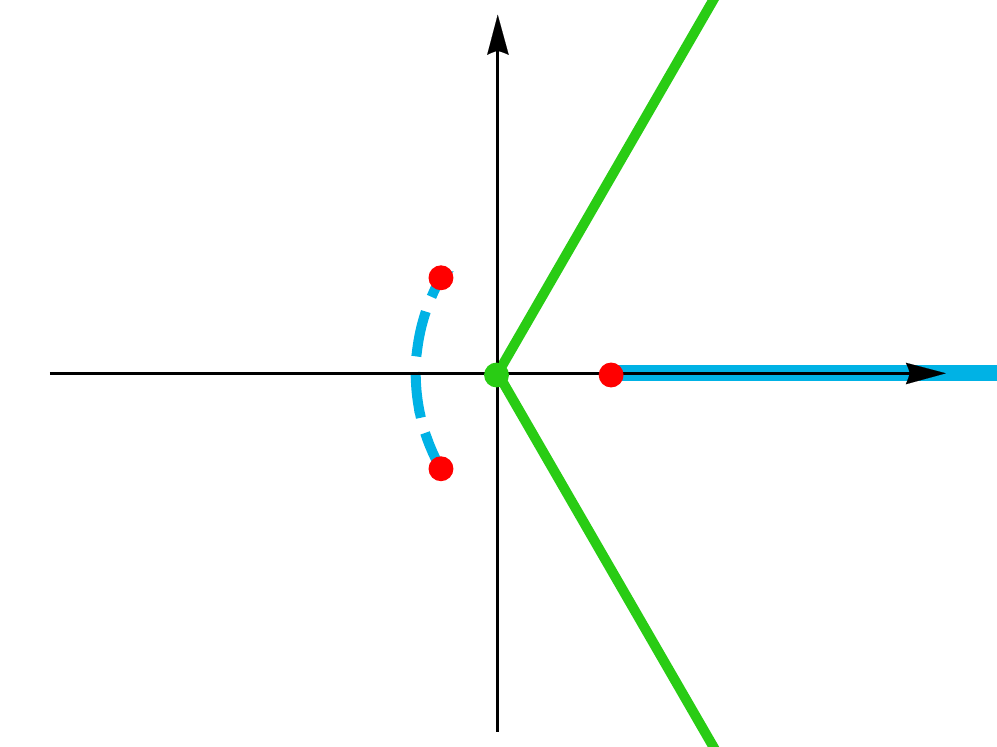}  \begin{picture}(0,0)\put(-76,90){${\rm Im}(x)$}   \put(-15,32){${\rm Re}(x)$} \put(-38,66){$\color{CycleGreen} {\cal C}^0$}  \put(-20,50){\color{myblue3} ${\bf bI^0}$}  \put(-102,46){\color{myblue3} ${\bf bIII^0}$}   \put(-50,35){\color{red} $\epsilon_3$}  \put(-70,58){\color{red} $\epsilon_2$} \put(-70,24){\color{red} $\epsilon_4$}  \end{picture}
\end{tabular}
\caption{Graphical representation of the union cycle ${\cal C}^0 =${\bf bII}$^+ \cup${\bf bII}$^-$ on a hexagonal torus lattice (left) and on an elliptic curve embedded in $\P_{112}$ (right). The cycle ${\cal C}^0$ is drawn in the $x$-plane $(v=1)$ by virtue of Weierstrass' elliptic function. \label{Fig:C0Picture}}
\end{center}
\end{figure}

Next, we focus on the deformation $\varepsilon_3^{(1)}$ for which it suffices to impose the exceptional cycle condition $x_2 = \ov x_3$ on equation (\ref{Eq:HyperZ2Z6Full}) to extract the real hypersurface equation describing the exceptional cycle volume. The points $(33)$, $(24)$ and $(42)$ in the $\Z_3$-invariant orbit $e_3^{(1)}$ are simultaneously deformed for a non-vanishing $\varepsilon_3^{(1)}$, such that the exceptional cycle consists initially of three distinct $S^2$'s resolving each of the three $\Z_2^{(1)}$ singularities, as shown in the left plot of figure~\ref{Fig:EvolutionExcCycleE3}. 
In order to extract the volume-dependence of a single $S^2$ as a function of the deformation parameter, we depict one third of the exceptional cycle volume in the left plot of figure~\ref{Fig:DeformedCycleVolumeZ21e3}, for which we observe a similar qualitative behaviour as for the exceptional cycle~$e_0^{(1)}$. More precisely, we notice a square-root type functional dependence of Vol($e_3^{(1)}$) for small deformations, which goes over into a linear behaviour and ends in a quadratic dependence for larger deformations. A quantitive difference with respect to the cycle $e_0^{(1)}$ is the region of validity for the parameter~$\varepsilon_{3}^{(1)}$. For values of $\varepsilon_3^{(1)}\sim 0.37$ and higher, the three two-spheres $S^2$ merge together into one large exceptional three-cycle as depicted in the right panel of figure~\ref{Fig:EvolutionExcCycleE3}, at which point  we can no longer reliably describe the exceptional cycle through the hypersurface formalism. This is manifested in the horizontal plateau truncating the exceptional cycle volume for values $\varepsilon_3^{(1)}\geq 0.37$ in the left plot of figure~\ref{Fig:DeformedCycleVolumeZ21e3}. The other two plots in figure~\ref{Fig:DeformedCycleVolumeZ21e3} represent the (normalised) volumes of the fractional three-cycles $\Pi^{\text{frac}}_\pm = \frac{1}{2} \left( \Pi^{\rm bulk} \pm {\boldsymbol \epsilon}_3^{(1)} \right)$ as a function of $\varepsilon_{3}^{(1)}$ with bulk orbit parallel to {\bf aI}$\times${\bf bI}$^0\times${\bf bI}$^0$. The representant in the bulk homology class is, however, not the factorisable three-cycle {\bf aI}$\times${\bf bI}$^0\times${\bf bI}$^0$ itself, but a bulk three-cycle {\bf aI}$\times{\cal C}^0\times{\cal C}^0$ consisting of the union of one-cycles ${\cal C}^0 =$ {\bf bII}$^+ \cup$ {\bf bII}$^- $ along both two-tori $T_{(2)}^2$ and $T_{(3)}^2$. Once again, it suffices to subtract (twice) the exceptional cycle volume from the fractional cycle volume to uncover the dependence of the bulk cycle on the deformation parameter $\varepsilon_{3}^{(1)}$ and verify that this functional behaviour matches the one of the three-cycle {\bf aI}$\times{\cal C}^0\times{\cal C}^0$. A pictorial representation of the one-cycle ${\cal C}^0$ is offered in figure~\ref{Fig:C0Picture}, from which it is immediately clear that the one-cycle does not represent a {\it sLag} cycle, since its pull-back of the K\"ahler two-form on the two-torus does not vanish. Nonetheless, the three-cycle {\bf aI}$\times{\cal C}^0\times{\cal C}^0$ belongs to the same homology class as the bulk three-cycles parallel to {\bf aI}$\times${\bf bI}$^0\times${\bf bI}$^0$, such that its integrated volumes are equal to each other, as argued in more detail in~\cite{Blaszczyk:2015oia}.     

\begin{figure}[h]
\begin{center}
\begin{tabular}{c@{\hspace{0.6in}}c@{\hspace{0.6in}}c}
\includegraphics[scale=0.5]{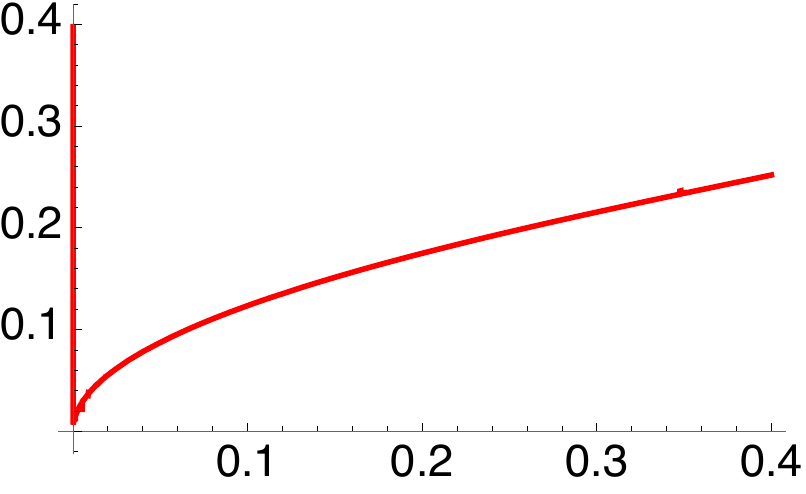} \begin{picture}(0,0) \put(0,0){$\varepsilon_1^{(1)}$} \put(-130,15){\begin{rotate}{90}Vol$_{\rm norm}(e_1)$\end{rotate}} \end{picture} & 
\includegraphics[scale=0.52]{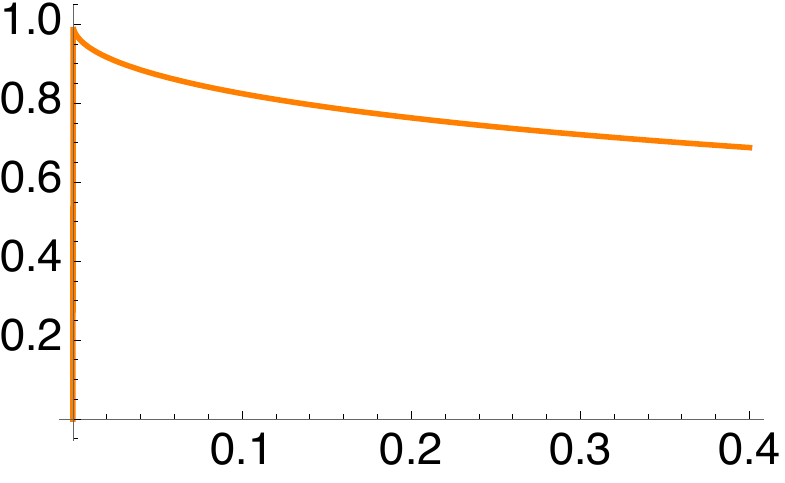} \begin{picture}(0,0) \put(0,0){$\varepsilon_1^{(1)}$} \put(-130,10){\begin{rotate}{90}Vol$_{\rm norm}(\Pi^{\rm frac}_-)$\end{rotate}} \end{picture} 
&
 \includegraphics[scale=0.52]{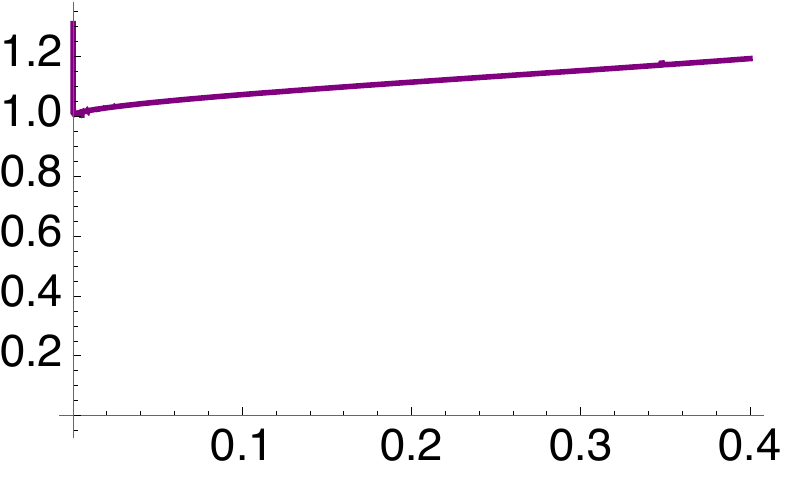} \begin{picture}(0,0) \put(0,0){$\varepsilon_1^{(1)}$} \put(-130,10){\begin{rotate}{90}Vol$_{\rm norm}(\Pi^{\rm frac}_+)$\end{rotate}} \end{picture}
\end{tabular}
\caption{Normalised volume of (one $S^2$ in) the resolved exceptional cycle $e^{(1)}_1$ on $T^4_{(1)}/\Z_6$ and the fractional cycles
 $\Pi^{\text{frac}}_\pm =\frac{ \Pi^{\text{bulk}} \pm {\boldsymbol \epsilon}^{(1)}_1}{2}$ on $T^2_{(1)} \times T^4_{(1)}/\Z_6$
as a function of the deformation parameter $\varepsilon^{(1)}_1$, with the bulk cycle $\Pi^{\text{bulk}}$ lying along the cycle ${\cal C}^0\times${\bf bIII}$^0$ on $T^4_{(1)}/\Z_6$. 
\label{Fig:DeformedCycleVolumeZ21e1}}
\end{center}
\end{figure}
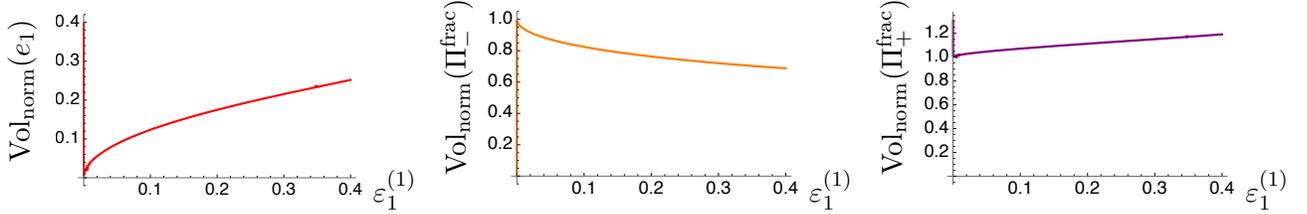

Discussing the global aspects of the exceptional cycle $e_{1}^{(1)}$ on $T^4_{(1)}/\Z_6$ follows a slightly different logic, as the geometric condition $x_2 = \ov{\lambda_3( x_3)}$ does not define a fixed set under the orientifold involution, i.e.~the resulting hypersurface equation is not real and therefore does not offer the desired direct access to the exceptional {\it sLag}. The intuition following from the study of the exceptional cycles $e_0^{(1)}$ and $e_3^{(1)}$ allows us, nonetheless, to express the functional dependence of $e_1^{(1)}$ on the deformation parameter $\varepsilon_1^{(1)}$ through a small detour: we first compute the normalised volume of the bulk cycle {\bf aI}$\times {\cal C}^0\times${\bf bIII}$^0$ as a function of the deformation parameter $\varepsilon_1^{(1)}$ and then subtract the normalised volume of the fractional cycle with integration contours completely along the real lines ${\rm Re}(x_{i=1,2,3})\geq 1$. The result of that computation is depicted in the left panel of figure~\ref{Fig:DeformedCycleVolumeZ21e1}, from which we can extract the square-root like functional dependence Vol$_{\rm norm}(e_1^{(1)})\sim\sqrt{\varepsilon_1^{(1)}}$.  
The plot does not contain information about a potential quadratic dependence on $\varepsilon_1^{(1)}$ for large deformations, as was the case for the exceptional cycles $e_0^{(1)}$ and $e_3^{(1)}$. It appears that this type of information can only be extracted explicitly from the hypersurface equation for the exceptional cycle $e_1^{(1)}$, whose form is constrained by the topology of the ambient $T^4$. When restricting to the real part of the hypersurface equation upon imposing the condition $x_2 = \ov{\lambda_3(x_3)}$, one can qualitatively see three distinct exceptional cycles growing out of the $\Z_2^{(1)}$ singularities $(31)$, $(21)$ and $(41)$ for non-vanishing $\varepsilon_1^{(1)}$, which merge together for larger deformation parameters analogously to the behaviour of the deformed exceptional cycle $e_3^{(1)}$ depicted in figure~\ref{Fig:EvolutionExcCycleE3}. In this respect, the $\Z_3$-action and the $T^4$ topology do qualitatively constrain the behaviour of the exceptional cycle $e_1^{(1)}$, even though their full effects cannot be extracted more quantitatively due to an indisputable imaginary component of the hypersurface equation for the exceptional cycle. 
The functional dependence of the (normalised) volumes for the fractional three-cycles $\Pi^{\rm frac}_\pm = \frac{1}{2} \left( \Pi^{\rm bulk} \pm {\boldsymbol \epsilon}_1^{(1)} \right)$ is given in the middle and right plot of figure~\ref{Fig:DeformedCycleVolumeZ21e1}, respectively. As expected, the volume of the fractional cycle $\Pi^{\rm frac}_-$ shrinks with growing deformation $\varepsilon_1^{(1)}$, while the volume of $\Pi^{\rm frac}_+$ grows with increasing deformation $\varepsilon_1^{(1)}$. Due to the exchange symmetry $T_{(2)}^2 \leftrightarrow T_{(3)}^2$, the discussion of the global description of the exceptional cycle $e_2^{(1)}$ is completely analogous to the one for $e_1^{(1)}$.  

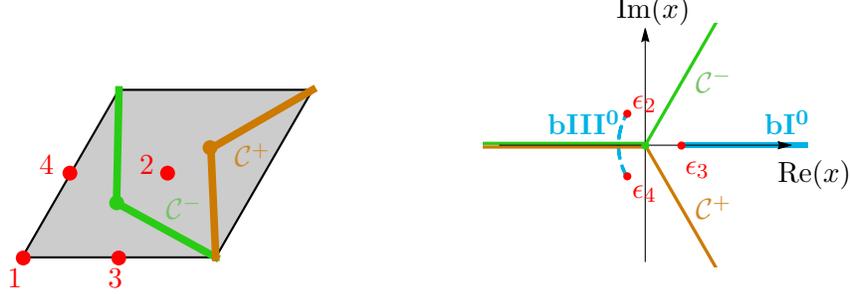
\begin{figure}[h]
\begin{center}
\begin{tabular}{c@{\hspace{0.8in}}c}
\includegraphics[scale=0.8]{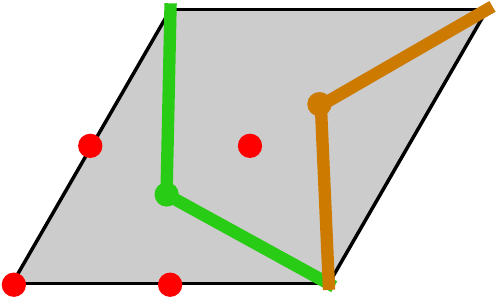}   \begin{picture}(0,0)  \put(-122,-8){$\color{red} 1$} \put(-110,35){$\color{red} 4$}  \put(-72,35){$\color{red} 2$}  \put(-84,-8){$\color{red} 3$}  \put(-62,18){$\color{CycleGreen} {\cal C}^-$} \put(-36,38){$\color{CycleBrown} {\cal C}^+$}\end{picture}
&\includegraphics[scale=0.4]{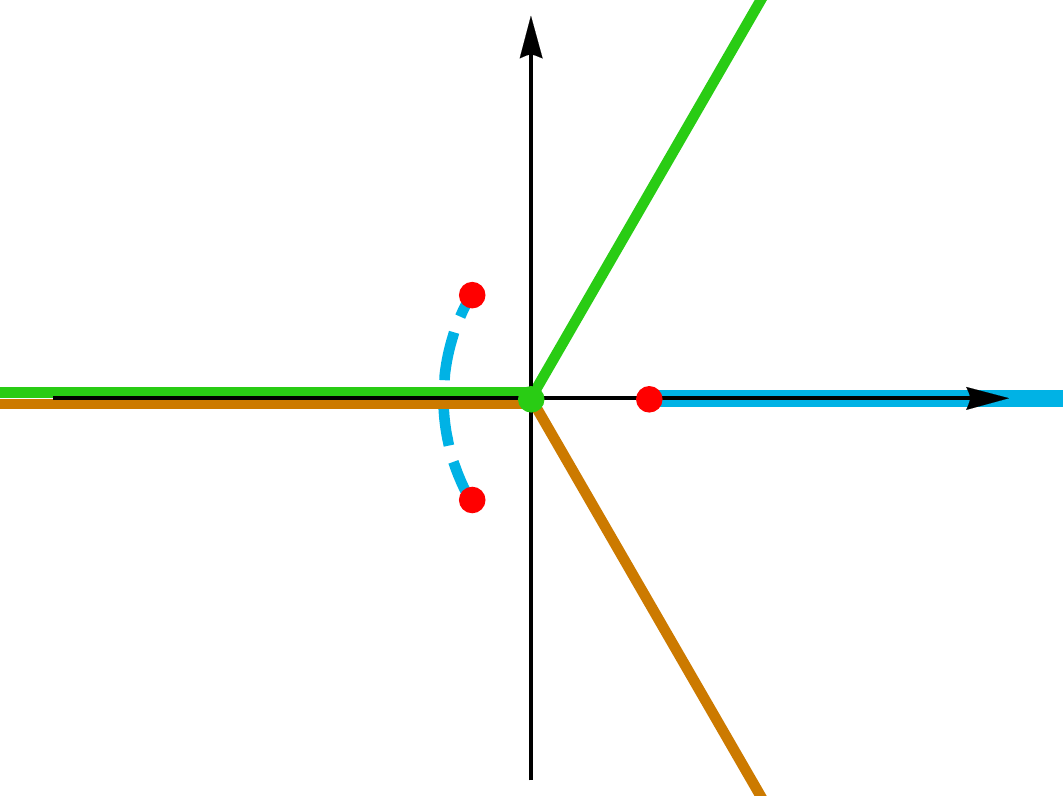}  \begin{picture}(0,0)\put(-76,94){${\rm Im}(x)$}   \put(-15,32){${\rm Re}(x)$} \put(-46,66){$\color{CycleGreen} {\cal C}^-$} \put(-46,18){$\color{CycleBrown} {\cal C}^+$}   \put(-20,50){\color{myblue3} ${\bf bI^0}$}  \put(-102,50){\color{myblue3} ${\bf bIII^0}$}   \put(-50,35){\color{red} $\epsilon_3$}  \put(-70,60){\color{red} $\epsilon_2$} \put(-70,26){\color{red} $\epsilon_4$}  \end{picture}
\end{tabular}
\caption{Graphical representation of the union cycles ${\cal C}^- =${\bf bII}$^0 \cup${\bf bII}$^-$ and ${\cal C}^+ =${\bf bII}$^0 \cup${\bf bII}$^+$ on a hexagonal torus lattice (left) and on an elliptic curve embedded in $\P_{112}$ (right). The cycles ${\cal C}^-$ and ${\cal C}^+$ are drawn in the $x$-plane $(v=1)$ by virtue of Weierstrass' elliptic function. \label{Fig:CMPicture}}
\end{center}
\end{figure}

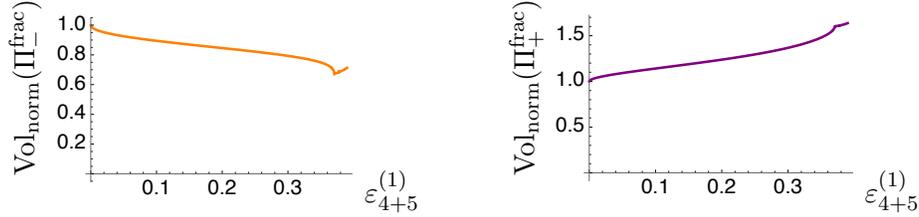
\begin{figure}[h]
\begin{center}
\begin{tabular}{c@{\hspace{1in}}c} 
\includegraphics[scale=0.6]{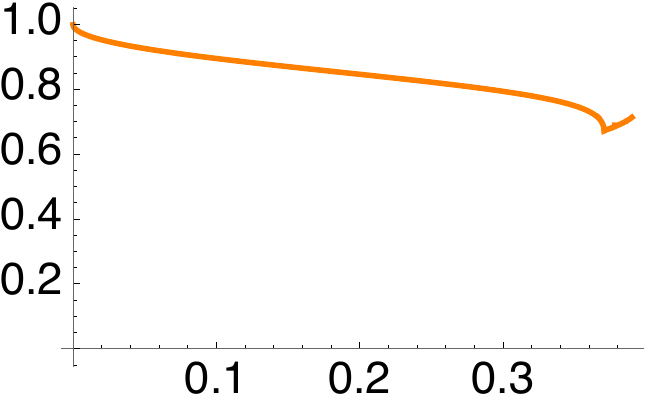} \begin{picture}(0,0) \put(0,0){$\varepsilon_{4+5}^{(1)}$} \put(-125,10){\begin{rotate}{90}Vol$_{\rm norm}(\Pi^{\rm frac}_-)$\end{rotate}} \end{picture} 
&
 \includegraphics[scale=0.6]{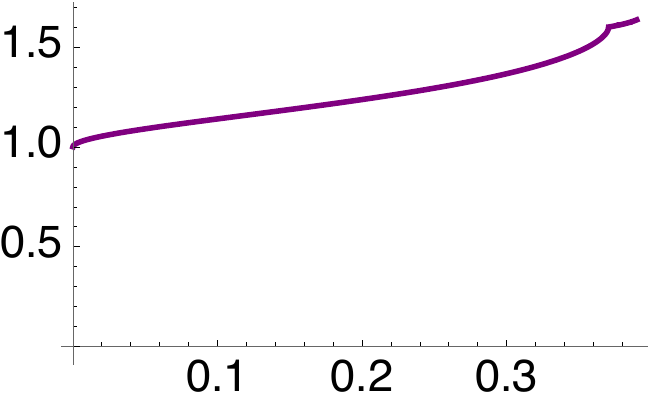} \begin{picture}(0,0) \put(0,0){$\varepsilon_{4+5}^{(1)}$} \put(-125,10){\begin{rotate}{90}Vol$_{\rm norm}(\Pi^{\rm frac}_+)$\end{rotate}} \end{picture}
\end{tabular}
\caption{Normalised volume of the fractional cycles
 $\Pi^{\text{frac}}_\pm =\frac{ \Pi^{\text{bulk}} \pm ({\boldsymbol \epsilon_4^{(1)}} + {\boldsymbol \epsilon_5^{(1)}})}{2}$ 
as a function of the deformation parameter $\varepsilon^{(1)}_{4+5}$, with the bulk cycle $\Pi^{\text{bulk}}$ lying along the cycle ${\cal C}^-\times{\cal C}^+$ on $T^4_{(1)}/\Z_6$. 
\label{Fig:DeformedCycleVolumeZ21e4}}
\end{center}
\end{figure}

This brings us finally to the global description of the exceptional cycles $e_{4}^{(1)}$ and $e_{5}^{(1)}$ on $T^4_{(1)}/\Z_6$, which are related to each other through the orientifold projection. In the hypersurface equation (\ref{Eq:HyperZ2Z6Full}) this relation under the $\OR$-projection is manifestly built in, such that a non-zero deformation parameter $\varepsilon_{4+5}^{(1)}$ resolves both $e_{4}^{(1)}$ and $e_{5}^{(1)}$ simultaneously and similarly for a non-zero deformation parameter $\varepsilon_{4-5}^{(1)}$. To extract the hypersurface equation for the exceptional cycles, we have to rotate the $x_{i=2,3}$-coordinates over an angle $\xi$ or $\xi^2$ (or use the M\"obius transformation $\lambda_2$ or $\lambda_4$), after which we can impose the algebraic condition $x_2 = \pm \ov x_3$. Unfortunately, the resulting hypersurface equation does not correspond to a fixed set under the orientifold involution, which is manifested by a purely imaginary contribution to the hypersurface equation. Hence, similarly to the exceptional cycle $e_1^{(1)}$, we are not able to directly access the exceptional cycles $e_{4}^{(1)}$ and $e_{5}^{(1)}$.\footnote{One can, however, focus on the real part of the hypersurface equation for $e_4^{(1)}$ and $e_5^{(1)}$ and compute the volume as a function of the respective deformation parameter. This offers a qualitative understanding of the geometry of $e_4^{(1)}$ and $e_5^{(1)}$ and shows that these orbits have a similar behaviour under deformation as the orbit $e_3^{(1)}$: for small deformations, the exceptional cycle volume exhibits a square-root type functional dependence, while the topology of the ambient $T^4_{(1)}$ enforces a quadratic behaviour for larger deformations. The two-spheres~$S^2$ at the resolved singularities in the orbit merge together into one large exceptional cycle for a sufficiently large deformation. This common behaviour is inherited from the isotropy between the orbits $e_3^{(1)}$, $e_4^{(1)}$ and $e_5^{(1)}$ on the parent toroidal orbifold $T^4_{(1)}/\Z_6$.}  The situation is even more complicated in this case as the fractional three-cycles wrapping one or more of the $\Z_2^{(1)}$-fixed points in the orbits $e_4^{(1)}$ and $e_5^{(1)}$ do not lie along the real axes in the $x_{2}$- and $x_{3}$-planes, such that we are not able to directly compute the fractional cycle volume as a function of $\varepsilon_{4+5}^{(1)}$ or $\varepsilon_{4-5}^{(1)}$ either. To understand the impact of the deformation $\varepsilon_{4+5}^{(1)}$ on the volume of a fractional cycle, one first has to 
establish that the resolved orbits $e_4^{(1)}$ and $e_5^{(1)}$ on the parent toroidal orbifold $T^6/(\Z_2\times\Z_6)$ with discrete torsion have the exact same structure as the resolved orbit~$e_3^{(1)}$ discussed above, upon respectively considering non-zero complex deformation parameters $\varepsilon_4^{(1)}$ and $\varepsilon_5^{(1)}$ individually. Taking afterwards the orientifold projection into account  implies - based on the calibration properties with respect to the volume three-form $\Omega_3$ - that the exceptional three-cycles ${\boldsymbol \epsilon}_4^{(1)} + {\boldsymbol \epsilon}_5^{(1)}$ and $\tilde{\boldsymbol \epsilon}_4^{(1)} + \tilde{\boldsymbol \epsilon}_5^{(1)}$ are resolved by a non-zero deformation parameter~$\varepsilon_{4+5}^{(1)}$, while the exceptional three-cycles ${\boldsymbol \epsilon}_4^{(1)} - {\boldsymbol \epsilon}_5^{(1)}$ and $\tilde{\boldsymbol \epsilon}_4^{(1)} - \tilde{\boldsymbol \epsilon}_5^{(1)}$ are resolved by a non-zero deformation parameter $\varepsilon_{4-5}^{(1)}$. To assess the impact of the deformation $\varepsilon_{4+5}^{(1)}$ on the volume of a fractional cycle wrapping $\Z_2^{(1)}$ singularities in the orbits $e_4^{(1)}$ and $e_5^{(1)}$, we exploit our intuition obtained from the other deformations in the $\Z_2^{(1)}$-twisted sector and propose the following method to compute the volume for e.g.~the fractional three-cycle {\bf aI}$\times${\bf bI}$^+\times${\bf bIII}$^-$: 

\begin{itemize}\label{It:Method}
\item Compute the (normalised) volume of the bulk cycle {\bf aI}$\times{\cal C}^-\times{\cal C}^+$, composed of the union one-cycles ${\cal C}^- =${\bf bII}$^0\cup${\bf bII}$^-$ and ${\cal C}^+ =${\bf bII}$^0\cup${\bf bII}$^+$ as drawn in figure~\ref{Fig:CMPicture}, as a function of the deformation parameter~$\varepsilon_{4+5}^{(1)}$;
\item Consider the volume of a single two-sphere $S^2$ obtained by deforming the exceptional cycle $e_3^{(1)}$, as presented in the left panel of figure~\ref{Fig:DeformedCycleVolumeZ21e3}, and re-interpret\footnote{This identification of the exceptional cycle volumes is supported by the $\Z_3$-symmetry among the orbits $e_3^{(1)}$, $e_4^{(1)}$ and $e_5^{(1)}$ on the ambient toroidal orbifold.} this volume as the volume of the resolved exceptional two-cycle in the $\Z_3$- and $\OR$-invariant exceptional three-cycle ${\boldsymbol \epsilon}_4^{(1)}+{\boldsymbol \epsilon}_5^{(1)}$;
\item Subtract or add the resulting exceptional cycle volume from the bulk cycle volume to obtain the volumes of the fractional cycles $\Pi_-^{\rm frac}$ and $\Pi_+^{\rm frac}$ respectively:
\begin{equation}
{\rm Vol}_{\rm norm}(\Pi^{\rm frac}_\pm) = {\rm Vol}_{\rm norm}(\Pi^{\rm bulk}) \pm {\rm Vol}_{\rm norm}({\boldsymbol \epsilon}_4^{(1)}+{\boldsymbol \epsilon}_5^{(1)}).
\end{equation}
\end{itemize}     
The proposed method does not allow us to obtain any quantitative information about the fractional cycle volume for a given deformation $\varepsilon_{4+5}^{(1)}\neq 0$, but it does enable us to envision the qualitative behaviour of the volumes of the fractional cycles $\Pi_\pm^{\rm frac}$ parallel to e.g.~the three-cycle {\bf aI}$\times${\bf bI}$^+\times${\bf bIII}$^-$ as presented in figure~\ref{Fig:DeformedCycleVolumeZ21e4}. There, we see that the volumes of the fractional three-cycles $\Pi_\pm^{\rm frac}$ exhibit the expected behaviour under deformation: the volume Vol$(\Pi_-^{\rm frac})$ decreases with growing $\varepsilon_{4+5}^{(1)}$, while the volume Vol$(\Pi_+^{\rm frac})$ increases for growing deformation~$\varepsilon_{4+5}^{(1)}$. The numerical noise for large deformations ($\varepsilon_{4+5}^{(1)} \geq 0.37$) is a reflection of the merging of the two-spheres $S^2$ at the resolved singularities into one large exceptional two-cycle.  

One can easily repeat this method for fractional three-cycles wrapping any of the other $\Z_2^{(1)}$ singularities in the orbits $e_4^{(1)}$ and $e_5^{(1)}$, or apply this method to probe the effect of a non-vanishing deformation $\varepsilon_{4-5}^{(1)}$ on the volume of such fractional three-cycles, provided one chooses the appropriate bulk three-cycle. In all of these cases, the qualitative functional behaviour of the fractional three-cycles can be brought back to the case presented in figure~\ref{Fig:DeformedCycleVolumeZ21e4}, namely Vol$(\Pi^{\rm frac}_\pm) \sim {\rm Vol}(\Pi^{\rm bulk}) \pm \sqrt{\varepsilon_{i}^{(1)}}$ with $i=4+5$ or $i=4-5$.

\subsubsection{\textit{\textbf{sLags}} in the deformed $\Z_2^{(3)}$-twisted sector}\label{Sss:sLagsZ23}
\begin{figure}[h]
\begin{center}
\begin{tabular}{c@{\hspace{0.4in}}c@{\hspace{0.4in}}c@{\hspace{0.4in}}c}
(a) & (b) & (c) & (d) \\ 
\includegraphics[scale=0.6]{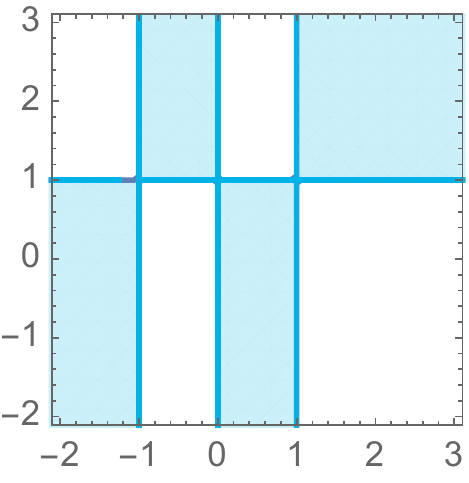} \begin{picture}(0,0)  \put(-45,-5){$x_1$}  \put(-96,45){$x_2$} \put(-32,64){\scriptsize \bf aI$\times$bI }  \put(-38,14){\begin{rotate}{90}\scriptsize \bf aIV$\times$bII \end{rotate}}  \put(-52,46){\begin{rotate}{90}\scriptsize \bf aIII$\times$bI \end{rotate}}  \put(-64,16){\begin{rotate}{90}\scriptsize \bf aII$\times$bII \end{rotate}}  \end{picture} &  \includegraphics[scale=0.6]{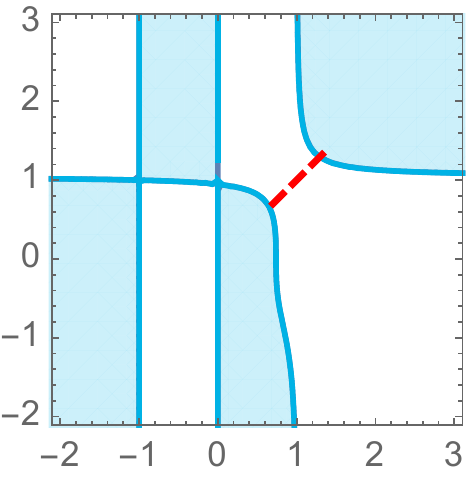} \begin{picture}(0,0)  \put(-45,-5){$x_1$}  \put(-96,45){$x_2$} \put(-30,60){\scriptsize (23)}  \end{picture}  &  \includegraphics[scale=0.6]{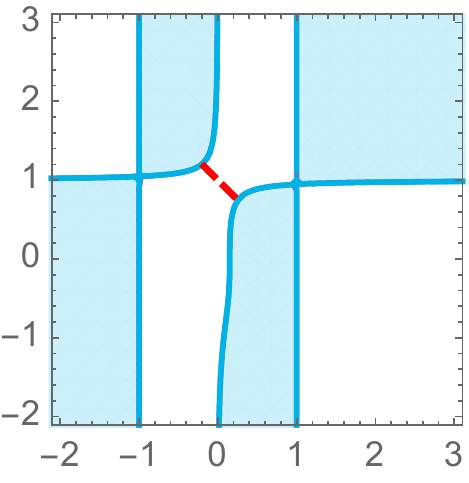} \begin{picture}(0,0)  \put(-45,-5){$x_1$}  \put(-96,45){$x_2$} \put(-63,60){\scriptsize (33)}   \end{picture}   &  \includegraphics[scale=0.6]{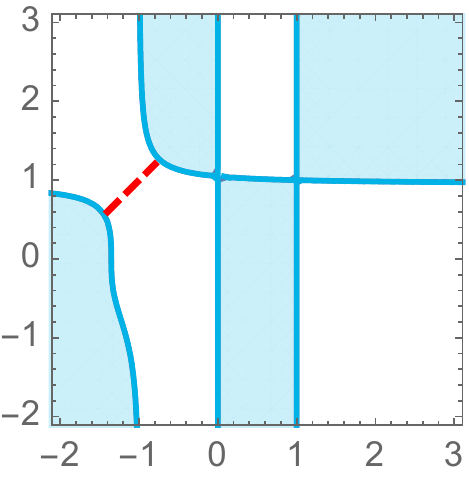}  \begin{picture}(0,0)  \put(-45,-5){$x_1$}  \put(-96,45){$x_2$}  \put(-62,58){\scriptsize (43)}    \end{picture}  \\
 &  $\varepsilon_2^{(3)}> 0$ & $\varepsilon_3^{(3)}> 0$ &  $\varepsilon_4^{(3)}> 0$  \\
 & \includegraphics[scale=0.6]{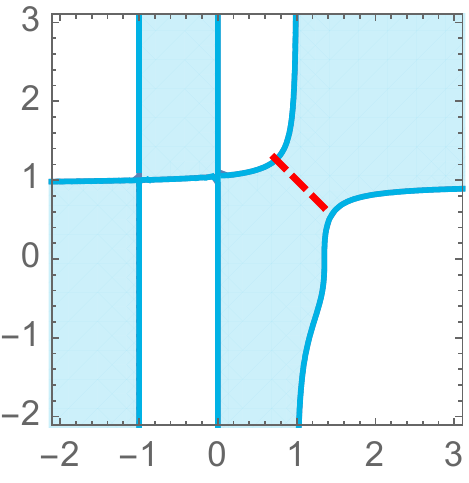} \begin{picture}(0,0)  \put(-45,-5){$x_1$}  \put(-96,45){$x_2$} \put(-32,56){\scriptsize (23)}  \end{picture}  & & \\
 &  $\varepsilon_2^{(3)}< 0$ & & 
\end{tabular}
\caption{Hypersurface equation (\ref{Eq:HyperZ2Z6Full}) in the ($\R$-projected) $(x_1,x_2)$ plane for various $\Z_2^{(3)}$ deformation parameters: all deformations are set to zero in (a); $|\varepsilon_2^{(3)}| = 0.05$ deforms the singularity (23) in (b); $\varepsilon_3^{(3)} = 0.05$ deforms the singularity (33) in (c); $\varepsilon_4^{(3)} = 0.05$ deforms the singularity (43) in (d). The contour lines correspond to zeros of $y$, while the blue-coloured regions represent {\it sLag} two-cycles calibrated with respect to ${\rm Re}(\Omega_2)$. The exceptional cycles with non-zero volume arising at the deformed singularities are indicated by a red dashed line. 
\label{Fig:DefZ23hypersurfaceEq}}
\end{center}
\end{figure}
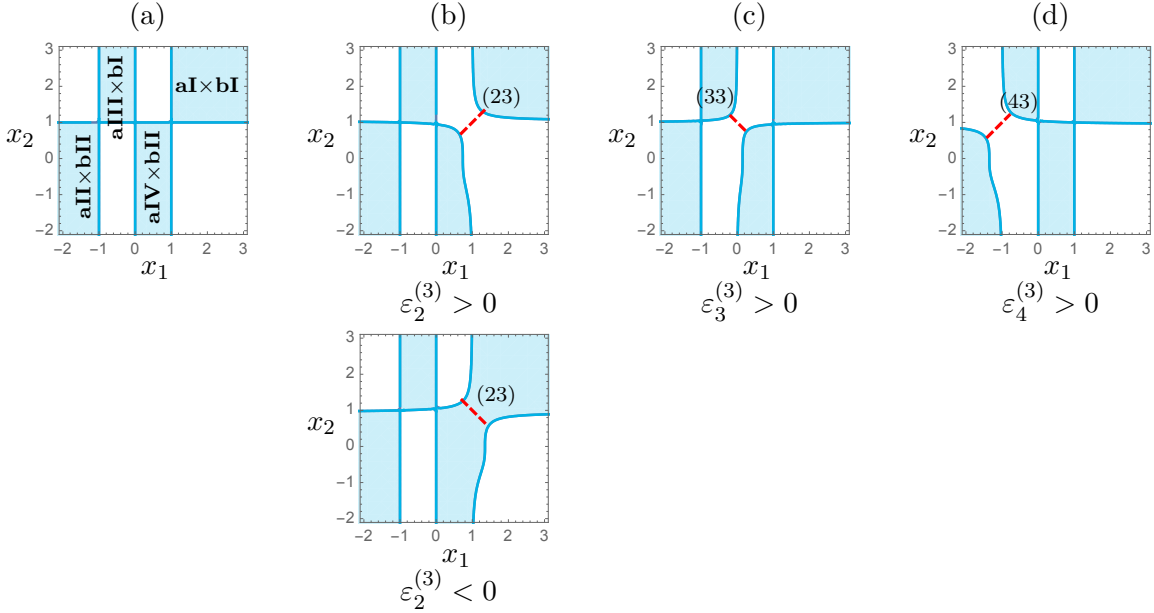

To investigate the deformation effects in the  $\Z_2^{(3)}$-twisted sector, we turn to the $v_i=1$ patch so that we can describe the {\it Lag} lines in terms of the homogeneous coordinates $x_i$ as in section~\ref{Ss:LCHF}. Real hypersurfaces at the orbifold point are represented in this coordinate by the {\it Lag} lines {\bf aI}, {\bf aII}, {\bf aIII} and {\bf aIV} on the two-torus $T_{(1)}^2$ and by {\bf bI}$^0$ and {\bf bII}$^0$ on $T_{(2)}^2$. Combining these {\it Lag} lines, we can construct a set of {\it sLag} two-cycles with topology $T^2/\Z_2$ on ${\rm Def}\left(T_{(3)}^4/\Z_2^{(3)}\right)$ 
and calibrated with respect to ${\rm Re}(\Omega_2)$, represented by the blue-coloured regions in figure~(\ref{Fig:DefZ23hypersurfaceEq}) (a), i.e.~the two-cycles {\bf aI}$\times${\bf bI}$^0$, {\bf aIV}$\times${\bf bII}$^0$, {\bf aIII}$\times${\bf bI}$^0$ and {\bf aII}$\times${\bf bII}$^0$. The white regions correspond to {\it sLag} two-cycles calibrated by ${\rm Im}(\Omega_2)$: {\bf aI}$\times${\bf bII}$^0$, {\bf aIV}$\times${\bf bI}$^0$, {\bf aIII}$\times${\bf bII}$^0$ and {\bf aII}$\times${\bf bI}$^0$. The blue contour-lines in the $\R$-projected $(x_1, x_2)$ plane represent the zero locus $y=0$ and intersect at the real $\Z_2^{(3)}$ fixed points (23), (33) and (43). In order to obtain fractional three-cycles calibrated w.r.t.~${\rm Re}(\Omega_3)$ on $(T^4/\Z_2^{(3)} \times T^2_{(3)})/\Z_6$, the two-cycles calibrated w.r.t.~${\rm Re}(\Omega_2)$ and ${\rm Im}(\Omega_2)$
on $T^4_{(3)}/\Z_2^{(3)}$ should be paired with a one-cycle {\bf bI}$^0$/{\bf bIII}$^0$ and {\bf bII}$^0$/{\bf bIV}$^0$, respectively, on $T_{(3)}^2$.

By turning on the $\Z_2^{(3)}$ deformation parameters $\varepsilon_\alpha^{(3)}$ one by one, figure~\ref{Fig:DefZ23hypersurfaceEq} shows exactly which singularities (or singular orbits under the $\Z_3 \subset \Z_6$ symmetry) are deformed and which singularities are displaced, in agreement with the lower part of table~\ref{tab:OverviewDeformperZ2Sector}. Statements about singularities $(1\alpha)$ 
cannot be made in this coordinate patch $v_i=1$ as they are located here at $x_1 = \infty$, and hence they require us instead to describe the hypersurface equation in terms of the homogeneous coordinates $v_i$ in the coordinate patch $x_i=1$. At the deformed singularities, an exceptional two-cycle with non-vanishing volume appears, as indicated by the red dashed lines in figure~\ref{Fig:DefZ23hypersurfaceEq}. We can study the effects of the deformation parameters more qualitatively by studying the zero locus of the hypersurface equation (\ref{Eq:HyperZ2Z6Full}) in a local patch around the singular point (23), for which the hypersurface equation reduces {\bf locally} to the form (after rescalings):
\begin{equation}\label{Eq:HyperLocalZ23}
\tilde y^2 = \tilde x_1 \tilde x_2 - 2 \varepsilon_{\alpha}^{(3)}, \qquad \qquad \alpha = 1,2,3,4.
\end{equation}
For the deformations $\varepsilon_1^{(3)}$, $\varepsilon_3^{(3)}$ and $\varepsilon_4^{(3)}$ we have to perform an appropriate M\"obius transformation from appendix~\ref{A:MT} to mould the hypersurface equation (\ref{Eq:HyperZ2Z6Full}) into this specific form, corresponding locally to a $\C^2/\Z_2$-type singularity. The two-cycles passing through the singularity~(23) are given by {\bf aI}$\times${\bf bI}$^0$ and {\bf aIV}$\times${\bf bII}$^0$, associated to the torus wrapping numbers $(n^1,m^1;n^2,m^2)=(1,0;1,0)$ and $(0,1;1,-2)$ on $T_{(1)}^2\times T_{(2)}^2$, respectively. Combining for instance the first two-cycle  {\bf aI}$\times${\bf bI}$^0$
with a one-cycle {\bf bI}$^0$ on $T_{(3)}^2$ yields a fractional three-cycle as defined in section~\ref{Ss:RemZ2Z6}, calibrated w.r.t. ${\rm Re}(\Omega_3)$. Its overall $\Z_2^{(3)}$ exceptional three-cycle is given  in terms of the basis exceptional three-cycles as:
\begin{equation}\label{Ex:Z2-3+13}
\Pi^{\Z_2^{(3)}} = (-)^{\tau^{\Z_2^{(3)}}} \left( (-)^{\tau^3} {\boldsymbol \epsilon}_{1}^{(3)} + (-)^{\tau^2 + \tau^3} {\boldsymbol \epsilon}_{2}^{(3)} \right).
\end{equation} 
By turning on the deformation parameter $\varepsilon_2^{(3)}$, an exceptional two-cycle $e_2^{(3)}$ with non-vanishing volume grows out of the singular point (23) on $T^4_{(3)}/\Z_2^{(3)}$, and depending on the sign of the deformation the volumes of the two-cycles {\bf aI}$\times${\bf bI}$^0$ and {\bf aIV}$\times${\bf bII}$^0$ will shrink or grow:
\begin{itemize} 
\item \vspace{-0.1in} \underline{$\varepsilon_{2}^{(3)}>0$:} {\bf aI}$\times${\bf bI}$^0$ and {\bf aIV}$\times${\bf bII}$^0$ still form two separate two-cycles with reduced size as an exceptional two-cycle emerges out of the singular point (23) in the region $\tilde y^2 <0$, as depicted in the upper diagram of figure~\ref{Fig:DefZ23hypersurfaceEq} (b). The local hypersurface equation (\ref{Eq:HyperLocalZ23}) reduces to the equation of a two sphere $S^2$ with radius $\sqrt{2 \varepsilon_2^{(3)}}$, when we impose the algebraic condition $\tilde x_1 = \ov{\tilde x_2}$ for the exceptional cycle. The exceptional two-cycle $e_2^{(3)}$ is calibrated with respect to ${\rm Re}(\Omega_2)$, a feature supported by the fact that the cycles {\bf aI}$\times${\bf bI}$^0$ and {\bf aIV}$\times${\bf bII}$^0$ remain {\it sLag} for positive deformations. Translating these considerations to the fractional three-cycle with exceptional part displayed in equation~\eqref{Ex:Z2-3+13}, we find that the exceptional three-cycle ${\boldsymbol \epsilon}_{2}^{(3)}$ is calibrated with respect to ${\rm Re}(\Omega_3)$ and that there should be a relative minus sign between the bulk three-cycle $\Pi^{\rm bulk}$ and its contribution to $\Pi^{\Z_2^{(3)}}$, i.e. $(-)^{\tau^{\Z_2^{(3)}} + \tau^2+ \tau^3}=-1$, in order for the volume of the fractional three-cycle to decrease for positive deformations.

The cycles {\bf aI}$\times${\bf bII}$^0$ and {\bf aIV}$\times${\bf bI}$^0$ on the other hand are no longer {\it sLag} on their own and melt together to one big two-cycle, still calibrated with respect to ${\rm Im} (\Omega_2)$. This bigger cycle is described by the union two-cycle {\bf aI}$\times${\bf bII}$^0$$\oplus${\bf aIV}$\times${\bf bI}$^0$ from which the exceptional cycle $e_2^{(3)}$ has been eliminated. 
\item \underline{$\varepsilon_{2}^{(3)}<0$:} we observe the opposite picture for negative deformations, namely the two-cycles {\bf aI}$\times${\bf bI}$^0$ and {\bf aIV}$\times${\bf bII}$^0$ have merged together to one big two-cycle as shown in the lower diagram of figure~\ref{Fig:DefZ23hypersurfaceEq} (b). This can be traced back to the fact that the exceptional two-cycle is now calibrated with respect to ${\rm Im}(\Omega_2)$. By taking the union two-cycle {\bf aI}$\times${\bf bI}$^0$$\oplus${\bf aIV}$\times${\bf bII}$^0$, we can ensure that the exceptional two-cycle drops out, such that the union two-cycle remains a {\it sLag} two-cycle. 

For negative deformation parameters, the sizes of the two-cycles {\bf aI}$\times${\bf bII}$^0$ and {\bf aIV}$\times${\bf bI}$^0$ shrink, in line with the consideration that  the exceptional two-cycle $e_2^{(3)}$ is now calibrated by the same two-form as both two-cycles. The $S^2$ topology of the exceptional cycle with radius $\sqrt{-2\varepsilon_{2}^{(3)}}$ follows from equation (\ref{Eq:HyperLocalZ23}) by restricting to the slice $\tilde x_1 = - \ov {\tilde{x_2}}$. Combining the two-cycles with the one-cycle {\bf bII}$^0$ on $T_{(3)}^2$ allows for the construction of fractional three-cycles, whose volumes are now decreasing for increasing $|\varepsilon_{2}^{(3)}|$. This implies a relative minus sign between $\Pi^{\rm bulk}$ and 
the contribution of ${\boldsymbol \epsilon}_{2}^{(3)}$
to $\Pi^{\Z_2^{(3)}}$ for the fractional three-cycle.
\end{itemize}

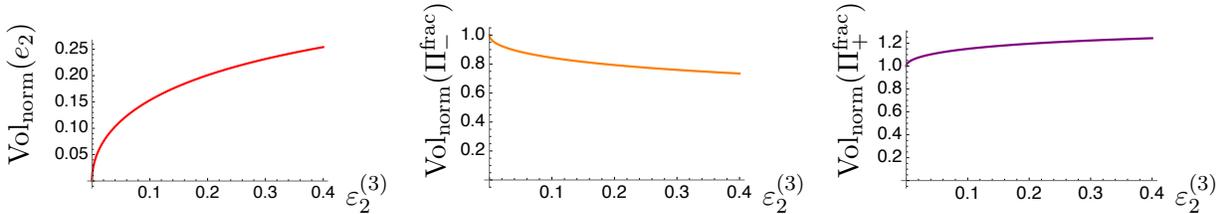
\begin{figure}[h]
\begin{center}
\begin{tabular}{c@{\hspace{0.6in}}c@{\hspace{0.6in}}c}
\includegraphics[scale=0.5]{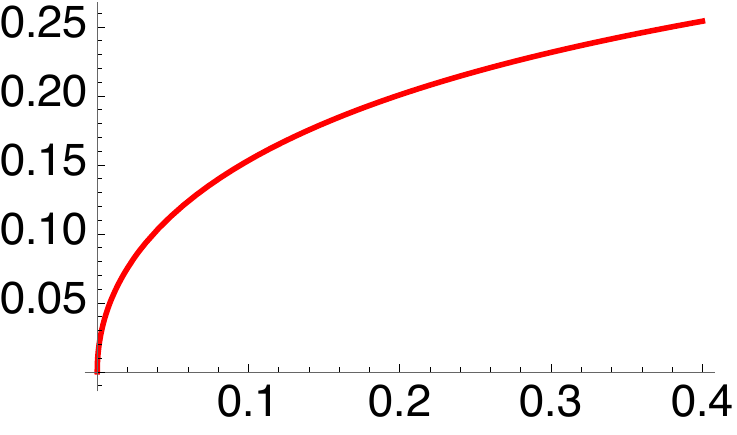} \begin{picture}(0,0) \put(0,0){$\varepsilon_2^{(3)}$} \put(-120,15){\begin{rotate}{90}Vol$_{\rm norm}(e_2)$\end{rotate}} \end{picture} & 
\includegraphics[scale=0.52]{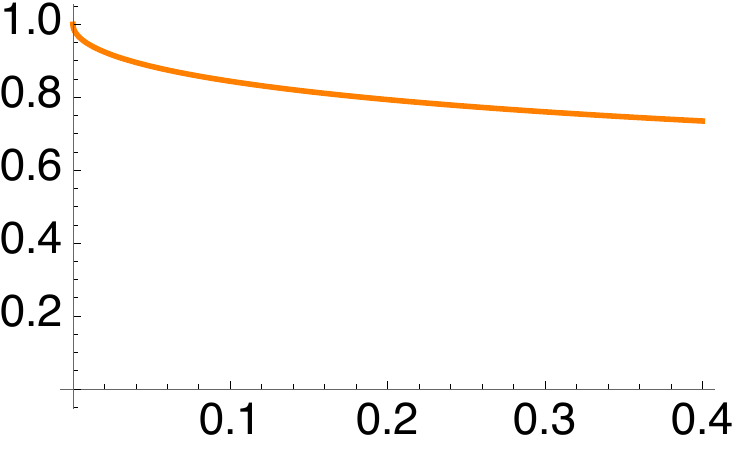} \begin{picture}(0,0) \put(0,0){$\varepsilon_2^{(3)}$} \put(-120,10){\begin{rotate}{90}Vol$_{\rm norm}(\Pi^{\rm frac}_-)$\end{rotate}} \end{picture} 
&
 \includegraphics[scale=0.52]{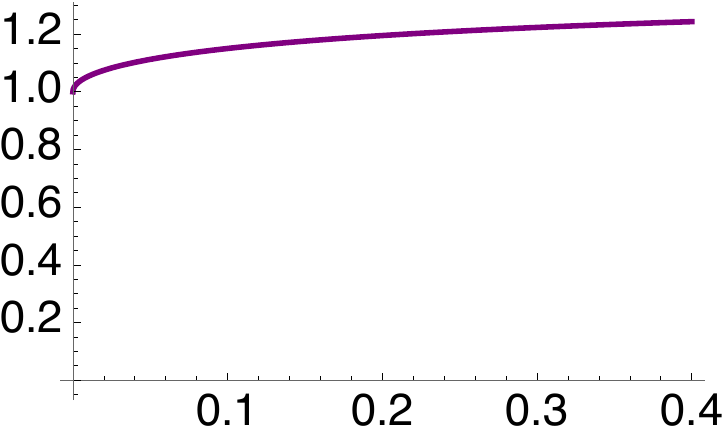} \begin{picture}(0,0) \put(0,0){$\varepsilon_2^{(3)}$} \put(-120,10){\begin{rotate}{90}Vol$_{\rm norm}(\Pi^{\rm frac}_+)$\end{rotate}} \end{picture}
\end{tabular}
\caption{Normalised volume of the exceptional cycle $e^{(3)}_2$ on $T^4_{(3)}/\Z_2^{(3)}$ and the fractional cycles
 $\Pi^{\text{frac}}_\pm =\frac{ \Pi^{\text{bulk}} \pm {\boldsymbol \epsilon}_{2}^{(3)}}{2}$ 
as a function of the deformation parameter $\varepsilon^{(3)}_2$, with the bulk cycle $\Pi^{\text{bulk}}$ lying along the cycle {\bf aIII}$\times{\cal C}^0$ on $T^4_{(3)}/\Z_2^{(3)}$. \label{Fig:DeformedCycleVolumeZ23e2}}
\end{center}
\end{figure}

If we want to turn to a {\bf global} description of the fractional three-cycles located at the deformed singularities, we have to consider the full hypersurface equation (\ref{Eq:HyperZ2Z6Full}) and impose the algebraic conditions $x_1= \pm \ov{x}_2$ (beyond a neighbourhood of the original singular point and possibly after acting with an appropriate M\"obius transformation on the coordinates), allowing us to determine the fixed loci of the orientifold projection. Note, however, that the resulting equation does not reduce to a real hypersurface equation, not even for vanishing deformations. The inability to describe $\Z_2^{(3)}$ deformations globally is an immediate consequence of the different complex structures on $T_{(1)}^2$ and $T_{(2)}^2$, which prevent the conditions $x_1= \pm \ov{x}_2$ to represent the fixed loci of the orientifold projection globally. Nevertheless, we are able to extract information about the functional dependence of the exceptional cycle volume on e.g.~the deformation parameter $\varepsilon_{2}^{(3)}$ by using the following strategy: compute the (normalised) volume of the bulk three-cycle {\bf aIII}$\times{\cal C}^0\times${\bf bI}$^0$ under non-vanishing deformation~$\varepsilon_{2}^{(3)}$ and subtract it from the (normalised) volume of the fractional three-cycle with integration contours completely along the real lines ${\rm Re}(x_i)\geq 1$ in the complex $x_{i=1,2,3}$-plane. The result of this computation is shown in the left panel of figure~\ref{Fig:DeformedCycleVolumeZ23e2} and exhibits a square-root like dependence on the parameter $\varepsilon_{2}^{(3)}$ for the (normalised) exceptional cycle volume. The behaviour of the (normalised) fractional three-cycle volumes Vol$(\Pi^{\rm frac}_\pm)$ under non-zero deformation $\varepsilon_{2}^{(3)}$ is shown in the middle and right panel of figure~\ref{Fig:DeformedCycleVolumeZ23e2}. As expected, the three-cycle $\Pi^{\rm frac}_-$ is characterised by a shrinking volume for increasing $\varepsilon_{2}^{(3)}$, while the volume of the three-cycle $\Pi^{\rm frac}_+$ increases for growing $\varepsilon_{2}^{(3)}$. Given that the other ($\Z_3$-orbits of)
$\Z_2^{(3)}$ exceptional divisors are related to $e_2^{(3)}$ by virtue of a M\"obius transformation, the volumes of the other exceptional cycles reproduce the same structure under their respective deformation as the one presented in figure~\ref{Fig:DeformedCycleVolumeZ23e2}.

Due to the exchange symmetry $T_{(2)}^2\leftrightarrow T_{(3)}^2$, we can straightforwardly transpose the entire analysis into the $\Z_2^{(2)}$ sector, where the same conclusions can be drawn for the $\Z_2^{(2)}$ exceptional and fractional three-cycle volumes. One subtle difference arises between the $\Z_2^{(2)}$- and $\Z_2^{(3)}$-twisted sector when the choice of the exotic O6-plane is taken into account. If either the $\OR\Z_2^{(2)}$- or $\OR\Z_2^{(3)}$-plane are taken to be the exotic O6-plane, their respective RR-charges are opposite, i.e.~$\eta_{(2)} = - \eta_{(3)}$, resulting in a different decomposition into $\OR$-even and -odd cycles for both sectors, according to table~\ref{tab:Z2Z6OrientifoldExceptionalCycles}, as will be discussed in the following section in terms of prototypical global D6-brane models.

\section{Deformation Moduli in Global D6-Brane Models}\label{S:DefModuliGlobalModels}

In this section, we apply the findings of geometric deformations in section~\ref{S:DefOrbS} to global D6-brane models of phenomenological interest
and discuss which moduli are stabilised at the orbifold point or constitute flat directions of the global model affecting physical gauge couplings,
either by a direct tree-level dependence or only via higher order and/or non-perturbative effects.

\subsection{Some generic considerations}\label{Ss:GenCon}

In section~\ref{Ss:RemZ2Z6}, the different types of three-cycles on the $T^6/(\Z_2 \times \Z_6 \times \OR)$ orientifold with discrete torsion were briefly reviewed.   
While a D6-brane $a$ by itself wraps a fractional three-cycle $\Pi_a$ of the form~\eqref{Eq:Pi-frac}, it is generically accompanied by its orientifold image $a'$ with associated three-cycle $\Pi_{a'}$. The global model remains inert under the exchange \mbox{$\Pi_a \leftrightarrow \Pi_{a'}$} when simultaneously changing the gauge representations for their conjugates, e.g. \mbox{$\N_a \leftrightarrow \ov{\N}_a$}.
The scalar potential only depends on the sum of the two~\cite{Blumenhagen:2002wn},
\begin{equation}\label{Eq:V-scalar}
{\cal V}_{\text{scalar}}^{\text{NS-NS}} \supset  \sum_a N_a \bigl[ \text{Vol}(\Pi_a) +  \text{Vol}(\Pi_{a'}) \bigr] - \text{Vol}(\Pi_{O6})
\quad
\left\{\begin{array}{lc}
=0 & \text{all D$6_a$-branes are {\it sLag}}
\\
> 0 & \text{else}
\end{array}
\right.
,
\end{equation}
which leads to the following qualitative situations observed first in the context of \mbox{$T^6/(\Z_2 \times \Z_2 \times \OR)$} and \mbox{$T^6/(\Z_2 \times \Z_6' \times \OR)$} models 
with discrete torsion in~\cite{Blaszczyk:2014xla,Blaszczyk:2015oia}, see also~\cite{Honecker:2015qba,Koltermann:2015oyv,Honecker:2016gyz}:
\begin{enumerate}
\item
The D6-brane $a$ couples to the $\Z_2^{(i)}$-twisted deformation modulus $\zeta^{(i)}_{\alpha}$ such that the {\it sLag} condition is violated for a non-vanishing
{\it vev} $\langle \zeta^{(i)}_{\alpha} \rangle \sim \sqrt{\varepsilon^{(i)}_{\alpha}}$.
 The deformation modulus $\zeta^{(i)}_{\alpha}$ itself can be seen as the period associated to the $\Z_2^{(i)}$ exceptional three-cycle $\tilde{\boldsymbol  \delta}_\alpha^{(i)}$:
\begin{equation}
\zeta^{(i)}_{\alpha} = \int_{\tilde {\boldsymbol \delta}_\alpha^{(i)}} \Omega_3,
\end{equation} 
where the Calabi-Yau three-form is defined in equation (\ref{Eq:Holo3FormZ2Z2}) and the three-cycle $\tilde{\boldsymbol  \delta}_\alpha^{(i)}$ is an $\OR$-odd linear combination of the exceptional three-cycles $({\boldsymbol  \epsilon}_\alpha^{(i)},\tilde{\boldsymbol  \epsilon}_\alpha^{(i)})$ in line with table~\ref{tab:Z2Z6OrientifoldExceptionalCycles}.
\item
The D6-brane $a$ couples to the $\Z_2^{(i)}$-twisted deformation modulus $\zeta^{(i)}_{\alpha}$ and stays {\it sLag} for arbitrary values of $\langle \zeta^{(i)}_{\alpha} \rangle \sim \sqrt{\varepsilon^{(i)}_{\alpha}}$.  In this case the deformation modulus $\zeta^{(i)}_{\alpha}$ corresponds to the period associated to an $\OR$-even linear combination of the  exceptional three-cycles $({\boldsymbol  \epsilon}_\alpha^{(i)},\tilde{\boldsymbol  \epsilon}_\alpha^{(i)})$ following table~\ref{tab:Z2Z6OrientifoldExceptionalCycles}. 
\item
The D6-brane $a$ does not couple directly to the $\Z_2^{(i)}$-twisted deformation modulus $\zeta^{(i)}_{\alpha}$.
\end{enumerate}
In the first case, it is argued that - from a low-energy field theory point of view - the stack of $N_a$ D$6_a$-branes supports a $U(N_a)$ gauge group, and the $U(1)_a$ factor within accounts for a D-term potential with Fayet-Iliopoulos term $D_a \propto \sqrt{\varepsilon^{(i)}_{\alpha}}$, whose numerical prefactor is fixed by the associated orientifold-odd combination of the exceptional wrapping numbers $(x^{(i)}_{\alpha,a}, y^{(i)}_{\alpha,a})$, cf. table~\ref{tab:DefParZ2Z6Complete} for $T^6/(\Z_2 \times \Z_6 \times \OR)$ with discrete torsion.\footnote{
As discussed in the previous section fractional three-cycles loose their {\it sLag} property under deformation, when one of its resolved exceptional three-cycles is no longer calibrated with respect to the same volume-form as the bulk three-cycle (and thereby also the orientifold fixed planes). According to equation~(\ref{Eq:V-scalar}) this results in a positive contribution to the total tension of the D6-branes and O6-planes, upon pursuing the dimensional reduction of the corresponding DBI-actions. Extracting the functional dependence of the FI-parameter on the deformation parameter then follows by computing the volume of the D6-brane on the resolved background and subtracting the tension of the O6-planes. In case the O6-planes are calibrated w.r.t.~${\rm Re}(\Omega_3)$, the positive contribution to the NS-NS scalar potential will scale as $\left( \int_{\Pi_a} {\rm Im}(\Omega_3) \right)^2$ for small deformations, which is understood from a four-dimensional perspective as (the square of) a FI-parameter~\cite{Blumenhagen:2002wn,Blumenhagen:2006ci,Ibanez:2012zz}.}
The appearance of the Fayet-Iliopoulos D-term leads to the stabilisation of the deformation modulus at the singular orbifold point, i.e. $\langle \zeta^{(i)}_{\alpha} \rangle =0$.\footnote{There exists in principle the possibility that
the {\it vev}s of charged scalars belonging to a vector-like pair in the bifundamental representation compensate the Fayet-Iliopoulos term $\langle \zeta^{(i)}_{\alpha} \rangle$, but due to the 
 form of the scalar potential ${\cal V}_{\text{scalar}} \sim \sum_z D_z^2$, such a {\it vev} can only atone for 
 the stabilisation of the deformation modulus if the two gauge factors have equal rank, i.e.~instead 
 a gauge symmetry breaking $SU(N) \times SU(N) \stackrel{\langle(\N,\ov{\N})\rangle \text{ or } \langle(\N,\N)\rangle \neq 0}{\longrightarrow} SU(N)_{\text{diag}}$
 occurs~\cite{Blaszczyk:2015oia}. \label{Footnote:SUNSUNFI}}
 \newline
In the second case, the D$6_a$-brane stack has only an orientifold-even exceptional wrapping number,  whereas in the last case, both orientifold-even and -odd wrapping numbers of the associated exceptional cycle vanish.
The scalar potential~\eqref{Eq:V-scalar} 
possesses a flat direction in the deformation modulus $\zeta^{(i)}_{\alpha}$ if for {\it all} D6-brane stacks in a given global model the second or third case applies,
as we will first demonstrate further in a global model with $USp(2)^4$ gauge group in section~\ref{Ss:ExUSp2-4} .

The four-dimensional gauge couplings at tree-level are obtained by dimensionally reducing the (6+1)-dimensional Dirac-Born-Infeld (DBI) action along 
the compact cycle $\Pi_a$ and its orientifold image $\Pi_{a'}$ of the D$6_a$-brane worldvolume~\cite{Klebanov:2003my,Blumenhagen:2003jy,Honecker:2011sm},
\begin{equation}\label{Eq:g2-tree}
\frac{4\pi}{g_{a,\text{tree}}^2} = \frac{1}{16 \sqrt{2} k_a} \frac{M_{\text{Planck}}}{M_{\text{string}}}  \; 
\frac{\text{Vol}(\Pi_a + \Pi_{a'})}{\sqrt{\text{Vol}_6}} \quad
\text{and } k_a = \left\{\begin{array}{cc}
1 & SU(N_a) \\ 2 & SO/USp(2N_a)
\end{array}\right.
,
\end{equation}
with the toroidal three-cycle volume at the orbifold point of the {\bf aAA} lattice of \mbox{$T^6/(\Z_2 \times \Z_6 \times \OR)$} given by:
\begin{equation}\label{Eq:Vol-Orb}
\begin{aligned}
\frac{\text{Vol}(\Pi_a)}{\sqrt{\text{Vol}_6}}\Big|_{\text{orb}} =
\frac{\text{Vol}(\Pi_{a'})}{\sqrt{\text{Vol}_6}} \Big|_{\text{orb}} =& \prod_{i=1}^3 L_a^{(i)} = 
 \sqrt{\frac{R_1^{(1)}}{R_2^{(1)}} (n^1)^2 + \frac{R_2^{(1)}}{R_1^{(1)}} (m^1)^2} \prod_{i=2}^3 \sqrt{ \frac{2}{\sqrt{3}} \left((n^i)^2 + n^i m^i + (m^i)^2 \right) }
\\
=& 
 \sqrt{\frac{R_1^{(1)}}{R_2^{(1)}}}
\times  \left\{\begin{array}{cl}
\frac{2}{\sqrt{3}} & a || \OR
\\
2 \sqrt{3} & a || \OR\Z_2^{(1)}
\end{array}\right.
,
\end{aligned}
\end{equation}
where $R_i^{(1)}$ are the radii associated to the one-cycles $\pi_{i=1,2}$ of the {\bf a}-type torus $T^2_{(1)}$ and $r_j$ the length scales of the hexagonal {\bf A}-type tori $T^2_{(j), j\in \{2,3\}}$ appearing in the definition of the two-torus volumes in equation~\eqref{Eq:2-torus-volume}.
\\
In the second case, the flat direction affects the strength of the gauge coupling, whereas in the third case the gauge coupling only feels the flat direction via (higher-order 
and non-perturbative) backreactions of the deformation on the toroidal cycles.
\\
In sections~\ref{Ss:ExMSSM},~\ref{Ss:ExLRSM} and~\ref{Ss:ExPS}, we will present how the three different cases appear in several globally consistent protoptype MSSM, L-R symmetric and PS-models, respectively.

For the $T^6/(\Z_2 \times \Z_6 \times \OR)$ orientifold with discrete torsion, {\bf aAA} lattice background and the choice of $\OR\Z_2^{(3)}$ as the exotic O6-plane orbit, 
the orientifold-even and -odd wrapping numbers  are obtained from
\begin{equation}\label{Eq:Z2Z6FractCyclesEven}
\mbox{\resizebox{0.92\textwidth}{!}{
$
\begin{aligned}\hspace{-4mm}
\Pi^{\rm frac}_a + \Pi^{\rm frac}_{a'} =& \frac{1}{4} \left({\color{blue} [2P_a+Q_a]} \rho_1 + {\color{blue} V_a} [-\rho_3 +2\rho_4]\right)
+ \frac{1}{4} \sum_{\alpha=0}^3 {\color{blue} 2 \, y ^{(1)}_{\alpha, a}} \, \tilde{\boldsymbol \epsilon}_\alpha^{(1)}
+ \frac{1}{4} \left({\color{blue} [x^{(1)}_{4, a} - x^{(1)}_{5, a}]} [{\boldsymbol \epsilon}_4^{(1)} - {\boldsymbol \epsilon}_5^{(1)}] 
+ {\color{blue} [y^{(1)}_{4, a} + y^{(1)}_{5, a}]} [\tilde{\boldsymbol \epsilon}_4^{(1)} + \tilde{\boldsymbol \epsilon}_5^{(1)}]
\right)
\\
& +\frac{1}{4}  \sum_{\alpha=1}^4 {\color{blue} y ^{(2)}_{\alpha, a}} \,[  2\,  \tilde{\boldsymbol \epsilon}_\alpha^{(2)} -{\boldsymbol \epsilon}_\alpha^{(2)}]
+ \frac{1}{4}  \sum_{\alpha=1}^4 {\color{blue} \left( 2\, x^{(3)}_{\alpha, a} + y ^{(3)}_{\alpha, a}\right)} {\boldsymbol \epsilon}_\alpha^{(3)} 
\end{aligned}
$}}
,
\end{equation}
and
\begin{equation}\label{Eq:Z2Z6FractCyclesOdd}
\mbox{\resizebox{0.92\textwidth}{!}{
$
\begin{aligned}\hspace{-4mm}
\Pi^{\rm frac}_a - \Pi^{\rm frac}_{a'} =&\frac{1}{4} \left({\color{red}  Q_a} [-\rho_1 + 2 \rho_2] + {\color{red} [2 U_a + V_a ]} \rho_3\right)
+ \frac{1}{4} \sum_{\alpha=0}^3 {\color{red} 2 \, x^{(1)}_{\alpha, a}} \,    {\boldsymbol \epsilon}_\alpha^{(1)}
+ \frac{1}{4} \left(   
{\color{red} [y^{(1)}_{4, a} - y^{(1)}_{5, a}] }
[\tilde{\boldsymbol \epsilon}_4^{(1)} - \tilde{\boldsymbol \epsilon}_5^{(1)}] 
+
{\color{red} [x^{(1)}_{4, a} + x^{(1)}_{5, a}] }
[{\boldsymbol \epsilon}_4^{(1)} + {\boldsymbol \epsilon}_5^{(1)}] 
\right)
\\
& +\frac{1}{4}  \sum_{\alpha=1}^4 {\color{red} \left( 2 \,x^{(2)}_{\alpha, a} + y ^{(2)}_{\alpha, a} \right)} {\boldsymbol \epsilon}_\alpha^{(2)}
+ \frac{1}{4}  \sum_{\alpha=1}^4 {\color{red} y ^{(3)}_{\alpha, a} }\left[ 2 \,  \tilde{\boldsymbol \epsilon}_\alpha^{(3)} - {\boldsymbol \epsilon}_\alpha^{(3)} 
\right]
\end{aligned}
$}}
,
\end{equation}
where $(P_a,Q_a,U_a,V_a)$ correspond to the four different products of toroidal wrapping numbers appearing in equation~\eqref{Eq:Bulk-Wrapping} and $(x^{(i)}_{\alpha,a}, y^{(i)}_{\alpha,a})$ to the exceptional wrapping numbers in equation~\eqref{Eq:Pi-frac}.
The latter have also been associated to the deformation parameters $\varepsilon^{(i)}_{\alpha}$ in table~\ref{tab:DefParZ2Z6Complete}.

At the singular orbifold point, only {\it relative} $\Z_2 \times \Z_2$ eigenvalues among pairs of D6-brane stacks $a$ and $b$ are of physical importance; e.g.
for the counting of chiral fermions the intersection numbers \mbox{$\Pi_a^{\Z_2^{(i)}} \circ \Pi_b^{\Z_2^{(i)}} \propto (-)^{\tau^{\Z_2^{(i)}}_a + \tau^{\Z_2^{(i)}}_b}$}
need to be computed. 
For illustrative purposes, we choose here the D$6_a$-brane stack supporting the strong interaction $SU(3)_a$ (or $SU(4)_a$ for PS models)
 to have $(\Z_2^{(1)}, \Z_2^{(2)}, \Z_2^{(3)})$ eigenvalues $(+++)$.
The details of a prototype MSSM example, four different L-R symmetric examples and two PS examples are provided in sections~\ref{Ss:ExMSSM}, \ref{Ss:ExLRSM} and~\ref{Ss:ExPS}, respectively. 
The D6-brane models presented in these sections form prototypes for the respective supersymmetric extensions of the Standard Model with three chiral generations of quarks and leptons and a minimal amount of (chiral) exotic matter (in the adjoint, symmetric and/or antisymmetric representation under the strong and weak interactions) constructed with $\varrho$-independent - i.e. all D6-branes along the one-cycle $\pi_1$ of $T^2_{(1)}$ -  supersymmetric {\it sLags} at the orbifold point of $T^6/(\Z_2 \times \Z_6 \times \OR)$ with discrete torsion. The discrete parameter choice in the D6-brane configurations is not unique, as different choices for the discrete Wilson lines, displacements, or $\Z_2^{(i)}$ eigenvalues may also yield global D6-brane models satisfying the same stringent conditions as the prototype D6-brane configurations. Nonetheless, such a global D6-brane model (with different discrete parameters) will always be characterised by the same massless (open and closed) string spectrum as one of the prototype D6-brane models. In other words, all global intersecting D6-brane models with three chiral generations of quarks and leptons and a minimal amount of exotic matter constructed with $\varrho$-independent supersymmetric {\it sLags} at the orbifold point will be classifiable by the massless string spectra of one of the prototype models considered here.

At this point, we anticipate that the bulk cycles of all D6-branes in the prototype models are either parallel to the $\OR$- or the $\OR\Z_2^{(1)}$-invariant (orbit of) O6-plane(s), and orientifold image branes
- such as $a$ and $a'$- thus only differ in their $\Z_2 \times \Z_2$ eigenvalues, with the precise relation depending on the discrete Wilson lines and displacements $(\vec{\tau},\vec{\sigma})$, 
see table~\ref{Tab:Ex-Z2-eigenvalues} for a compact summary.
\begin{table}
\begin{equation*}\hspace{-15mm}
\begin{array}{|c|c||c|c||c|}\hline
\muc{5}{|c|}{\text{\bf $\Z_2 \times \Z_2$ eigenvalues of D6-branes \& couplings to $\Z_2^{(i)}$ twisted sectors}}
\\\hline\hline
\text{brane stack } z & (\Z_2^{(1)},\Z_2^{(2)},\Z_2^{(3)}) & 
\OR\text{-image } z' &   (\Z_2^{(1)},\Z_2^{(2)},\Z_2^{(3)}) &
\begin{array}{c}
 \Pi_z+\Pi_{z'} \sim 
 \\(\Pi^{\Z_2^{(1)}},\Pi^{\Z_2^{(2)}},\Pi^{\Z_2^{(3)}}) \end{array}
\\\hline\hline
a_{\text{\bf all}}, h_{\text{\bf PS-II}} & (+++) 
& a_{\text{\bf all}}^{\prime} , h_{\text{\bf PS-II}}^{\prime} & (-+-)
& (\emptyset , \ast, \emptyset)
\\
h_{1,2}^{\text{\bf L-R II}}, h_{1,2}^{\text{\bf L-R IIb},\prime} & (+\pm\pm)
& h_{1,2}^{\text{\bf L-R II},\prime},h_{1,2}^{\text{\bf L-R IIb}} & (-\pm \mp)
& (\emptyset , \ast, \emptyset)
\\
d_{\text{\bf L-R}}, h_{\text{\bf MSSM}}^{\prime}, h_{\text{\bf PS-I}}^{\prime}   & (+--)
& d_{\text{\bf L-R}}^{\prime}, h_{\text{\bf MSSM}}, h_{\text{\bf PS-I}}  & (--+)
& (\emptyset , \ast, \emptyset)
\\\hline
c_{\text{\bf MSSM}} & (-+-)
& c_{\text{\bf MSSM}}^{\prime} & (+--)
& (\emptyset, \emptyset, \ast)
\\
h_{1,2}^{\text{\bf L-R I}} & (+\pm\pm)
& h_{1,2}^{\text{\bf L-R I},\prime} & (-\mp\pm)
& (\emptyset, \emptyset, \ast)
\\\hline
d_{\text{\bf MSSM}} , h_{1,2}^{\text{\bf L-R IIc},\prime}  & (+--)
& d_{\text{\bf MSSM}}^{\prime} ,h_{1,2}^{\text{\bf L-R IIc}} & (+++)
& (\ast, \emptyset, \emptyset)
\\\hline\hline
b_{\targ2{MSSM \, \& \, PS-I}{L-R \, \& \, PS-II}}
& (\mp \mp +) & \muc{2}{|c|}{} & (\ast,\ast,\ast)
\\\hline
c_{\targ2{L-R \, \& \, PS-II}{PS-I}}
& ( \mp  \pm-) &  \muc{2}{|c|}{} & (\ast,\ast,\ast)
\\\hline
\end{array}
\end{equation*}
\caption{$\Z_2 \times \Z_2$ eigenvalues of D6-brane stacks $z$ and their orientifold images $z'$ 
in the phenomenologically appealing MSSM, L-R symmetric and PS models with 
D6-brane configurations displayed in tables~\ref{tab:5stackMSSMaAAPrototypeI},~\ref{tab:6stackLRSaAA} and~\ref{tab:PatiSalamaAAPrototypeI} \&~\ref{tab:PatiSalamaAAPrototypeII}, respectively,
together with the dependence of gauge couplings in equation~\protect\eqref{Eq:g2-tree} 
on twisted sectors, e.g.~via $\Pi_{a_{\text{all}}} +\Pi_{a_{\text{all}}'} = \frac{1}{2} (\Pi_a^{\text{bulk}} + \Pi_a^{\Z_2^{(2)}})$.
The $USp(2)_{b,c}$ factors experience flat directions in all three $\Z_2$-twisted sectors and are thus sensitive to all {\it absolute} choices of $\Z_2 \times \Z_2$ eigenvalues upon deformations, whereas each $U(N_z)$ gauge factor is only sensitive to flat directions in one $\Z_2^{(i)}$-twisted sector; e.g.~the gauge coupling of the D$6_a$-stack is only affected by the $\Z_2^{(2)}$ eigenvalue.
}
\label{Tab:Ex-Z2-eigenvalues}
\end{table}
Further anticipating 
the decomposition of bulk and exceptional wrapping numbers into orientifold-even and -odd parts as summarised in table~\ref{Tab:Ex-Wrappings},
\begin{table}
{\footnotesize
\begin{equation*}\hspace{-20mm}
\begin{array}{|c||c|c|c|c|c|c|c|c|c|c|c|c||c|}\hline
\text{brane }z & \!\!\!\!\begin{array}{c} a_{\text{\bf all}} \\ d_{\text{\bf L-R}}  \end{array}\!\!\!\!
& \!\!\!\!\begin{array}{c} b_{\targ2{MSSM}{\& PS-I}} \\ b_{\targ2{L-R}{\& PS-II}} \end{array}\!\!\!\! & c_{\text{\bf MSSM}} & 
 \!\!\!\!\begin{array}{c}c_{\targ2{L-R \, \&}{PS-II}} \\ c_{\text{\bf PS-I}}  \end{array}\!\!\!\!  & d_{\text{\bf MSSM}} 
 &  \!\!\!\!\begin{array}{c} h_{\text{\bf MSSM}} \\ h_{\text{\bf PS-I}} \end{array}\!\!\!\!   & h_{\text{\bf PS-II}}
& \!\!\!\!\begin{array}{c} h_1^{\text{\bf L-R I}}  \\ h_2^{\text{\bf L-R I}} \end{array}\!\!\!\! 
& \!\!\!\!\begin{array}{c}  h_1^{\text{\bf L-R II}}  \\ h_2^{\text{\bf L-R II}} \end{array}\!\!\!\! 
& \!\!\!\!\begin{array}{c}  h_1^{\text{\bf L-R IIb}}  \\ h_2^{\text{\bf L-R IIb}} \end{array}\!\!\!\! 
& \!\!\!\!\begin{array}{c}  h_1^{\text{\bf L-R IIc}}  \\ h_2^{\text{\bf L-R IIc}} \end{array}\!\!\!\! 
\\\hline\hline
\muc{12}{|c|}{\text{\bf Orientifold-even wrapping numbers}}
\\\hline\hline
{\color{blue} [2P_z+Q_z]} & 2 & 6  & \muc{2}{|c|}{ 6 } & 6 
& 2 
& 6 & 2 & 6 & 6 &  6 
\\\hline
& \muc{11}{|c|}{ {\color{blue} V_z}=0}
\\\hline\hline
& \muc{11}{|c|}{ {\color{blue} y ^{(1)}_{\alpha \in \{0,1,2,3\}, z}}=0= {\color{blue} [y^{(1)}_{4, z} + y^{(1)}_{5, z}]}}
\\\hline
  {\color{blue} [x^{(1)}_{4, z} - x^{(1)}_{5, z}]} & 0 & \mp 2 & 0 & \mp 2 & 2 & 0  & 0 & 0 & 0 &  0 & \pm 2  
 \\\hline\hline
    {\color{blue} y ^{(2)}_{1, z} = y ^{(2)}_{2, z}} & \mp 2 & \mp 2 & 0 & \pm 2 & 0 & 2 & 2 & 0 & \pm 2 & 0 & 0  
    \\\hline
{\color{blue} y ^{(2)}_{3, z} = y ^{(2)}_{4, z} }& \muc{9}{|c|}{ 0 }&  \pm 2  & 0 
\\\hline\hline
 \!\!\!\!\targ2{\color{blue} [ 2\, x^{(3)}_{1, z} + y ^{(3)}_{1, z} ] }{={\color{blue} [ 2\, x^{(3)}_{2, z} + y ^{(3)}_{2, z}]}} \!\!\!\!
& 0 & -6 & \muc{2}{|c|}{6}  & 0 & 0 & 0 & \pm 2  & 0 &  0 & 0  
\\\hline
& \muc{11}{|c|}{ {\color{blue} [ 2\, x^{(3)}_{3, z} + y ^{(3)}_{3, z}]} =0 =  {\color{blue}  [ 2\, x^{(3)}_{4, z} + y ^{(3)}_{4, z}]}
}
\\\hline\hline
\muc{12}{|c|}{\text{\bf Orientifold-odd wrapping numbers}}
\\\hline\hline
& \muc{11}{|c|}{ {\color{red}  Q_z}  =0= {\color{red} [2 U_z + V_z ]} }
\\\hline\hline
{\color{red} x^{(1)}_{\alpha\in\{0,1,2\}, z}} & \muc{6}{|c|}{ 0} & 1 & 1  & 1  & -1 & 0
\\\hline
{\color{red} x^{(1)}_{3, z}} & -2 & 0 & 2 & 0 & 0 &  2 & 1 & 1  & 1  & -1 & 0
\\\hline
{\color{red} [x^{(1)}_{4, z} + x^{(1)}_{5, z}] } & 2 & 0 & -2 & 0 & 0 &  -2 & \muc{5}{|c|}{ 0 }
\\\hline
&  \muc{11}{|c|}{  {\color{red} [y^{(1)}_{4, z} - y^{(1)}_{5, z}] }   =0  }
\\\hline\hline
\targ2{\color{red} [ 2 \,x^{(2)}_{1, z} + y ^{(2)}_{1, z}]}{ ={\color{red} [ 2 \,x^{(2)}_{2, z} + y ^{(2)}_{2, z}]} } 
& 0 & 0 & 6 & 0 & -6 & 0 & 0  & \pm 2  &  0 &  0 & \targ2{6}{ 0} 
\\\hline
\targ2{\color{red} [ 2 \,x^{(2)}_{3, z} + y ^{(2)}_{3, z} ] }{= {\color{red} [ 2 \,x^{(2)}_{4, z} + y ^{(2)}_{4, z} ]} } & \muc{10}{|c|}{  0 }   & \targ2{0}{6}
\\\hline\hline
 {\color{red} y ^{(3)}_{1, z} } = {\color{red} y ^{(3)}_{2, z} } 
& -2 & 0 & 0 & 0 & 2  & -2 & -2 &  0  & \mp 2  &  0 & \targ2{-2}{0}
\\\hline
{\color{red} y ^{(3)}_{3, z} }   = {\color{red} y ^{(3)}_{4, z} }& \muc{9}{|c|}{   0  }  & \pm 2  & \targ2{0}{2} 
\\\hline
\end{array}
\end{equation*}
}
\caption{Orientifold-even and -odd wrapping numbers of the phenomenologically appealing global MSSM, L-R symmetric
and PS models on $T^6/(\Z_2 \times \Z_6 \times \OR)$ with discrete torsion discussed in sections~\ref{Ss:ExMSSM},~\ref{Ss:ExLRSM} and~\ref{Ss:ExPS}, respectively.
A non-vanishing orientifold-odd wrapping number stabilises the associated deformation modulus according to
table~\protect\ref{tab:DefParZ2Z6Complete}. If some orientifold-odd wrapping number vanishes, but the orientifold-even number with identical fixed point labels
is non-zero, the associated tree-level gauge coupling  in equation~\protect\eqref{Eq:g2-tree} experiences a flat direction in the deformation modulus, see 
table~\protect\ref{Tab:Stab-vs-Flat} for details.
}
\label{Tab:Ex-Wrappings}
\end{table}
we can summarise the counting of stabilised moduli per model as well as flat directions affecting the gauge couplings according to the relation~\eqref{Eq:g2-tree}
in table~\ref{Tab:Stab-vs-Flat}.
\begin{table}
\begin{equation*}\hspace{-15mm}
\begin{array}{|c||c||c|c|c|c||c|c||c|c||c|c||c|c|c|}\hline
\muc{11}{|c|}{\text{\bf Counting of stabilised complex structure moduli \& flat directions with $g^{-2}_{\text{D6}_{x \in \{a,b,c,d,h_{1,2}\} }}$ dependence}}
\\\hline\hline
& \varrho
& \varepsilon^{(1)}_{0,1,2} & \varepsilon^{(1)}_{3} & \varepsilon^{(1)}_{4+5} & \varepsilon^{(1)}_{4-5} 
& \varepsilon^{(2)}_{1,2} & \varepsilon^{(2)}_{3,4}
& \varepsilon^{(3)}_{1,2} & \varepsilon^{(3)}_{3,4}
& \#_\text{stab}^{\text{max}}
\\\hline\hline
{\bf MSSM} &  \muc{2}{|c|}{\text{none}} &\muc{2}{|c|}{ a,c,h } & [b,d]_{\text{flat}} & c,d & \text{none} & a,d,h & \text{none}
& 6 
\\\hline\hline
\text{\bf L-R I}  &  \text{none} & h_{1,2} & a,d,h_{1,2} & a,d & [b,c]_{\text{flat}} & h_{1,2} & \text{none} & a,d & \text{none}
& 9
\\\hline
\text{\bf L-R II} &  \text{none} & h_{1,2} & a,d,h_{1,2} &  a,d &  [b,c]_{\text{flat}}& [a,b,c,d,h_{1,2}]_{\text{flat}} & \text{none} & a,d, h_{1,2} & \text{none}
& 7
\\\hline
\text{\bf L-R IIb}  &  \text{none} & h_{1,2} & a,d,h_{1,2} & a,d &  [b,c]_{\text{flat}} & [a,b,c,d]_{\text{flat}} &  [h_{1,2}]_{\text{flat}} & a,d & h_{1,2}
& 9
\\\hline
\text{\bf L-R IIc}  &   \muc{2}{|c|}{\text{none}} & a,d & a,d &  [b,c,h_{1,2}]_{\text{flat}} & h_1 & h_2 & a,d, h_1 & h_2
& 10
\\\hline\hline
\text{\bf PS I} &  \muc{2}{|c|}{\text{none}} &\muc{2}{|c|}{a,h} & [b,c]_{\text{flat}} & [a,b,c,h]_{\text{flat}} & \text{none} & a,h &  \text{none}
& 4
\\\hline
\text{\bf PS II}  &  \text{none}& h & a,h & a & [b,c]_{\text{flat}} & [a,b,c,h]_{\text{flat}} & \text{none} &  a,h & \text{none} 
& 7
\\\hline
\end{array}
\end{equation*}
\caption{Counting of stabilised deformation moduli according to the orientifold-odd wrapping numbers displayed in table~\protect\ref{Tab:Ex-Wrappings}.
The L-R symmetric model of prototype IIc presented in section~\protect\ref{Ss:ExLRSM} is expected to have the maximal number of ten out of $14_{\Z_2} (+4_{\Z_{3,6}})$ twisted complex structure moduli 
 stabilised at the singular orbifold point.
The tree-level gauge couplings of the gauge factors $USp(2)_b \times USp(2)_c \times \prod_{i=1}^2 U(1)_{h_i}^{\text{massive}}$ in this L-R symmetric example IIc
experience a flat direction in the 
deformation parameterised by $\varepsilon^{(1)}_{4-5}$ with the prefactor fixed by the orientifold-even wrapping number ${\color{blue} [x^{(1)}_{4,z} - x^{(1)}_{5,z}]_{z \in \{b,c,h_1,h_2 \}}}
\in \{2, -2\}$ in table~\protect\ref{Tab:Ex-Wrappings}.
}
\label{Tab:Stab-vs-Flat}
\end{table}

Details of deformations of the global MSSM, L-R symmetric and PS models will be discussed in sections~\ref{Ss:ExMSSM},~\ref{Ss:ExLRSM} and~\ref{Ss:ExPS}, respectively.

\clearpage
\subsection{Deformations in a global $USp(2)^4$ model}\label{Ss:ExUSp2-4}

Before discussing phenomenologically interesting particle physics models on D6-branes with all their intricacies of stabilised moduli, here we first 
consider deformations in a global model with $USp(2)^4$ gauge group, which {\it a priori} is expected to contain only flat supersymmetric directions. To obtain gauge group enhancement 
on all D6-brane stacks and consistency with the bulk RR tadpole cancellation conditions~\cite{Forste:2010gw,Ecker:2014hma} for the choice of the $\OR\Z_2^{(3)}$-orbit
as exotic O6-plane, the bulk part of each fractional three-cycle has to be either parallel to the $\OR$- or the $\OR\Z_2^{(1)}$-invariant orbit.
In either case, only gauge group enhancement of the type $U(N) \hookrightarrow USp(2N)$ occurs for arbitrary choices
of $\Z_2 \times \Z_2$ eigenvalues and arbitrary discrete Wilson line and displacement parameters  $(\tau^1,\sigma^1)$ on the {\bf a}-type torus $T^2_{(1)}$, see table~10 of~\cite{Ecker:2014hma} for details.
With the discrete Wilson lines and displacements $(\targ2{\tau^2}{\tau^3};\targ2{\sigma^2}{\sigma^3})$ along $T^4_{(1)}$ listed as lower index,
such three-cycles have the form:
{\footnotesize
\begin{equation}
\begin{aligned}
{}\hspace{-20mm}
\Pi^{|| \OR}_{(\targ2{0}{1};\targ2{1}{1})} = & \frac{\rho_1}{4} 
+ \frac{(-1)^{\tau^{\Z_2^{(1)}}}}{4} \left({\boldsymbol \epsilon}^{(1)}_5  - {\boldsymbol \epsilon}^{(1)}_4  \right)
 + \frac{(-1)^{\tau^{\Z_2^{(2)}}}}{4}\left(  [ {\boldsymbol \epsilon}^{(2)}_{\kappa_1}  -2 \, \tilde{\boldsymbol \epsilon}^{(2)}_{\kappa_1} ]
+ (-1)^{\tau^1} \,  [ {\boldsymbol \epsilon}^{(2)}_{\kappa_2}  -2 \, \tilde{\boldsymbol \epsilon}^{(2)}_{\kappa_2} ]\right)
 - \frac{(-1)^{\tau^{\Z_2^{(3)}}}}{4}\left(  {\boldsymbol \epsilon}^{(3)}_{\kappa_1}   + (-1)^{\tau^1} \, {\boldsymbol \epsilon}^{(3)}_{\kappa_2}  
 \right)
 ,
\\
{}\hspace{-20mm}
\Pi^{|| \OR\Z_2^{(1)}}_{(\targ2{1}{\tau^3};\targ2{1}{0})} =&
\frac{3 \, \rho_1}{4} 
+ \frac{(-1)^{\tau^{\Z_2^{(1)}}+\tau^3}}{4} \left(  {\boldsymbol \epsilon}^{(1)}_4 - {\boldsymbol \epsilon}^{(1)}_5 \right)
-  \frac{(-1)^{\tau^{\Z_2^{(2)}}+\tau^3}}{4} \left(
 [{\boldsymbol \epsilon}^{(2)}_{\kappa_1} - 2 \tilde{\boldsymbol \epsilon}^{(2)}_{\kappa_1} ]
+ (-1)^{\tau^1} [{\boldsymbol \epsilon}^{(2)}_{\kappa_2} - 2 \tilde{\boldsymbol \epsilon}^{(2)}_{\kappa_2} ]
\right)
- \frac{3 \, (-1)^{\tau^{\Z_2^{(3)}}}}{4}\left(  {\boldsymbol \epsilon}^{(3)}_{\kappa_1} +(-1)^{\tau^1} \,  {\boldsymbol \epsilon}^{(3)}_{\kappa_2} \right)
,
\end{aligned}
\end{equation}
}
\noindent and the remaining two possibilities of gauge group enhancement, $\Pi^{|| \OR}_{(\targ2{\tau^2}{1};\targ2{0}{1})}$ with arbitary discrete Wilson line \mbox{$\tau^2 \in \{0,1\}$}
 and $\Pi^{|| \OR\Z_2^{(1)}}_{(\targ2{1}{0};\targ2{1}{1})}$, 
are obtained from those by choosing identical values of $(\vec{\tau};\vec{\sigma})$ along $(T^2)^3$ and replacing the $\Z_2 \times \Z_2$ eigenvalues
$(\vec{\tau}^{\Z_2})$ as follows:
\begin{equation}
\begin{aligned}
(\tau^{\Z_2^{(1)}}+\tau^2+1, \tau^{\Z_2^{(2)}} ,\tau^{\Z_2^{(3)}}+\tau^2+1 )^{|| \OR}_{(\targ2{\tau^2}{1};\targ2{0}{1})}
& \leftrightarrow (\vec{\tau}^{\Z_2})^{|| \OR}_{(\targ2{0}{1};\targ2{1}{1})}
,
 \\
(\tau^{\Z_2^{(1)}}+ \tau^3 + 1, \tau^{\Z_2^{(2)}} + \tau^3 +1 , \tau^{\Z_2^{(3)}} )^{|| \OR\Z_2^{(1)}}_{(\targ2{1}{\tau^3};\targ2{1}{0})} 
& \leftrightarrow (\vec{\tau}^{\Z_2})^{|| \OR\Z_2^{(1)}}_{(\targ2{1}{0};\targ2{1}{1})}
.
\end{aligned}
\end{equation}
$\Pi^{|| \OR}_{(\targ2{0}{1};\targ2{1}{1})}$ with the choice of $\Z_2 \times \Z_2$ eigenvalues $(+++)$ and discrete data $\tau^1=0=\sigma^1$ on $T^2_{(1)}$
only differs from the QCD stack $a$ of all phenomenologically appealing models discussed later on in this article
in the choice of one discrete Wilson line, $\tau^2$, cf.~e.g.~the MSSM D6-brane configuration in table~\ref{tab:5stackMSSMaAAPrototypeI}. 
$\Pi^{|| \OR\Z_2^{(1)}}_{(\targ2{1}{\tau^3};\targ2{1}{0})}$ with the choice $\tau^3=0$ on the other hand corresponds to the left- and right-symmetric
D6-branes $b$ and $c$, respectively, of all L-R symmetric and 
the PS prototype II model, cf. tables~\ref{tab:6stackLRSaAA} and~\ref{tab:PatiSalamaAAPrototypeII}, 
for the choice of $\Z_2 \times \Z_2$ eigenvalues $(+++)$ and $(-+-)$, respectively.
Moreover,  $\Pi^{|| \OR\Z_2^{(1)}}_{(\targ2{1}{\tau^3};\targ2{1}{0})}$ with the choice $\tau^3=0$ corresponds to
stack $b$ (and $c$) of the MSSM (and the PS prototype I), cf. table~\ref{tab:5stackMSSMaAAPrototypeI} (and~\ref{tab:PatiSalamaAAPrototypeI}),
for a different choice of $\Z_2 \times \Z_2$ eigenvalues $(--+)$ (and $(+--)$).

The D6-brane data of a global model with $USp(2)^4$ gauge group enhancement
satisfying all bulk and twisted RR tadpole cancellation conditions\footnote{Notice that the K-theory constraints are trivially satisfied in this model.} 
is presented in table~\ref{tab:4stackUSp2-4},
\mathtabfix{
\begin{array}{|c||c|c||c|c|c||c|}\hline 
\muc{7}{|c|}{\text{\bf D6-brane configuration of a global $USp(2)^4$ model on the {\bf aAA} lattice of }T^6/(\Z_2 \times \Z_6 \times \OR)_{\eta=-1}}
\\\hline \hline
&\text{\bf wrapping numbers} &\frac{\rm Angle}{\pi}&\text{\bf $\Z_2^{(i)}$ eigenvalues}  & (\vec \tau) & (\vec \sigma)& \text{\bf gauge group}\\
\hline \hline
 a &(1,0;1,0;1,0)&(0,0,0)& (+++)
 & (0,0,1) & (0,1,1) & USp(2)
\\
d  &(1,0;1,0;1,0)&(0,0,0)& (-+-)
& (0,0,1) & (0,1,1) & USp(2)
\\\hline
 b&(1,0;-1,2;1,-2)&(0,\frac{1}{2},-\frac{1}{2})&(+++)&(0,1,0) & (0,1,0)&USp(2)\\
 c&(1,0;-1,2;1,-2)&(0, \frac{1}{2},-\frac{1}{2})&(-+-)&(0,1,0) & (0,1,0)&USp(2)\\ 
 \hline
\end{array}
}{4stackUSp2-4}{D6-brane configuration of a global model on the {\bf aAA} lattice of $T^6/(\Z_2 \times \Z_6 \times \OR)$ with discrete torsion and 
with $\OR\Z_2^{(3)}$ as the exotic O6-plane orbit. The D6-branes $b$ and $c$ are chosen identical to all L-R symmetric models and the PS II model in tables~\ref{tab:6stackLRSaAA} and~\ref{tab:PatiSalamaAAPrototypeII}, respectively.}
and the non-vanishing bulk and exceptional wrapping numbers are displayed here in table~\ref{Tab:Ex-USp2-4-wrappings} for convenience and for comparison with those of the phenomenologically interesting models in table~\ref{Tab:Ex-Wrappings}.
\begin{table}
\begin{equation*}\hspace{-20mm}
\begin{array}{|c||c|c||c|c|}\hline
\muc{5}{|c|}{\text{\bf Wrapping numbers of a global $USp(2)^4$ model}}
\\\hline\hline
z & a & d & b & c 
\\\hline\hline
{\color{blue} [2P_z+Q_z]}  & 2 & 2 & 6 & 6 
\\\hline
 {\color{blue} [x^{(1)}_{4, z} - x^{(1)}_{5, z}]} & -2 & 2  & 2 & -2
 \\\hline
     {\color{blue} y ^{(2)}_{1, z} = y ^{(2)}_{2, z}} & -2 & -2 & 2 & 2 
   \\\hline
    {\color{blue} [ 2\, x^{(3)}_{1, z} + y ^{(3)}_{1, z} ] }{={\color{blue} [ 2\, x^{(3)}_{2, z} + y ^{(3)}_{2, z}]}} 
    & -2 & 2  & -6 & 6 
\\\hline
\end{array}
\end{equation*}
\caption{Non-vanishing wrapping numbers of the global $USp(2)^4$ model with D6-brane data specified in table~\protect\ref{tab:4stackUSp2-4}. The 
$\Z_2 \times \Z_2$ eigenvalues are chosen such that the D6-branes $b$ and $c$ agree with those of all L-R symmetric models and the PS II model, cf.
 table~\protect\ref{Tab:Ex-Wrappings}.}
\label{Tab:Ex-USp2-4-wrappings}
\end{table}
They are obtained from the expansion of the following fractional three-cycles:
\begin{equation}\label{Eq:USp2-4-model-cycles}
\begin{aligned}
\Pi_{a,d} =& \frac{\rho_1}{4} 
\mp  \frac{1}{4} \left({\boldsymbol \epsilon}^{(1)}_4  - {\boldsymbol \epsilon}^{(1)}_5  \right)
 + \frac{1}{4}\left(  [ {\boldsymbol \epsilon}^{(2)}_{1}  -2 \, \tilde{\boldsymbol \epsilon}^{(2)}_{1} ] +   [ {\boldsymbol \epsilon}^{(2)}_{2}  -2 \, \tilde{\boldsymbol \epsilon}^{(2)}_{2} ]\right)
 \mp \frac{1}{4}\left( {\boldsymbol \epsilon}^{(3)}_{1}  + {\boldsymbol \epsilon}^{(3)}_{2}  \right)
 ,
\\
\Pi_{b,c} =& \frac{3 \, \rho_1}{4} 
\pm \frac{1}{4} \left(  {\boldsymbol \epsilon}^{(1)}_4 - {\boldsymbol \epsilon}^{(1)}_5 \right)
 -  \frac{1}{4} \left( [{\boldsymbol \epsilon}^{(2)}_{1} - 2 \tilde{\boldsymbol \epsilon}^{(2)}_{1} ] +  [{\boldsymbol \epsilon}^{(2)}_{2} - 2 \tilde{\boldsymbol \epsilon}^{(2)}_{2} ] \right)
\mp \frac{3}{4}\left(  {\boldsymbol \epsilon}^{(3)}_{1} +   {\boldsymbol \epsilon}^{(3)}_{2} \right)
.
\end{aligned}
\end{equation}
The massless matter spectrum of this D6-brane configuration is displayed in table~\ref{Tab:Ex-USp2-4-spectrum} and agrees in the $bb$, $bc$ and $cc$ sectors by construction with 
that of the L-R symmetric models and the PS II model displayed in  table~\ref{tab:6stackLRSMaAAVisibleSpectrum}.
\begin{table}
{\footnotesize
\begin{equation*}
\begin{array}{|c|c||c|c|c|}\hline
\muc{4}{|c|}{\text{\bf Massless matter spectrum of a global $USp(2)^4$ model on the {\bf aAA} lattice of } T^6/(\Z_2 \times \Z_6 \times \OR)_{\eta=-1}}
\\\hline\hline
\text{sector} & USp(2)_a \times USp(2)_d \times USp(2)_b \times USp(2)_c & \text{sector} & USp(2)_a \times USp(2)_d \times USp(2)_b \times USp(2)_c
\\\hline\hline
aa & (\1_{\Anti},\1;\1,\1) 
& bb & 5 \times (\1,\1;\1_{\Anti},\1)
\\
dd &  (\1,\1_{\Anti};\1,\1) 
& cc & 5 \times (\1,\1;\1,\1_{\Anti})
\\
ad & 2 \times (\2,\2;\1,\1)
& bc & 10 \times (\1,\1;\2,\2)
\\\hline
ab & 2 \times (\2,\1;\2,\1)
& dc &  2 \times (\1,\2;\1,\2)
\\\hline
\end{array}
\end{equation*}
}
\caption{Massless matter spectrum of the global $USp(2)^4$ model with rigid D6-brane configuration specified in table~\protect\ref{tab:4stackUSp2-4}
on the {\bf aAA} lattice of $T^6/(\Z_2 \times \Z_6 \times \OR)$ with discrete torsion, which is chosen such that the $bb$, $bc$ and $cc$ sectors agree with those of the L-R symmetric models and the PS I model in 
table~\protect\ref{tab:6stackLRSMaAAVisibleSpectrum}. The $ac$ and $db$ sectors do not contain any massless state.}
\label{Tab:Ex-USp2-4-spectrum}
\end{table}

Here, it is noteworthy that at the orbifold point only the {\it relative} $\Z_2 \times \Z_2$ eigenvalues among the D6-branes $a$, $b$, $c$ and $d$ are of physical relevance.
The {\it absolute} $\Z_2 \times \Z_2$ eigenvalues in all three $\Z_2^{(i)}$-twisted sectors will, however, become important when switching on deformations since 
the gauge couplings scale with the cycle volume, $\nicefrac{1}{g^2_{z,\text{tree}}} \propto \text{Vol}(\Pi_z)$ for $\Pi_z = \Pi_{z'}$,
 and the change in $\text{Vol}(\Pi_z)$ is e.g. proportional to $y^{(2)}_{1,z} \cdot \sqrt{\varepsilon^{(2)}_1}$ with the {\it absolute} sign of $y^{(2)}_{1,z} =\pm 2$ specified in table~\ref{Tab:Ex-USp2-4-wrappings} for the choice of discrete D6-brane data in table~\ref{tab:4stackUSp2-4}.
Exchanging $(+++) \leftrightarrow (-+-)$ in the D6-brane configuration of  table~\ref{tab:4stackUSp2-4} will obviously only pairwise permute the D6-brane labels $a \leftrightarrow d$ and $b \leftrightarrow c$,
while the sign flip $\tari{(+++)}{(-+-)} \leftrightarrow \tari{(--+)}{(+--)}$ provides a physically distinct model once deformations along the directions of non-vanishing exceptional wrapping numbers
listed in table~\ref{Tab:Ex-USp2-4-wrappings} are switched on. The  D6-branes $b$ and $c$ in this latter case agree with those of the PS prototype I model specified in table~\ref{tab:PatiSalamaAAPrototypeI},
whose $b$-brane in turn agrees with the MSSM stack $b$ of the weak interactions in table~\ref{tab:5stackMSSMaAAPrototypeI}.

Since the non-vanishing wrapping numbers in  table~\ref{Tab:Ex-USp2-4-wrappings} are only of orientifold-even type - as expected for a model with just $USp$-gauge factors -
each gauge coupling feels flat directions along the untwisted complex structure $\varrho$ of the {\bf a}-type two-torus $T^2_{(1)}$ and the 
twisted complex strucuture moduli $\zeta^{(i)}_{\alpha}$ associated to the deformation parameters $\sqrt{\varepsilon^{(i)}_{\alpha}} \sim \langle \zeta ^{(i)}_{\alpha}\rangle$
with $(i,\alpha) \in \{(1, 4-5),(2,1),(2,2),(3,1),(3,2) \}$.
All other twisted complex structure moduli only affect the gauge couplings through higher-order, e.g. field redefinitions at loop level, or non-perturbative effects.

It is also important to stress that, while the $\{b,c\}$ sector of the $USp(2)^4$ model {\it locally} agrees with the L-R symmetric models in section~\ref{Ss:ExLRSM}
and the PS II model  in section~\ref{Ss:ExPS} (or the PS I model in section~\ref{Ss:ExPS} upon flipping the $\Z_2^{(1)} \times \Z_2^{(2)}$ eigenvalues), the {\it global} models differ significantly:
All phenomenogically appealing models contain at least two $U(N_1) \times U(N_2)$ gauge factors with $N_1 \neq N_2$. The QCD stack with either $U(3)_a$ or $U(4)_a$ 
gauge symmetry e.g.~couples by an orientifold-odd exceptional wrapping number to the two $\Z_2^{(3)}$-twisted deformation moduli $\zeta^{(3)}_{1,2}$, which leads to their stabilisation, i.e.
$\varepsilon^{(3)}_1 = 0 = \varepsilon^{(3)}_2$. Except for the L-R symmetric model of prototype I, the spectrum does not contain any other $U(3)$ gauge factor, and even in the L-R symmetric prototype I
 the two hidden stacks with $U(3)_{h_1} \times U(3)_{h_2}$ symmetry do not possess orientifold-odd wrapping numbers along the directions $\zeta^{(3)}_{1,2}$, as can be read off from table~\ref{Tab:Ex-Wrappings}. 
 In all phenomenologically appealing prototypes presented in this article, the Fayet-Iliopoulos term(s) generated by some non-vanishing deformation parameter(s) $\varepsilon^{(3)}_{1,2}$ can thus not be compensated by any supersymmetry preserving {\it vev} of some charged scalar matter field.
The same argument involving the QCD stack $a$ applies to two deformations, $\varepsilon^{(1)}_3$ and $\varepsilon^{(1)}_{4+5}$, from the $\Z_2^{(1)}$-twisted sector.
Further stabilisations of deformation moduli involving the remaining stacks with unitary gauge groups will be discussed below on a model-by-model basis.
As can be read off from the summary in table~\ref{Tab:Stab-vs-Flat}, all phenomenologically appealing examples share the property that the D6-brane $b$ of weak interactions -- and in all L-R symmetric and PS models also the right-symmetric D6-brane $c$ -- experience a flat direction in the deformation parameter $\varepsilon^{(1)}_{4-5}$ changing the value of the respective gauge coupling(s).

The wrapping numbers and discrete displacements of the D6-brane configuration in table~\ref{tab:4stackUSp2-4} indicate that the branes $a,d$ are represented by the three-cycle {\bf aI}$\times${\bf bIII}$^0\times${\bf bIII}$^0$ and the branes $b,c$ by the three-cycle {\bf aI}$\times${\bf bIV}$^0\times${\bf bII}$^0$ in the hypersurface formalism, following the dictionary at the singular orbifold point in section~\ref{Sss:sLagsOrbi}. The deformation $\varepsilon_{4-5}^{(1)}$ in the $\Z_2^{(1)}$-twisted sector allows for resolved exceptional three-cycles calibrated with respect to the same ${\rm Re}(\Omega_3)$ as the bulk parts of the four fractional three-cycles in the parameter region $\varepsilon_{4-5}^{(1)} \leq 0$, and similarly the deformations $\varepsilon_{1,2}^{(i)}$ in the $\Z_2^{(i=2,3)}$-twisted sectors yield correctly calibrated (resolved) exceptional three-cycles in the parameter regions $\varepsilon_{1,2}^{(i)} \leq 0$. In both cases, this can be verified by applying the M\"obius transformation $\lambda_4$ on the cycles {\bf bIII}$^0$ and {\bf bIV}$^0$, after which one can study the ($\R$-projected) countour plots that form the equivalent to figures~\ref{Fig:DefZ21hypersurfaceEq} and~\ref{Fig:DefZ23hypersurfaceEq} in section~\ref{Ss:DefsLagS}. To obtain the functional behaviour of the fractional three-cycle volumes under the deformations, we first determine which bulk three-cycles and which exceptional cycles at $\Z_2^{(i)}$ singularities are wrapped by the fractional three-cycles in table~\ref{tab:4stackUSp2-4}, cf.~equation~\eqref{Eq:USp2-4-model-cycles}. Then, we reconstruct the (normalised) volume of the fractional three-cycles using the results for the exceptional three-cycle volumes from sections~\ref{Sss:sLagsZ21} and~\ref{Sss:sLagsZ23} as primary building blocks, in the same spirit as the method presented on page~\pageref{It:Method}. 
More precisely, we compute the fractional three-cycle volumes directly as the sum or difference of (normalised) bulk three-cycle volumes and (normalised) exceptional three-cycle volumes as computed in the aforementioned sections.

\begin{figure}[h]
\begin{center}
\begin{tabular}{c@{\hspace{1in}}c} 
\includegraphics[scale=0.62]{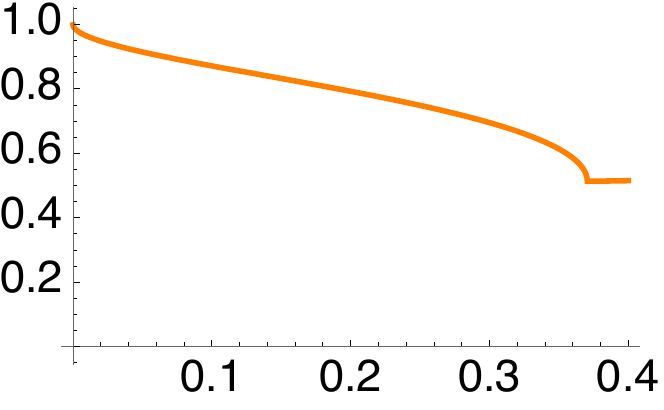} \begin{picture}(0,0) \put(0,0){$\big|\varepsilon_{4-5}^{(1)}\big|$} \put(-130,10){\begin{rotate}{90}Vol$_{\rm norm}(\Pi^{\rm frac}_a)$\end{rotate}} \end{picture} 
&
 \includegraphics[scale=0.62]{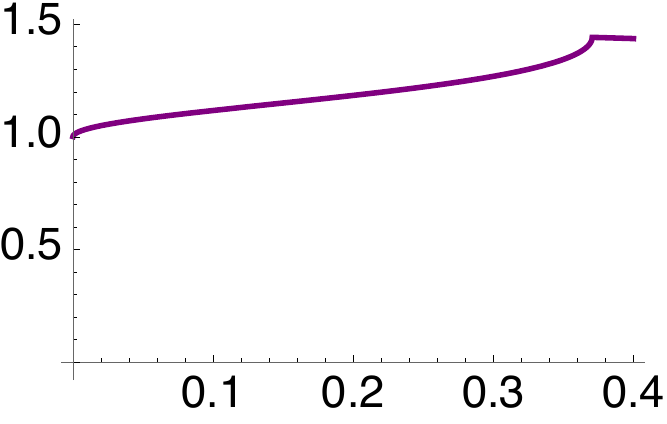} \begin{picture}(0,0) \put(0,0){$\big|\varepsilon_{4-5}^{(1)}\big|$} \put(-130,10){\begin{rotate}{90}Vol$_{\rm norm}(\Pi^{\rm frac}_d)$\end{rotate}} \end{picture}
\end{tabular}
\caption{Normalised volume of the fractional cycles
 $\Pi^{\text{frac}}_a =\frac{\rho_1 - \big( {\boldsymbol \epsilon}_4^{(1)}  - {\boldsymbol \epsilon}_5^{(1)} \big) }{4}$ and $\Pi^{\text{frac}}_d  =\frac{\rho_1 + \big( {\boldsymbol \epsilon}_4^{(1)}  - {\boldsymbol \epsilon}_5^{(1)} \big) }{4}$ 
as a function of the deformation parameter $\big|\varepsilon^{(1)}_{4-5}\big|$, with the bulk cycle $\Pi^{\text{bulk}}$ lying along the cycle ${\cal C}^0\times{\cal C}^0$ on $T^4_{(1)}/\Z_6$ as argued in section~\ref{Sss:sLagsZ21}. 
\newline
These plots  serve as templates for describing the fractional three-cycle volumes as a function of some deformation along the flat direction in the twisted complex structure modulus $\zeta_{4-5}^{(1)}$ in the subsequent subsections. In those cases, the plots will represent the normalised volume of a fractional three-cycle $\Pi_x + \Pi_{x'} =\frac{\rho_1 \pm \big( {\boldsymbol \epsilon}_4^{(1)}  - {\boldsymbol \epsilon}_5^{(1)} \big) }{2}$.
\label{Fig:DeformedCycleADToyZ1}}
\end{center}
\end{figure}
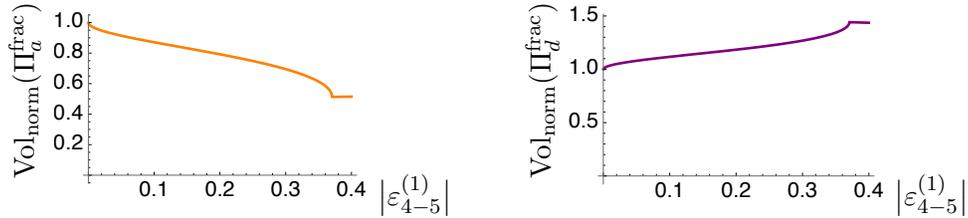

\begin{figure}[h]
\begin{center}
\begin{tabular}{c@{\hspace{1in}}c} 
\includegraphics[scale=0.62]{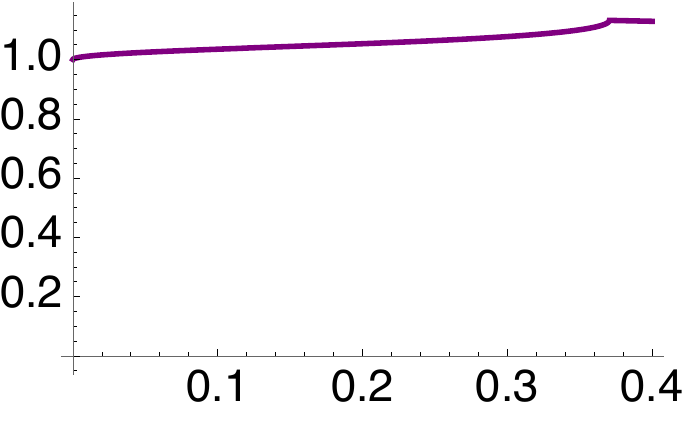} \begin{picture}(0,0) \put(0,0){$\big|\varepsilon_{4-5}^{(1)}\big|$} \put(-135,10){\begin{rotate}{90}Vol$_{\rm norm}(\Pi^{\rm frac}_b)$\end{rotate}} \end{picture} 
&
 \includegraphics[scale=0.62]{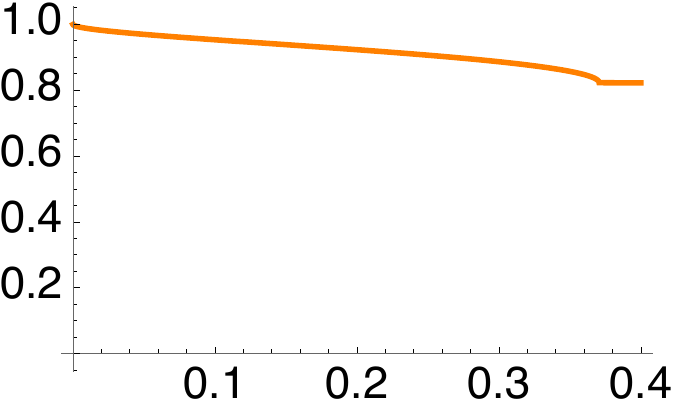} \begin{picture}(0,0) \put(0,0){$\big|\varepsilon_{4-5}^{(1)}\big|$} \put(-135,10){\begin{rotate}{90}Vol$_{\rm norm}(\Pi^{\rm frac}_c)$\end{rotate}} \end{picture}
\end{tabular}
\caption{Normalised volume of the fractional cycles
 $\Pi^{\text{frac}}_b =\frac{3\rho_1 + \big( {\boldsymbol \epsilon}_4^{(1)}  - {\boldsymbol \epsilon}_5^{(1)} \big) }{4}$ and $\Pi^{\text{frac}}_c  =\frac{3\rho_1 - \big( {\boldsymbol \epsilon}_4^{(1)}  - {\boldsymbol \epsilon}_5^{(1)} \big) }{4}$ 
as a function of the deformation parameter $\big|\varepsilon^{(1)}_{4-5}\big|$. 
\newline
These plots serve as templates for describing the fractional three-cycle volumes as a function of some deformation along the flat direction  in the twisted complex structure modulus $\zeta_{4-5}^{(1)}$  discussed in subsequent subsections. In those cases, the plots will represent the normalised volume of a fractional three-cycle $\Pi_x + \Pi_{x'} =\frac{3\rho_1 \pm \big( {\boldsymbol \epsilon}_4^{(1)}  - {\boldsymbol \epsilon}_5^{(1)} \big) }{2}$.
\label{Fig:DeformedCycleBCToyZ1}}
\end{center}
\end{figure}

\begin{figure}[h]
\begin{center}
\begin{tabular}{c@{\hspace{0.8in}}c@{\hspace{0.8in}}c} 
\hspace*{-0.5in}\includegraphics[scale=0.62]{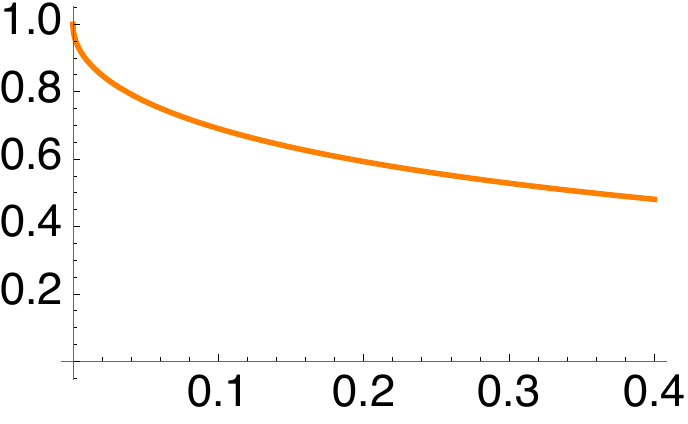} \begin{picture}(0,0) \put(0,0){$\big|\varepsilon_{2}^{(2)}\big|$} \put(-135,10){\begin{rotate}{90}Vol$_{\rm norm}(\Pi^{\rm frac}_{a|d})$\end{rotate}} \end{picture} 
&
 \includegraphics[scale=0.62]{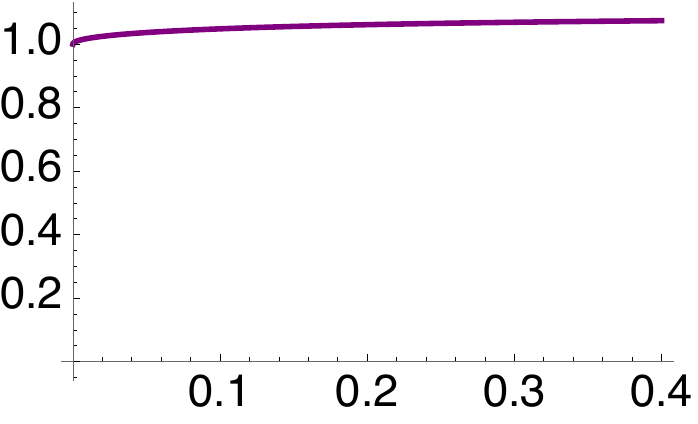} \begin{picture}(0,0) \put(0,0){$\big|\varepsilon_{2}^{(2)}\big|$} \put(-138,10){\begin{rotate}{90}Vol$_{\rm norm}(\Pi^{\rm frac}_{b|c})$\end{rotate}} \end{picture}
 &
 \includegraphics[scale=0.62]{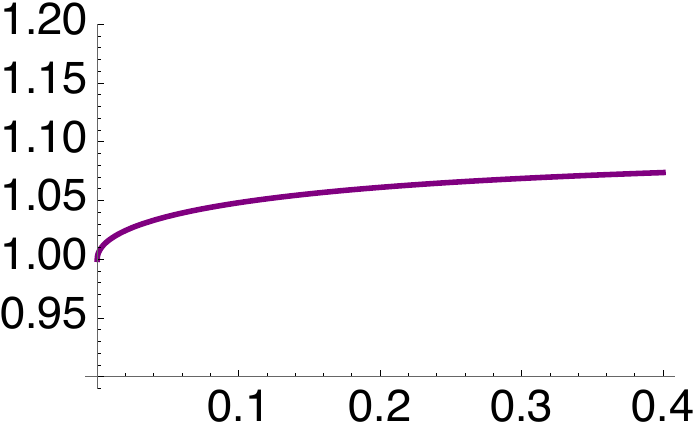} \begin{picture}(0,0) \put(0,0){$\big|\varepsilon_{2}^{(2)}\big|$} \put(-138,10){\begin{rotate}{90}Vol$_{\rm norm}(\Pi^{\rm frac}_{b|c})$\end{rotate}} \end{picture}
\end{tabular}
\caption{Normalised volume of the fractional cycles
 $\Pi^{\text{frac}}_{a|d} =\frac{\rho_1 - \big( 2 \, \tilde{\boldsymbol \epsilon}^{(2)}_{2}   -{\boldsymbol \epsilon}^{(2)}_{2}  \big) }{4}$ and $\Pi^{\text{frac}}_{b|c} =\frac{3\rho_1 +  \big( 2 \, \tilde{\boldsymbol \epsilon}^{(2)}_{2}  -{\boldsymbol \epsilon}^{(2)}_{2}   \big) }{4}$ with $z\in \{ b,c\}$ 
as a function of the deformation parameter $\big|\varepsilon^{(2)}_{2}\big|$. The right panel offers a close-up of the middle panel in the range $[0.9-1.2]$ of the fractional cycle volume for the D6-brane stacks $b$ and $c$.  
\newline
These plots serve as templates for the fractional three-cycle volumes as a function of deformations along the flat directions in the twisted complex structure moduli
$\zeta_{1,2}^{(2)}$ appearing in the next subsections, in which case the plots represent the normalised volume of a fractional three-cycle $\Pi^{\text{frac}}_{x} + \Pi^{\text{frac}}_{x'} =\frac{\rho_1 - \big( 2 \, \tilde{\boldsymbol \epsilon}^{(2)}_{2}   -{\boldsymbol \epsilon}^{(2)}_{2}  \big) }{2}$ or $\Pi^{\text{frac}}_x +\Pi^{\text{frac}}_{x'} =\frac{3\rho_1 +  \big( 2 \, \tilde{\boldsymbol \epsilon}^{(2)}_{2}  -{\boldsymbol \epsilon}^{(2)}_{2}   \big) }{2}$, respectively.  
\label{Fig:DeformedCycleADBCToyZ2}}
\end{center}
\end{figure}

The results of these computations are presented in figures~\ref{Fig:DeformedCycleADToyZ1} and~\ref{Fig:DeformedCycleBCToyZ1} for branes $\{a,d\}$ and $\{b,c\}$, respectively, for
 the deformation parameter $\varepsilon_{4-5}^{(1)}$ in the $\Z_2^{(1)}$-twisted sector and in figure~\ref{Fig:DeformedCycleADBCToyZ2} for all branes $\{a,b,c,d\}$ for 
  the deformation parameter $\varepsilon_{2}^{(2)}$ in the $\Z_2^{(2)}$-twisted sector. Through the relationship (\ref{Eq:g2-tree}) between the tree-level gauge coupling of the gauge theory living on a D6-brane and the volume of the three-cycle wrapped by the D6-brane internally, we can deduce the qualitative behaviour of the gauge coupling when going away from the singular orbifold point by switching on deformations along flat directions. More explicitly, in this toy model we observe in figures~\ref{Fig:DeformedCycleADToyZ1} and~\ref{Fig:DeformedCycleBCToyZ1} that the volumes of the fractional cycles wrapped by branes $a$ and~$c$ decrease for increasing deformation parameter $\big|\varepsilon^{(1)}_{4-5}\big|$ in the $\Z_2^{(1)}$-twisted sector, implying that the gauge couplings of their respective gauge theories increase when going away from the orbifold singular point. The gauge couplings of branes $b$ and~$d$ on the other hand are weaker at points in the moduli space away from the orbifold point, as the fractional cycle volumes wrapped by branes $b$ and~$d$ grow for non-zero deformation along the flat direction in the $\Z_2^{(1)}$-twisted sector, as depicted in figures~\ref{Fig:DeformedCycleADToyZ1} and~\ref{Fig:DeformedCycleBCToyZ1}. For the $\Z_2^{(2)}$ deformations, we observe in figure~\ref{Fig:DeformedCycleADBCToyZ2} that branes $a$ and~$d$ obtain a stronger gauge coupling, whereas branes $b$ and $c$ have a weaker gauge coupling away from the singular  orbifold point. The computations for the deformation parameter $\varepsilon_{1}^{(2)}$ in the $\Z_2^{(2)}$-twisted sector follow the same logic and reproduce the exactly same respective functional dependence as in figures~\ref{Fig:DeformedCycleADBCToyZ2} for the various D6-brane volumes, as can be inferred from the exceptional wrapping numbers in table~\ref{Tab:Ex-USp2-4-wrappings}.

The deformations in the $\Z_2^{(3)}$-twisted sector require the exactly same method as the one used for the $\Z_2^{(2)}$-twisted sector, which leads (qualitatively) to similar functional dependences of the fractional cycle volumes on the parameters $\varepsilon_{1,2}^{(3)}$ as presented in figure~\ref{Fig:DeformedCycleADBCToyZ2}: the volumes of branes $a$ and $b$ exhibit the functional behaviour of the left panel, while the volumes of the branes $c$ and $d$ increase for growing deformations $\big|\varepsilon_{1,2}^{(3)}\big|$ as in the middle and right panel of figure~\ref{Fig:DeformedCycleADBCToyZ2}. Hence, the gauge groups $USp(2)_a$ and $USp(2)_b$ acquire a stronger gauge coupling at the string scale $M_{\text{string}}$, whereas the gauge groups $USp(2)_c$ and $USp(2)_d$ a weaker gauge coupling under deformations along the two flat directions in the $\Z_2^{(3)}$-twisted sector.

\subsection{Deformations in a global MSSM model}\label{Ss:ExMSSM}

In~\cite{Ecker:2015vea}, we had performed a systematic search for MSSM-like models on $T^6/(\Z_2 \times \Z_6 \times \OR)$ with discrete torsion
and found that there is a unique choice of {\it bulk} parts supporting the
Standard Model gauge group $SU(3)_a \times SU(2)_b \times U(1)_Y \subset U(3)_a \times USp(2)_b \times U(1)_c \times U(1)_d$, and that all seemingly different 
choices of discrete (Wilson line, displacement and $\Z_2 \times \Z_2$ eigenvalue)
parameters lead to the same massless matter spectrum when completing to a {\it global} five-stack model with one `hidden' stack, see the discussion in section 4.1 of~\cite{Ecker:2015vea} for details.
The D6-brane configuration displayed in table~\ref{tab:5stackMSSMaAAPrototypeI} agrees with the previous example from~\cite{Ecker:2015vea} in all {\it relative} $\Z_2 \times \Z_2$
eigenvalues and {\it absolute} values of Wilson line and displacement parameters and is thus identical at the orbifold point. For the sake of comparing the different phenomenologically appealing models, we choose here, however, the {\it absolute} $\Z_2 \times \Z_2$ eigenvalues $(+++)$ for the (up to rank) common QCD stack $a$.
\mathtabfix{
\begin{array}{|c||c|c||c|c|c||c|}\hline 
\muc{7}{|c|}{\text{\bf D6-brane configuration of a global 5-stack MSSM model on the {aAA} lattice}}
\\\hline \hline
&\text{\bf wrapping numbers} &\frac{\rm Angle}{\pi}&\text{\bf $\Z_2^{(i)}$ eigenvalues}  & (\vec \tau) & (\vec \sigma)& \text{\bf gauge group}\\
\hline \hline
 a&(1,0;1,0;1,0)&(0,0,0)&(+++)&(0,1,1) & (0,1,1)& U(3)\\
 b&(1,0;-1,2;1,-2)&(0,\frac{1}{2},-\frac{1}{2})&(--+)&(0,1,0) & (0,1,0)&USp(2)\\
 c&(1,0;-1,2;1,-2)&(0, \frac{1}{2},-\frac{1}{2})&(-+-)&(0,1,1) & (0,1,1)&U(1)\\ 
  d&(1,0;-1,2;1,-2)&(0, \frac{1}{2},-\frac{1}{2})&(+--)&(0,0,1) & (0,0,1)& U(1)\\
  \hline
    h&(1,0;1,0;1,0)&(0,0,0)&(--+)&(0,1,1) & (0,1,1)& U(4)\\
 \hline
\end{array}
}{5stackMSSMaAAPrototypeI}{D6-brane configuration of a global five-stack model with gauge group $SU(3)_a \times SU(2)_b \times U(1)_Y \times SU(4)_h \times \Z_3$ after Green-Schwarz mechanism, which leads to the orientifold-even and -odd wrapping numbers labelled by MSSM in table~\protect\ref{Tab:Ex-Wrappings}.}
The massless open string spectrum of this MSSM example is for convenience displayed in table~\ref{tab:5stackMSSMaAAPrototypeISpectrum}.
\mathtabfix{
\begin{array}{|c|c|c|c||c|c|c|c|}
\hline \multicolumn{8}{|c|}{\text{\bf Overview of the massless matter spectrum of a global 5-stack MSSM on the {aAA} lattice of } T^6/(\Z_2 \times \Z_6 \times \OR)}\\
\hline \hline
\!\!\!\text{sector}\!\!\!  &(U(3)_a \times USp(2)_b \times U(4)_h)_{U(1)_c\times U(1)_d}& Q_Y  & \Z_3
& \!\!\!\text{sector}\!\!\!  &(U(3)_a \times USp(2)_b \times U(4)_h)_{U(1)_c\times U(1)_d}& Q_Y  & \Z_3
 \\
\hline \hline
ab &3 \times  (\3, \2, \1)_{(0,0)} & \nicefrac{1}{6} & 0  
&  aa'&  2\times[ ({\bf \ov{\3}_{\Anti}},\1,\1)_{(0,0)} + h.c.] & \pm \nicefrac{1}{3}  &0
\\
 ac&6 \times (\ov \3, \1,\1)_{(1,0)}  & \nicefrac{1}{3}  &1 
 &  bb &  5 \times  (\1,\1_{\Anti},\1)_{(0,0)}& 0  &0
 \\
 ad  & 3 \times (\3, \1, \1)_{(0,-1)}& \nicefrac{-1}{3} & 1 
 & cc & 4 \times (\1,\1,\1)_{(0,0)}& 0  &0
  \\
 ad' & 3 \times  (\ov \3, \1, \1)_{(0,-1)} & \nicefrac{-2}{3} &1 
 & dd & 5 \times (\1,\1,\1)_{(0,0)} & 0  &0
 \\
 bc   &3 \times (\1, \2,\1)_{(1,0)} + 3 \times \left[ (\1, \2,\1)_{(1,0)} + h.c. \right] &\nicefrac{1}{2},\pm \nicefrac{1}{2}   & 1,1||2   
 & dd'  & [(\1,\1,\1)_{(0,2)} + h.c.]   & \pm 1  &1||2
 \\\cline{5-8}
   bd    &6 \times (\1,\2,\1)_{(0,-1)}+2 \times \left[ (\1,\2,\1)_{(0,1)}  + h.c. \right]   & \nicefrac{-1}{2}, \nicefrac{\pm 1}{2}  & 1,2||1  
   & bh  &3 \times (\1,\2,\4)_{(0,0)} &0 &  2 
  \\
  cd &3 \times (\1,\1,\1)_{(-1,1)} +3 \times \left[ (\1,\1,\1)_{(-1,1)} + h.c. \right] & 0  & 1,1||2 
  &  ch' &6 \times (\1,\1,\ov\4)_{(-1,0)}  & \nicefrac{-1}{2} &  0
  \\
     cd'  &3 \times (\1,\1,\1)_{(1,1)}+3 \times \left[ (\1,\1,\1)_{(1,1)} + h.c. \right] & 1, \pm 1  &0 
    &  dh &3 \times (\1,\1,\ov\4)_{(0,1)} & \nicefrac{1}{2}&  0 
     \\\cline{1-4}  
     ah &2 \times \left[ (\3,\1,\ov\4)_{(0,0)}  +h.c.\right] & \nicefrac{\pm1}{6}  & 1||2
     & dh'&3 \times (\1,\1,\4)_{(0,1)}  & \nicefrac{1}{2}& 1 
   \\
  ah'  & [(\3,\1,\4)_{(0,0)}  +h.c.] & \nicefrac{\pm1}{6} &  2 ||1 
&     hh' &2 \times [ (\1,\1,{\bf 6_{\Anti}})_{(0,0)} + h.c.] & 0  & 1||2\\
 \hline
\end{array}
}{5stackMSSMaAAPrototypeISpectrum}{Chiral and non-chiral massless (open string) matter spectrum of the five-stack D6-brane model from table~\ref{tab:5stackMSSMaAAPrototypeI}
with gauge group $SU(3)_a\times USp(2)_b \times U(1)_Y \times SU(4)_h \times \Z_3$
after the Green-Schwarz mechanism has been taken into account.
For vector-like states, different charges under the discrete $\Z_3 \subset U(1)_c + 2 U(1)_d + 2 U(1)_h$ symmetry  are denoted using the logic symbol~$||$.
\newline
The closed string sector for this model - as well as all other explicit examples discussed in this article -  contains $(h^{11}_+,h^{11}_-,h^{21})=(4,15,19)$ vectors, K\"ahler moduli and complex structure moduli, respectively.
}

Ignoring the possibility of {\it vev}s for charged scalars as suggested in footnote~\ref{Footnote:SUNSUNFI} on page~\pageref{Footnote:SUNSUNFI} for the moment, we can read off from table~\ref{Tab:Stab-vs-Flat}
that the QCD stack $a$ and the `hidden' stack $h$ each couple with orientifold-odd wrapping numbers to the deformation moduli $\zeta^{(1)}_{3,4+5}$ 
and $\zeta^{(3)}_{1,2}$, while the D6-brane $c$ couples in an orientifold-odd way to $\zeta^{(1)}_{3,4+5}$ and $\zeta^{(2)}_{1,2}$ and 
the D6-brane $d$ to $\zeta^{(2)}_{1,2}$ and $\zeta^{(3)}_{1,2}$. This MSSM example thus allows to stabilise at most six of the 14 $\Z_2$-twisted deformation moduli 
as displayed in table~\ref{Tab:Stab-vs-Flat}. Out of the remaining eight $\Z_2$-twisted deformation moduli, only $\zeta^{(1)}_{4-5}$ couples in an orientifold-even 
way to the two D6-branes $b$ and $d$. The associated gauge couplings of $SU(2)_b \simeq USp(2)_b$ and $U(1)_Y = \nicefrac{U(1)_a}{6} + \nicefrac{[U(1)_c+U(1)_d]}{2}$ thus experience one flat direction at tree-level.
The remaining seven $\Z_2$-twisted deformation moduli $\zeta^{(1)}_{0,1,2}$ and $\zeta^{(2,3)}_{(3,4)}$ 
on the other hand do not couple directly to any D6-brane in this global model and can at most change the couplings in the low-energy effective MSSM Lagrangian via 
non-perturbative or higher order corrections. 
\newline
Let us emphasise here that at the singular orbifold point, the volume of a given fractional three-cycle and thereby the tree-level
gauge coupling only depends on the {\it untwisted} complex structure modulus $\varrho \propto \nicefrac{R^{(1)}_2}{R^{(1)}_1}$ as stated in equations~\eqref{Eq:g2-tree} and~\eqref{Eq:Vol-Orb}, while the {\it untwisted} K\"ahler moduli $v_{i,i\in \{1,2,3\}}$ defined in equation~\eqref{Eq:2-torus-volume} influence the gauge couplings once the one-loop gauge threshold corrections are taken into account, as will be discussed in section~\ref{S:GKFDefModuli}.

Before turning to the technicalities of executing the different deformations, let us briefly discuss possible caveats in the counting of the maximal number of stabilised deformation moduli
when allowing for {\it vev}s of charged matter fields. While the massless matter spectrum in table~\ref{tab:5stackMSSMaAAPrototypeISpectrum} provides all charged (open string) scalars that might possibly 
trigger some D6-brane recombination process while compensating the Fayet-Iliopoulos term generated by the {\it vev} of some (closed string) deformation modulus, a more detailed analysis of the origin of each 
matter state per D6-brane intersection sector $x(\omega^k y^{(\prime)})$ in table~\ref{tab:Z2Z65stackMSSMTotalSpectrum} is required to deduce allowed terms in the low-energy effective action, in particular the {\it necessary} selection rule of a closed polygon with $n$ edges along the partaking D6-bane directions per $n$-point coupling.

Let us start by considering the chiral sector of the spectrum in table~\ref{tab:5stackMSSMaAAPrototypeISpectrum} only. The only states in bifundamental representations of two unitary gauge factors of equal rank
are the right-handed {\it s}electrons and {\it s}neutrinos\footnote{Notice that $U(1)_{\text{massive}} = U(1)_c - U(1)_d$ in this example acts as {\it perturbative} global Peccei-Quinn symmetry in the low-energy effective action with the right-handed {\it s}neutrinos naturally identified as QCD axions of a generalised `stringy' DFSZ model~\cite{Honecker:2013mya,Honecker:2015ela,Honecker:2015qba,Honecker:2016gyz}.} in the $cd'$ and $cd$ sectors, respectively, charged under $U(1)_c \times U(1)_d$. 
As summarised in table~\ref{Tab:Stab-vs-Flat}, both D6-branes $c$ and $d$ couple to the twisted deformation moduli $\zeta^{(2)}_{1,2}$, and table~\ref{Tab:Ex-Wrappings} shows that the corresponding 
orientifold-odd exceptional wrapping numbers are opposite, $[2 x^{(2)}_{\alpha,c} + y^{(2)}_{\alpha,c} ]_{\alpha=1,2} = - [2 x^{(2)}_{\alpha,d} + y^{(2)}_{\alpha,d} ]_{\alpha=1,2}$.
The D-terms can therefore only be compensated by some {\it vev} of the right-handed {\it s}neutrinos due to their opposite $U(1)_c$ and $U(1)_d$ charges.
In~\cite{Ecker:2015vea}, three-point functions involving {\it vev}s of this type were shown to be suitable for generating mass terms e.g. of the vector-like down-type quark pairs originating from the $ac$ and $ad$ sectors as well as of the vector-like left-handed lepton (or Higgs) pairs from the $bc$ and $bd$ sectors.
At this point, we however, notice two caveats: on the one hand, our discussion so far up to now only includes the {\it relative} sign among the $U(1)_c \times U(1)_d$ charges, whereas in a more thorough field theoretical study in remains to be seen if the {\it s}neutrino/axion or its hermitian conjugate representation are suitable for canceling a Fayet-Iliopoulos term; on the other hand, the naive geometric intuition of $\Pi_c + \Pi_d = \frac{2 \, \Pi^{\text{bulk}}_{c=d} + \Pi^{\Z_2^{(1)},\text{odd}}_c + \Pi^{\Z_2^{(1)},\text{even}}_d + \Pi^{\Z_2^{(3)},\text{even}}_c + \Pi^{\Z_2^{(3)},\text{odd}}_d}{4}$ as a merging of cycles in the $\Z_2^{(2)}$-twisted sector is contrasted by the naive merging of orientifold image cycles $\Pi_c + \Pi_{c'}= \frac{\Pi^{\text{bulk}}_c + \Pi^{\Z_2^{(3)}}_c}{2}$ and $\Pi_d + \Pi_{d'}= \frac{\Pi^{\text{bulk}}_d + \Pi^{\Z_2^{(1)}}_d}{2}$ leaving only one $\Z_2$-twisted sector.

One might wonder if taking the vector-like matter states into consideration as well can produce additional flat directions, in particular if (some of) the antisymmetric representations of $SU(3)_a$ or $SU(4)_h$
receive a {\it vev}. The former is clearly undesirable since it would break the part $SU(3)_a \times U(1)_Y$ of the Standard Model gauge group. On the other hand, if the vector-like pairs
in the $aa'$ and $hh'$ sectors of table~\ref{tab:5stackMSSMaAAPrototypeISpectrum} were originating from ${\cal N}=2$
supersymmetric sectors, their {\it vev}s would not be protected by the ${\cal N}=1$ $SU(N)$ D-term argument of~\cite{Blaszczyk:2015oia} but would instead be expected to constitute flat directions
associated to the recombination of orientifold image cycles of the type $\Pi_z+ \Pi_{z'} = \frac{\Pi^{\text{bulk}}_z + \Pi^{\Z_2^{(2)}}_z }{2}$ for $z \in \{a,h\}$ such that the deformation moduli $\zeta^{(1)}_{3,4+5}$
and $\zeta^{(3)}_{1,2}$ in the $\Z_2^{(1)}$- and $\Z_2^{(3)}$-twisted sectors were to be at most stabilised by the existence of stacks $c$ and $d$, respectively.
However, the state-per-sector list in table~\ref{tab:Z2Z65stackMSSMTotalSpectrum} shows that usually a chiral multiplet in one sector $x(\omega^k y)$ is paired with an anti-chiral multiplet in some $x(\omega^{l \neq k} y)$
sector. Even the vector-like pair of states in the antisymmetric representation in the $aa'$ (or $hh'$) sector with the orientifold image D6-branes parallel along all three two-tori $T^2_{(i)}$ does not constitute a genuine ${\cal N}=2$ supersymmetric sector due to the different {\it relative}  $\Z_2 \times \Z_2$ eigenvalues  entering the orbifold projection on the {\it a priori} three chiral multiplets containing e.g. the massless scalar $\psi^{i \text{ or } \ov{i}}_{-\nicefrac{1}{2}}|0 \rangle_{\text{NS}}$.
A much more detailed and dedicated computation of $n$-point couplings appearing in the low-energy effective action is thus needed to completely settle this issue. But this goes way beyond the project discussed here, since not even 3-point Yukawa couplings have been computed for the cases at hand of a vanishing intersection angle along $T^2_{(1)}$.
The vanishing arguments of~\cite{Abel:2003vv,Abel:2003yx,Cvetic:2003ch,Lust:2004cx,Abel:2004ue} rely on ${\cal N}=2$ supersymmetry on the factorisable six-torus $T^6=(T^2)^3$ or its $\Z_2 \times \Z_2$ orbifold {\it without} discrete torsion and are clearly not applicable for the models {\it with} discrete torsion discussed in this article.

The only element left to discuss at this point is the $\Z_2^{(1)}$-twisted deformation $\varepsilon_{4-5}^{(1)}$, to which the D6-branes $b$ and $d$ couple through the orientifold-even exceptional three-cycle. Hence, this deformation represents a $\Z_2^{(1)}$-twisted modulus with a flat direction as indicated in table~\ref{Tab:Stab-vs-Flat}. A closer look at the three-cycle configuration in table~\ref{tab:5stackMSSMaAAPrototypeI} for branes $b$ and $d$ reveals that the functional dependence of their volumes in terms of $\varepsilon_{4-5}^{(1)}$ is equivalent to the branes supporting the gauge groups $USp(2)_c$ and $USp(2)_b$, respectively, in the toy model of the previous section: the $b$-brane volume behaves as the righthand part of figure~\ref{Fig:DeformedCycleBCToyZ1} under a deformation by $\varepsilon_{4-5}^{(1)} \neq 0$, while the $d$-brane cycle exhibits the behaviour of the lefthand side of figure~\ref{Fig:DeformedCycleBCToyZ1} under the deformation parameter~$\varepsilon_{4-5}^{(1)}$. This implies that the weak gauge coupling becomes stronger when the $\Z_2^{(1)}$-twisted modulus $\zeta_{4-5}^{(1)}$ acquires a non-zero {\it vev} and we move away from the singular orbifold point along the flat direction. The massless hypercharge on the other hand is characterised by a smaller gauge coupling away from the orbifold point. Compatibility between the calibration forms for the bulk and (resolved) exceptional three-cycles limits the parameter space of the deformations to the half-line $\varepsilon_{4-5}^{(1)}\leq 0$, similarly to the toy model of the previous section.

\subsection{Deformations in global Left-Right symmetric models}\label{Ss:ExLRSM}

In this section, we study how deformations affect each of the prototype L-R symmetric models classified in~\cite{Ecker:2015vea}.
The common {\it local} D6-brane configuration of the observable sector is displayed in the upper part of table~\ref{tab:6stackLRSaAA}, with the four different types of {\it global} completion by two `hidden' sector D6-branes provided in the lower part of the same table.
\mathtabfix{
\begin{array}{|c||c|c||c|c|c||c|}\hline 
\muc{7}{|c|}{\text{\bf D6-brane configurations of global L-R symmetric models on }T^6/(\Z_2 \times \Z_6 \times \OR)}
\\\hline \hline
&\text{\bf wrapping numbers} &\frac{\rm Angle}{\pi}&\text{\bf $\Z_2^{(i)}$ eigenvalues}  & (\vec \tau) & (\vec \sigma)& \text{\bf gauge group}\\
\hline \hline
\muc{7}{|c|}{\text{\bf Universal observable sector}}
\\\hline \hline
 a&(1,0;1,0;1,0)&(0,0,0)&(+++)&(0,1,1) & (0,1,1)& U(3)\\
 b&(1,0;-1,2;1,-2)&(0,\frac{1}{2},-\frac{1}{2})&(+++)&(0,1,0) & (0,1,0)&USp(2)\\
 c&(1,0;-1,2;1,-2)&(0, \frac{1}{2},-\frac{1}{2})&(-+-)&(0,1,0) & (0,1,0)&USp(2)\\ 
  d&(1,0;1,0;1,0)&(0, 0,0)&(+--)&(0,1,1) & (0,1,1)& U(1)\\
  \hline\hline\hline
  \muc{7}{|c|}{\text{\bf Global completion of prototype I}}
  \\\hline\hline
    h_1&(1,0;1,0;1,0)&(0,0,0)&(+++)&(0,0,0) & (0,0,0)& U(3)\\
        h_2&(1,0;1,0;1,0)&(0,0,0)&(+--)&(0,0,0) & (0,0,0)& U(3)\\
\hline\hline\hline
  \muc{7}{|c|}{\text{\bf Global completion of prototype II}}
  \\\hline\hline
    h_1&(1,0;-1,2;1,-2)&(0,\frac{1}{2},-\frac{1}{2})&(+++)&(0,0,0) & (0,0,0)& U(1)\\
        h_2&(1,0;-1,2;1,-2)&(0,\frac{1}{2},-\frac{1}{2})&(+--)&(0,0,0) & (0,0,0)& U(1)\\
\hline\hline\hline
  \muc{7}{|c|}{\text{\bf Global completion of prototype IIb}}
  \\\hline\hline
   h_1&(1,0;-1,2;1,-2)&(0,\frac{1}{2},-\frac{1}{2})&(-+-)&(0,0,0) & (1,0,0)& U(1)\\
        h_2&(1,0;-1,2;1,-2)&(0,\frac{1}{2},-\frac{1}{2})&(--+)&(0,0,0) & (1,0,0)& U(1)\\
\hline\hline\hline
  \muc{7}{|c|}{\text{\bf Global completion of prototype IIc}}
  \\\hline\hline
  h_1&(1,0;-1,2;1,-2)&(0,\frac{1}{2},-\frac{1}{2})&(+++)&(0,0,1) & (0,0,1)& U(1)\\
        h_2&(1,0;-1,2;1,-2)&(0,\frac{1}{2},-\frac{1}{2})&(+++)&(0,0,1) & (1,1,1)& U(1)\\
 \hline
\end{array}
}{6stackLRSaAA}{{\it Local} observable L-R symmetric D6-brane sector in the first block with four different {\it global} completions by two `hidden' branes $h_1$, $h_2$ listed below.
Observe that for prototype IIb, we have $(+++)_{h_1'}$ and $(+--)_{h_2'}$, and thus the only difference w.r.t. prototype II is the different choice of the discrete displacement parameter $\sigma^1_{h_i^{\text{\bf IIb}}}=1$.
}
The common observable part of the matter spectrum is displayed in table~\ref{tab:6stackLRSMaAAVisibleSpectrum},
\mathtabfix{ 
\begin{array}{|c|c|c|c||c|c|c|c|}
\hline \multicolumn{8}{|c|}{\text{\bf Visible spectrum of all L-R symmetric \& the PS I models on }T^6/(\Z_2 \times \Z_6 \times \OR)}\\
\hline \hline
\text{sector} & (U(3)_a \times USp(2)_b \times USp(2)_c )_{U(1)_d}  & \Z_3^{\text{\bf I}} & \widetilde{U(1)}_{B-L}^{\text{\bf II+IIb+IIc}}
& \text{sector} & (U(3)_a \times USp(2)_b \times USp(2)_c )_{U(1)_d}  & \Z_3^{\text{\bf I}} & \widetilde{U(1)}_{B-L}^{\text{\bf II+IIb+IIc}}\\
\hline\hline
\muc{8}{|c|}{\text{\bf L-R symmetric models \& PS I model (with $\3 \to \4$, ${\bf \ov{3}}_{\Anti} \to {\bf 6}_{\Anti}$ of $U(3)_a \to U(4)_a$)}}
\\\hline
ab &3 \times  (\3,\2,\1)_{(0)} & 1 & \nicefrac{1}{3} 
&  aa'&  2\times[ ({\bf \ov{\3}_{\Anti}},\1,\1)_{(0)} + h.c.] &2||1 &  \nicefrac{\pm 2}{3}
\\
ac &3 \times  (\ov \3,\1,\2)_{(0)} &2 & \nicefrac{-1}{3} 
&  bb &  5 \times  (\1,\1_{\Anti},\1)_{(0)}&0 & 0
\\
bc  &10 \times (\1,\2,\2)_{(0)} &0 & 0
&  cc &  5 \times  (\1,\1,\1_{\bf A})_{(0)}&0 & 0
\\\hline\hline
\muc{8}{|c|}{\text{\bf L-R symmetric models only}}
\\\hline\hline
ad & (\3,\1,\1)_{(-1)} + h.c.  & 1||2& \nicefrac{\pm 4}{3} 
& bd &3 \times (\1,\2,\1)_{(-1)} &0 & 1
\\ 
ad' &2 \times \left[ (\3,\1,\1)_{(1)} + h.c.\right] &1||2 &\nicefrac{ \mp  2}{3}
& cd &3 \times (\1,\1,\2)_{(1)} &0 & -1\\
\hline
\end{array}
}{6stackLRSMaAAVisibleSpectrum}{Common visible spectrum of the L-R symmetric models. Prototype I has the low-energy gauge group $SU(3)_a \times USp(2)_b \times USp(2)_c \times SU(3)_{h_1} \times SU(3)_{h_2} \times \Z_3$ 
with the $\Z_3$ charge displayed in the third and seventh column. The prototype II model has the low-energy gauge group $SU(3)_a \times USp(2)_b \times USp(2)_c \times \widetilde{U(1)}_{B-L}$ with the 
$\widetilde{U(1)}_{B-L} \equiv \frac{1}{3} U(1)_a - U(1)_d - U(1)_{h_1} + U(1)_{h_2}$ charge displayed in the fourth and eighth column. For the prototype IIb \& IIc models, the $\widetilde{U(1)}_{B-L}$ symmetry is massive and thus only a {\it perturbative} global symmetry.
The massless matter spectrum is completed by the `hidden' sectors displayed in table~\protect\ref{tab:6stackLRSMaAAHiddenSpectrum} for each prototype I, II, IIb and IIc.
The massless closed string sector is for each case identical to the MSSM-like model, i.e.  it contains $(h^{11}_+,h^{11}_-,h^{21})=(4,15,19)$ vectors, K\"ahler moduli and complex structure moduli, respectively.
}
and the individual `hidden' spectra are given in  table~\ref{tab:6stackLRSMaAAHiddenSpectrum}.
\mathtabfix{
\begin{array}{|c|c|c||c|c|c|}
\hline \multicolumn{6}{|c|}{\text{\bf Overview of the `hidden' spectra for the L-R symmetric models on }T^6/(\Z_2 \times \Z_6 \times \OR)}\\
\hline \hline
 \multicolumn{6}{|c|}{\text{\bf  Prototype I}}\\
\hline \hline
\text{sector}  & (U(3)_a \times USp(2)_b \times USp(2)_c \times U(3)_{h_1} \times U(3)_{h_2} )_{U(1)_d}  & \Z_3
& \text{sector}  & (U(3)_a \times USp(2)_b \times USp(2)_c \times U(3)_{h_1} \times U(3)_{h_2} )_{U(1)_d}  & \Z_3
\\\hline \hline
ah_1&2 \times  (\ov\3,\1,\1,\3,\1)_{(0)}  &0
& dh_1 & 2 \times (\1,\1,\1,\ov\3,\1)_{(1)} &2
\\
ah_2 &2 \times (\3,\1,\1,\1,\ov\3)_{(0)}  &0
& dh_2 &2 \times(\1,\1,\1,\1,\3)_{(-1)}  &1
\\
bh_1  &  (\1,\2,\1,\ov\3,\1)_{(0)}  +[ (\1,\2,\1,\ov\3,\1)_{(0)} +h.c. ] &2, 2||1
& h_1 h_2 &(\1,\1,\1,\3,\ov\3)_{(0)} + h.c.  &0
\\
bh_2  &  (\1,\2,\1,\1,\3)_{(0)} + [(\1,\2,\1,\1,\3)_{(0)} + h.c. ] &1, 1||2 
& h_1 h_2'& 2\times\left[ (\1,\1,\1,\3,\3)_{(0)} + h.c.  \right] &2||1
\\
ch_1 & (\1,\1,\2,\ov\3,\1)_{(0)} + [(\1,\1,\2,\ov\3,\1)_{(0)} +h.c. ] &2 , 2||1
&    h_1 h_1' & 2\times[ (\1,\1,\1,{\bf \ov{\3}_{\Anti}},\1)_{(0)} + h.c.] &2||1
\\
ch_2 &(\1,\1,\2,\1,\3)_{(0)} + [ \1,\1,\2,\1,\3)_{(0)} + h.c.] &1,1||2
&   h_2 h_2' & 2\times[ (\1,\1,\1,\1,{\bf \ov{\3}_{\Anti}})_{(0)} + h.c.] &2||1
    \\\hline\hline \multicolumn{6}{|c|}{\text{\bf Prototype II}}\\
\hline \hline
\text{sector} & (U(3)_a \times USp(2)_b \times USp(2)_c)_{U(1)_d\times U(1)_{h_1} \times U(1)_{h_2}}  & \widetilde{U(1)}_{B-L}
& \text{sector} & (U(3)_a \times USp(2)_b \times USp(2)_c)_{U(1)_d\times U(1)_{h_1} \times U(1)_{h_2}}  & \widetilde{U(1)}_{B-L}
\\\hline \hline
ah_1 & 2 \times \left[ (\3,\1,\1)_{(0,-1,0)} + h.c. \right] & \nicefrac{\pm 4}{3}
& dh_1  & 2 \times \left[ (\1,\1,\1)_{(1,-1,0)} + h.c. \right]  & 0
 \\
ah_1' & (\3,\1,\1)_{(0,1,0)} + h.c.& \nicefrac{ \mp 2}{3}  
& dh_1'  & (\1,\1,\1)_{(1,1,0)} + h.c.  & \mp 2
\\
ah_2 & 2 \times \left[ (\3,\1,\1)_{(0,0,-1)} + h.c. \right] & \nicefrac{ \mp 2}{3} 
& dh_2 & 2\times \left[(\1,\1,\1)_{(1,0,-1)} + h.c. \right]  & \mp 2
\\
ah_2' & (\3,\1,\1)_{(0,0,1)} + h.c.& \ \nicefrac{\pm 4}{3}
& dh_2' & (\1,\1,\1)_{(1,0,1)} + h.c.  & 0
\\
bh_1  & 3 \times  (\1,\2,\1)_{(0,1,0)} + 3 \times \left[ (\1,\2,\1)_{(0,-1,0)} +h.c. \right]& -1, \pm 1 
& h_1 h_2 & 5 \times \left[ (\1,\1,\1)_{(0,1,-1)} + h.c. \right]   & \mp 2
 \\
bh_2  & 3 \times  (\1,\2,\1)_{(0,0,-1)} + 3 \left[ \times  (\1,\2,\1)_{(0,0,1)} + h.c. \right] & -1,  \pm 1 
& h_1 h_2' & 6\times \left[ (\1,\1,\1)_{(0,1,1)} + h.c. \right] &0
\\
ch_1  & 3 \times (\1,\1,\2)_{(0,-1,0)} + 3 \times \left[  (\1,\1,\2)_{(0,1,0)} +h.c. \right]& 1, \mp 1  
&     h_1h_1 & 4 \times (\1,\1,\1)_{(0,0,0)} &0
 \\
ch_2  & 3 \times (\1,\1,\2)_{(0,0,1)} + 3 \left[ \times (\1,\1,\2)_{(0,0,1)} + h.c. \right]& 1, \pm 1
&     h_2h_2 &4 \times (\1,\1,\1)_{(0,0,0)}& 0 
  \\\hline \hline \multicolumn{6}{|c|}{\text{\bf Prototype IIb}}\\
\hline \hline
h_1 h_2 &5 \times \left[ (\1,\1,\1)_{(0,1,-1)} + h.c. \right]   & \mp 2
&    h_1h_1 & 4 \times (\1,\1,\1)_{(0,0,0)} &0 
 \\
h_1 h_2' &6\times \left[ (\1,\1,\1)_{(0,1,1)} + h.c. \right] &0
&  h_2h_2 &4 \times (\1,\1,\1)_{(0,0,0)}& 0
    \\\hline\hline \multicolumn{6}{|c|}{\text{\bf Prototype IIc}}\\
\hline \hline
ah_1 &3 \times (\ov\3,\1,\1)_{(0,1,0)}&  \nicefrac{-4}{3}
& dh_1 &3\times (\1,\1,\1)_{(1,-1,0)}  & 0 
 \\
ah_1' &3 \times (\3,\1,\1)_{(0,1,0)}& \nicefrac{-2}{3} 
& dh_1' &3\times (\1,\1,\1)_{(-1,-1,0)}  & -2 
\\
bh_1  &4 \times \left[ (\1,\2,\1)_{(0,-1,0)} +h.c. \right]& \pm 1 
&  h_1h_1 & 5 \times (\1,\1,\1)_{(0,0,0)} &0 
 \\
ch_1 &6 \times (\1,\1,\2)_{(0,-1,0)} +2 \times \left[  (\1,\1,\2)_{(0,1,0)} +h.c. \right]  & 1,   \mp 1
&  h_2h_2 &5 \times (\1,\1,\1)_{(0,0,0)}& 0
    \\\hline
\end{array}
}{6stackLRSMaAAHiddenSpectrum}{`Hidden' massless spectrum per L-R symmetric model completing the common observable sector  displayed in table~\protect\ref{tab:6stackLRSMaAAVisibleSpectrum}.}

The QCD stack $a$ agrees by construction with the one of the MSSM example discussed in section~\ref{Ss:ExMSSM}, while the left- and right-symmetric groups $USp(2)_b \times USp(2)_c$ 
have identical bulk cycles and discrete Wilson line and displacement parameters, but differ in their $\Z_2 \times \Z_2$ eigenvalues from the D6-brane $b$ of the MSSM example.
The D6-brane $d$ only differs from the QCD stack $a$ in the $\Z_2 \times \Z_2$ eigenvalues and stack size $N_d=1$ vs. $N_a=3$.

The orientifold-even and -odd wrapping numbers are summarised table~\ref{Tab:Ex-Wrappings} and lead to the naive counting of the maximal number of stabilised deformation moduli in table~\ref{Tab:Stab-vs-Flat}. We observe the following differences:
\begin{itemize}
\item L-R IIc has the maximal number of ten stabilised deformation moduli due to the different choices of displacements $(\sigma^1,\sigma^2)=(0,0)_{h_1}, (1,1)_{h_2}$ of the two `hidden' D6-branes.
\item L-R I is the only model with non-Abelian `hidden' sector and thus states in bifundamental representations of gauge groups of identical rank, $SU(3)_a \times SU(3)_{h_1} \times SU(3)_{h_2}$. While 
the twisted deformation modulus $\langle \zeta^{(1)}_3 \rangle =0$ is stabilised at the orbifold point by the presence of D6-brane $d$, {\it vev}s of $\zeta^{(1)}_{0,1,2}$ and $ \zeta^{(2)}_{1,2}$ could potentially be compensated by {\it vev}s of scalars in the $h_1h_2^{(\prime)}$ sectors, which would simultaneously break $SU(3)_{h_1} \times SU(3)_{h_2} \to SU(3)_h^{\text{diag}}$.\footnote{ More explicitly, one can consider e.g.~the vector-like pair $(\1,\1,\1,\3,\ov\3)_{(0)} + h.c.$ from the $h_1 h_2$ sector and solve the Abelian and non-Abelian D-term constraints in terms of non-vanishing {\it vev}s of the scalar components in both bifundamental representations of $U(3)_{h_1} \times U(3)_{h_2}$. Such a vacuum configuration would break the non-Abelian gauge groups to the diagonal gauge group $U(3)_h^{\rm diag}$, which would correspond geometrically to the recombination of the two D6-brane stacks $h_1,h_2$ into a single stack $h$ wrapping $[\nicefrac{\Pi^{\text{bulk}}_h + \Pi^{\Z_2^{(1)}}_h]}{2}$, cf. also the microscopic origin of the vector-like matter states from the $h_1(\omega^0 h_2)$ sector according to table~\ref{tab:Z2Z6-LRS-hidden}. }
\item L-R II, IIb, IIc contain Abelian `hidden' D6-branes and thus three gauge groups of equal rank, $U(1)_d \times U(1)_{h_1} \times U(1)_{h_2}$. 
\\
In the prototypes II and IIb, some {\it vev} of the type $\langle \zeta^{(1)}_{0,1,2} \rangle$ might potentially be compensated by {\it vev}s associated to charged scalar fields belonging to vector-like pairs in the $h_1h_2^{(\prime)}$ sector. For prototype IIb, the same considerations apply also to $\langle \zeta^{(3)}_{3,4}\rangle$. Details of these potential compensations among {\it vev}s of closed and open string scalars depend on the microscopic origin of the latter as detailed in  table~\ref{tab:Z2Z6-LRS-hidden}.
\item
Prototypes I and IIc have according to the na\"ive counting in table~\ref{Tab:Stab-vs-Flat} only one flat direction, which affects the tree level gauge couplings, in the $\Z_2^{(1)}$-twisted sector, while prototypes II and IIb additionally have two flat directions of direct physical consequence in the $\Z_2^{(2)}$-twisted sector.
\end{itemize}
The counting of matter states per intersection sector is displayed in tables~\ref{tab:Z2Z6-LRS+PSII-visible} and~\ref{tab:Z2Z6-LRS-hidden} for the universal observable and individual `hidden' sectors, respectively, 
and -- just as for the MSSM example -- a dedicated derivation of the low-energy effective action is needed to determine if {\it vev}s of matter states can indeed allow for potentially flat directions in the deformation moduli space as stated above.

Looking closer at the $\Z_2^{(i)}$-twisted moduli with a flat direction, we notice first of all that the discussion for the $USp(2)_b \times USp(2)_c$ sector
can be brought back to the analysis presented in section~\ref{Ss:ExUSp2-4} for the global $USp(2)^4$ toy model.
A common calibration w.r.t. ${\rm Re}(\Omega_3)$ for bulk and resolved exceptional three-cycles then constrains the deformation parameter to lie on the half-line $\varepsilon_{4-5}^{(1)}\leq 0$. Going away from the orbifold point along the flat direction of the $\Z_2^{(1)}$-twisted modulus $\zeta^{(1)}_{4-5}$ implies a weaker gauge coupling for the left stack $USp(2)_b$ and a stronger gauge coupling for the right stack $USp(2)_c$.\footnote{For the prototype IIc model, the modulus $\zeta^{(1)}_{4-5}$ also couples to the `hidden' D6-brane stacks $h_1$ and $h_2$, such that a deformation along its flat direction also affects their respective $U(1)$ gauge couplings 
at the string scale $M_{\text{string}}$, before these Abelian gauge groups are spontaneously broken at the KK-scale
 by virtue of the St\"uckelberg mechanism. 
That is to say, the ($\Z_2^{(1)}$-twisted sector of the) fractional three-cycle for $h_1/h_2$ can be brought back to the ($\Z_2^{(1)}$-twisted sector of the) fractional cycle supporting the $USp(2)_{b/c}$ gauge group, such that the gauge coupling for $U(1)_{h_1}$ decreases and the one for $U(1)_{h_2}$ increases for non-zero deformation $\varepsilon_{4-5}^{(1)}$. Realising a strongly coupled anomalous $U(1)$ gauge theory at the string scale through these geometric deformations of $\Z_2$ singularities opens up avenues for D6-brane model building scenarios~\cite{Shiu:2017ta} realising Nambu-Jona-Lasinio type models upon integrating out the massive $U(1)$.} 

\begin{figure}[h]
\begin{center}
\begin{tabular}{c@{\hspace{0.8in}}c@{\hspace{0.8in}}c} 
\hspace*{-0.4in}\includegraphics[scale=0.65]{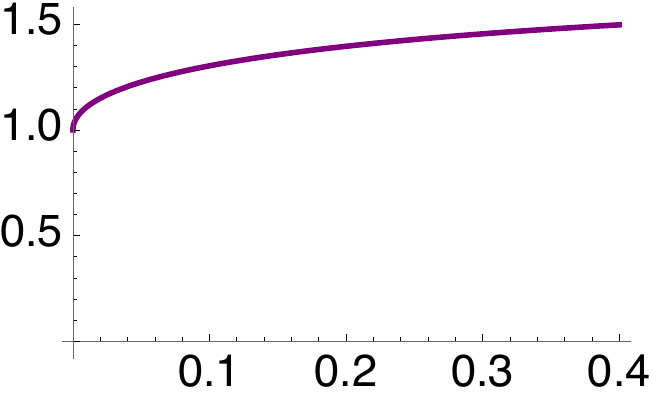} \begin{picture}(0,0) \put(0,0){$\big|\varepsilon_{2}^{(2)}\big|$} \put(-135,10){\begin{rotate}{90}Vol$_{\rm norm}(\Pi^{\rm frac}_{d})$\end{rotate}} \end{picture} 
&
 \includegraphics[scale=0.62]{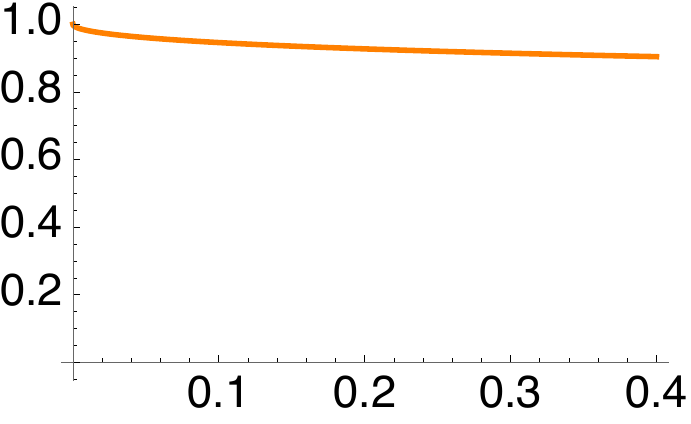} \begin{picture}(0,0) \put(0,0){$\big|\varepsilon_{2}^{(2)}\big|$} \put(-138,10){\begin{rotate}{90}Vol$_{\rm norm}(\Pi^{\rm frac}_{h_2})$\end{rotate}} \end{picture}
 &
 \includegraphics[scale=0.62]{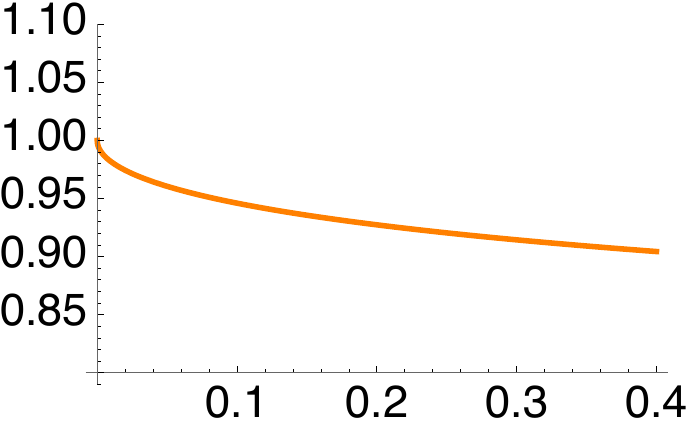} \begin{picture}(0,0) \put(0,0){$\big|\varepsilon_{2}^{(2)}\big|$} \put(-138,10){\begin{rotate}{90}Vol$_{\rm norm}(\Pi^{\rm frac}_{h_2})$\end{rotate}} \end{picture}
\end{tabular}
\caption{Normalised volume of the fractional cycles
 $\Pi^{\text{frac}}_d + \Pi^{\text{frac}}_{d'} =\frac{\rho_1 + \big( 2 \, \tilde{\boldsymbol \epsilon}^{(2)}_{2}   -{\boldsymbol \epsilon}^{(2)}_{2}  \big) }{2}$ and $\Pi^{\text{frac}}_{h_2} +\Pi^{\text{frac}}_{h_2'} =\frac{3\rho_1 -  \big( 2 \, \tilde{\boldsymbol \epsilon}^{(2)}_{2}  -{\boldsymbol \epsilon}^{(2)}_{2}   \big) }{2}$ for the prototype II + IIb models  
as a function of the deformation parameter $\big|\varepsilon^{(2)}_{2}\big|$. The right panel offers a close-up of the middle panel in the range $[0.8-1.1]$ of the fractional cycle volume of the hidden D6-brane stack $h_2$ in the prototype II + IIb models. \label{Fig:DeformedCycleLRSZ2}}
\end{center}
\end{figure}

The deformation parameters $\varepsilon_{1,2}^{(2)}$ in the $\Z_2^{(2)}$-twisted sector allow for a mutually compatible calibration w.r.t. ${\rm Re}(\Omega_3)$ between bulk and exceptional three-cycles provided the parameters lie on the half-line $\varepsilon_{1,2}^{(2)}\leq 0$. The ($\Z_2^{(2)}$ sector of the) fractional three-cycle for the {\it strong} D6-brane stack~$a$ is identical to the geometry of the three-cycle (in the $\Z_2^{(2)}$ sector) supporting the $USp(2)_a$ gauge group in the toy model of section~\ref{Ss:ExUSp2-4}. 
This implies that the {\it QCD} gauge group acquires a stronger gauge coupling when either the twisted modulus $\zeta_1^{(2)}$ or $\zeta_2^{(2)}$ acquires a non-zero {\it vev}. The opposite occurs for the left stack $b$ and right stack $c$, whose gauge couplings decrease for non-zero deformations along the flat directions of $\zeta_{1,2}^{(2)}$ 
as detailed in the context of the $USp(2)^4$ toy model in section~\ref{Ss:ExUSp2-4}. 
Also the $d$-brane couples to the deformation moduli $\zeta_{1,2}^{(2)}$ along flat directions, yet its geometric properties cannot be reduced to a situation discussed in previous sections. Using the same modus operandi as in section~\ref{Ss:ExUSp2-4}, we can compute the normalised volume of the fractional three-cycle of the $d$-brane, which is shown on the left-hand side of figure~\ref{Fig:DeformedCycleLRSZ2}. From this figure, the qualitative picture obviously exhibits the $U(1)_d$ gauge coupling decreasing for non-zero deformations in the $\Z_2^{(2)}$-twisted sector along the flat directions $\zeta_{1,2}^{(2)}$. Due to the relative factor $1/3$ in the definition of the generalised $\widetilde{U(1)}_{B-L}$ symmetry
of prototype II, the behaviour of the $U(1)_d$ gauge coupling is expected to be dominant with respect to the behaviour of the $U(1)_a$ gauge coupling under $\Z_2^{(2)}$ deformation, implying that the $\widetilde{U(1)}_{B-L}$ symmetry will also be more weakly coupled for non-zero {\it vev}s in the $\zeta_{1,2}^{(2)}$-directions.

According to table~\ref{Tab:Stab-vs-Flat}, the hidden D6-brane stacks $h_1$ and $h_2$ in the prototype II and IIb models couple to deformations with flat directions too, which requires us to investigate their fractional cycle volume under deformation. In both prototype models the (relevant part of) fractional three-cycle for $h_1$ can be recast into the fractional three-cycle supporting the $USp(2)_b$ gauge group in the global toy model of section~\ref{Ss:ExUSp2-4}. This implies that the fractional cycle volume for hidden stack $h_1$ exhibits the same behaviour as depicted in the middle and right panel of figure~\ref{Fig:DeformedCycleADBCToyZ2} and that the $U(1)_{h_1}$ gauge coupling becomes more weakly coupled for non-zero deformations $\langle \zeta_{1,2}^{(2)} \rangle \neq0$ in case of prototype II models and $\langle \zeta_{3,4}^{(2)} \rangle \neq0$ in case of prototype IIb models. The (relevant part of the) fractional three-cycle for $h_2$ has not yet been encountered before in this article,
 but using the same techniques as in section~\ref{Ss:ExUSp2-4} we can compute its fractional three-cycle volume as a function of the $\Z_2^{(2)}$ deformations, yielding the plots in the middle and right panel of figure~\ref{Fig:DeformedCycleLRSZ2}. Hence, the $U(1)_{h_2}$ gauge coupling is expected to increase when going away from the singular orbifold point along the flat directions $\zeta_{1,2}^{(2)}$ for the prototype II models and along the flat directions $\zeta_{3,4}^{(2)}$ for the prototype IIb models.

\subsection{Deformations in global Pati-Salam models}\label{Ss:ExPS}

In~\cite{Ecker:2014hma}, a systematic computer scan led to two prototype PS models. 
In order to streamline the discussion of deformations and of one-loop corrections to the gauge couplings, we here present the prototype I with (up to the size $N_a$) identical stacks $\{a,b\}$ as in the MSSM example of section~\ref{Ss:ExMSSM}. The D6-brane configuration is displayed in table~\ref{tab:PatiSalamaAAPrototypeI} and differs from the original one in~\cite{Ecker:2014hma} by a swap of the last two two-tori, $T^2_{(2)} \leftrightarrow T^2_{(3)}$, and by a flip of the {\it absolute} $\Z_2 \times \Z_2$ eigenvalues. 
\mathtabfix{
\begin{array}{|c||c|c||c|c|c||c|}\hline 
\muc{7}{|c|}{\text{\bf D6-brane configuration of a global Pati-Salam model on the {aAA} lattice: prototype I}}
\\\hline \hline
&\text{\bf wrapping numbers} &\frac{\rm Angle}{\pi}&\text{\bf $\Z_2^{(i)}$ eigenvalues}  & (\vec \tau) & (\vec \sigma)& \text{\bf gauge group}\\
\hline \hline
 a&(1,0;1,0;1,0)&(0,0,0)&(+++)&(0,1,1) & (0,1,1)& U(4)\\
 b&(1,0;-1,2;1,-2)&(0,\frac{1}{2},-\frac{1}{2})&(--+)&(0,1,0) & (0,1,0)&USp(2)\\
 c&(1,0;-1,2;1,-2)&(0, \frac{1}{2},-\frac{1}{2})&(+--)&(0,1,0) & (0,1,0)&USp(2)\\ 
 \hline
  h&(1,0;1,0;1,0)&(0,0,0)&(--+)&(0,1,1) & (0,1,1)& U(6)\\
 \hline
\end{array}
}{PatiSalamaAAPrototypeI}{Pati-Salam model, prototype I, on the $T^6/(\Z_2 \times \Z_6 \times \OR)$ orientifold with discrete torsion. The three-cycles wrapped by the D6-branes $a$ and $b$ are identical to those of the MSSM example in
table~\protect\ref{tab:5stackMSSMaAAPrototypeI}.}
The spectrum of prototype I is displayed in table~\ref{tab:PatiSalamaAAPrototypeISpectrum}.
\mathtab{
\begin{array}{|c|c||c|c|}
\hline \multicolumn{4}{|c|}{\text{\bf Matter spectrum of the prototype I global Pati-Salam model on aAA}}\\
\hline
\hline
 \text{sector} & U(4)_a\times USp(2)_b\times USp(2)_c \times U(6)_h
& \text{sector} & U(4)_a\times USp(2)_b\times USp(2)_c \times U(6)_h
\\
\hline 
 ab & 3 \times (\4, \2, \1; \1) 
 & aa'& 2 \times \left[  ({\bf 6}_\Anti,\1,\1;\1) + h.c. \right]  
 \\
ac & 3 \times (\ov \4, \1, \2;\1)
&    bb'& 5 \times  (\1,\1_\Anti, \1;\1)
  \\
bc &  10 \times (\1,\2,\2; \1)
&     cc'& 5 \times  (\1,\1, \1_\Anti;\1)
      \\\hline
   ah & 2 \times [ (\4,\1,\1;\ov{\6}) + h.c. ]  
&      bh & 3 \times (\1, \2, \1;{\bf 6})
\\ 
ah' &   (\4,\1,\1;\6) + h.c.   
&      ch & 3 \times (\1, \1, \2; \ov{\bf 6})\\
      hh'& 2 \times [ (\1,\1,\1;{\bf15}_\Anti) + h.c.  ] & &
      \\
    \hline
\end{array}
}{PatiSalamaAAPrototypeISpectrum}{Chiral and vector-like massless  (open string) matter spectrum of the prototype I global PS model on the $T^6/(\Z_2 \times \Z_6 \times \OR)$ orientifold with D6-brane configuration given in table~\ref{tab:PatiSalamaAAPrototypeI}. The universal closed string spectrum of all global models in this article can e.g. be found in the caption of table~\protect\ref{tab:5stackMSSMaAAPrototypeISpectrum}.
 Both $U(1)_a\times U(1)_d$ gauge factors acquire  masses through the St\"uckelberg mechanism and survive as {\it perturbative} global symmetries at low energies.
}

The D6-brane configuration of the prototype II in table~\ref{tab:PatiSalamaAAPrototypeII} is up to the analogous swaps in two-torus indices and {\it absolute} choice of $\Z_2 \times \Z_2$ eigenvalues plus a swap in $h \leftrightarrow h'$ also identical to the original one from~\cite{Ecker:2014hma} with the massless matter spectrum displayed in table~\ref{tab:6stackLRSMaAAVisibleSpectrum} for the observable part 
\mathtabfix{
\begin{array}{|c||c|c||c|c|c||c|}\hline 
\muc{7}{|c|}{\text{\bf D6-brane configuration of a global Pati-Salam model on the {aAA} lattice: prototype II}}
\\\hline \hline
&\text{\bf wrapping numbers} &\frac{\rm Angle}{\pi}&\text{\bf $\Z_2^{(i)}$ eigenvalues}  & (\vec \tau) & (\vec \sigma)& \text{\bf gauge group}\\
\hline \hline
 a&(1,0;1,0;1,0)&(0,0,0)&(+++)&(0,1,1) & (0,1,1)& U(4)\\
 b&(1,0;-1,2;1,-2)&(0,\frac{1}{2},-\frac{1}{2})&(+++)&(0,1,0) & (0,1,0)&USp(2)\\
 c&(1,0;-1,2;1,-2)&(0, \frac{1}{2},-\frac{1}{2})&(-+-)&(0,1,0) & (0,1,0)&USp(2)\\ 
 \hline
  h&(1,0;-1,2;1,-2)&(0, \frac{1}{2},-\frac{1}{2})&(+++)&(0,0,0) & (0,0,0)& U(2)\\
 \hline
\end{array}
}{PatiSalamaAAPrototypeII}{Pati-Salam model, prototype II, on the $T^6/(\Z_2 \times \Z_6 \times \OR)$ orientifold with discete torsion. The three-cycles wrapped by D6-branes $\{a,b,c\}$ are identical to those of the L-R symmetric models in  table~\protect\ref{tab:6stackLRSaAA}.
Moreover, the `hidden' brane $h$ of this PS II model wraps the same three-cycle as $h_1$ of the L-R symmetric II model.}
and in table~\ref{tab:PatiSalamaAAPrototypeIISpectrum} for the `hidden' part. All discrete data are chosen such that the sector $\{a,b,c\}$ of the PS II model agrees (up to the stack size $N_a$) with that of the L-R symmetric models in section~\ref{Ss:ExLRSM}.
\mathtab{
\begin{array}{|c||c|c|c|}
\hline \multicolumn{4}{|c|}{\text{\bf `Hidden' spectrum of the prototype II global Pati-Salam model on aAA}}\\
\hline
\hline
 \text{sector} & U(4)_a\times USp(2)_b\times USp(2)_c \times U(2)_h
 & \text{sector} & U(4)_a\times USp(2)_b\times USp(2)_c \times U(2)_h
\\
\hline 
    ah & 2 \times [ (\4,\1,\1;\ov{\2}) + h.c. ]  
    &  bh & 3 \times (\1, \2, \1;{\2})  + 3\times [(\1, \2, \1;{\2}) + h.c. ] 
\\ 
 ah' &   (\4,\1,\1;\2) + h.c.    
 &  ch & 3 \times (\1, \1, \2; \ov{\2}) + 3\times [(\1, \1, \2;{\2}) + h.c. ] \\
      hh'& 6 \times [ (\1,\1,\1;\1_\Anti) + h.c.  ] & & 
      \\
    \hline
\end{array}
}{PatiSalamaAAPrototypeIISpectrum}{Chiral and vector-like massless `hidden' matter spectrum of the prototype II global Pati-Salam model with D6-brane configuration in table~\protect\ref{tab:PatiSalamaAAPrototypeII}. The observable spectrum is (up to the rank of stack $a$) identical to the $\{a,b,c\}$ sector of the L-R symmetric models, cf. 
table~\protect\ref{tab:6stackLRSMaAAVisibleSpectrum}.
Also in this prototype, both $U(1)_a\times U(1)_d$ gauge factors acquire a St\"uckelberg mass, turning them into {\it perturbative} global symmetries in the low-energy effective field theory.
}

The orientifold-odd and -even wrapping numbers are given in table~\ref{Tab:Ex-Wrappings} and lead to the naive counting of the maximal number of stabilised moduli 
of four and seven for the PS I and PS II model, respectively, in table~\ref{Tab:Stab-vs-Flat}. Even though the models differ in the sectors $\{b,c,h\}$, in both cases
the gauge couplings of branes $b,c$ are sensitive to flat directions in the deformation moduli $\zeta^{(1)}_{4-5}$ and $\zeta^{(2)}_{1,2}$, with also the gauge couplings of $a$ and $h$
sensitive to $\zeta^{(2)}_{1,2}$. 

Looking more closely at table~\ref{Tab:Stab-vs-Flat}, we observe that the deformation moduli $\zeta_{0,1,2}^{(1)}$ only couple to the hidden stack $h$ in the prototype PS II model. This opens up the possibility to stabilise one of the {\it vev}s $\langle \zeta_{0,1,2}^{(1)} \rangle$ by compensating it with the {\it vev}s of the scalar components in a vector-like pair $(\1, \2, \1;{\2}) + h.c.$ in the bifundamental representation under $USp(2)_b\times U(2)_h$ or a vector-like pair $(\1, \1,\2;{\2}) + h.c.$ in the bifundamental representation under $USp(2)_c\times U(2)_h$. In case that the necessary terms indeed appear in the low-energy effective action,
 the gauge group will be spontaneously broken from $USp(2)_{x\in \{ b,c \} } \times U(2)$ to a diagonal $SU(2)^{\rm diag}$, corresponding to a recombination of two D-brane stacks,
 which, however, does not bear an interpretation as a new fractional cycle only wrapping $\Z_2^{(i)}$ singularities in a single sector $i$, cf. the overview over the couplings to $\Z_2^{(i)}$-twisted sectors in table~\ref{Tab:Stab-vs-Flat}. 
 Whether or not such a recombination can happen, has to be studied in more detail from a geometric perspective in the future and cannot be solely assessed through the study of the D-term equations.
Furthermore, a more detailed study of the full scalar potential is required as well for these prototype models, but goes well beyond the scope of this article as well. Yet, we offer the listing of matter states per intersection sector in table~\ref{tab:Z2Z64StackPSITotalSpectrum} for the prototype I example and in table~\ref{tab:Z2Z6-LRS+PSII-visible} for the prototype II examples, from which a field theoretic study of the scalar potential could initiate in terms of the {\it necessary} selection rule of the existence of closed polygons.

Let us instead analyse the $\Z_2^{(i)}$-twisted moduli representing flat directions for the PS models, as anticipated in table~\ref{Tab:Stab-vs-Flat}. In the $\Z_2^{(1)}$-twisted sector, the D6-branes $b$ and $c$ couple to the twisted modulus $\zeta_{4-5}^{(1)}$ and (the relevant part of) their fractional three-cycles can be brought back to the fractional three-cycles supporting the $USp(2)_{b,c}$ gauge groups in the toy model of section~\ref{Ss:ExUSp2-4}. There is a substantial qualitative difference between the prototype I and prototype II models: for the prototype I models the fractional three-cycle volume for the $b$-brane is given by the right panel of figure~\ref{Fig:DeformedCycleBCToyZ1} and the one for the $c$-brane is depicted on the left-hand side of figure~\ref{Fig:DeformedCycleBCToyZ1}, whereas the prototype II models have the reverse identifications with respect to figure~\ref{Fig:DeformedCycleBCToyZ1}. This implies that the {\it left}-symmetric $USp(2)_b$ gauge group in the prototype I models acquires a stronger gauge coupling when we go away from the singular orbifold point along the flat direction $ \zeta_{4-5}^{(1)}$, while the $USp(2)_b$ gauge group in the prototype II models receives a weaker gauge coupling. The gauge coupling of the {\it right}-symmetric $USp(2)_c$ gauge group exhibits the exact opposite behaviour w.r.t.~the one of the {\it left}-symmetric stack for non-zero deformations $\langle \zeta_{4-5}^{(1)}\rangle \neq0$. This qualitative difference between the two prototype PS models can be traced back to the relative difference in $\Z_2\times \Z_2$ eigenvalues for the $b$- and $c$-stack between both prototypes. 

The $\Z_2^{(2)}$ deformations affect the gauge coupling of the hidden gauge group in the prototype~I and II PS models, as the hidden D6-brane stack $h$ in both cases
couples to the twisted deformation moduli~$\zeta_{1,2}^{(2)}$ along flat directions. The (relevant part) of the fractional three-cycle wrapped by the hidden stack $h$ in the prototype I model is characterised by the same geometry as the $d$-brane in the L-R symmetric model, such that its volume exhibits the behaviour of the left panel of figure~\ref{Fig:DeformedCycleLRSZ2} under non-zero deformation parameters $\varepsilon_{1,2}^{(2)}$. Hence, the hidden $U(6)_h$ gauge group in the prototype I model has a smaller gauge coupling at the string scale $M_{\text{string}}$ for resolved $\Z_2^{(2)}$ singularities. The $\Z_2^{(2)}$ exceptional three-cycle of the hidden stack $h$ in the prototype II model on the other hand takes the same form as the one of the $USp(2)_b$ stack in the toy model of section~\ref{Ss:ExUSp2-4}.
Hence, its fractional three-cycle volume is characterised by the middle and right panel of figure~\ref{Fig:DeformedCycleADBCToyZ2}, and the hidden gauge group $U(2)_h$ has a smaller gauge coupling in the prototype II model as well when we consider non-zero deformations $\varepsilon_{1,2}^{(2)}$ along flat directions. Note that compatibility of the common calibration w.r.t. ${\rm Re}(\Omega_3)$ of the resolved exceptional three-cycles constrains the parameter space of the deformations to the half-lines~$\varepsilon_{4-5}^{(1)}\leq0$ and~$\varepsilon_{1,2}^{(2)}\leq0$ in both prototype PS models.

\section{Gauge Couplings at One-Loop, Geometric Moduli and $M_{\text{string}}$}\label{S:GKFDefModuli}

Up to this point, changes in the volumes of D6-branes have been discussed, which are related to the tree-level gauge couplings according to equation~\eqref{Eq:g2-tree}.
At the orbifold point, the volumes stem solely from the toroidal cycles as detailed in equation~\eqref{Eq:Vol-Orb}.
For the MSSM example of section~\ref{Ss:ExMSSM}, the relation~\eqref{Eq:Vol-Orb} amounts to the following relation among the tree-level gauge couplings at the orbifold point:
\begin{equation}\label{Eq:g2-tree-relations}
\frac{1}{g^2_{SU(3)_a}} = \frac{1}{g^2_{SU(4)_h}} = \frac{2}{3} \frac{1}{g^2_{USp(2)_b}} = \frac{6}{19} \frac{1}{g^2_{U(1)_Y}},
\end{equation}
which obviously disagrees with the proposal of gauge coupling unification at $M_{\text{string}} \simeq M_{GUT} \sim 10^{16}$GeV~\cite{Blumenhagen:2003jy}.

As discussed in section~\ref{S:DefModuliGlobalModels}, 
deformations of exceptional cycles along directions of only orientifold-even wrapping numbers change the volumes of fractional cycles such that 
the degeneracy of e.g.~the gauge coupling strength of $USp(2)_b$ and $USp(2)_c$ in the L-R symmetric and PS models is lifted, and also the relations in~\eqref{Eq:g2-tree-relations} might be ablished. 

On the other hand, the degeneracy of gauge couplings is already at the orbifold point lifted when including one-loop corrections $\Delta_{G_x}$
 to the gauge couplings of any gauge factor $G_x$, 
\begin{equation}
\frac{ 8 \pi^2}{g^2_{G_x}(\mu)} = \frac{ 8 \pi^2}{g^2_{G_x,\text{tree}}} + \frac{b_{G_x}}{2} \ln \frac{M^2_{\text{string}}}{\mu^2} + \frac{\Delta_{G_x}}{2} ,
\end{equation}
which depend on the discrete D6-brane
data such as $\Z_2 \times \Z_2$ eigenvalues and discrete Wilson lines and displacement parameters. In section~\ref{Ss:GKF1Loop}, we will therefore present the 
one-loop gauge thresholds $\Delta_{G_x}$ and discuss in section~\ref{Ss:MstringVsCouplings} impacts on the low-energy phenomenology of the global particle physics vacua
with D6-brane data specified in tables~\ref{tab:5stackMSSMaAAPrototypeI},~\ref{tab:6stackLRSaAA},~\ref{tab:PatiSalamaAAPrototypeI} and~\ref{tab:PatiSalamaAAPrototypeII}.
In particular, the class of models at hand shares the unusual feature that all D6-branes wrap the $\OR(\Z_2^{(1)})$-invariant one-cycle along $T^2_{(1)}$ such that 
not only one-loop gauge thresholds can - at least in principle - 
 contribute to lowering the string scale $M_{\text{string}}$ well beyond the Planck scale $M_{\text{Planck}}$~\cite{Honecker:2012qr}, but the weakness of 
gravity can already be generated at tree level by a large hierarchy between the two radii $R_2^{(1)}$ and $R_1^{(1)}$ of the two-torus $T^2_{(1)}$.

\subsection{Gauge couplings at one-loop}\label{Ss:GKF1Loop}

Let us briefly summarise the results of the formalities of one-loop gauge threshold corrections for the case at hand of fractional D6-branes parallel along at least one two-torus~\cite{Blumenhagen:2007ip,Gmeiner:2009fb,Honecker:2011sm,Honecker:2011hm} before applying these general results to the stringy particle physics vacua on the  
\mbox{$T^6/(\Z_2 \times \Z_6 \times \OR)$} background with exotic O6-plane orbit $\OR\Z_2^{(3)}$. The K\"ahler metrics and contributions to the $SU(N_x)$ beta function coefficients,
\begin{equation}
\begin{aligned}
b_{SU(N_a)} =& \underbrace{N_a \left(-3+\varphi^{\Adj_a}\right)}+ \underbrace{\frac{N_a}{2} \, \left( \varphi^{\Sym_a} + \varphi^{\Anti_a} \right)}
 + \underbrace{\left( \varphi^{\Sym_a} -\varphi^{\Anti_a} \right)}
 +\underbrace{\sum_{b\neq a} \frac{N_b}{2} \left( \varphi^{ab} + \varphi^{ab'}\right) } 
 \\
 \equiv & \qquad \sum_{k=0}^2
 b_{a(\omega^k a)}^{\cal A} \quad + \; \qquad \sum_{k=0}^2 b_{a(\omega^k a')}^{\cal A} \qquad\quad 
 + \quad \sum_{k=0}^2 b_{(\omega^k a)}^{\cal M} \qquad +\quad  \sum_{b\neq a} \sum_{k=0}^2 \left(b_{a(\omega^k b)}^{\cal A} + b_{a(\omega^k b')}^{\cal A} \right) 
,
\end{aligned}
\end{equation}
 per open string sector are displayed in table~\ref{tab:beta_coeffs_Kaehler_metrics}
\mathtabfix{
\begin{array}{|c||c|c|c|}\hline
\muc{4}{|c|}{\text{\bf  K\"ahler metrics and $SU(N_x)$ beta function coefficients on $T^6/(\Z_2 \times \Z_{2M} \times \OR)$ with discrete torsion for vanishing angle } \phi_{xy}^{(1)}=0}
\\\hline\hline
(\phi^{(1)}_{xy},\phi^{(2)}_{xy},\phi^{(3)}_{xy}) & K_{{\bf R}_x} 
& b^{\cal A}_{xy}  = 
\begin{array}{c}
\sum_{i=1}^3 \delta_{\sigma^i_x}^{\sigma^i_y} \; \delta_{\tau^i_x}^{\tau^i_y} \tilde{b}^{{\cal A},(i)}_{xy} \\
\text{ or } \; \delta_{\sigma^1_x}^{\sigma^1_y} \; \delta_{\tau^1_x}^{\tau^1_y}  \tilde{b}^{\cal A}_{xy}  \end{array}
& b^{\cal M}_{x} (= \sum_{i=1}^3 b^{{\cal M},(i)}_{x} \text{ for } (\vec{\phi})=(\vec{0}))
 \; \text{ for } \eta_{\OR\Z_2^{(3)}}=-1 \text{ (only for $y=x'$)} 
\\\hline\hline
(0,0,0) & \frac{g_{\text{string}}}{v_1v_2v_3} \, \sqrt{2\pi} L_x^{(i)} &  \sum_{i=1}^3 \delta_{\sigma^i_x}^{\sigma^i_y} \; \delta_{\tau^i_x}^{\tau^i_y} \frac{ ( -N_y) \, I_{xy}^{\Z_2^{(i)},(j \cdot k)}}{4}
&  
\begin{array}{cl}
  -2 \, \left( 1 + (-1)^{\tau^2_x \sigma^2_x}  - (-1)^{\tau^3_x \sigma^3_x} \right) & \text{for } x \pp \OR
\\
  -2 \, \left( 1 - (-1)^{\tau^2_x \sigma^2_x}  + (-1)^{\tau^3_x \sigma^3_x} \right) & \text{for } x \pp \OR\Z_2^{(1)}
\end{array}
\\\hline
(0,\phi,-\phi) &   \frac{g_{\text{string}}}{v_1v_2v_3} \, \sqrt{2\pi} L_x^{(1)}
& \delta_{\sigma^1_x}^{\sigma^1_y} \; \delta_{\tau^1_x}^{\tau^1_y}  \frac{ (- N_y )\;  (I_{xy}^{(2 \cdot 3)} + I_{xy}^{\Z_2^{(1)},(2 \cdot 3)} )}{4}  & 
 - \frac{1}{2} ( |\tilde{I}_x^{\OR, (2, \cdot 3)}| + |\tilde{I}_x^{\OR\Z_2^{(1)},(2 \cdot 3)}|)
\\\hline
\end{array}
}{beta_coeffs_Kaehler_metrics}{K\"ahler metrics $ K_{{\bf R}_x}$ with two-torus volumes $v_i$ defined in~\protect\eqref{Eq:2-torus-volume} and one-cycle lengths $L_x^{(i)}$in~\protect\eqref{Eq:Vol-Orb}
 for matter representations ${\bf R}_x \in \{ (\N_x,\ov{\N}_y), (\N_x,\N_y), (\Anti_x), (\Sym_x) \}$
and contributions to the beta function coefficients from annulus and M\"obius strip topologies, $b^{\cal A}_{xy}$ , $b^{\cal M}_{x}$. 
For details on the computation of $\Z_2$-invariant intersection numbers, the interested reader is referred to e.g.~\cite{Honecker:2012qr,Ecker:2014hma}.
The factor $(-1)^{2 b_i \, \sigma^i_x \tau^i_x}$ appearing in the M\"obius strip contribution to the beta function coefficient is required for consistency of spectra on tilted tori $(b_i=\frac{1}{2})$
as first noted in the caption of table 49 in~\cite{Forste:2010gw}, but the exact shape of the corresponding M\"obius strip amplitude with $b_i \, \sigma^i_x \tau^i_x \neq 0$
- needed for any known phenomenologically appealing model with rigid D6-branes - 
 is to the present day not known, see appendix B.1 of~\cite{Honecker:2012qr} for an extended discussion. 
}
for the cases at hand. For the generic case with three non-vanishing angles and/or non-rigid D6-branes, the interested reader is referred to~\cite{Gmeiner:2009fb,Honecker:2011sm,Honecker:2011hm,Honecker:2012qr}.
The associated one-loop gauge threshold corrections per open string sector with at least one vanishing angle
are collected in table~\ref{tab:1-loop-thresholds},
\mathtabfix{ \begin{array}{|c||c|c|}\hline
\muc{3}{|c|}{\text{\bf One-loop corrections to gauge couplings on $T^6/(\Z_2 \times \Z_{2M} \times \OR)$ with discrete torsion for } \phi_{xy}^{(1)}=0}
\\\hline\hline
(\phi^{(1)}_{xy},\phi^{(2)}_{xy},\phi^{(3)}_{xy}) &  \Delta^{\cal A}_{xy} = N_y \tilde{\Delta}^{\cal A}_{xy}  \subset \Delta^{\cal A}_{SU(N_x)}  & \Delta^{\cal M}_x \subset \Delta^{\cal M}_{SU(N_x)}
\\\hline\hline
(0,0,0) & \sum_{i=1}^3 \tilde{b}^{{\cal A},(i)}_{xy} \, \Lambda_{\tau^i_{xy},\sigma^i_{xy}} (v_i)
& \sum_{i=1}^3 \tilde{b}^{{\cal M},(i)}_x \, \widehat{\Lambda}_{b_i,\tau^i_x,\sigma^i_x} (\widehat{v}_i)
\\\hline
(0,\phi,-\phi) & \tilde{b}^{{\cal A}}_{xy} \, \Lambda_{\tau^1_{xy},\sigma^1_{xy}} (v_1) 
+ \frac{N_y \, \ln 2}{2} \left( I_{xy}^{\Z_2^{(2)}} -  I_{xy}^{\Z_2^{(3)}} \right) \left(\frac{\text{sgn}(\phi)}{2} -\phi \right)
& \tilde{b}^{{\cal M}}_{x} \, \widehat{\Lambda}_{0,\tau^1_x,\sigma^1_x} (\widehat{v}_1)
+ \left( |I^{\OR\Z_2^{(2)}}_x| -  |I^{\OR\Z_2^{(3)}}_x|
\right)\frac{\ln 2}{2}
\\\hline
\end{array}
}{1-loop-thresholds}{One-loop corrections to the gauge couplings of fractional D6-branes at some vanishing angle on $T^6/(\Z_2 \times \Z_{2M} \times \OR)$ orientifolds with discrete torsion. 
The annulus lattice sums only depend on {\it relative} Wilson lines  $\tau^i_{xy} \equiv |\tau^i_x - \tau^i_y| \in \{0, 1\}$ and displacements $\sigma^i_{xy} = |\sigma^i_x - \sigma^i_y|\in \{0, 1\}$.
To complete the picture, the gauge thresholds for rigid D6-branes at three non-vanishing angles can be found in~\cite{Blumenhagen:2007ip,Honecker:2011sm}, where also the conversion from $\Delta_{G_x}$ to the 
holomorphic gauge kinetic function ${\rm f}_{G_x}$ using the K\"ahler metrics from table~\protect\ref{tab:beta_coeffs_Kaehler_metrics} is discussed in detail. 
For all D6-brane examples parallel to the $\OR(\Z_2^{(1)})$-invariant planes on the {\bf aAA} lattice of $T^6/(\Z_2 \times \Z_6 \times \OR)$ discussed in this article,
the constant contribution from the M\"obius strip topology vanishes due to  $|I^{\OR\Z_2^{(2)}}_x| =  |I^{\OR\Z_2^{(3)}}_x|$. The standard Annulus lattice sums are defined 
in~\protect\eqref{Eq:Def-Lambdas-Annulus}, while the
hatted lattice sums in the one-loop M\"obius corrections are defined in equation (\ref{Eq:Formal-MS-LatticeSum}).
}
where the following abbreviations of lattice sums for the annulus topology are used,
\begin{equation}\label{Eq:Def-Lambdas-Annulus}
\begin{aligned}
\Lambda_{0,0}(v) =& - \ln \left( 2 \pi L^2 \eta^4 (i v) \right),
\\
\Lambda_{\tau,\sigma \neq 0,0}(v) =&  -  \ln \left( e^{-\pi (\sigma)^2 v/4} \frac{| \vartheta_1 (\frac{\tau - i \sigma \, v}{2}, iv) |}{\eta(iv)}  \right)^2 ,
\end{aligned}
\end{equation}
with the two-torus volume $v$ defined by
\begin{equation}\label{Eq:2-torus-volume}
v = \left\{\begin{array}{cc}  
\frac{R_1R_2}{\alpha'} & \Z_2 ({\bf a})\\
\frac{\sqrt{3}}{2} \frac{r^2}{\alpha'} & \Z_3 ({\bf A})
\end{array}\right.
,
\end{equation}

and the one-cycle length $L$ as given in equation~\eqref{Eq:Vol-Orb}.

For later use in section~\ref{Ss:MstringVsCouplings}, we already provide the asymptotic behaviour of the lattice sums here for two-torus volumina larger than $\alpha'$:
\begin{equation}\label{Eq:Asymptotics-LatticeSums}
\begin{aligned}\hspace{-7mm}
\Lambda_{\tau,\sigma}(v)  \stackrel{v \gg 1}{\longrightarrow} &
\left\{\begin{array}{cr}
\frac{\pi  v}{3} -\ln (2\pi L^2) & (\tau,\sigma)=(0,0)\\
\frac{[3 (1-\sigma)^2 -1] \, \pi v}{6} - 2\, \delta_{\sigma,0} \, \ln[2 \sin(\frac{\pi\tau}{2})]
 & 
 \neq (0,0)
\end{array}\right\}
\stackrel{\tau,\sigma \in \{0,1\}}{=}  \left\{\begin{array}{cr}
\frac{\pi v}{3} - \ln (2\pi L^2) & (0,0)
\\
\frac{\pi v}{3} -2 \ln2  & (1,0)
\\
-\frac{\pi v}{6} & (\tau,1)
\end{array}\right.
,
\end{aligned}
\end{equation}
which turns out to be already an excellent approximation for $v \gtrsim 1$, cf. figure~2 of~\cite{Honecker:2012qr}.

Let us discuss for example the {\bf MSSM model} with D6-brane data specified in table~\ref{tab:5stackMSSMaAAPrototypeI}. The massless matter spectrum per sector is provided in table~\ref{tab:Z2Z65stackMSSMTotalSpectrum}, from which the beta function coefficients $b^{{\cal A},(i)}$
-- or whenever vanishing the reduced numbers $\tilde{b}^{{\cal A},(i)}$ -- can be read off. 
Combining this information with the discrete D6-brane data in table~\ref{tab:5stackMSSMaAAPrototypeI}, we can derive 
the full {\bf annulus contributions} to the gauge thresholds, which are  given by (with $R_1 \equiv R_1^{(1)}$):
\begin{equation}\hspace{-15mm}\label{Eq:MSSM-thresholds}
\begin{aligned}
\Delta_{SU(3)_a}^{\cal{A}, \text{\bf MSSM}} =&  3 \times \left( \tilde{\Delta}_{aa} + \tilde{\Delta}_{aa'} \right)
+ 2 \times  \tilde{\Delta}_{ab}
+ \left( \tilde{\Delta}_{ac} +  \tilde{\Delta}_{ac'} \right)
+ \left( \tilde{\Delta}_{ad} +  \tilde{\Delta}_{ad'} \right)
+ 4 \times \left(  \tilde{\Delta}_{ah} +  \tilde{\Delta}_{ah'} \right)
\\
=& 16 \, \Lambda_{0,0}^{|| \OR(\Z_2^{(1)})} (v_1) + 2 \, \Lambda_{0,0}^{|| \OR} (v_2) - \frac{4}{3} \, \ln 2
\\
& \stackrel{v \gg 1}{\longrightarrow} \;
\frac{16 \pi}{3} v_1 + \frac{2 \pi}{3} v_2 - \ln \left(2^{4/3} (2\pi)^{18} \Bigl( \frac{R_1^2}{\alpha'} \Bigr)^{16} \Bigl( \frac{r_2^2}{\alpha'} \Bigr)^2 \right)
,
\\
\\
\Delta_{USp(2)_b}^{\cal{A}, \text{\bf MSSM}} =& 3 \times \tilde{\Delta}_{ab} 
+ 2 \times \tilde{\Delta}_{bb} 
+ \tilde{\Delta}_{bc}
+ \tilde{\Delta}_{bd}
+ 4 \times \tilde{\Delta}_{bh}
\\
=& 28 \,  \Lambda_{0,0}^{|| \OR(\Z_2^{(1)})}(v_1) -2 \,  \Lambda_{0,0}^{|| \OR\Z_2^{(1)}}(v_2) -2 \,   \Lambda_{0,0}^{|| \OR\Z_2^{(1)}}(v_3) +  \Lambda_{1,1}(v_3) + \frac{11}{3} \, \ln 2
\\
& \stackrel{v \gg 1}{\longrightarrow} \;
\frac{28 \pi}{3} v_1  - \frac{2\pi}{3} v_2 - \frac{5\pi}{6} v_3 - \ln \left(2^{-11/3} (2\pi)^{24} \Bigl( \frac{R_1^2}{\alpha'}\Bigr)^{28}  \Bigl(\frac{r_2^2}{\alpha'}\Bigr)^{-2} \Bigl(\frac{r_3^2}{\alpha'}\Bigr)^{-2}  \right)
,
\\
\\
\Delta_{U(1)_Y}^{\cal{A}, \text{\bf MSSM}} =& 
\frac{1}{36} \, \Delta_{U(1)_a}^{\cal{A}, \text{\bf MSSM}} 
+ \frac{1}{4} \, \Delta_{U(1)_c}^{\cal{A}, \text{\bf MSSM}}
+ \frac{1}{4} \, \Delta_{U(1)_d}^{\cal{A}, \text{\bf MSSM}}
+ \left( - \tilde{\Delta}_{ac} +  \tilde{\Delta}_{ac'} \right)
+  \left( - \tilde{\Delta}_{ad} +  \tilde{\Delta}_{ad'} \right)
+ \left( - \tilde{\Delta}_{cd} + \tilde{\Delta}_{cd'} \right)
\\
=& \frac{152}{3} \, \Lambda_{0,0}^{|| \OR(\Z_2^{(1)})} (v_1)  + \frac{1}{3} \,  \Lambda_{0,0}^{|| \OR}  (v_2) + 2 \, \Lambda_{0,0}^{|| \OR\Z_2^{(1)}}  (v_2) -2 \, \Lambda_{1,1}(v_2) + \Lambda_{0,0}^{|| \OR}  (v_3) +\Lambda_{1,1}(v_3)
+ \frac{47}{18} \, \ln 2
\\
& \stackrel{v \gg 1}{\longrightarrow} \;
\frac{152 \pi }{3} v_1 + \frac{10 \pi}{9} v_2 + \frac{\pi}{6} v_3
- \ln \left( 9 \cdot 2^{-47/18} (2\pi)^{54}
\Bigl( \frac{R_1^2}{\alpha'}\Bigr)^{152/3}  
\Bigl(\frac{r_2^2}{\alpha'}\Bigr)^{7/3} 
\Bigl( \frac{r_3^2}{\alpha'} \Bigr)
\right)
,
\\
\\
\Delta_{SU(4)_h}^{\cal{A}, \text{\bf MSSM}} =& 3 \times  \Bigl(  \tilde{\Delta}_{ah} +  \tilde{\Delta}_{ah'} \Bigr)
+ 2 \times \tilde{\Delta}_{bh}
+ \left(  \tilde{\Delta}_{ch} + \tilde{\Delta}_{ch'} \right)
+ \left( \tilde{\Delta}_{dh} +  \tilde{\Delta}_{dh'} \right)
+ 4 \times  \Bigl(  \tilde{\Delta}_{hh} +  \tilde{\Delta}_{hh'} \Bigr)
\\
=& 16 \, \Lambda_{0,0}^{|| \OR(\Z_2^{(1)})}(v_1)  - 2 \,  \Lambda_{0,0}^{|| \OR}(v_2) + 2 \, \ln 2
\\
& \stackrel{v \gg 1}{\longrightarrow}
\frac{16 \pi}{3} v_1 - \frac{2 \pi}{3} v_2 
- \ln \left( 2^{-2} (2\pi)^{14} \Bigl( \frac{R_1^2}{\alpha'}\Bigr)^{16}  \Bigl(\frac{r_2^2}{\alpha'}\Bigr)^{-2}
\right)
,
\end{aligned}
\end{equation}
where we have made use of the same notation $\tilde{\Delta}_{xy}=\tilde{\Delta}_{yx}$ as in e.g.~\cite{Honecker:2011sm}.
We will come back to the impact of these one-loop correction in section~\ref{Ss:MstringVsCouplings} after having also determined the M\"obius strip contributions,
but already point out here that all threshold contributions have a positive dependence on the two-torus volume $v_1$, while $v_2$ and $v_3$ appear with negative prefactors  
in the one-loop correction to $USp(2)_b$ and $SU(4)_h$.

The annulus contributions to the gauge thresholds of the {\bf L-R symmetric models} are analogously computed using the D6-brane data in table~\ref{tab:6stackLRSaAA} and the resulting state-per-sector counting in tables~\ref{tab:Z2Z6-LRS+PSII-visible} and~\ref{tab:Z2Z6-LRS-hidden}
with the following results:
\begin{equation}
\begin{aligned}
\Delta_{SU(3)_a}^{\cal{A}, \text{\bf L-R}} =&   \left\{\begin{array}{lr}
16 \,  \Lambda_{0,0}^{|| \OR(\Z_2^{(1)})}(v_1)    & \text{\bf L-R I \& II}   \\
10 \,   \Lambda_{0,0}^{|| \OR(\Z_2^{(1)})}(v_1) + 6 \,  \Lambda_{0,1}(v_1) & \text{\bf L-R IIb}   \\
13 \,   \Lambda_{0,0}^{|| \OR(\Z_2^{(1)})}(v_1) + 3 \, \Lambda_{0,1}(v_1)  & \text{\bf L-R IIc}
    \end{array}\right\}
    -4 \, \Lambda_{0,0}^{|| \OR}(v_2) - \frac{10}{3} \, \ln 2
 \,    ,
 \\
\\
\Delta_{USp(2)_b}^{\cal{A}, \text{\bf L-R}} =&  \left\{\begin{array}{lr} 
33 \,  \Lambda_{0,0}^{|| \OR(\Z_2^{(1)})}(v_1)   & \text{\bf L-R I \& II} \\
24\,  \Lambda_{0,0}^{|| \OR(\Z_2^{(1)})}(v_1) + 9 \,  \Lambda_{0,1}(v_1) & \text{\bf L-R IIb} \\
28\,  \Lambda_{0,0}^{|| \OR(\Z_2^{(1)})}(v_1) + 5 \,  \Lambda_{0,1}(v_1)  & \text{\bf L-R IIc}
\end{array}\right\}
- 4 \,  \Lambda_{0,0}^{|| \OR\Z_2^{(1)}}(v_2) -\frac{20}{3}\, \ln 2
 \,    ,
\\
\\
\Delta_{USp(2)_c}^{\cal{A}, \text{\bf L-R}} =&   \left\{\begin{array}{lr} 
33 \,  \Lambda_{0,0}(v_1)^{|| \OR(\Z_2^{(1)})}  & \text{\bf L-R I \& II} \\
24\,  \Lambda_{0,0}(v_1)^{|| \OR(\Z_2^{(1)})} + 9 \,  \Lambda_{0,1}(v_1) & \text{\bf L-R IIb} \\
29\,  \Lambda_{0,0}(v_1)^{|| \OR(\Z_2^{(1)})} + 4 \,  \Lambda_{0,1}(v_1)  & \text{\bf L-R IIc}
\end{array}\right\}
- 4 \,  \Lambda_{0,0}^{|| \OR\Z_2^{(1)}}(v_2) -\frac{20}{3}\, \ln 2
\,     .
\end{aligned}
\end{equation}
The asymptotic behaviour is again easily extracted, producing again only positive contributions from the two-torus volume $v_1$.

For the {\bf prototype I L-R symmetric model}, the non-Abelian hidden gauge groups experience the following one-loop gauge threshold correction,
\begin{equation}
\begin{aligned}
\Delta_{SU(3)_{h_1/h_2}}^{\cal{A}, \text{\bf L-R I only}} =&  16 \,  \Lambda_{0,0}^{|| \OR(\Z_2^{(1)})}(v_1) \mp  \frac{20}{3}\, \ln 2
\, ,
\end{aligned}
\end{equation}
and finally the generalised massless $\widetilde{U(1)}_{B-L}$ gauge group of the {\bf prototype II L-R symmetric model} receives the following one loop correction:
\begin{equation}
\begin{aligned}
\Delta_{\widetilde{U(1)}_{B-L}}^{\cal{A}, \text{\bf L-R II}} =&
\frac{704}{3} \, \Lambda_{0,0}^{|| \OR(\Z_2^{(1)})}(v_1) 
+ \frac{16}{3} \,  \Lambda_{0,0}^{|| \OR} (v_2)
+ 16 \, \Lambda_{0,0}^{|| \OR}(v_3) 
+ 16  \, \Lambda_{0,0}^{|| \OR\Z_2^{(1)}}(v_3)  
+ \frac{674}{27} \, \ln 2
\, .
\end{aligned}
\end{equation}

Finally, for the {\bf PS models}, the analogous computations using the D6-brane data in tables~\ref{tab:PatiSalamaAAPrototypeI} and~\ref{tab:PatiSalamaAAPrototypeII} and the state-per-sector counting in tables~\ref{tab:Z2Z64StackPSITotalSpectrum} and~\ref{tab:Z2Z6-LRS+PSII-visible} leads to:
\begin{equation}
\begin{aligned}
\Delta_{SU(4)_a}^{\cal{A}, \text{\bf PS}} =& 16 \, \Lambda_{0,0}^{|| \OR(\Z_2^{(1)})}(v_1) + \left\{\begin{array}{ll}
4 \,  \Lambda_{0,0}^{|| \OR}(v_2) - \frac{4}{3} \, \ln 2  & \text{\bf PS I}  \\
-8 \,  \Lambda_{0,0}^{|| \OR}(v_2) - \frac{16}{3} \, \ln 2 & \text{\bf PS II} 
\end{array}\right.
\, ,
\\
\Delta_{USp(2)_b}^{\cal{A}, \text{\bf PS}} = \Delta_{USp(2)_c}^{\cal{A}, \text{\bf PS}} =&  33 \,  \Lambda_{0,0}^{|| \OR(\Z_2^{(1)})}(v_1) - 4\, \Lambda_{0,0}^{|| \OR\Z_2^{(1)}}(v_2) +  \left\{\begin{array}{ll} 
 \frac{8}{3} \, \ln 2 & \text{\bf PS I} \\
-2 \,  \Lambda_{1,1}(v_2) - \frac{26}{3} \, \ln 2 & \text{\bf PS II} \end{array}\right.
\, ,
\\
\Delta_{SU(6_{\text{\bf I}}/2_{\text{\bf II}})_h}^{\cal{A}, \text{\bf PS}} =& \left\{\begin{array}{ll}
16 \, \Lambda_{0,0}^{|| \OR(\Z_2^{(1)})}(v_1) - 4 \, \Lambda_{0,0}^{|| \OR}(v_2) + 128 \, \ln 2 & \text{\bf PS I} \\
48 \, \Lambda_{0,0}^{|| \OR(\Z_2^{(1)})}(v_1)  - 4 \, \Lambda_{0,0}^{|| \OR\Z_2^{(1)}}(v_2)  -4 \, \Lambda_{1,1}(v_2)- \frac{1280}{3} \, \ln 2 & \text{\bf PS II}
\end{array}\right.
\, .
\end{aligned}
\end{equation}
As for the MSSM and L-R symmetric examples, any of these gauge threshold contributions grows (asymptotically) linearly with $v_1$.

For the {\bf M\"obius strip} topology, modified lattice sums,
\begin{equation}\label{Eq:Formal-MS-LatticeSum}
\widehat{\Lambda}_{b,\tau,\sigma}(\hat{v})
\qquad
\text{with}
\qquad
\hat{v} \equiv \frac{v}{1-b},
\end{equation}
with $b \in \{0,\frac{1}{2}\}$ appear. 
For the {\it untilted torus} with $b=0$, i.e. the two-torus $T^2_{(1)}$ in the $T^6/(\Z_2 \times \Z_6 \times \OR)$ examples at hand, and only discrete Wilson lines and displacements $\tau,\sigma\in \{0,1\}$,  
the lattice sum on $T^2_{(1)}$ is simply given by
\begin{equation}\label{Eq:Lambda-Hat-for-b0}
\widehat{\Lambda}_{0,\tau,\sigma}(\hat{v}) \stackrel{\tau,\sigma \in \{0,1\}}{\equiv}  \widehat{\Lambda}_{0}(\hat{v}) =\Lambda_{0,0}(v) - 2 \, \ln 2
,
\end{equation}
where the constant term $-2 \, \ln 2$ stems from the replacement~\cite{Gmeiner:2009fb,Honecker:2011sm} $L^2 \to 2 \, \hat{L}^2$ with $\hat{L}^2=2 \, L^2$ in the first line of equation~\eqref{Eq:Def-Lambdas-Annulus}.

However, for $(\tau,\sigma)=(1,1)$ on tilted tori $b=\nicefrac{1}{2}$, the sign factor $(-1)^{\tau\sigma}$ in the beta function coefficients in table~\ref{tab:beta_coeffs_Kaehler_metrics}
indicates that also the lattice sum for the hexagonal two-tori  $T^2_{(2)} \times T^2_{(3)}$ needs to be modified in a yet unknown way.
We will therefore write the formal expression~\eqref{Eq:Formal-MS-LatticeSum} for the lattice sums throughout the computation and only estimate their size via the ansatz for the asymptotics $\widehat{\Lambda}_{\frac{1}{2},\tau,\sigma}(\hat{v})
\stackrel{v \gg 1}{\longrightarrow} \frac{\pi c^{\frac{1}{2}}_{\tau,\sigma}}{3} v$ with coefficients $c^{\frac{1}{2}}_{\tau,\sigma} = {\cal O}(1)$ when discussing hierarchies among compact direction in section~\ref{Ss:MstringVsCouplings}.

Using the same D6-brane data as for the annulus amplitudes, we obtain as joint expressions for {\bf all models},
\begin{equation}\label{Eq:Eq:MS-contribtions-all}
\begin{aligned}
\Delta^{{\cal M}, {\text{\bf all}}}_{SU(3/4)_a} = \Delta^{{\cal M}, {\text{\bf MSSM}}}_{SU(4)_h} = \Delta^{{\cal M}, {\text{\bf PS I}}}_{SU(6)_h}
=& -4  \, \widehat{\Lambda}_{0,0,0}^{|| \OR(\Z_2^{(1)})}(\hat{v}_1)
 +   2  \,  \widehat{\Lambda}_{\frac{1}{2},1,1} (\hat{v}_2)
-2 \,  \widehat{\Lambda}_{\frac{1}{2},1,1} (\hat{v}_3)
\,  ,
\\
\Delta^{{\cal M}, {\text{\bf all}}}_{USp(2)_b} = \Delta^{{\cal M}, {\text{\bf PS + L-R}}}_{USp(2)_c} 
=& - 6 \, \widehat{\Lambda}_{0,0,0}^{|| \OR(\Z_2^{(1)})}(\hat{v}_1)
-  \widehat{\Lambda}_{\frac{1}{2},1,1} (\hat{v}_2)
-   \widehat{\Lambda}_{\frac{1}{2},0,0}^{|| \OR\Z_2^{(1)}} (\hat{v}_3)
\, ,
 \end{aligned}
\end{equation}
since M\"obius strip contributions are independent of the $\Z_2 \times \Z_2$ eigenvalues.
The remaining massless and anomaly-free gauge symmetries of the {\bf MSSM}, {\bf PS II} and {\bf L-R symmetric II}  model receive the following 
M\"obius strip contributions to the one-loop gauge thresholds,
\begin{equation}\label{Eq:MS-contribtions-individuals}
\begin{aligned}
\Delta^{{\cal M}, {\text{\bf MSSM}}}_{U(1)_Y} =& - \frac{38}{3} \,  \widehat{\Lambda}_{0,0,0}^{|| \OR(\Z_2^{(1)})}(\hat{v}_1)
 +   \widehat{\Lambda}_{\frac{1}{2},0,0}^{\OR\_2^{(1)}} (\hat{v}_2) -   \frac{2}{3}  \,  \widehat{\Lambda}_{\frac{1}{2},1,1} (\hat{v}_2)
 +    \frac{5}{3}  \, \widehat{\Lambda}_{\frac{1}{2},1,1} (\hat{v}_3)
\,  ,
\\
\Delta^{{\cal M}, {\text{\bf L-R II}}}_{\widetilde{U(1)}_{B-L}} =&
- \frac{176}{3}  \, \widehat{\Lambda}_{0,0,0}^{|| \OR(\Z_2^{(1)})}(\hat{v}_1) 
 + 8  \,  \widehat{\Lambda}_{\frac{1}{2},0,0}^{\OR\Z_2^{(1)}}  (\hat{v}_2)
+ \frac{16}{3} \, \widehat{\Lambda}_{\frac{1}{2},1,1} (\hat{v}_2)
 -8 \,  \widehat{\Lambda}_{\frac{1}{2},0,0}^{\OR\Z_2^{(1)}}  (\hat{v}_3)
- \frac{16}{3} \,  \widehat{\Lambda}_{\frac{1}{2},1,1} (\hat{v}_3)
\, ,
\\
\Delta^{{\cal M}, {\text{\bf PS II}}}_{SU(2)_h} =& 
-12 \,  \widehat{\Lambda}_{0,0,0}^{|| \OR(\Z_2^{(1)})}(\hat{v}_1)
 + 2 \,  \widehat{\Lambda}_{\frac{1}{2},0,0}^{\OR\Z_2^{(1)}} (\hat{v}_2)
-2 \,  \widehat{\Lambda}_{\frac{1}{2},0,0}^{\OR\Z_2^{(1)}}  (\hat{v}_3)
\, .
 \end{aligned}
\end{equation}
At this point, it is noteworthy that the absolute values of the negative coefficients  in equations~\eqref{Eq:Eq:MS-contribtions-all} and~\eqref{Eq:MS-contribtions-individuals}
of the lattice sum $\widehat{\Lambda}_{0,0,0}^{|| \OR(\Z_2^{(1)})}(\hat{v}_1)$ on $T^2_{(1)}$ are always smaller than the positive coefficients in the corresponding Annulus contributions, and thus the one-loop threshold correction of each gauge group still grows (asymptotically) linearly with the two-torus volume $v_1$ in any model considered in this article.
Moreover, the coefficients of $\widehat{\Lambda}_{\frac{1}{2},1,1} (\hat{v}_2)$ and $\widehat{\Lambda}_{\frac{1}{2},1,1} (\hat{v}_3)$ have identical absolute value and opposite sign for $SU(3/4)_a^{\text{\bf all}}$, $SU(4)_h^{\text{\bf MSSM}}$, $SU(6_{\bf I}/2_{\bf II})_h^{\text{\bf PS}}$. When choosing isotropic torus volumes $v_2=v_3$, these yet unknown contributions thus cancel.

\subsection{Balancing $M_{\text{string}}$ with gravitational and gauge couplings in four dimensions}\label{Ss:MstringVsCouplings}

Dimensional reduction of the ten-dimensional Einstein-Hilbert term leads to the following relation among the four-dimensional Planck scale $M_{\text{Planck}}$, the string scale $M_{\text{string}}$, the string coupling $g_{\text{string}}$ and the  K\"ahler moduli $v_i$ inherited from the underlying torus $T^6=(T^2_{(i)})^3$,
\begin{equation}\label{Eq:Scales+Volumes}
\frac{M_{\text{Planck}}^2}{M_{\text{string}}^2} = \frac{4\pi}{g^2_{\text{string}}} \, v_1v_2v_3
\quad
\text{with }
g_{\text{string}} \equiv e^{\phi_{10}},
\end{equation}
which in conventional D6-brane models on tori or toroidal orbifolds  with phenomenologically acceptable sizes of gauge couplings at tree-level is only consistent
with a high string scale $M_{\text{string}} \lesssim M_{\text{GUT}}$~\cite{Klebanov:2003my,Blumenhagen:2003jy}, since usually the O6-plane tensions are cancelled by D6-branes at non-trivial angles on all three two-tori.
However, the situation in all global D6-brane models discussed in this article is special since the existence of the exotic O6-plane orbit $\OR\Z_2^{(3)}$ on $T^6/(\Z_2 \times \Z_6 \times \OR)$ with discrete torsion 
 enforces all D6-branes to lie along the $\OR(\Z_2^{(1)})$-invariant direction on the first two-torus $T^2_{(1)}$ with length scale $R_1^{(1)}$, while gravitational forces also propagate along the perpendicular direction and are thus sensitive to $v_1 =\frac{R_1^{(1)} R_2^{(1)}}{\alpha'}$.
The tree-level gauge coupling in equation~\eqref{Eq:g2-tree} of $SU(3)_a$ and $SU(4)_h$ of the MSSM example can e.g.~be rewritten using the relation~\eqref{Eq:Scales+Volumes},
\begin{equation}\label{Eq:g2-via-Mstring}
\frac{8\pi^2}{g^2_{SU(3)_a/SU(4)_h,\text{tree}}} = \frac{\pi}{2\sqrt{6}} \sqrt{\frac{R_1^{(1)}}{R_2^{(1)}}}   \frac{M_{\text{Planck}}}{M_{\text{string}}}
,
\end{equation}
which means that a large hierarchy between $M_{\text{Planck}}$ and $M_{\text{string}}$ can - at least in principle - be compensated by a large hierarchy between $R_2^{(1)}$ and $R_1^{(1)}$ to arrive at some phenomenologically appealing order of magnitude of the gauge couplings as already briefly sketched in~\cite{Honecker:2016gyz}. 
Such large hierarchies in the D6-brane models at hand would provide explicit examples for the low-string scale scenario, see e.g.~\cite{Kiritsis:2003mc,Lust:2008qc,Anchordoqui:2009mm,Anchordoqui:2011eg,Cvetic:2011iq,Lebed:2011fw,Buckley:2011vc,Anastasopoulos:2011hj,Berenstein:2014wva}.

We will now also take into account the one-loop gauge threshold corrections, which for the MSSM example of section~\ref{Ss:ExMSSM}
 have the following asymptotic behaviour,
\begin{equation}\label{Eq:Delta-A+MS-MSSM}
\begin{aligned}
\frac{\Delta^{{\cal A}+{\cal M}, \text{\bf MSSM}}_{SU(3)_a}}{2} \stackrel{v \gg 1}{\longrightarrow} & \quad 
2\pi \, v_1 +\frac{\pi}{3} \left(v_2 + c^{\frac{1}{2}}_{1,1} \bigl(v_2 - v_3 \bigr)  \right) - \ln \left( \Bigl(\frac{R_1^{(1)}}{R_2^{(1)}} \, v_1 \Bigr)^{6}  v_2 \right) - 12
,
\\
\frac{\Delta^{{\cal A}+{\cal M}, \text{\bf MSSM}}_{USp(2)_b}}{2} \stackrel{v \gg 1}{\longrightarrow} & \quad 
\frac{11 \pi}{3} \, v_1 - \frac{\pi}{3} \left(1 + \frac{c^{\frac{1}{2}}_{1,1}}{2}  \right)\, v_2 - \frac{3\pi}{4} \, v_3
- \ln \left(\Bigl(\frac{R_1}{R_2} \, v_1 \Bigr)^{11} (v_2)^{-1} v_3)^{-\nicefrac{3}{2}}  \right) - 11
,
\\
\frac{\Delta^{{\cal A}+{\cal M}, \text{\bf MSSM}}_{U(1)_Y}}{2} \stackrel{v \gg 1}{\longrightarrow} & \quad 
\frac{209 \pi}{9} \, v_1 + \frac{(8-  c^{\frac{1}{2}}_{1,1} )\pi}{9} \, v_2
+ \frac{ (3 + 10 \, c^{\frac{1}{2}}_{1,1}) \pi}{36} \, v_3
- \ln \left( \Bigl( \frac{R_1}{R_2} \, v_1\Bigr)^{19}   v_2^{\frac{5}{3}} v_3^{\frac{1}{2}}  \right)
 - 36
,
\\
\frac{\Delta^{{\cal A}+{\cal M}, \text{\bf MSSM}}_{SU(4)_h}}{2} \stackrel{v \gg 1}{\longrightarrow} & \quad 
2\pi \, v_1 +\frac{\pi}{3} \left( - v_2 + c^{\frac{1}{2}}_{1,1} \bigl(v_2 - v_3 \bigr)  \right)
 - \ln \left( \Bigl( \frac{R_1}{R_2} \, v_1 \Bigr)^{6} v_2^{-1} \right) - 7 
,
\end{aligned}
\end{equation}
where we  have rewritten the logarithmic terms in terms of the complex structure parameter $R_2^{(1)}/R_1^{(1)}$
 on $T^2_{(1)}$ and the K\"ahler moduli $v_i$ using the definition~\eqref{Eq:2-torus-volume} and evaluated the constant contributions, e.g. $ \ln \left(2^{4/3} (2\pi)^{18}  \Bigl( \frac{2}{\sqrt{3}} \Bigr)^2 \right) \approx 34.29$ in the annulus contribution $\Delta^{{\cal A}, \text{\bf MSSM}}_{SU(3)_a}$.

Let us stress here again that in all MSSM, L-R symmetric and PS models of section~\ref{S:DefModuliGlobalModels},
the gauge coupling of the QCD stack is weakened by the positive (asymptotically) linear one-loop contribution of the volume $v_1$, at least for $v_1 \gtrsim 2$.
Due to the lack of knowledge of the exact shape of the M\"obius strip contribution for D6-branes with bulk part along some orientifold-invariant direction and
non-vanishing Wilson line and displacement, $(\tau,\sigma)=(1,1)$, along some tilted two-torus $T^2_{(2 \text{ or } 3)}$, we have to distinguish two cases, where our (asymptotic) field theory results in
equation~\eqref{Eq:Delta-A+MS-MSSM} are classified as reliable:
\begin{enumerate}
\item
Isotropic volumes of the last two tori, i.e. $v_2=v_3$: in this case, the unknown M\"obius strip contributions within the gauge thresholds of $SU(3)_a$ and $SU(4)_h$ of the MSSM example cancel.
\item
A much larger first two-torus, i.e. $v_1 \gg v_2, v_3$: the one-loop gauge threshold corrections are expected to be dominated by the asymptotics linear in $v_1$, and all gauge couplings 
will be weakened due to the positive prefactor of $v_1$ for every single gauge group in each MSSM, L-R symmetric and PS model of section~\ref{S:DefModuliGlobalModels}.
\end{enumerate}
Let us discuss each of these two cases further: in the first case of two isotropic tori \mbox{$T^2_{(2)} \times T^2_{(3)}$}, for generic volumes $v_1 > 2$ the (asymptotic) linear dependence on $v_1$ surpasses the negative constant contributions to all four one-loop gauge thresholds in~\eqref{Eq:Delta-A+MS-MSSM}, and at least for \mbox{$v_{2,3}> 2.5$} the QCD stack will have a weaker coupling at one-loop due to the (asymptotic) linear dependence on $v_2$, while the coupling of the hidden stack becomes stronger. 
This behaviour facilitates the formation of a gaugino condensate on the hidden stack, which will in turn lead to supersymmetry breaking mediated to the observable sector
on the one hand via gravitational couplings and on the other hand through the messenger particles with $USp(2)_b \times U(1)_Y \times SU(4)_h$ charges in the 
last block of table~\ref{tab:5stackMSSMaAAPrototypeISpectrum}.
Going to the edge of the validity of the geometric regime, $v_1 \gtrsim 1$ and $6 \gtrsim v_{2,3} \gtrsim 1$, one can read off from equation~\eqref{Eq:Delta-A+MS-MSSM}
that the one-loop corrections to the inverse of (the squared of) the $SU(3)_a \times SU(4)_h$ gauge couplings will be negative, while for $USp(2)_b \times U(1)_Y$ 
the contributions of $v_1$ cancel the negative constant contributions. To achieve $v_1 \gtrsim 1$, we further assume $R_1^{(1)} \sim R_2^{(1)} \sim \sqrt{\alpha'}$, and equation~\eqref{Eq:g2-via-Mstring}
favours a high string scale $M_{\text{string}}$. Such a choice of scales is in turn consistent with very small volumes $v_i \sim {\cal O}(1)$ in~\eqref{Eq:Scales+Volumes} and a not too weak string coupling  $g_{\text{string}}$.

In the second case of one two-torus significantly larger than the other two, $v_1 \gg v_2,v_3$, we can e.g. make the ansatz of $M_{\text{string}} \sim 10^{12}$ GeV and $g_{\text{string}} \sim 0.1$ in 
equation~\eqref{Eq:Scales+Volumes} leading to $v_1v_2v_3 \sim 10^{11}$. In order to achieve $\alpha^{-1}_{QCD} \sim {\cal O}(1)$, equation~\eqref{Eq:g2-via-Mstring} then requires $\nicefrac{R_1^{(1)}}{R_2^{(1)}} \sim 10^{-12}$, which is problematic if we require $v_{i \in \{1,2,3\}} \gtrsim 1$ and $\nicefrac{R_{1,2}^{(1)}}{\sqrt{\alpha'}} \gtrsim 1$ to be in the geometric regime where the supergravity approximation is expected to be reliable.
Even when we choose a larger string coupling of  $g_{\text{string}} \sim 0.3$, $v_1 \sim 10^{12}$ and $v_2 = v_3 \gtrsim 1$, all one-loop corrections in equation~\eqref{Eq:Delta-A+MS-MSSM} 
- as well as the analogous expressions for all L-R symmetric and PS models -
receive exponentially large positive contributions from $v_1$, thereby exponentially suppressing all gauge couplings.  

Let us now assume a high string scale $M_{\text{string}} \sim M_{\text{GUT}}$ in equation~\eqref{Eq:Scales+Volumes} and discuss the possibility of gauge coupling unification.
The tree-level relation~\eqref{Eq:g2-tree-relations} together with the fact that in the MSSM example only $USp(2)_b$ and $U(1)_d \subset U(1)_Y$ possess one flat direction along the deformation modulus $\varepsilon^{(1)}_{4-5}$, boils down to the requirement that the combination of a deformation along this direction with the one-loop gauge threshold correction reduces the inverse of the gauge coupling squared 
by $50 \%$ for $USp(2)_b \simeq SU(2)_{\text{weak}}$ and by roughly a factor of three for $U(1)_Y$.
Figure~\ref{Fig:DeformedCycleBCToyZ1} shows that the volume ${\rm Vol}(\Pi_b) \propto \nicefrac{1}{g_{USp(2)_b}^2}$ decreases by at most $20 \%$ when reaching the upper bound $|\varepsilon^{(1)}_{4-5}| \approx 0.4$, while ${\rm Vol}(\Pi_d)$  increases at the same time by about $10 \%$.
Thus, the one-loop corrections of $USp(2)_b \times U(1)_Y$ both have to be sufficiently negative compared to the one-loop correction of $SU(3)_a$. 
Using equation~\eqref{Eq:Delta-A+MS-MSSM},  we find that for the unknown constant $c^{\frac{1}{2}}_{1,1} < - \frac{3}{22}$ at least the prefactor of $v_3$ is negative for both gauge threshold differences
$\nicefrac{[\Delta^{{\cal A}+{\cal M}, \text{\bf MSSM}}_{USp(2)_b} - \Delta^{{\cal A}+{\cal M}, \text{\bf MSSM}}_{SU(3)_a}]}{2}$
and $\nicefrac{[\Delta^{{\cal A}+{\cal M}, \text{\bf MSSM}}_{U(1)_Y} - \Delta^{{\cal A}+{\cal M}, \text{\bf MSSM}}_{SU(3)_a}]}{2}$ under consideration, while for 
$\frac{5}{4} < c^{\frac{1}{2}}_{1,1}$, at least the prefactor of $v_2$ is negative.
In both cases, either $v_2$ or $v_3$ has to be the largest K\"ahler modulus, and only the missing computation of the M\"obius strip contribution for tilted tori and non-vanishing Wilson line
and displacement parameters, i.e. $2b_i\tau^i \sigma^i =1$, can settle the question if gauge coupling unification in the MSSM example is feasible.

\section{Conclusions and Outlook}\label{S:conclu}

This article forms the third part in a tryptic about deforming $\Z_2$ singularities of toroidal orbifolds $T^6/(\Z_2\times \Z_{2M})$ with discrete torsion in Type IIA superstring theory, focusing here on the - technically most involved but at the same time of greatest phenomenological interest~- 
  $T^6/(\Z_2 \times \Z_6 \times \OR)$ orientifold. To study the deformations of such singularities, we employ the techniques of the hypersurface formalism developed for the $\Z_2 \times \Z_2$ toroidal orbifold with discrete torsion and extend them to the $\Z_2 \times \Z_6$ toroidal orbifold, in a similar spirit as was done for the $\Z_2 \times \Z_6'$ toroidal orbifold before. The extension boils down to modding out an additional $\Z_3 \subset \Z_6$ action along a four-torus in the hypersurface formalism for the $\Z_2 \times \Z_2$ toroidal orbifold. 
  Since the $\Z_6$ action does not constrain the geometry of the first two-torus, the deformations in the $\Z_2^{(1)}$-twisted sector are structurally different from the ones in the $\Z_2^{(i=2, 3)}$-twisted sectors. This finding agrees with the difference in the relevant Hodge numbers as follows:
 we can identify six deformation moduli (four real and one complex) in the $\Z_2^{(1)}$-twisted sector and four (real) deformation moduli in the $\Z_2^{(i=2,3)}$-twisted sector each, as presented in the hypersurface equation (\ref{Eq:HyperZ2Z6Full}). This inherent difference among the $\Z_2^{(i)}$-twisted sectors presents new challenges that where absent for the orbifold groups $\Z_2 \times \Z_2$ and $\Z_2 \times \Z_6'$, which act isotropically on all three two-tori. It also forces us to discuss the deformations in the $\Z_2^{(1)}$-twisted sector in section~\ref{Sss:sLagsZ21}
 separately from the ones in the $\Z_2^{(i=2,3)}$-twisted sectors in section~\ref{Sss:sLagsZ23}.  

Introducing an anti-holomorphic involution - a necessary geometric part of the Tye IIA orientifold projection - in the hypersurface formalism opens the door for identifying {\it sLag} three-cycles, albeit only a minimal subset of {\it sLag} three-cycles defined as the fixed loci under this involution. By virtue of Weierstrass' elliptic function, one can easily divide this subset into bulk three-cycles and fractional three-cycles for vanishing deformations. For non-zero deformations, exceptional three-cycles are characterised in the hypersurface formalim by (real) algebraic equations whose local form reduces to the ones of a (resolved) $\C^2/\Z_2$ singularity. Their global description on the other hand is for most deformations not attainable due to the compact topology of the ambient $T^4$ and/or different complex structures of the two-torus lattices within the factorisable ambient $T^4_{(2 \text{ or } 3)}$ in case of the $\Z_2^{(2 \text{ or } 3)}$-twisted sector. Only the exceptional three-cycles associated to the $\Z_2^{(1)}$ deformation parameters $\varepsilon_{0}^{(1)}$ and $\varepsilon_3^{(1)}$ can be fully described globally, with the first deforming only the singularity at the origin of $T^4_{(1)}/\Z_2^{(1)}$ and the latter deforming a $\Z_3$-invariant orbit of three singularities on $T^4_{(1)}/\Z_6$. By computing the volume-dependence on these deformation parameters for bulk and fractional three-cycles parallel to the $\OR$-invariant orientifold plane we are able to cross-check and validate the results of the exceptional three-cycle volume. Moreover, by studying the effects of deformations $\varepsilon_{0}^{(1)}$ and $\varepsilon_3^{(1)}$ on simple fractional three-cycles in detail we obtain the necessary intuition to investigate the effects of other deformations, for which the exceptional three-cycles cannot be accessed directly due to the absence of a global description. This also allows to propose a method for assessing qualitatively any fractional three-cycle volume as a function of a particular deformation parameter, as explained on page~\pageref{It:Method}.

When applying these techniques to global intersecting D6-brane models, we observe in first instance that $\OR$-even exceptional three-cycles give rise to deformation moduli with a flat direction, while D6-branes wrapping $\OR$-odd exceptional three-cycles couple to deformation moduli which ought to be stabilised at {\it vev} to avoid the presence of non-vanishing Fayet-Iliopoulos terms, which in turn would signal a breakdown of supersymmetry at the string or Kaluza-Klein scale. A loophole to this consideration emerges when the massless open string spectrum contains suitably charged states whose scalar components can develop non-zero {\it vev}s to compensate the non-zero FI-term(s) in all D-term equations simultaneously.
 If present, this phenomenon would result in a spontaneous breaking of the gauge group supported by the D6-branes and would correspond geometrically to a recombination of two separate D-brane stacks (or a separation of D-branes within a single stack). In order to fully understand whether the loophole is realisable, it is of uttermost importance to acquire a better handle on the low-energy effective action from first principles through CFT computations~\cite{Akerblom:2007np,Akerblom:2007uc,Blumenhagen:2007ip,Honecker:2011sm} or through dimensional reduction~\cite{Douglas:1996sw}. We hope to resolve and report on this issue in future work.

Bypassing the loophole we can count the maximal number of stabilised deformation moduli $\langle \zeta^{(i)}_{\lambda} \rangle = 0$, i.e.~those $\zeta^{(i)}_{\lambda}$ coupling to D6-branes via the respective $\OR$-odd exceptional wrapping number.
 An overview of this counting for the previously constructed and phenomenologically interesting global D6-brane models on $T^6/(\Z_2 \times \Z_6 \times \OR)$ is provided in table~\ref{Tab:Stab-vs-Flat}, which shows that a maximum of ten out of 14 twisted complex structure moduli can be stabilised in the L-R symmetric prototype IIc model. For the other prototype models, the maximal number of stabilised deformation moduli is lower, with the lowest number being four in case of the global prototype I PS model. In table~\ref{Tab:Stab-vs-Flat}, we also observe a number of twisted moduli with a flat direction affecting the low-energy effective field theory. By going away from the singular orbifold point along these flat directions, the (inverse squared of the) tree-level gauge coupling of some D6-brane acquires a (square root-like) dependence on the non-zero twisted moduli, provided the D6-brane wraps only the associated $\OR$-even exceptional three-cycle. Depending on the relative sign between the bulk three-cycle and the exceptional three-cycle, the gauge theory on such a D6-brane can become more weakly or more strongly coupled at the string scale, as discussed e.g.~for the {\it left-symmetric} gauge group $SU(2)_L$ in all global prototype models, for the {\it strong} gauge group $SU(3)_{QCD}$ and the generalised $B-L$ symmetry in global L-R symmetric models and for the hidden gauge group in global PS models. An important observation to be made here concerns the $\Z_2\times\Z_2$ eigenvalues of the respective D6-branes: at the orbifold point, the physics only depends on the {\it relative} $\Z_2\times \Z_2$ eigenvalues among the D6-branes (e.g.~in the computation of the particle spectra), whereas the {\it absolute} $\Z_2\times \Z_2$ eigenvalues enter explicitly in the fractional three-cycle volumes on the deformed toroidal orbifold. This last feature can be useful to improve the matching of the gauge coupling strength at the string scale, or might allow for further subdivisions in the global prototype models. Finally, in each of the global models there exists also a set of deformation moduli to which none of the D6-branes couples, implying that deforming the associated $\Z_2$ singularities is only expected to impact the fractional three-cycle volumes in subleading order through higher order corrections such as field redefinitions of the moduli or instanton corrections.  
 
Additional {\it untwisted} moduli-dependent corrections to the tree-level gauge couplings result from the one-loop gauge threshold corrections, which exhibit a linear and logarithmic dependence on the K\"ahler moduli 
associated to the two-torus volumes (in units of $\alpha'$) in the geometric regime ($v_i  \gtrsim 1$). In the global D6-branes models considered here, all fractional three-cycles have a bulk one-cycle part along $T_{(1)}^2$ parallel to the $\OR(\Z_2^{(1)})$-invariant plane, such that the gauge threshold corrections depend in each intersection sector on the K\"ahler modulus $v_1$ representing the area of the first two-torus $T_{(1)}^2$.
This dependence with an overall positive prefactor presents various phenomenological challenges for compactifications with an intermediate string scale $M_{\text{string}}$ and LARGE internal volumes as discussed in section~\ref{Ss:MstringVsCouplings}: 
a large hierarchy between the Planck scale $M_{\text{Planck}}$ and the string scale $M_{\text{string}}$ can yield a reasonable tree-level coupling when compensated by a hierarchically large complex structure modulus $\nicefrac{R_2^{(1)}}{R_1^{(1)}}$ of $T_{(1)}^2$.
Taking into account the $v_1$-dependent one-loop gauge threshold corrections makes it, however, difficult to stay in the geometric regime and implies  an exponential suppression of the running gauge coupling, excluding the possibility of a strong coupling regime of the QCD stack. Thus, the phenomenologically interesting, global D6-brane models at hand prefer a high string scale, while gauge unification is neither easily realised  at tree-level, nor when including the one-loop gauge threshold corrections.

Each global D6-brane prototype model at hand forms a realisation of a supersymmetric Minkowski vacuum with all internal RR-fluxes or NS-NS-fluxes set to zero. Despite the underlying mathematical consistency of these global D6-brane models, there are two important elements missing in these vacuum configurations: a mechanism to stabilise {\it all} moduli {\it vev}s including the dilaton
(and thereby also the moduli masses) and an alternative mechanism for supersymmetry breaking beyond the formation of a gaugino condensate in some hidden sector - which is of particular relevance to those models with only Abelian `hidden' gauge bosons such as the L-R symmetric prototypes II, IIb and IIc -
producing soft supersymmetry terms lifting the mass degeneracy of the massless open string matter states. In Type II superstring theory, one can (at least in principle) kill these two birds with one stone by turning on internal RR- and NS-NS-fluxes generating a non-vanishing F-term for one (or more) of the moduli multiplets. This consideration begs the question whether one can consistently switch on twisted internal NS-NS-fluxes supported along resolved exceptional three-cycles and discuss moduli stabilisation for the deformation moduli (or twisted complex structure moduli) perturbatively, in contrast to the moduli stabilisation scheme for K\"ahler moduli in Type IIB string theory through solely non-perturbative effects as discussed e.g.~in~\cite{Denef:2005mm,Lust:2005dy,Lust:2006zg}. This question is being addressed in a separate research project and we hope to answer it positively in the near future. Moreover, when switching on twisted NS-NS-fluxes one generally also expects to generate a scalar potential for the axionic (or CP-odd) partners of the twisted complex structure moduli, and it begs the question whether the shape and symmetries of this scalar potential exhibit the generalised Kaloper-Sorbo structure, as recently established~\cite{Bielleman:2015ina,Carta:2016ynn} for Type IIA orientifold compactifications on smooth Calabi-Yau backgrounds. Similar to the potential cancellation of FI-terms by virtue of {\it vev}s associated to charged matter states, this question should be settled at tree-level through a dimensional reduction of the Chern-Simons action and the Dirac-Born-Infeld action an effective four-dimensional ${\cal N}=1$ supergravity following the methods in~\cite{Kerstan:2011dy,Grimm:2011dx}.

\vspace{5mm}

\noindent
{\bf Acknowledgements:} 
W.S.~would like to thank I\~naki Garc\'ia-Etxebarria for enlightening conversations.
This work is partially supported by the {\it Cluster of Excellence `Precision Physics, Fundamental Interactions and Structure of Matter' (PRISMA)} DGF no. EXC 1098,
the DFG research grant HO 4166/2-2 and the DFG Research Training Group {\it `Symmetry Breaking in Fundamental Interactions'} GRK 1581. W.S. is supported by the ERC Advanced Grant SPLE under contract ERC-2012-ADG-20120216-320421, by the grant FPA2012-32828 from the MINECO, and the grant SEV-2012-0249 of the ``Centro de Excelencia Severo Ochoa" Programme.

\appendix

\section{M\"obius Transformations}\label{A:MT}
In section~\ref{Ss:LCHF}, a two-torus $T^2$ was described as an elliptic curve $E$ in the weighted projective space $\P^2_{112}$ through eq.~(\ref{Eq:TEHypersurface}) with a built-in $\Z_2$ symmetry acting on the homogeneous coordinate $y$ of weight 2. The fixed points of this $\Z_2$ action correspond to the roots of the polynomial $F(x,v)$ in equation~(\ref{Eq:GenFEC}). The position of these roots in the $x$-coordinate or in the $v$-coordinate are tied to the complex structure of the two-torus. In table~\ref{tab:OverviewZ2FixedPointsPerT2} we provide a list of the roots in all coordinate patches for a square two-torus and a hexagonal two-torus.

\mathtabfix{
\begin{array}{|c|c|c||c|c|c|}
\hline
\multicolumn{6}{|c|}{\text{\bf $\Z_2^{(i)}$ fixed points per $T_{(i)}^2$ in various coordinate patches}}\\
\hline \hline
\multicolumn{3}{|c||}{\text{\bf $\Z_2^{(i)}$ fixed point on square $T_{(1)}^2$}} & \multicolumn{3}{|c|}{\text{\bf $\Z_2^{(i)}$ fixed point on hexagonal $T_{(l=2,3)}^2$}} \\
\hline \hline
\text{fixed point $\alpha$} & x_1\text{-coordinate } & v_1\text{-coordinate} &\text{fixed point $\alpha$}  & x_l\text{-coordinate }& v_l\text{-coordinate}  \\
\hline
1 & x_1 = \infty & v_1 = 0 & 1 & x_l = \infty & v_l = 0 \\
2 & x_1 = 1& v_1 = 1 & 3& x_l = 1 & v_l = 1  \\
3 & x_1 = 0& v_1 = \infty & 2& x_l = \xi & v_l = \xi^2  \\
4 & x_1 = -1& v_1 = -1 & 4& x_l = \xi^2 & v_l = \xi  \\
\hline
\end{array}
}{OverviewZ2FixedPointsPerT2}{Overview of the $\Z_2^{(i)}$ fixed points on the square torus $T_{(1)}^2$ and the hexagonal tori $T_{(l=2,3)}^2$ in the homogeneous coordinates $x_i$ ($v_i=1$ patch) for the second and fifth column, and in the homogeneous coordinates $v_i$ ($x_i=1$ patch) for the third and last column. The labelling of the fixed points matches the one in figure~\ref{Fig:T2LatticesSquareHexa}.
}

There exist automorphisms $\lambda_\alpha:E\rightarrow E$ of the elliptic curve~\cite{Blaszczyk:2015oia,Koltermann:2016oyv} interchanging the coordinates $x$ and $v$ through the action:
\begin{equation}
\lambda_\alpha: \left( \begin{array}{c} x \\ v \end{array}\right) \mapsto \lambda_\alpha \left( \begin{array}{c} x \\ v \end{array}\right) = \frac{1}{\sqrt{2 \epsilon_\alpha^2 + \epsilon_\beta \epsilon_\gamma}} \left( \begin{array}{cc} \epsilon_\alpha & \epsilon_\alpha^2 + \epsilon_\beta \epsilon_\gamma \\1 & - \epsilon_\alpha  \end{array}\right) \left( \begin{array}{c} x \\ v \end{array}\right), \qquad  y \mapsto y,
\end{equation} 
where the $\epsilon_\alpha$ correspond to the roots entering in equation~(\ref{Eq:GenFEC}). These automorphisms $\lambda_\alpha$ interchange the $\Z_2$ fixed points: 
\begin{equation}
\lambda_2:  1 \leftrightarrow 2, \, 3 \leftrightarrow 4, \quad \lambda_3: 1 \leftrightarrow 3, \, 2 \leftrightarrow 4, \quad  \lambda_4: 1 \leftrightarrow 4, \, 2 \leftrightarrow 3,
\end{equation}
in line with the coordinate transformation, and are further constrained by the tiltedness of the torus lattice:
\begin{equation}
\text{untilted: } \lambda_\alpha = \ov{\lambda_\alpha}, \qquad \text{tilted: } \lambda_2 = \ov{\lambda_4}, \, \lambda_3 = \ov{\lambda_3}.
\end{equation}
In a coordinate patch where $v=1$ (or $x=1$), the automorphisms $\lambda_\alpha$ coincide with M\"obius transformations. 
Recall that for a square two-torus the roots are given by $\epsilon_2 = 1$, $\epsilon_3 = 0$ and $\epsilon_4=-1$, while a hexagonal two-torus is characterised by the roots $\epsilon_2 = \xi$, $\epsilon_3=1$ and $\epsilon_4 = \xi^2$. The automorphisms $\lambda_\alpha$ also allow to exchange {\it Lag} lines among each other. For a square $T^2$ the {\it Lag} lines {\bf aI}-{\bf aIV} from table~\ref{tab:LagSquareT2} behave as follows under M\"obius transformations:  $\lambda_2$ swaps the {\it Lag} lines {\bf aI} $\leftrightarrow$ {\bf aIV} leaving {\bf aII} and {\bf aIII} invariant, $\lambda_3$ exchanges {\bf aI} $\leftrightarrow$ {\bf aIV} and {\bf aII} $\leftrightarrow$ {\bf aIII}, and finally $\lambda_4$ swaps {\bf aII} $\leftrightarrow$ {\bf aIII} leaving {\bf aI} and {\bf aIV} invariant. 
For a hexagonal $T^2$, the {\it Lag} lines from table~\ref{tab:LagHexaT2} behave as follows under M\"obius transformations: 
 $\lambda_3$ keeps all the cycles {\bf bX} invariant, whereas  $\lambda_{2}$ and $\lambda_4$ exchange the {\it Lag} lines {\bf bI}$^0 \leftrightarrow$ {\bf bIII}$^0$ and  {\bf bII}$^0 \leftrightarrow$ {\bf bIV}$^0$.

\section{Correction Terms for $\Z_2^{(1)}$-Deformations}\label{A:CorrectionTerms}
In section~\ref{Sss:sLagsZ21} we observed that certain deformations in the $\Z_2^{(1)}$-twisted sector deform too many singular orbits. The deformations with parameters $\varepsilon_{4+5}^{(1)}$ and $\varepsilon_{4-5}^{(1)}$ for instance deform the singularity (33), inadvertently. Such a feature was for the first time observed in~\cite{Blaszczyk:2015oia} and appears to be intrinsic to $\Z_2$-twisted sectors subject to an additional $\Z_3$ action on the full covering four-torus $T^4$, as testified in table~\ref{tab:OverviewDeformperZ2Sector}. In order to overcome the unwanted deformation of singular points, which are spatially separated from the one where the deformation parameter is localised, 
we have to add a counter-term which eliminates the effect of the initial deformation there. Building further on the $\varepsilon_{4+5}^{(1)}$ example, we have to switch on a correction term $\varepsilon_3^{(1)}$ as a function of $\varepsilon_{4+5}^{(1)}$. In order to determine this functional dependence, we consider a {\it sLag} two-cycle on $T^4_{(1)}/\Z_6$, represented by the algebraic condition $x_2 = \ov x_3$, going through the singular point (33) and enforce that there is no exceptional two-cycle emerging from that point for a deformation parameter $\varepsilon_{4+5}^{(1)}\neq 0$ elsewhere. This allows us to find a series expansion for $\varepsilon_3^{(1)}$ as a function of $\varepsilon_{4+5}^{(1)} \equiv \varepsilon_{\rm in}$:
\begin{align}\label{Eq:Defe4Corre3}
\varepsilon_3^{(1)} \big(\varepsilon_{\rm in}\big) =& - \frac{1}{4} \varepsilon_{\rm in}^2 - \frac{5}{4 \times 3!} \varepsilon_{\rm in}^3- \frac{31}{4 \times (3!)^2} \varepsilon_{\rm in}^4   - \frac{13 \times 17}{4 \times (3!)^3} \varepsilon_{\rm in}^5 - \frac{1753}{4 \times (3!)^4} \varepsilon_{\rm in}^6   - \frac{11\times13\times53}{2\times (3!)^5} \varepsilon_{\rm in}^7  \nonumber\\
& \quad - \frac{167\times421}{2 \times (3!)^6} \varepsilon_{\rm in}^8  - \frac{137 \times 3331}{8 \times (3!)^6} \varepsilon_{\rm in}^9 - \frac{13\times 17 \times 41\times 43}{4\times(3!)^6} \varepsilon_{\rm in}^{10}    + {\cal O}(\varepsilon_{\rm in}^{11} ).
\end{align}
We have to point out that this approach only works for small (initial) deformations $\varepsilon_{\rm in}$, as long as the correction term $\varepsilon_3^{(1)} (\varepsilon_{\rm in})$ remains smaller than the original deformation. Applying this requirement for the (conservative) assumption $\big|\varepsilon_3^{(1)}\big| \sim \frac{1}{10} \varepsilon_{\rm in}$ implies the range $\big|\varepsilon_{\rm in}\big|\lesssim 0.3$ for the initial deformation.

Similar considerations have to be made for other deformation parameters, as summarised in table \ref{tab:SummDefCorrZ21}. In each particular case, we apply the same method as described above and construct a Taylor expansion in terms of the initial deformation. 
\mathtab{
\begin{array}{|c||c|c|c|}
\hline \multicolumn{4}{|c|}{\text{\bf Deformations \& correction terms in the $\Z_2^{(1)}$ sector }} \\
\hline \hline
& \multicolumn{3}{|c|}{\text{\bf Deformation parameters}} \\
\hline \hline
\text{\bf Correction Terms}  & \varepsilon_3^{(1)} &  \varepsilon_{4+5}^{(1)} &  \varepsilon_{4-5}^{(1)}     \\
\hline 
\varepsilon_3^{(1)}  & - & \checkmark  \text{ eq. (\ref{Eq:Defe4Corre3})} &  \checkmark  \text{ eq. (\ref{Eq:Defe5Corre3})}   \\
\varepsilon_{4+5}^{(1)}& \checkmark   \text{ eq. (\ref{Eq:Defe3Corre4})} & - & -  \\
\hline
\end{array}
}{SummDefCorrZ21}{Overview of correction terms for various $\Z_2^{(1)}$ deformation parameters with the corresponding Taylor expansion indicated by the equations in the main text.
}
More explicitly, a non-vanishing deformation parameter $\varepsilon_3^{(1)}\equiv \varepsilon_{\rm in}$ deforms the $\Z_2^{(1)}$-fixed point orbits $e_4^{(1)}$ and $e_5^{(1)}$ unintentionally, as pointed out in table~\ref{tab:OverviewDeformperZ2Sector}. This effect can be undone by turning on a correction terms $\varepsilon_{4+5}^{(1)}$, as a function of $\varepsilon_3^{(1)}$. Following the same logic as presented above, one finds the following series expansion for $\varepsilon_{4+5}^{(1)}$:
\begin{align} \label{Eq:Defe3Corre4}
\varepsilon_{4+5}^{(1)} (\varepsilon_{\rm in}) &= - \frac{1}{2} \varepsilon_{\rm in}^2 -  \frac{1}{2\times 3} \varepsilon_{\rm in}^3  -  \frac{19}{2^5\times 3^2} \varepsilon_{\rm in}^4 -  \frac{1}{2^5} \varepsilon_{\rm in}^5  -  \frac{1}{2^6\times 3^2} \varepsilon_{\rm in}^6  -  \frac{5^2 \times 17}{2^9\times 3^3} \varepsilon_{\rm in}^7 \nonumber\\
& -  \frac{149 \times 463}{2^{13}\times 3^5} \varepsilon_{\rm in}^8   -  \frac{131 \times 223}{2^{12}\times 3^5}\varepsilon_{\rm in}^9 -  \frac{266401}{2^{14}\times 3^6} \varepsilon_{\rm in}^{10} + {\cal O}( \varepsilon_{\rm in}^{11}).
\end{align}
Turning on the deformation parameter $\varepsilon_{4-5}^{(1)}\equiv \varepsilon_{\rm in}$ on the other hand deforms the $\Z_2^{(1)}$-fixed point orbit $e_3^{(1)}$ unwillingly, as can be seen from table~\ref{tab:OverviewDeformperZ2Sector}. This effect can be undone by a correction term $\varepsilon_3^{(1)}$ depending on the initial deformation parameter $\varepsilon_{4-5}^{(1)}$, whose functional dependence follows from repeating the same logic as above: 
\begin{align}\label{Eq:Defe5Corre3}
\varepsilon_3^{(1)} (\varepsilon_{\rm in}) &= - \frac{1}{2^2 \times 3} \varepsilon_{\rm in}^2 + \frac{1}{2^3 \times 3 \sqrt{3}} \varepsilon_{\rm in}^3 + \frac{1}{2^4 \times 3^2 } \varepsilon_{\rm in}^4 - \frac{11}{2^5 \times 3^3 \sqrt{3}}\varepsilon_{\rm in}^5 - \frac{7}{2^6 \times 3^4}\varepsilon_{\rm in}^6 \nonumber \\
&  + \frac{11}{2^3 \times 3^5 \sqrt{3}} \varepsilon_{\rm in}^7 + \frac{59}{2^7 \times 3^7} \varepsilon_{\rm in}^8 - \frac{3019}{2^9 \times 3^6 \sqrt{3}} \varepsilon_{\rm in}^9 - \frac{25}{2^6 \times 3^9} \varepsilon_{\rm in}^{10} + {\cal O}(\varepsilon_{\rm in}^{11}).
\end{align}
After having determined the relevant correction term for each of the deformations, one can also explicitly verify how the correction terms restore the singular nature of the orbits, as depicted in the lower plots of figure~\ref{Fig:DefZ21hypersurfaceEq}~(f) and~(g).

\section{Tables of Matter States per D6-Brane Intersection Sector}\label{A:Tables-MatterSector}

In this appendix, we summarise the massless open string spectrum per sector for each of the global prototype D6-brane models with interesting phenomenological features as discussed in section~\ref{S:DefModuliGlobalModels}. Table~\ref{tab:Z2Z65stackMSSMTotalSpectrum} contains the massless open string spectrum of the global prototype MSSM model 
discussed in section~\ref{Ss:ExMSSM}, and
 table~\ref{tab:Z2Z64StackPSITotalSpectrum} contains the counting of massless open string states per intersection sector for the global PS  prototype I model discussed in section~\ref{Ss:ExPS}. 
 Table~\ref{tab:Z2Z6-LRS+PSII-visible} displays the counting of the observable massless open string sector of all L-R symmetric models discussed in section~\ref{Ss:ExLRSM} as well as of the PS prototype II model.
 Finally, table~\ref{tab:Z2Z6-LRS-hidden} provides the sector-per-state counting for the different `hidden' completions of the massless matter spectrum per L-R symmetric model.

 The presentation in terms of matter states per sector is indispensable for the computation of the gauge threshold corrections in section~\ref{S:GKFDefModuli} and is useful for the discussion about potentially flat directions for twisted moduli in section~\ref{Ss:ExMSSM} to~\ref{Ss:ExPS}.

\begin{sidewaystable}[h]
\begin{minipage}{21cm}
\begin{center}
\hspace*{-14mm}
{\small
$\begin{array}{|c||c|c|c|c|c|c|c|c|c|c|}
\hline  \multicolumn{10}{|c|}{\text{\bf Total amount of massless matter per sector for a 5-stack MSSM model on the {aAA} lattice of $T^6/(\Z_2 \times \Z_6 \times \OR)$ with discrete torsion}}\\
\hline \hline 
(\chi^{x (\omega^{k \in \{0,1,2\}} y)})
& y=a & y=a'
& y=b& y=c & y=c'
& y=d & y=d'
& y=h & y=h'
\\  
\hline
x= a & (0,0,0) &  \!\!\tarh{([2]_{2-2+2},-1,1)}{(0,0,0)}\!\!
& (2,1,0) & (-4,-1,-1) & (0,0,0)
& (2,1,0)  & (-2,-1,0)
& ([2],1,-1) & ([2]_{-2+2+2},0,0)
\\\cline{2-3}
x= b & \muc{2}{|c|}{}
&  \!\!\tarh{(0,\frac{3 +[2]}{2},\frac{-3+[2]}{2})}{(0,0,0)}\!\!  & (0,-3 + [2],[4]) & -
& (0,3+[2],3+[2]) & -
&(-2,-1,0) & -
\\\cline{4-4}
x= c & \muc{3}{|c|}{} & (0,\frac{[4]}{2},\frac{[4]}{2})  &\!\!\tarh{([2]_{2+2-2},3+[2],-3+[2])}{(0,0,0)}
& (0,[4],-3+[2]) &(0,[4],3+[2])
& (0,0,0) &  (-4,-1,-1) 
\\\cline{5-6}
x= d & \muc{5}{|c|}{} & (0,\frac{-3 +|2|}{2},\frac{3+|2|}{2})  & \!\!\tarh{(0,3+[2],-3+[2])}{([2]_{-2+2+2},0,0)}\!\!
& (2,1,0) &  (2,0,1) 
\\\cline{7-8}
x= h & \muc{7}{|c|}{} & (0,0,0) &  \!\!\tarh{([2]_{2-2+2},-1,1)}{(0,0,0)}\!\! 
\\
\hline
\end{array}$
}
\caption{Overview of the total amount of chiral and vector-like massless (open string) matter per sector $x(\omega^k y)$ for the global five-stack MSSM-like model with fractional D6-brane configuration given in table~\ref{tab:5stackMSSMaAAPrototypeI}. If the net-chirality $|\chi^{x(\omega^ky)}| < \varphi^{x(\omega^ky)}$, the sector $x(\omega^ky)$ comes with a set of vector-like pairs of matter states, whose multiplicity corresponds to $n^{x(\omega^ky)}_{NC} \equiv\varphi^{x(\omega^ky)} - |\chi^{x(\omega^ky)}|$, e.g. $n^{a(\omega^0h)}_{NC}=[2]$ denotes one vector-like pair of bifundamental states in the sector $a(\omega^0h)$. 
The diagonal entries $\varphi^{x(\omega x)} = \varphi^{x(\omega^2 x)} = \frac{\varphi^{\Adj_x}}{2}$ count the number of states in the adjoint representation for the $x$-stack, e.g. the entry $(0,\frac{[4]}{2},\frac{[4]}{2})$
in the $c(\omega^k c)_{k=0,1,2}$ sectors corresponds to four multiplets in the adjoint representation arising at intersections of $c$ with its orbifold images $(\omega^k c)_{k=1,2}$.
The upper and lower entries in the $x(\omega^k x')$ sectors count the number of states in the antisymmetric and symmetric representation, respectively. The entries $\tarh{([2]_{2-2+2},-1,1)}{(0,0,0)}$ for the $a(\omega^k a')$ 
sectors e.g. amount to one vector-like pair of antisymmetric reprentations from the $aa'$ sector plus a second pair spread over the $a(\omega a')+a(\omega^2 a')$ sectors.
The lower index of e.g. $2-2+2$ in the $aa'$ sector, moreover, determines the decomposition of $\sum_{i=1}^3 \delta^0_{\sigma^i_{xy}} \delta^0_{\tau^i_{xy}} \tilde{b}^{{\cal A},(i)}_{xy}$ introduced in table~\protect\ref{tab:beta_coeffs_Kaehler_metrics}, which is required as input for the computation of the one-loop gauge threshold $\Delta_{aa'}^{\cal A}$ according to 
table~\protect\ref{tab:1-loop-thresholds}.
\label{tab:Z2Z65stackMSSMTotalSpectrum}}\end{center}
\end{minipage}
\\
\\ \\ \\
\begin{minipage}{21cm}\begin{center}
$\begin{array}{|c||c|c|c|c|c|c|c|}
\hline  \multicolumn{7}{|c|}{\text{\bf Total amount of matter per sector for the Pati-Salam model prototype I on } T^6/(\Z_2 \times \Z_6 \times \OR)}\\
\hline \hline 
(\chi^{x (\omega^{k\in\{0,1,2\}} y)})& y=a  & y=a'
& y=b & y=c
& y=h & y=h'
\\  
\hline
x= a &   (0,0,0) &  \!\!\tarh{([2]_{2-2+2},-1,1)}{(0,0,0)}\!\!
& (2,1,0) & (-2,0,-1) 
& ([2]_{2+2-2},1,-1) & ([2]_{-2+2+2},0,0)  
   \\
\cline{2-3}
x= b & \muc{2}{|c|}{}
&  \!\!\tarh{(0,\frac{3 +[2]}{2},\frac{-3+[2]}{2})}{(0,0,0)}\!\!  & ([2]_{2-2+2},[4],[4]) & (-2,-1,0) & -
\\\cline{4-4}
x= c &  \muc{3}{|c|}{} & \!\!\tarh{(0,\frac{3+[2]}{2},\frac{-3+[2]}{2})}{(0,0,0)} & (2,0,1) & - 
  \\\cline{5-5}
x= h &  \muc{4}{|c|}{} & (0,0,0) & \!\!\tarh{([2]_{2-2+2},-1,1)}{(0,0,0)}\!\!
 \\
\hline
\end{array}
$
\caption{Overview of the chiral and vector-like matter spectrum per sector $x(\omega^k y)$ of the PS I model with D6-brane data specified in table~\protect\ref{tab:PatiSalamaAAPrototypeI}. 
For details of the notation see the caption of table~\protect\ref{tab:Z2Z65stackMSSMTotalSpectrum}.
\label{tab:Z2Z64StackPSITotalSpectrum}}\end{center}
\end{minipage}
\end{sidewaystable}


\begin{sidewaystable}[h]
\begin{minipage}{21cm}
\begin{center}
\hspace*{-14mm}
$\begin{array}{|c||c|c|c|c||c|c||c|c|}
\hline  \multicolumn{9}{|c|}{\text{\bf Total amount of matter per sector: universal visible part for all L-R symmetric models and Pati-Salam II on } T^6/(\Z_2 \times \Z_6 \times \OR)}\\
\hline \hline 
\!\!(\chi^{x (\omega^{k\in \{0,1,2\}} y)})\!\!
& y=a & y=a' 
& y=b& y=c & y=d_{\text{\bf L-R}}  & y=d_{\text{\bf L-R}}'
& y=h_{\text{\bf PS II}} &  y=h_{\text{\bf PS II}}^{\prime}
\\\hline
x= a & (0,0,0) &  \!\!\tarh{([2]_{2-2+2},-1,1)}{(0,0,0)} \!\!
& (2,0,1) & (-2,-1,0) &  ([2]_{-2+2+2},0,0)   & ([2]_{2+2-2},1,-1)
& ([2],-1,1) & ([2],0,0)
\\\cline{2-3}
x= b & \muc{2}{|c|}{}
&  \!\!\tarh{(0,\frac{3 +[2]}{2},\frac{-3+[2]}{2})}{(0,0,0)}\!\!  & ([2]_{2-2+2},[4], [4]) &  (2,1,0)    & -  & (0,[4],-3+[2]) & - 
\\\cline{4-4}
x= c & \muc{3}{|c|}{}
& \!\!\tarh{(0,\frac{3 +[2]}{2},\frac{-3+[2]}{2})}{(0,0,0)} &  (-2,0,-1)     & - & (0,3+[2],[4]) & -
\\\hline\hline
x= d_{\text{\bf L-R}} & \muc{4}{|c||}{} & (0,0,0) &   \!\!\tarh{([2]_{2-2+2},-1,1)}{(0,0,0)} \!\! & \muc{2}{|c|}{}
\\\hline\hline
x=h_{\text{\bf PS II}} & \muc{6}{|c||}{} & (0,\frac{[4]}{2},\frac{[4]}{2}) & \!\!\tarh{([2]_{2-2+2},3+[2],-3+[2])}{(0,0,0)}\!\!
\\\hline
\end{array}
$
\caption{Overview of the massless visible matter states per intersection sector of the L-R symmetric models and the massless visible plus `hidden' matter of
the PS II model with D6-brane data specified in tables~\protect\ref{tab:6stackLRSaAA} and~\protect\ref{tab:PatiSalamaAAPrototypeII}, respectively. 
For the sector-per-state counting of the `hidden' matter in the L-R symmetric models see table~\protect\ref{tab:Z2Z6-LRS-hidden}.
Details of the notation are given in the caption of table~\protect\ref{tab:Z2Z65stackMSSMTotalSpectrum}.
\label{tab:Z2Z6-LRS+PSII-visible}}
\end{center}
\end{minipage}
\\
\\ \\ \\
\begin{minipage}{21cm}
{\footnotesize
\begin{center}
\hspace*{-22mm}
$\begin{array}{|c||c|c|c|c||c|c|c|c|}
\hline  \multicolumn{9}{|c|}{\text{\bf Total amount of `hidden' matter per sector for  the L-R symmetric models on $T^6/(\Z_2 \times \Z_6 \times \OR)$ with discrete torsion}}\\
\hline \hline 
& \muc{4}{|c||}{\text{\bf L-R I}} &\muc{4}{|c|}{\text{\bf L-R II}} 
\\\hline
\!\!(\chi^{x (\omega^k y)})\!\! & y=h_1 & y=h_1'  & y=h_2 & y=h_2'
 & y=h_1 & y=h_1'  & y=h_2 & y=h_2'
 \\  
\hline
x= a & \!\!\!(0,-1,-1)\!\!\!  &(0,0,0)  & (0,1,1)  & (0,0,0) 
& ([2],-1,1) & ([2],0,0) & ([2],1,-1) & ([2],0,0)
\\
x=b & (2,0,-1) & - & (-2,0,1)  & -
& \!\!\!(0_{0,-2,0},[4],-3+[2])\!\!\! & - & \!\!\!(0_{0+2+0},[4],3+[2])\!\!\! & -
\\
x=c & (2,-1,0) & - & (-2,1,0) & -  
& \!\!\!\!(0_{0,-2,0},3+[2],[4])\!\!\!\! & - & \!\!\!(0_{0+2+0},-3+[2],[4])\!\!\! & -
\\
x= d &  (0,1,1)  & (0,0,0) & (0,-1,-1) & (0,0,0)
& ([2],1,-1) & ([2],0,0) & ([2],-1,1) & ([2],0,0)
 \\
x= h_1 &  (0,0,0)  &\!\!\!\!\tarh{([2]_{2+2-2},-1,1}{(0,0,0)}\!\!\!\! & \!\!\!([2]_{-2+2+2},0,0)\!\!\! & \!\!\!([2]_{2-2+2},-1,1)\!\!\!
& (0,\frac{[4]}{2},\frac{[4]}{2}) & \!\!\!\tarh{([2]_{2-2+2},3+[2],-3+[2])}{(0,0,0)}\!\!\! & \!\!\!([2]_{-2+2+2},[4],[4])\!\!\! & \!\!\!([2]_{2+2-2},-3+[2],3+[2])\!\!\!
\\\cline{2-3}\cline{6-7}
x= h_2 & \muc{2}{|c|}{} & (0,0,0)  & \!\!\tarh{([2]_{2+2-2},-1,1)}{(0,0,0)}\!\!
& & & (0,\frac{[4]}{2},\frac{[4]}{2}) & \!\!\tarh{([2]_{2-2+2},3+[2],-3+[2])}{(0,0,0)}\!\!
\\\hline \hline 
& \muc{4}{|c||}{\text{\bf L-R IIb}} &\muc{4}{|c|}{\text{\bf L-R IIc}} 
\\\hline
\!\!(\chi^{x (\omega^k y)})\!\! & y=h_1 & y=h_1'  & y=h_2 & y=h_2'
 & y=h_1 & y=h_1'  & y=h_2 & y=h_2'
 \\  
\hline
x= a & (0_2,0,0) & (0_2,0_1,0_1) & (0_2,0,0) & (0_2,0_1,0_1) 
& (-2,-1,0) & (2,1,0) & (0_2,0,0_1)  & (0_2,0,0_1)
\\
x=b & (0,0_5,0_4) & - &  (0,0_5,0_4) & - 
& (0,[4],[4]) & - & (0,0_5,0_5) & - 
\\
x=c & (0,0_4,0_5) & - & (0,0_4,0_5) & -
& (0,3+[2],3+[2]) & - & (0,0_4,0_4) & - 
\\
x= d &  (0_2,0,0) & (0_2,0_1,0_1) & (0_2,0,0) & (0_2,0_1,0_1) 
& (2,1,0) & (-2,-1,0) & (0_2,0,0) & (0_2,0_1,0_1)  
 \\
x= h_1 & \!\!\!(0,\frac{[4]}{2},\frac{[4]}{2})\!\!\! & \!\!\!\tarh{([2]_{2-2+2},3+[2],-3+[2] )}{(0,0,0)}\!\!\! & \!\!\!([2]_{-2+2+2},[4],[4])\!\!\! & \!\!\!([2]_{2+2-2},-3+[2],3+[2])\!\!\!
& (0,\frac{[5]}{2},\frac{[5]}{2}) & \!\!\tarh{([2]_{-2+2+2},[5],[5]}{}\!\! & (0,0_5,0_5) & (0,0_5,0_5)
\\\cline{2-3}\cline{6-7}
x= h_2 & \muc{2}{|c|}{} & (0,\frac{[4]}{2},\frac{[4]}{2}) &  \!\!\tarh{([2]_{2-2+2},3+[2],-3+[2] )}{(0,0,0)}\!\!
& \muc{2}{|c|}{} & (0,\frac{[5]}{2},\frac{[5]}{2})
& \!\!\tarh{([2]_{-2+2+2}, [5],[5]}{(0,0,0)}\!\!
\\\hline
\end{array}
$
\caption{Overview of the total amount of `hidden' chiral and vector-like matter per sector $x(\omega^k h_i^{(\prime)})$ and $h_i(\omega^k h_j^{(\prime)})$
in the L-R symmetric models with D6-brane data specified in table~\protect\ref{tab:6stackLRSaAA}.
The counting of the associated visible matter states per sector is displayed in table~\protect\ref{tab:Z2Z6-LRS+PSII-visible}, and the notation is explained in the caption of table~\protect\ref{tab:Z2Z65stackMSSMTotalSpectrum}.
\label{tab:Z2Z6-LRS-hidden}}
\end{center}
}
\end{minipage}
\end{sidewaystable}

\clearpage

\addcontentsline{toc}{section}{References}
\bibliographystyle{ieeetr}
\bibliography{refs_Z2Z6-Deform}

\end{document}